\documentclass[times,final]{elsarticle}

\usepackage{jcomp}
\usepackage{framed,multirow}

\usepackage[utf8]{inputenc} 				
\usepackage[main=english, ngerman]{babel}	
\usepackage[T1]{fontenc} 					
\usepackage{textcomp} 					
\usepackage{amsmath}					
\usepackage{amssymb} 					
\usepackage{cases}                      
\usepackage[hidelinks]{hyperref}		
\usepackage[noabbrev]{cleveref}			
\usepackage[final]{microtype}			
\usepackage{csquotes}					
\usepackage{graphicx}                   
\usepackage{grffile}					
\usepackage{subcaption}					
\usepackage{lmodern}		            
\usepackage{multirow}                   
\usepackage{array}						
\usepackage{makecell}					
\usepackage{booktabs}					
\usepackage{placeins}					
\usepackage{xcolor}						
\usepackage{pgfplots}					

\pgfplotsset{compat=newest}
\usepgfplotslibrary{groupplots,dateplot,fillbetween}
\usetikzlibrary{calc,angles,quotes}
\usetikzlibrary{arrows,arrows.meta,patterns,shapes.arrows,calc,angles,quotes,babel}
\DeclareUnicodeCharacter{2212}{−}
\newlength\figureheight
\newlength\figurewidth
\newlength\mdist 	
\newlength\radius 	

\newcommand{\cpp}{C\nolinebreak\hspace{-.05em}\raisebox{.4ex}{\tiny\textbf{+}}\nolinebreak\hspace{-.02em}\raisebox{.4ex}{\tiny\textbf{+}}} 	

\usepackage{mathtools}
\newcommand{\lbmpy}{{{\em lbmpy}}}
\newcommand{\walberla}{\textsc{waLBerla}}

\usepackage{url}
\usepackage{xcolor}
\definecolor{newcolor}{rgb}{.8,.349,.1}

\journal{Journal of Computational Physics}

\begin{document}

\verso{C.\ Schwarzmeier \textit{et al.}}
	
\begin{frontmatter}
	
\title{Comparison of free-surface and conservative Allen--Cahn phase-field lattice Boltzmann method}

\author[1]{Christoph \snm{Schwarzmeier}\corref{cor1}}
\ead{christoph.schwarzmeier@fau.de}
\cortext[cor1]{Corresponding author}

\author[1,2]{Markus \snm{Holzer}}

\author[3]{Travis \snm{Mitchell}}

\author[4]{Moritz \snm{Lehmann}}

\author[4]{Fabian \snm{Häusl}}

\author[1,2]{Ulrich \snm{Rüde}}

\address[1]{Chair for System Simulation, Friedrich-Alexander-Universität Erlangen-Nürnberg, Cauerstraße 11, 91058 Erlangen, Germany}

\address[2]{CERFACS, 42 Avenue Gaspard Coriolis, 31057 Toulouse Cedex 1, France}

\address[3]{School of Mechanical and Mining Engineering, The University of Queensland, St Lucia, QLD 4072, Australia}

\address[4]{Biofluid Simulation and Modeling - Theoretische Physik VI, University of Bayreuth, 95447 Bayreuth, Germany}

\received{...}
\finalform{...}
\accepted{...}
\availableonline{...}
\communicated{...}


\begin{abstract}
This study compares the free-surface lattice Boltzmann method (FSLBM) with the conservative Allen--Cahn phase-field lattice Boltzmann method (PFLBM) in their ability to model two-phase flows in which the behavior of the system is dominated by the heavy phase.
Both models are introduced and their individual properties, strengths and weaknesses are thoroughly discussed.
Six numerical benchmark cases were simulated with both models, including (i) a standing gravity and (ii) capillary wave, (iii) an unconfined rising gas bubble in liquid, (iv) a Taylor bubble in a cylindrical tube, and (v) the vertical and (vi) oblique impact of a drop into a pool of liquid.
Comparing the simulation results with either analytical models or experimental data from the literature, four major observations were made.
Firstly, the PFLBM selected was able to simulate flows purely governed by surface tension with reasonable accuracy.
Secondly, the FSLBM, a sharp interface model, generally requires a lower resolution than the PFLBM, a diffuse interface model.
However, in the limit case of a standing wave, this was not observed.
Thirdly, in simulations of a bubble moving in a liquid, the FSLBM accurately predicted the bubble's shape and rise velocity with low computational resolution.
Finally, the PFLBM's accuracy is found to be sensitive to the choice of the model's mobility parameter and interface width.
\end{abstract}

\end{frontmatter}
	


\section{Introduction}
Multiphase flows are important in a wide range of natural and industrial applications. 
For instance, they can manifest as undesired foam in the food industry~\cite{leuner2020OvercomingUnwantedFoam,thunnesen2021UltrasonicEffectsFoam}, in the transport of hydrocarbons from subsurface environments~\cite{WU2017350}, or in the transfer of micro-particles into the environment during rainfall~\cite{lehmann2021ejection}.
The laboratory experiments associated with studying the fundamental dynamics of multiphase flows can be expensive and time-consuming, while only supplying limited insight into the governing fluid mechanics.
With advances of computational infrastructure, it has now become common to supplement physical with numerical experiments through the use of computational fluid dynamics (CFD).
This tends to provide a cheaper, time-efficient solution to flow problems and allows direct insights to the flow field, as every arbitrary location inside the fluid can be monitored.\par

This work aims to present and analyze a numerical method that can be used to supplement experiments by numerical simulations of immiscible two-phase flows, in which the flow dynamics of the lighter phase are assumed to have a negligible influence on the heavier phase, and overall dynamics of the system.
In such cases, the flow in the lighter phase is commonly neglected, reducing the two-phase flow to a flow with a free boundary, or more commonly referred to as free-surface flow~\cite{scardovelli1999DirectNumericalSimulation}.
It has been previously shown that such simplification is valid in simulations for e.g., single gas bubbles rising in a liquid~\cite{bogner2017DirectSimulationLiquid,donath2011VerificationSurfaceTension} and has been applied to simulate foaming~\cite{korner2005LatticeBoltzmannModel}.\par

While the flow in the lighter phase is assumed to be negligible, the simulation of the flow in the heavier phase often requires a highly resolved computational grid to capture all the relevant flow structures.
Therefore, the numerical methods presented here are designed and targeted for massively parallel computing environments.
For efficient numerical fluid simulations on such hardware, the lattice Boltzmann method (LBM) has been established as a modern alternative to classical approaches for CFD that are based on the discretization of the Navier--Stokes equations.
As all operations require only information of a local neighborhood, the LBM is inherently suitable for parallel computing and has been extended with models for simulating a variety of different physics including multiphase flows~\cite{wohrwag2018TernaryFreeEnergyEntropic,shan1994SimulationNonidealGases,chiu2011ConservativePhaseField}, particulate flows~\cite{rettinger2022EfficientFourwayCoupled,kuron2016MovingChargedParticles}, thermal effects~\cite{mitchell_majidi_rahimian_leonardi_2021} and others.\par

There are several multiphase LBM models available in the literature that can be distinguished by the representation of the interface between the phases.
Models having a sharp interface representation include the free-surface lattice Boltzmann method (FSLBM)~\cite{korner2005LatticeBoltzmannModel}, the level-set method~\cite{becker2009CombinedLatticeBGK}, the front-tracking approach~\cite{lallemand2007LatticeBoltzmannFronttrackinga}, and the color gradient model~\cite{gunstensen1991LatticeBoltzmannModel}.
In contrast, the interface is represented in a diffuse manner in the pseudopotential model~\cite{shan1994SimulationNonidealGases}, the free-energy model~\cite{swift1995LatticeBoltzmannSimulation}, and phase-field models.
The latter are either based on solving the Cahn--Hilliard~\cite{inamuro2004LatticeBoltzmannMethod,zheng2005LatticeBoltzmannInterface} or Allen--Cahn~\cite{Fakhari2017} equation to model the interfacial dynamics.
In this article, the comparative study is restricted to the FSLBM and the conservative Allen--Cahn phase-field LBM (PFLBM)~\cite{Fakhari2017,Mitchell2018}.
Both of these models have well-optimized parallel implementations, and have been shown to be capable of simulating systems with high density and viscosity contrasts corresponding to liquid--gas systems.\par

The FSLBM extends the LBM with a volume-of-fluid approach~\cite{hirt1981VolumeFluidVOF} where the sharp interface between the two phases is captured by an indicator field~\cite{korner2005LatticeBoltzmannModel}.
The fluid dynamics of the lighter phase are entirely neglected, and only the effect of pressure forces at the interface is modeled.
It is implicitly assumed that the density and viscosity ratio between the two fluid phases is infinite.
The sharp interface formulation and avoiding computations in the lighter phase lead to high computational efficiency with low memory requirements.
Although the algorithm's implementation is challenging, it is also well suited for parallel hardware like graphics processing units (GPUs)~\cite{ lehmann2021ejection,janssen2011free}.\par

The conservative Allen--Cahn equation~\cite{chiu2011ConservativePhaseField, SUN2007626} is the basis of the conservative Allen--Cahn phase-field LBM~\cite{Fakhari2017,Mitchell2018}, a model designed to simulate two-phase flow problems with high density and viscosity contrasts.
The algorithm is simpler than that of the FSLBM, where different equations must be solved depending on the type of cell.
In contrast, the PFLBM can be purely formulated via the standard lattice Boltzmann equation with additional force terms~\cite{Fakhari2017}.
As in the single-phase LBM, all operations are restricted to a local cell-neighborhood making the PFLBM well-suited for parallel computing.
While prior phase-field models were not capable of simulating multiphase flows with large density and viscosity ratios~\cite{He1998,Premnath2005,FAKHARI20093046}, the PFLBM has been successfully used in simulations with density ratios of up to $10^3$ and viscosity ratios of up to $10^2$~\cite{mitchell_majidi_rahimian_leonardi_2021,Mitchell2019,Mitchell2020_pof,Mitchell2020103376,Mitchell2021}.
This is equivalent to an air--water system and makes the model a possible alternative for free-surface flows where the dynamics are governed by the heavier phase.
The Allen--Cahn phase-field equation tracks the dynamics of the interface.
The diffusivity of the interface suggests that a PFLBM simulation must have a higher resolution than an FSLBM simulation.
On the other hand, due to its algorithmic simplicity, it is easier to optimize the implementation for different architectures, including accelerator hardware like GPUs~\cite{Holzer2021}.\par

In this work, the models are compared with respect to their algorithmic properties and ability to simulate two-phase flows in which the lighter phase has negligible impact on the flow dynamics.
First, the numerical foundations of the LBM, FSLBM and PFLBM are introduced in more detail.
Then, the models are compared with respect to methodology and numerical implementation.
Based on six numerical experiments, the accuracy and the required computational grid resolution of the models are compared.
For all numerical experiments, each model's simulation results were cross-validated with independent implementations from other code bases, as listed below.
The choice of these tests is discussed, as the test cases must be reasonably applicable to both models.
The initial test case features a standing surface wave governed by gravitational forces and is referred to as a gravity wave. 
Surface tension effects are not modeled in this test case.
In the second benchmark, the same standing wave setup is used, however, the flow is governed by surface tension rather than gravitational forces.
With respect to the gravity wave test case, the capillary wave allows one to exclusively evaluate the models' capability to capture the effects of surface tension.
For both the gravity and capillary wave, there exist analytical models that can be used to validate the simulation accuracy.
In the third and fourth test case, an unconfined single rising gas bubble in liquid, and a confined Taylor bubble in a cylindrical tube are simulated and compared with experimental data from the literature.
Finally, the fifth and sixth benchmark case feature dynamic coalescence, i.e., the formation of a splash crown caused by the impact of a droplet into liquid.
The results are qualitatively compared with experimental data from the literature.
Both models are regularly applied to simulate capillary waves~\cite{donath2011VerificationSurfaceTension,korner2005LatticeBoltzmannModel,Kumar2019}, rising bubbles~\cite{bogner2017DirectSimulationLiquid,donath2011VerificationSurfaceTension, Kumar2019, mitchell2019DevelopmentEvaluationMultiphase,fakhari_geier_lee_2016,FAKHARI201722}, and drop impacts~\cite{lehmann2021ejection,Kumar2019,fakhari_geier_lee_2016,FAKHARI201722,thurey2007PhysicallyBasedAnimation}, however, a direct comparison between them is missing from the literature.
Finally, it is concluded that the PFLBM is more accurate in simulating flows governed purely by surface tension forces compared to the FSLBM used in this article.
However, in flow problems governed by surface tension and gravitational acceleration, the FSLBM required less computational resolution than the PFLBM while having more accuracy in the tests performed here.
Additionally, the PFLBM was sensitive to the choice of the model's mobility parameter and interface width, affecting accuracy and numerical stability.\par

In this work, general properties related to computational performance such as the grid's resolution and memory requirements are discussed, while quantitative data are not presented.
Data such as these would only represent the state of the implementations used here and would not allow a general conclusion to be made.

The FSLBM simulations were performed using the open source \cpp~framework \walberla{}~\cite{waLBerla} and cross-validated with FluidX3D~\cite{lehmann2021ejection, lehmann2022AnalyticSolutionPiecewise}.
The PFLBM simulations were also performed using \walberla{} together with the code generation framework \lbmpy{}~\cite{lbmpy}.
These simulations were cross-validated using TCLB~\cite{TCLB}.
The implementations used in this work and all simulation setups are freely available online, as described in Appendix~\ref{subsec:app-implementation-links}.\par


\section{Numerical methods}
The first part of this section briefly introduces the LBM, before presenting the numerical foundations of the FSLBM and the PFLBM.
The section is concluded by comparing both models focusing on their computational properties.

\subsection{The lattice Boltzmann method} \label{subsec:LBM}
The classical approach to CFD is to simulate the evolution of a flow problem via the discretization of the Navier--Stokes equations.
Contrary to this, the LBM is based on the lattice Boltzmann equation (LBE) and has gained popularity in the last few decades.
The LBE is given by,
\begin{align}\label{eq:LBE}
f_i \left(\boldsymbol{x} + \boldsymbol{c}_i \Delta t, t + \Delta t\right) - f_i \left(\boldsymbol{x}, t\right) = \Omega_i \left(\boldsymbol{x}, t\right) + F_{i}\left(\boldsymbol{x}, t\right),
\end{align}
with $f_i \left(\boldsymbol{x}, t\right) \in \mathbb{R}$ being a discrete particle distribution function (PDF) that describes the probability that there exists a virtual fluid particle at position $\boldsymbol{x} \in \mathbb{R}^{d}$ and time $t \in \mathbb{R^{+}}$ traveling with discrete lattice velocity $\boldsymbol{c}_i \in \Delta x / \Delta t \, \{-1,0,1\}^{d}$~\cite{lbm_book}.
The domain is discretized using a uniformly spaced Cartesian grid with spacing $\Delta x \in \mathbb{R^{+}}$ where the macroscopic fluid velocity in each cell is discretized using a D$d$Q$q$ velocity set such that  $i \in \{0,1,\dots,q-1\}$.
Here, $d \in \mathbb{N}$ refers to the number of dimensions in space and $q \in \mathbb{N}$ refers to the number of discrete lattice velocities.
In each velocity set, the so-called lattice speed of sound, $c_{s}= \sqrt{1/3} \, \Delta x / \Delta t$, defines the relation between density, $\rho(\boldsymbol{x}, t) \in \mathbb{R^{+}}$, and pressure, $p(\boldsymbol{x}, t) = c_{s}^{2} \rho(\boldsymbol{x}, t)$, with $\Delta t \in \mathbb{R^{+}}$ denoting the temporal resolution.
External forces are included in the LBM by $F_{i}(\boldsymbol{x}, t) \in \mathbb{R}$.\par

The collision operator, $\Omega_i \left(\boldsymbol{x}, t\right) \in \mathbb{R}$, models particle collisions and redistributes PDFs. 
In this study, collision operators are based on the multiple relaxation time (MRT) scheme~\cite{raw_moments} that can be written as,
\begin{align} \label{eq:mrt}
\boldsymbol{\Omega} \left(\boldsymbol{x}, t\right) = \boldsymbol{M}^{-1} \cdot \boldsymbol{\hat{S}} \cdot \boldsymbol{M} \cdot \left(\boldsymbol{f}^{\text{eq}} \left(\rho, \boldsymbol{u}\right) - \boldsymbol{f} \left(\boldsymbol{x}, t\right)\right),
\end{align}
where $\boldsymbol{M} \in \mathbb{R}^{q\times q}$ denotes a $q \times q$ Matrix, constructed from a set of $q$ moments, that transforms the PDFs to the moment space~\cite{raw_moments}.
In the moment space, the collision is resolved by subtracting the PDFs' equilibria, $\boldsymbol{f}^{\text{eq}}(\rho,\boldsymbol{u}) \in \mathbb{R}^{q}$, from the PDFs and applying the diagonal relaxation matrix $\boldsymbol{\hat{S}}\in \mathbb{R}^{q\times q}$.
It contains the relaxation rate, $\omega_{i} < 2 / \Delta t$, the inverse of which is referred to as the relaxation time, $\tau_{i} = 1/\omega_{i}$.
For the MRT employed here, the relaxation time corresponding to second-order moments, $\tau$, is directly related to the kinematic viscosity of the fluid through,
\begin{equation}\label{eq:kin-viscosity}
\nu = c_{s}^{2} \left(\tau -\frac{\Delta t}{2} \right).
\end{equation}
The equilibrium PDF is given as,
\begin{align}\label{eq:feq}
f_i^{\text{eq}}(\rho,\boldsymbol{u}) = \rho w_i +  \rho_0 w_i \left( \frac{\boldsymbol{c}_i \cdot\boldsymbol{u}}{c_s^2}+\frac{(\boldsymbol{c}_i \cdot\boldsymbol{u})^2}{2c_s^4}-\frac{\boldsymbol{u}\cdot\boldsymbol{u}}{2c_s^2} \right),
\end{align}
and can be derived from the continuous Maxwell--Boltzmann distribution~\cite{Bauer_Silva_2020} using the macroscopic velocity, $\boldsymbol{u} \equiv \boldsymbol{u}\left(\boldsymbol{x}, t\right) \in \mathbb{R}^{d}$, density, $\rho \equiv \rho\left(\boldsymbol{x}, t\right)$, and lattice weight, $w_i \in \mathbb{R}$.
When setting the LBM reference density $\rho_0 = 1$ in \Cref{eq:feq}, the incompressible LBM formulation is obtained, whereas $\rho_0 = \rho$ reveals the LBM in compressible form~\cite{Luo97}.\par

If the collision operator's moment set is constructed with the so-called raw moments and all moments are relaxed with the same relaxation rate, $\omega = 1/\tau$, the commonly used Bhatnagar–Gross–Krook (BGK), also referred to as single relaxation time (SRT) collision operator is obtained~\cite{lbm_book},
\begin{align} \label{eq:BGK}
\boldsymbol{\Omega} \left(\boldsymbol{x}, t\right) = \omega \left(\boldsymbol{f}^{\text{eq}} \left(\rho, \boldsymbol{u}\right) - \boldsymbol{f} \left(\boldsymbol{x}, t\right)\right).
\end{align}\par

A major contributor to the LBM's popularity is its formulation as an explicit time-stepping scheme and the fact that all non-linear operations (collision) are local to a computational cell, while the advection (streaming) is linear~\cite{GEIER2015507}.
This means that \Cref{eq:LBE} can be separated into the subsequent steps of collision and streaming denoted by,
\begin{align}\label{eq:collision}
f^{\star}_i \left(\boldsymbol{x}, t\right) &= f_i \left(\boldsymbol{x}, t\right) + \Omega_i \left(\boldsymbol{x}, t\right), \\
\label{eq:streaming}
f_i \left(\boldsymbol{x} + \boldsymbol{c}_i \Delta t, t + \Delta t\right) &= f^{\star}_i \left(\boldsymbol{x}, t\right),
\end{align}
where $f^{\star}_i (\boldsymbol{x}, t)$ indicates the post-collision status of the PDFs.
This illustrates that the resulting scheme can be parallelized well and is therefore excellently-suited for massively parallel, large-scale simulations~\cite{Holzer2021}. In practice usually both steps are combined, as shown in \Cref{eq:LBE}, to achieve the best parallel performance \cite{lehmann2021accuracy}.

\subsection{Free-surface lattice Boltzmann method} \label{subsec:fslbm}
The free-surface lattice Boltzmann method (FSLBM) used in this work is based on the approach from Körner et al.~\cite{korner2005LatticeBoltzmannModel}. 
It allows the simulation of a moving interface between two immiscible fluids and assumes that the heavier phase completely governs the flow dynamics of the system.
As the flow dynamics of the lighter phase are ignored, the problem reduces to a single-phase flow with a free boundary.
This assumption applies to two-phase systems with substantial density and viscosity ratios between the phases.
In the following, the heavier and lighter phases will be called liquid and gas phases, respectively.\par

The boundary between the two phases is treated in a volume-of-fluid-like approach~\cite{hirt1981VolumeFluidVOF}.
As such, the fill level, $\varphi(\boldsymbol{x}, t)$, in a cell is defined as the ratio of its liquid volume to its total volume, and acts as an indicator to describe the affiliation to a phase.
Using this definition, all cells belonging to the fluid domain are either categorized as liquid ($\varphi(\boldsymbol{x}, t)=1$), gas ($\varphi(\boldsymbol{x}, t)=0$) or interface ($\varphi(\boldsymbol{x}, t) \in (0,1)$).
The latter type assembles a sharp interface, i.e., a closed layer of single interface cells that separates liquid from gas cells.
In terms of the LBM implementation, liquid and interface cells are treated as normal cells that contain PDFs and participate in the collision and streaming described in \Cref{subsec:LBM}.
As opposed to this, gas cells neither contain PDFs nor participate in the LBM update.\par

The fill level, $\varphi\left(\boldsymbol{x}, t\right)$, fluid density, $\rho\left(\boldsymbol{x}, t\right)$, and volume, $\Delta x^{3}$, of a cell are used to define its liquid mass as, 
\begin{equation}
m\left(\boldsymbol{x}, t\right) = \varphi\left(\boldsymbol{x}, t\right) \rho\left(\boldsymbol{x}, t\right) \Delta x^{3}.
\end{equation}
The mass flux, $\Delta m_{i}\left(\boldsymbol{x}, t\right)$, is tracked for interface cells and computed from the LBM streaming step as,
\begin{equation} \label{eq:fslbm-mass-flux}
\frac{\Delta m_{i}\left(\boldsymbol{x}, t\right)}{\Delta x^{3}}\left(\boldsymbol{x}, t\right) = 
\begin{cases}
0 & \boldsymbol{x} + \boldsymbol{c}_{i}\Delta t \in \text{gas} \\

f_{\overline{i}}^{\star}\left(\boldsymbol{x} + \boldsymbol{c}_{i}\Delta t, t\right) - f_{i}^{\star}\left(\boldsymbol{x}, t\right) & \boldsymbol{x} + \boldsymbol{c}_{i}\Delta t \in \text{liquid}\\

\frac{1}{2}\left(\varphi\left(\boldsymbol{x}, t\right) + \varphi\left(\boldsymbol{x} + \boldsymbol{c}_{i}\Delta t, t\right) \right)
\left(f_{\overline{i}}^{\star}\left(\boldsymbol{x} + \boldsymbol{c}_{i}\Delta t, t\right) - 
f_{i}^{\star}\left(\boldsymbol{x}, t\right)\right) & \boldsymbol{x} + \boldsymbol{c}_{i}\Delta t \in \text{interface}
\end{cases}
\end{equation}
where $\boldsymbol{c}_{\overline{i}} = -\boldsymbol{c}_{i}$ is the inversion of the lattice direction $i$.\par

An interface cell is converted to gas or liquid when it gets emptied, $\varphi(\boldsymbol{x}, t)< 0-\varepsilon_{\varphi}$, or filled, $\varphi(\boldsymbol{x}, t)>1+\varepsilon_{\varphi}$, with respect to the heuristically chosen threshold, $\varepsilon_{\varphi}=10^{-2}$, that is defined to prevent oscillatory conversions~\cite{pohl2008HighPerformanceSimulation}.
It is important to note that liquid or gas cells can not be converted directly into one another.
Instead, when converting an interface cell, surrounding liquid and gas cells are converted to interface cells to maintain a closed interface layer.
In the case of conflicting conversions, the separation of liquid and gas cells is prioritized.\par

In the course of the simulation, unnecessary interface cells may appear that either have no liquid or no gas neighbor.
In that case, the mass flux from \Cref{eq:fslbm-mass-flux} is modified as suggested in Reference~\cite{thurey2007PhysicallyBasedAnimation} to either force these cells to fill or empty.\par

When converting an interface cell with fill level, $\varphi^{\text{conv}}(\boldsymbol{x}, t)$, to liquid or gas, the fill level of the converted cell is set to $\varphi(\boldsymbol{x}, t)=1$ or $\varphi(\boldsymbol{x}, t)=0$, respectively.
This leads to an excess mass of, 
\begin{equation}
\frac{m_{\text{ex}}\left(\boldsymbol{x}, t\right)}{\rho\left(\boldsymbol{x}, t\right) \Delta x^{3}} = 
\begin{cases}
\varphi^{\text{conv}}\left(\boldsymbol{x}, t\right) - 1 & \text{ if } \boldsymbol{x} \text{ is converted to liquid} \\
\varphi^{\text{conv}}\left(\boldsymbol{x}, t\right) & \text{ if } \boldsymbol{x} \text{ is converted to gas}
\end{cases}
\end{equation}
that must be distributed to neighboring cells.
In this work, excessive mass is distributed evenly among surrounding interface cells, or evenly among surrounding interface and liquid cells in the implementation in FluidX3D.\par

A cell conversion from liquid to interface and vice-versa does not modify the PDFs of the cell. In contrast, the PDFs in cells converted from gas to interface are not yet available.
They are initialized using \Cref{eq:feq} with $\rho$, and $\boldsymbol{u}$ averaged from all surrounding liquid and non-newly converted interface cells.\par

The LBM collision as in \Cref{eq:collision} is applied to all liquid and interface cells with \Cref{eq:feq} being used in compressible form.
Unlike suggested in Reference~\cite{korner2005LatticeBoltzmannModel}, the gravitational force is not weighted with the fill level of an interface cell in the implementation used here.\par

During the LBM streaming step, according to \Cref{eq:streaming}, PDFs streaming from gas cells to interface cells do not exist and must be reconstructed.
This is accomplished using an anti-bounce-back pressure boundary condition at the interface,
\begin{equation} \label{eq:fslbm-anti-bounce-back}
f_{i}^{\star}\left(\boldsymbol{x} - \boldsymbol{c}_{i}\Delta t, t\right)
= f_{i}^\text{eq}\left(\rho^{\text{G}}, \boldsymbol{u}\right)
+ f_{\overline{i}}^\text{eq}\left(\rho^{\text{G}}, \boldsymbol{u}\right)
- f_{\overline{i}}^{\star}\left(\boldsymbol{x}, t\right) \quad \forall i: \boldsymbol{x} - \boldsymbol{c}_{i}\Delta t \in \text{gas}
\end{equation}
where $\boldsymbol{u} \equiv \boldsymbol{u}(\boldsymbol{x}, t)$ is the velocity of the interface cell and $\rho^{\text{G}} \equiv \rho^{\text{G}}(\boldsymbol{x}, t)=p^{\text{G}}(\boldsymbol{x}, t)/c_{s}^{2}$ is the gas density computed from the pressure of the gas phase, $p^{\text{G}}(\boldsymbol{x}, t)$.
In the original model~\cite{korner2005LatticeBoltzmannModel}, it was suggested to reconstruct PDFs based on their orientation with respect to the interface normal.
However, this approach overwrites existing information and was observed to lead to anisotropic artifacts~\cite{bogner2016CurvatureEstimationVolumeoffluid, lehmann2019high}.
Here, as suggested in Reference~\cite{bogner2016CurvatureEstimationVolumeoffluid}, only missing PDFs are reconstructed, and no information is dropped.\par

The gas pressure, 
\begin{equation}
p^{\text{G}}\left(\boldsymbol{x}, t\right) = p^{\text{V}}\left(t\right) - p^{\text{L}}\left(\boldsymbol{x}, t\right),
\end{equation}
consists of the volume pressure, $p^{\text{V}}(t) \in \mathbb{R^{+}}$, and the Laplace pressure, $p^{\text{L}}(\boldsymbol{x}, t) \in \mathbb{R^{+}}$.
The volume pressure can be either atmospheric in which $p^{\text{V}}(t)=\text{constant}$ or result from the change of the volume, $V(t) \in \mathbb{R^{+}}$, of an enclosed gas volume, i.e., bubble with,
\begin{equation}
p^{\text{V}}\left(t\right)=p^{\text{V}}\left(0\right) \frac{V\left(0\right)}{V\left(t\right)}.
\end{equation}
The Laplace pressure, 
\begin{equation} \label{eq:fslbm-laplace-pressure}
p^{\text{L}}\left(\boldsymbol{x}, t\right)=2 \sigma \kappa\left(\boldsymbol{x}, t\right),
\end{equation} 
incorporates the surface tension, $\sigma \in \mathbb{R^{+}}$, and the interface curvature, $\kappa\left(\boldsymbol{x}, t\right) \in \mathbb{R}$.
There exist different approaches for computing the interface curvature that are based on the finite difference method (FDM) or a local triangulation of the interface~\cite{lehmann2022AnalyticSolutionPiecewise,bogner2016CurvatureEstimationVolumeoffluid}.
The simulation results shown here are obtained using the FDM as described in Reference~\cite{bogner2016CurvatureEstimationVolumeoffluid}.
The interface normal, as required by the FDM curvature model, is modified near solid obstacle cells according to Reference~\cite{donath2011WettingModelsParallel}.
Other curvature computation models have been tested and will be discussed in \Cref{subsubsec:cap-wave}.\par

In applications where bubbles must be properly simulated, an additional bubble model extension is required for the FSLBM.
Since gas volumes can coalesce and divide, this algorithm must keep track of the volume pressure of individual bubbles and handle coalescence and segmentation accordingly.
Such algorithms are referred to as bubble models and are algorithmically challenging when applied in parallel computing environments.
Here, the bubble model from Reference~\cite{pohl2008HighPerformanceSimulation} is used to simulate bubble coalescence correctly and in parallel environments.
It is based on the combination of the interface normal and the \textit{seed-fill} algorithm~\cite{pavlidis1982AlgorithmsGraphicsImage}.
In contrast, in FluidX3D, the bubble model is based on the \textit{Hoshen--Kopelman} algorithm~\cite{haeusl2021SoftObjects}.\par

\subsection{Conservative Allen--Cahn model}
The conservative Allen--Cahn model is described in several other publications~\cite{Mitchell2018,Mitchell2021,fakhari_geier_lee_2016}.
Here, the governing equations and their discretization with the LBM are only briefly introduced.

\subsubsection{ Governing equations }
The phase-field model studied in this work is built on the following macroscopic equations,
\begin{align}
\label{eqn-continuity}
\frac{\partial \rho}{\partial t} + \nabla\cdot\rho\boldsymbol{u} &= 0, \\
\label{eqn-momentum}
\rho \left(\frac{\partial \boldsymbol{u}}{\partial t} + \boldsymbol{u} \cdot \nabla \boldsymbol{u} \right) &= -\nabla p + \nabla \cdot \left( \mu \left[\nabla \boldsymbol{u} + (\nabla \boldsymbol{u})^T\right]\right)+ \boldsymbol{F}_s + \boldsymbol{F}_b, \\
\frac{\partial \phi}{\partial t} + \boldsymbol{\nabla} \cdot (\phi \boldsymbol{u} ) &= \boldsymbol{\nabla} \cdot M \left( \boldsymbol{\nabla}\phi - \frac{1-(\phi - \phi_0)}{\xi}\boldsymbol{n}\right), \label{eqn-phasefield}
\end{align}
the first of which represents the continuity equation.
\Cref{eqn-momentum} is the momentum equation with the hydrodynamic pressure, $p \equiv p(\boldsymbol{x}, t)$, and \Cref{eqn-phasefield} is the Allen--Cahn equation used for the tracking of the interface. 
Here, the mobility is denoted by $M \in \mathbb{R^{+}}$, the interface width by $\xi \in \mathbb{N^{+}}$, $\boldsymbol{n}\equiv\boldsymbol{n}(\boldsymbol{x}, t)=\boldsymbol{\nabla}\phi/|\boldsymbol{\nabla}\phi|$ is the unit vector normal to the liquid--gas interface, and $\mu \in \mathbb{R^{+}}$ is the fluid's dynamic viscosity.\par
The principle behind phase-field models is to allocate an additional scalar field for the phase indicator parameter, $\phi \equiv \phi(\boldsymbol{x}, t)\in [0,1]$.
This phase indicator represents the fluid with higher density by $\phi_H=1$ and the lower density fluid by $\phi_{\text{L}}=0$. The bounds of $\phi_{\text{H}}$ and $\phi_{\text{L}}$ can be seen to vary in the literature, and is generally a point of contention. Nonetheless, the authors specify the bounds as $(0,1)$ to minimize issues that may otherwise arise in the light phase fluid.\par

The forces acting on the fluid include the body force associated with gravity, and the surface tension forces resulting from the liquid-gas interface.
These are given as,
\begin{align}
\boldsymbol{F}_b \equiv \boldsymbol{F}_b(\boldsymbol{x}, t) &= \rho(\boldsymbol{x}, t) \boldsymbol{g}, \\
\boldsymbol{F}_s \equiv \boldsymbol{F}_s(\boldsymbol{x}, t) &= \mu_{\phi}\nabla\phi(\boldsymbol{x}, t),
\end{align}
respectively, with gravitational acceleration, $\boldsymbol{g} \in \mathbb{R}^{d}$ and chemical potential, $\mu_{\phi} \in \mathbb{R}$.

\subsubsection{Lattice Boltzmann equations}
Discretizing the conservative Allen--Cahn equation with the LBM yields,
\begin{align}
\label{eq:lb_interface}
h_i (\boldsymbol{x} + \boldsymbol{c}_i \Delta t, t + \Delta t) - h_i(\boldsymbol{x},t) = \Omega_{i}^h \left[ h_i^{\text{eq}}(\phi,\boldsymbol{u}) -  h_i(\boldsymbol{x},t)\right]|_{(\boldsymbol{x}, t)},
\end{align}
where the collision operator of the phase-field LBE is given by $\Omega_{i}^h(\boldsymbol{x},t) \in \mathbb{R}$, the phase-field PDFs by $h_{i}(\boldsymbol{x},t) \in \mathbb{R}$, and the phase-field relaxation time by,
\begin{equation}\label{eq:pflbm-tau}
\tau_{\phi}=M/c_s^2.
\end{equation}
Thus, the mobility of the interface defines the behavior of the interface tracking LBM step.
The density
\begin{equation}
\rho(\boldsymbol{x}, t) = \rho(\phi) = \rho_{\text{L}} + (\phi(\boldsymbol{x}, t) - \phi_{\text{L}})(\rho_{\text{H}} - \rho_{\text{L}})
\end{equation} 
as used in the PDF equilibrium in \Cref{eq:feq}, is computed via interpolation from the phase indicator $\phi(\boldsymbol{x}, t)$ as suggested in Reference~\cite{Fakhari2017}.
Using this formulation of the LBM step, the zeroth-order moment,
\begin{align}
\label{eq:zeroth_moment_phase}
\phi\left(\boldsymbol{x}, t\right) = \sum_q h_q\left(\boldsymbol{x}, t\right),
\end{align}
computes $\phi\left(\boldsymbol{x}, t\right)$.
The conservative Allen--Cahn equation is recovered by applying,
\begin{align}
\boldsymbol{F}^{\phi}(\boldsymbol{x},t) = \frac{4\phi(1-\phi)}{\xi} \cdot \boldsymbol{n},
\end{align}
in the collision space according to Guo's forcing scheme~\cite{Mitchell2019,guo2002DiscreteLatticeEffects}.

The LBE for the hydrodynamics is given by,
\begin{align}
g_i(\boldsymbol{x} + \boldsymbol{c}_i \Delta t, t + \Delta t) - g_i(\boldsymbol{x},t) = \Omega_{i}^g \left[ g_i^{\text{eq}}(p^*,\boldsymbol{u}) - g_i(\boldsymbol{x},t)\right]|_{(\boldsymbol{x}, t)},
\end{align}
with collision operator, $\Omega_{i}^g(\boldsymbol{x},t) \in \mathbb{R}$, for the hydrodynamic PDFs, $g_i(\boldsymbol{x},t) \in \mathbb{R}$, and normalized pressure, $p^{*} \equiv p^{*}(\boldsymbol{x},t) = p(\boldsymbol{x},t)/(\rho(\boldsymbol{x},t) c_s^2)$. Note here that the LBE is formulated such that the zeroth-order moment recovers the normalized pressure,
\begin{align}
\label{eq:zeroth_moment_hydro}
p^{*}(\boldsymbol{x},t) = \sum_i g_i(\boldsymbol{x},t).
\end{align}
Additionally, it is important to notice that for $g_i^{\text{eq}}(p^*, \boldsymbol{u}) \in \mathbb{R}$, the incompressible formulation of the equilibrium PDFs is used.\par

The forcing term to recover the Navier--Stokes equation is,
\begin{align}
\boldsymbol{F}(\boldsymbol{x},t) = \boldsymbol{F}_s + \boldsymbol{F}_b + \boldsymbol{F}_p + \boldsymbol{F}_{\mu},
\end{align}
which consists of terms to recover the correct pressure gradient term, $\boldsymbol{F}_p \equiv \boldsymbol{F}_{p}(\boldsymbol{x},t) \in \mathbb{R}$, the viscous forces, $\boldsymbol{F}_{\mu} \equiv \boldsymbol{F}_{\mu}(\boldsymbol{x},t) \in \mathbb{R}$, the surface forces, $\boldsymbol{F}_{s} \equiv \boldsymbol{F}_{s}(\boldsymbol{x},t) \in \mathbb{R}$, and the body forces, $\boldsymbol{F}_{b} \equiv \boldsymbol{F}_{b}(\boldsymbol{x},t) \in \mathbb{R}$. The force vector is directly applied in the collision space according to Guo's forcing scheme~\cite{Mitchell2019,guo2002DiscreteLatticeEffects}.
The pressure and viscous forces are given as,
\begin{align}
\boldsymbol{F}_p (\boldsymbol{x},t) 	  &= -p^*c_s^2(\rho_{\mathrm{H}} - \rho_{\mathrm{L}})\boldsymbol{\nabla}\phi, \\
\boldsymbol{F}_{\mu}(\boldsymbol{x},t)  &= \nu(\rho_{\mathrm{H}}-\rho_{\mathrm{L}})\left[\boldsymbol{\nabla}\boldsymbol{u}+(\boldsymbol{\nabla}\boldsymbol{u})^T\right]\cdot\boldsymbol{\nabla}\phi,
\end{align}
where $\rho_{\text{H}} \equiv \rho_{\text{H}}(\boldsymbol{x},t)$, and $\rho_{\text{L}} \equiv \rho_{\text{L}}(\boldsymbol{x},t)$ denote the density in the heavy and light phase, respectively~\cite{fakhari_geier_lee_2016}.
The kinematic viscosity, $\nu \equiv \nu(\boldsymbol{x},t)$, is computed with \Cref{eq:kin-viscosity} using the linearly interpolated relaxation time
\begin{equation}
\tau(\boldsymbol{x},t) = \tau(\phi) = \tau_{\text{L}} + (\phi(\boldsymbol{x},t) - \phi_{\text{L}})(\tau_{\text{H}} - \tau_{\text{L}}),
\end{equation}
where $\tau_{\text{H}} \equiv \tau_{\text{H}}(\boldsymbol{x},t)$ is the relaxation time of the heavy phase and $\tau_{\text{L}} \equiv \tau_{\text{L}}(\boldsymbol{x},t)$ is the relaxation time of the light phase.
It is noted here that the deviatoric stress tensor can be obtained from moments of the non-equilibrium distribution to avoid the need for finite difference approximations in the velocity field.

\subsection{Comparison of methodology and numerical implementation}
In this section, the FSLBM and PFLBM are compared in terms of various aspects ranging from methodology to implementation.
This is done to illustrate the various assumptions made in each model, and provide an understanding for the quantitative comparisons made in the later sections.

\subsubsection{Treatment of the low density phase}
One of the major differences between the FSLBM and the PFLBM presented is the treatment of the lighter fluid phase.
Contrary to the PFLBM, the flow dynamics of the lighter phase are ignored in the FSLBM.
Although this allows the PFLBM to be applicable to a broader range of applications, this work focuses on flows where the lighter phase is believed to have negligible influence on the system.
In this case, the computations in the second phase are assumed to be unnecessary when using the PFLBM. Further to this, the lighter phase has a lower viscosity than the heavier phase.
Consequently, the flow is more likely to become turbulent, and $\Delta x$ and $\Delta t$ must be chosen to avoid instabilities in the lighter phase.
Concerning the heavier phase, $\Delta x$ and $\Delta t$ tend to be much smaller than necessary for stability. This impacts the efficiency of the simulation and is one of the driving motivations of the FSLBM. While not considered here, these drawbacks could be moderately compensated by using adaptive refinement of the computational grid~\cite{FAKHARI201722}.\par

\subsubsection{Representation of the interface}
Another significant difference between the two models is the representation of the interface between the phases.
In the FSLBM, the interface layer has a width of one cell and is therefore referred to as a sharp interface.
The fill level in the cell captures the interfacial movement.
On the other hand, the PFLBM represents the interface layer in a diffuse manner with a width of typically around five lattice cells~\cite{Mitchell2018}.
The Allen--Cahn equation describes the advection of the interface.
In general, it is preferential that the interface width is more than a magnitude smaller than the smallest characteristic length scale of the system~\cite{lbm_book}.

\subsubsection{Conservation of mass}
Both models in their originally proposed states are mass conserving~\cite{korner2005LatticeBoltzmannModel,chiu2011ConservativePhaseField}.
However, it was observed that single interface cells can become trapped in liquid or gas in the FSLBM~\cite{thurey2007PhysicallyBasedAnimation}.
It is argued that these artifacts do not perturb the fluid simulation but are visible as artifacts.
To resolve these artifacts, it is suggested to forcefully convert these cells to the cell type in their surrounding, leading to a loss in mass.
Following this approach, the current FSLBM implementation does not fully conserve mass.

\subsubsection{Numerical implementation}
This section focuses on implementation-related aspects of the FSLBM and PFLBM, such as their applicability to code generation, parallel computing, and memory requirements.

\paragraph{Code generation}
With metaprogramming techniques, it is possible to describe the complete PFLBM model in an abstract symbolic form embedded in a high-level programming language, e.g., Python~\cite{Holzer2021}.
Highly optimized code in a performance-oriented programming language, e.g., C or CUDA, is generated automatically. 
Furthermore, performance optimizations, including spatial blocking, common subexpression elimination, and simultaneous instructions on multiple data (SIMD) vectorization, are applied by the code generator.
This provides portability to different computing architectures, such as accelerator hardware like GPUs, and reduces code complexity while increasing the maintainability of the code base. 
The PFLBM consists of essentially only two continuous LBM steps, making it perfectly applicable for code generation.
Here, the entire model, including boundary conditions, forces, and inter-process communication, is implemented using the code generation framework \lbmpy~\cite{lbmpy}.\par

In contrast, the FSLBM is not expected to be as well suited to code generation directly.
While a compute kernel for the LBM step can be generated, many other model components are not inherently suitable to code generation. 
In various parts of the model, the type and direction of a neighboring cell define the operation.
For instance, in the mass exchange algorithm, \Cref{eq:fslbm-mass-flux}, different lattice directions have to be treated according to the type of the neighboring cell in this direction.
The abstract form of the code might then be similar to the direct implementation in a performance-oriented programming language.
Therefore, future work remains to evaluate the applicability of the FSLBM to code generation techniques.\par

\paragraph{Parallelization}
The common requirement of a highly resolved computational grid can often not be sufficiently computed on a single processor or compute node of a cluster for practically relevant simulations.
The PFLBM scales almost perfectly~\cite{Holzer2021} on parallel computing environments and inherently tracks coalescence and segmentation of gas volumes through the Allen--Cahn equation.\par

Without modeling bubble coalescence and segmentation, parallelization of the FSLBM is straightforward on any parallel hardware, scaling just as well as the single-phase LBM.
It must be remarked that this is sufficient for a wide variety of applications such as the standing wave and drop impact test cases presented in \Cref{sec:experiments}.
However, when tracking individual gas volumes, a bubble model is required that monitors information such as the bubble's identifiers, the gas pressure, and the identifier of the process on which parts of the bubble reside.
The parallel implementation of a bubble model is challenging and the models presented in Reference~\cite{pohl2008HighPerformanceSimulation} rely on either global all-to-all communication or global sequential communication in each LBM time step.
As an extension to that, Reference~\cite{donath2009LocalizedParallelAlgorithm} presented more complicated bubble models where regional all-to-all or sequential communication is sufficient.
However, Reference~\cite{donath2009LocalizedParallelAlgorithm} has shown that neither of the mentioned bubble models scale ideally on a parallel computing environment relying on inter-process communication.
In the implementation used in this study, the model from Reference~\cite{pohl2008HighPerformanceSimulation} with global all-to-all communication is used.\par

\paragraph{Memory requirements}
The FSLBM requires similar memory allocation to a single-phase LBM implementation, making it well suited for systems with limited memory like GPUs.
In contrast, the PFLBM requires a second LBM step with separate PDFs for the phase field, approximately doubling the amount of memory required to be stored at each lattice cell, making it less attractive for hardware with constrained memory.
In particular, in setups where high surface detail (e.g., a large number of small droplets) needs to be resolved, with the FSLBM, droplets can have a minimum diameter of three cells.
With the PFLBM on the other hand, the minimum droplet diameter is increased to at least 10 cells.
To match resolved surface details, the lateral increase in lattice resolution for the PFLBM combined with the higher memory requirements per cell lead to approximately $2\cdot(10/3)^3\approx74$-fold increase in required memory, making the FSLBM clearly the better choice in such use-cases.

\section{Numerical experiments} \label{sec:experiments}
This section compares the FSLBM and the PFLBM using numerical experiments.
Choosing the proper test case for comparing two distinct models in terms of accuracy and computational performance is a challenging task.
One must select test cases to which both models are applicable, and keep in mind that each model may be subjected to different forms of errors.
Additionally, it is crucial to select a benchmark where the correct solution is known \textit{a priori}, either from experimental data or analytical models to give a point of reference for the modeled results.\par

While many references provide experimental data for different kinds of multiphase flows, comparing two models based on only experimental measurements can be misleading.
Every experiment is subject to uncertainties that can not be considered in numerical simulations, and if both models disagree with experimental observations in contradicting form, no meaningful conclusion can be drawn.
Therefore, it is always preferable to base the initial comparison on test cases, for which the exact solution is known from analytical calculations.\par

Numerical tools used for fluid simulations generally consist of various, coupled models responsible for certain physical aspects.
For instance, each of the models discussed here has multiple approaches for including wetting effects~\cite{bogner2016CurvatureEstimationVolumeoffluid,donath2011WettingModelsParallel,Ding2007,FAKHARI2017620}.
In an initial comparison, a suitable test case should only include the minimally required components of the models to avoid drawing incorrect conclusions caused by a single component in one of the models.\par

Here, six numerical experiments were used to compare the FSLBM and PFLBM.
Citations have been provided to literature in which each of these models has been applied to the chosen test cases, arguably showing that they are both applicable modeling procedures for the cases.
Two of these tests simulated a standing wave with analytical models available in the literature.
The two cases differed by only the driving force in the flow.
While a gravity wave oscillates due to a body force, a capillary wave does so due to the forces resulting from surface tension.
In each of the test cases, the respective other force was neglected.
The third and fourth test cases featured unconfined and confined buoyancy driven flows.
That is, simulations of a gas bubble rising in a large pool of liquid and a Taylor bubble traveling through a cylindrical pipe, both of which were compared with experimental data from the literature.
In the final test cases, dynamic coalescence was investigated by simulating the impact of a vertical and oblique drop into a pool of liquid.
The results were qualitatively compared to photographs of the laboratory experiments from the literature.\par

In all simulations with the FSLBM, the SRT collision model from \Cref{eq:BGK} was used.
To improve numerical stability in the PFLBM, a weighted orthogonal MRT collision model according to \Cref{eq:mrt} was employed, and the individual moments in both LBM steps were relaxed according to Reference~\cite{Kumar2019}.
It is important to note here that also the second-order moments for the interface tracking LBM step were relaxed with $\tau_{\phi}$.
Within all test cases, the specified relaxation rates were constant across the various resolutions leading to what is also known as diffusive scaling in the LBM.
Setting the second-order moments directly to the equilibrium led to nonphysical results.
Both models used the D$2$Q$9$ velocity set for the standing wave simulations.
For all other simulations, a D$3$Q$19$ velocity set was employed by the FSLBM while the PFLBM was set up with two D$3$Q$27$ lattices for the two LBM steps. 
It is common to introduce the Cahn number, $\mathrm{Cn} = \xi / L$, to describe the PFLBM's interface width $\xi$.
It is highlighted in this work that for convergence assessments, the value of $\xi$ remained constant rather than $\mathrm{Cn}$, as solutions are desired to tend towards a sharp interface result.
In the FSLBM, body forces were modeled according to Guo et al.~\cite{guo2002DiscreteLatticeEffects}.
The forcing terms applied to the LBM steps in the PFLBM model were according to Reference~\cite{Fakhari2017}.
In the simulations of both models, no-slip boundary walls were realized through the bounce-back boundary condition~\cite{lbm_book}.
In agreement with the usual choice in the LBM literature, the reference density was chosen to be $\rho_{0} = \rho_{H} = 1$ in all simulations.\par

In the FSLBM, the fill level was initialized with a Monte Carlo-like sampling method.
A two-dimensional grid consisting of equally spaced, $101 \times 101$, sample points was created in each cell.
The ratio of samples within the specified initial profile to the total number of samples per cell gave the initial fill level.
In the PFLBM, the diffuse interface was initialized with, 
\begin{eqnarray}
\phi_x = \phi_0 \pm \frac{\phi_{\text{H}}-\phi_{\text{L}}}{2} \mathrm{tanh}\left(\frac{x-x_0}{\xi/2}\right),
\end{eqnarray}
in the direction normal to an interface located at $x_0$.\par

The surface meshes visualized for the bubbles and drop impacts were obtained using a marching cube algorithm with destination value $\varphi=0.5$ and $\phi=0.5$ for the FSLBM and PFLBM, respectively.
If not explicitly specified otherwise, all quantities but non-dimensional numbers are denoted in the lattice Boltzmann unit system.
All simulations shown in this article were performed with double-precision floating-point arithmetic.

\subsection{Standing waves} \label{subsec:waves}
In this section, both models' simulation results for a standing gravity and capillary wave are presented and compared with their analytical solutions.

\subsubsection{Gravity wave} \label{subsubsec:grav-wave}
A gravity wave is a standing wave that oscillates at the phase boundary between two immiscible fluids.
Its fluid dynamics are entirely governed by gravitational forces, with surface tension forces being negligible in comparison.

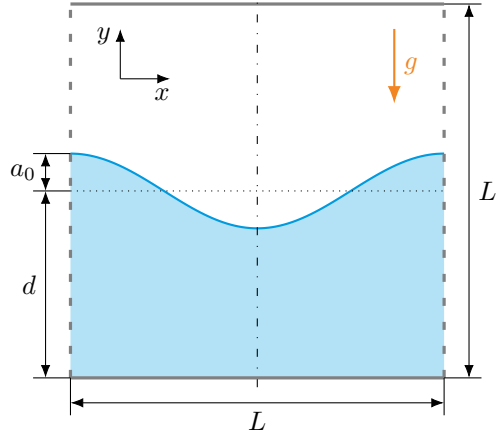
\begin{figure}[htbp]
	\centering
	\setlength{\figureheight}{0.3\textwidth}
	\setlength{\figurewidth}{0.3\textwidth}
	\setlength\mdist{0.02\textwidth}
	\begin{tikzpicture}
\definecolor{darkorange24213334}{RGB}{242,133,34}
\definecolor{dodgerblue0154222}{RGB}{0,154,222}

\begin{axis}%
[width=\figurewidth,
height=\figureheight,
xmin=0,
xmax=1,
ymin=0,
ymax=1,
ticks=none,
axis lines=none,
clip=false,
scale only axis
]
\addplot[thick, name path=f, domain=0:1,samples=50,smooth,dodgerblue0154222] {0.5+0.1*cos(deg(pi*x*2))};

\path[name path=axis] (axis cs:0,0) -- (axis cs:1,0);

\addplot [thick, color=dodgerblue0154222, fill=dodgerblue0154222!30] fill between[of=f and axis,	soft clip={domain=0:1}];
\end{axis}

\draw[very thick, loosely dashed, black!50] (0,0)--(0,\figureheight);
\draw[very thick,, loosely dashed, black!50] (\figurewidth,0)--(\figurewidth,\figureheight);

\draw[very thick,, black!50] (0,\figureheight)--(\figurewidth,\figureheight);
\draw[very thick,, black!50] (0,0)--(\figurewidth,0);

\draw[<->, >=Latex] (\figurewidth+\mdist,0)--(\figurewidth+\mdist,\figureheight) node [pos=0.5,right] {$L$};
\draw[-] (\figurewidth,0)--(\figurewidth+1.5\mdist,0);
\draw[-] (\figurewidth,\figureheight)--(\figurewidth+1.5\mdist,\figureheight);

\draw[<->, >=Latex] (0,-\mdist)--(\figurewidth,-\mdist) node [pos=0.5,below] {$L$};
\draw[-] (0,0)--(0,-1.5\mdist);
\draw[-] (\figurewidth,0)--(\figurewidth,-1.5\mdist);

\draw[dotted] (0,0.5\figureheight)--(\figurewidth,0.5\figureheight);
\draw[<->, >=Latex] (-\mdist,0)--(-\mdist,0.5\figureheight) node [pos=0.5,left] {$d$};
\draw[-] (-1.5\mdist,0)--(0,0);
\draw[-] (-1.5\mdist,0.5\figureheight)--(0,0.5\figureheight);

\draw[<->, >=Latex] (-\mdist,0.5\figureheight)--(-\mdist,0.6\figureheight) node [pos=0.5,left] {$a_{0}$};
\draw[-] (-1.5\mdist,0.6\figureheight)--(0,0.6\figureheight);

\draw[loosely dashdotted] (0.5\figurewidth,-0.025\figureheight)--(0.5\figurewidth,1.025\figureheight);

\draw[thick, ->, >=Latex, darkorange24213334] (\figurewidth-2\mdist,\figureheight-\mdist)--(\figurewidth-2\mdist,\figureheight-4\mdist) node [pos=0.5,right] {$g$};

\draw[->, >=Latex] (2\mdist,\figureheight-3\mdist)--(4\mdist,\figureheight-3\mdist) node [pos=0.5,below right] {$x$};
\draw[->, >=Latex] (2\mdist,\figureheight-3\mdist)--(2\mdist,\figureheight-1\mdist) node [pos=0.5,above left] {$y$};
\end{tikzpicture}%
	\caption{
		Simulation setup of the two-dimensional standing gravity and capillary wave with wavelength, $L$, liquid depth, $d$, and initial wave amplitude, $a_0$.
		There were periodic boundary conditions at the domain's side-walls in $x$-direction and no-slip boundary conditions at the top- and bottom walls in $y$-direction.
		The gravitational acceleration, $g$, was only present in the gravity wave test.}
	\label{fig:wave-setup}
\end{figure}

\paragraph{Simulation setup} \label{par:grav-wave-setup}
A gravity wave with wavelength, $L$, was simulated in a quadratic domain of size $L\times L \times 1$ ($x$-, $y$-, $z$-direction).
As illustrated in \Cref{fig:wave-setup}, a free boundary was initialized with the profile, $y(x) = d + a_{0} \cos\left(kx \right)$, with liquid depth, $d=0.5L$, initial amplitude, $a_{0}=0.01L$, and wavenumber, $k=2\pi/L$.
There were no-slip boundary conditions at both walls in the $y$-direction and periodic boundary conditions at all other domain walls.
Due to the gravitational acceleration, $g$, the initial profile evolved into a standing wave oscillating around the liquid depth, $d$, and dampened by viscous forces.
The Reynolds number,
\begin{equation} \label{eq:re-wave}
\mathrm{Re} = \frac{a_{0}\omega_{0}L}{\nu},
\end{equation}
is defined by the angular frequency of the wave,
\begin{equation}
\omega_{0} = \sqrt{g k \, \mathrm{tanh} \left( k d \right)}.
\end{equation}
In both models, the heavier phase was initialized with hydrostatic pressure according to $g$ such that the LBM pressure at $y(x)=d$ equaled the constant atmospheric volume pressure $p^{\text{V}}(t) = p_{0} = \rho_{0} c_{s}^{2} = 1/3$.\par

The surface elevation, $a^{*}(x, t) = a(x, t)/a_{0}$, and the time, $t^{*} = t\omega_{0}$, are non-dimensionalized to ease comparison.
The simulations were run until $t^{*} = 80$ and the surface elevation, i.e., amplitude $a(x,t)$, was monitored at $x=0$ every $t^{*} = 0.1$.
It was computed by the sum of all cells' fill levels in the $y$-direction at $x=0$ in the FSLBM.
In the PFLBM, the surface elevation was evaluated by interpolating the position at which the phase-field value is $\phi=0.5$.\par

The simulations were carried out with $\mathrm{Re}=10$ and $L \in \{50,100,200,400,800\}$ for the FSLBM and $L \in \{50,100,200,400\}$ for the PFLBM.
The FSLBM's gas phase was considered to be the atmosphere, having a constant atmospheric volume pressure of $p^{\text{V}}(t) = p_{0}$ defined by the LBM reference density $\rho_{0}=1$.
In the PFLBM, the density ratio, $\tilde{\rho} ~= \rho_{\text{H}} / \rho_{\text{L}} = 1000$, and kinematic viscosity ratio, $\tilde{\nu} ~= \nu_{\text{H}} / \nu_{\text{L}} = 1$, mimic a liquid--gas system and were chosen to conform with the analytical solution of the capillary wave in \Cref{par:cap-wave-model}.
The relaxation rate was set to $\omega = 1.8$ and $\omega_{\text{H}} = 1.99$, for the FSLBM and for the heavy phase in the PFLBM, respectively.
The mobility, $M = 0.02$, and interface width, $\xi = 5$, were chosen in the PFLBM conforming to usual choices in the literature~\cite{Mitchell2019}.\par

\paragraph{Analytical model} \label{par:grav-wave-model}
The analytical model for the gravity wave is derived by linearization of the continuity and Euler equations with a free-surface boundary condition~\cite{dingemans1997WaterWavePropagation}.
The surface elevation, i.e., the amplitude of the standing wave,
\begin{equation}
a(x,t) = a_{\text{D}}(t) \cos \left( kx - \omega_{0} t \right) + d,
\end{equation}
is obtained under the assumption of an inviscid fluid resulting in zero damping with $a_{\text{D}}(t)=a_{0}$.
Viscous damping is considered by,
\begin{equation}
a_{\text{D}}(t) = a_{0} \mathrm{e}^{-2 \nu k^{2} t},
\end{equation}
as provided in Reference~\cite{lamb1975Hydrodynamics}.
The model is valid for $k |a_{0}| \ll 1$ and $k |a_{0}| \ll kd$~\cite{dingemans1997WaterWavePropagation}, which is applicable in this study with $k |a_{0}| = 0.02\pi \ll 1 < kd = \pi$.

\begin{figure}[htbp]
	\centering
	\setlength{\figureheight}{0.5\textwidth}
	\setlength{\figurewidth}{0.8\textwidth}
	\input{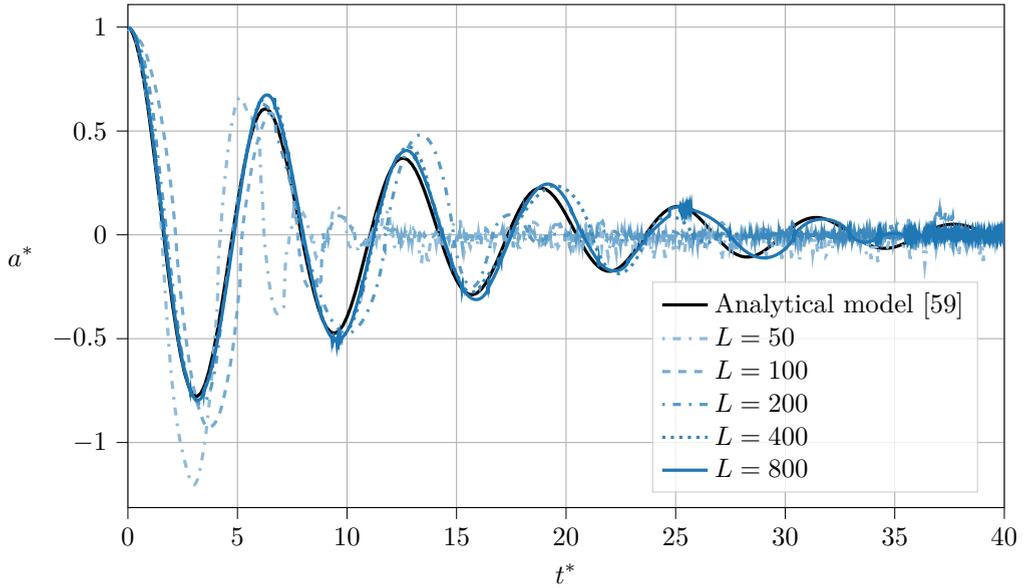}%
	\caption{
		Gravity wave as simulated by the FSLBM with $L \in \{50,100,200,400,800\}$ in terms of non-dimensional amplitude, $a^{*}(0,t^{*})$, and time, $t^{*}$.
		Due to the small initial amplitude, $a_{0} = 0.01L$, the FSLBM's resolution must be sufficiently high to capture the movement of the interface.
	}
	\label{fig:fslbm-gravity-wave-convergence}
\end{figure}

\paragraph{Results and discussion} \label{par:grav-wave-results}
\Cref{fig:fslbm-gravity-wave-convergence} shows the amplitude, $a^{*}(0,t^{*})$, over time, $t^{*}$, for different wavelengths, $L$, simulated with the FSLBM.
As immediately evident, the FSLBM could not reasonably simulate the gravity wave setup chosen here with small resolutions.
This is caused by the requirement of a small initial amplitude, $a_{0}=0.01L$, to be consistent with the analytical solution.
The surface of the wave moves only in a range of a few LBM cells or even purely within one cell.
This could not be simulated with sufficient accuracy with the FSLBM due to its sharp interface representation on a fixed Cartesian grid.
On the other hand, with higher resolution, the amplitudes span more cells and the FSLBM converged well with reasonable accuracy to the analytical model.
In particular, a resolution of $L=50$ did not allow a meaningful simulation, however, $L \in \{100,200,400,800\}$ allowed $2, 3, 4, 5$ periods to be simulated.\par

\begin{figure}[htbp]
	\centering
	\setlength{\figureheight}{0.5\textwidth}
	\setlength{\figurewidth}{0.8\textwidth}
	\input{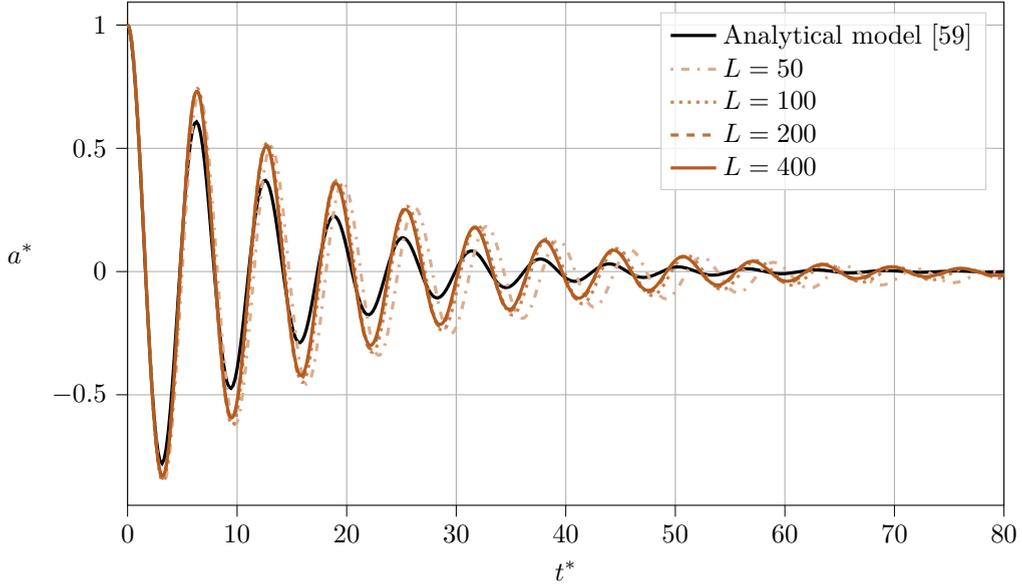}%
	\caption{Gravity wave as simulated by the PFLBM with $L \in \{50,100,200,400\}$ in terms of non-dimensional amplitude, $a^{*}(0,t^{*})$, and time, $t^{*}$. The model was able to capture small interface movement even with low resolution.}
	\label{fig:pf-gravity-wave-convergence}
\end{figure}

Due to the diffuse interface of the PFLBM, the model was capable of simulating even very small amplitudes as shown in \Cref{fig:pf-gravity-wave-convergence}.
The simulations converged well and from $L=100$ on, the phase of the wave was predicted accurately.
However, the model clearly underestimated the wave's damping for the parameters used in this study.\par

\begin{figure}[htbp]
	\centering
	\setlength{\figureheight}{0.5\textwidth}
	\setlength{\figurewidth}{0.8\textwidth}
	\input{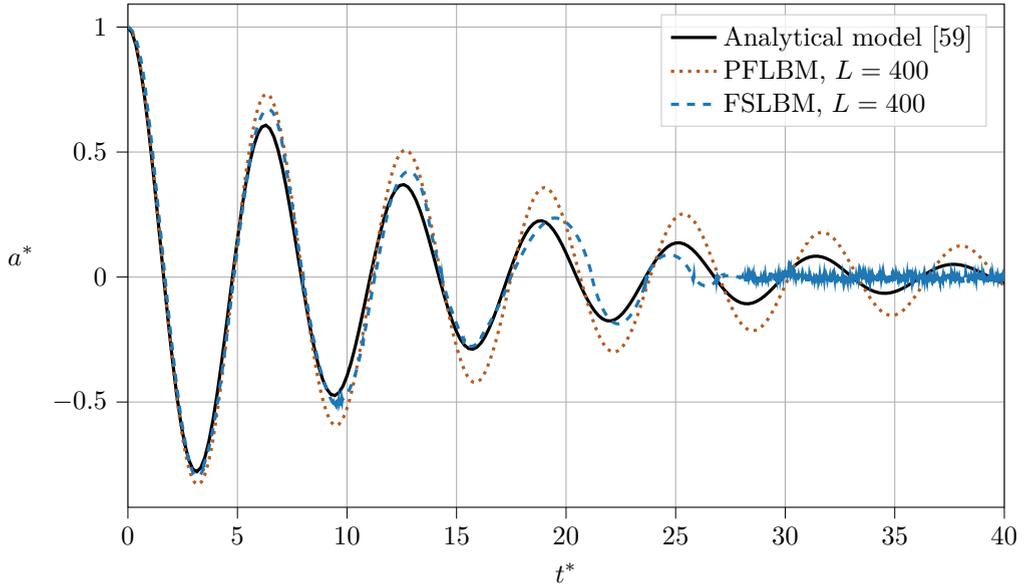}%
	\caption{
		Amplitude of the gravity wave as simulated by the FSLBM and PFLBM at $L=400$ in terms of non-dimensional amplitude, $a^{*}(0,t^{*})$, and time, $t^{*}$.
		The PFLBM was able to capture significantly smaller amplitudes.
	}
	\label{fig:comparison-gravity-wave}
\end{figure}

In \Cref{fig:comparison-gravity-wave}, the FSLBM and PFLBM are compared directly.
The resolution of the FSLBM was chosen such that a sufficient number of periods have been simulated to allow a meaningful comparison.
The computational grid had to be resolved to a very high level to have the amplitude to span over multiple cells (note that many fewer periods were resolved by the FSLBM for the same resolution).
However, in this test case, there was only a single fluid and a single gas domain divided by one interface with little curvature.
Therefore, the width of PFLBM's diffuse interface was less significant here and did not enforce a highly resolved computational grid.
While this is representative for a variety of applications, it is not for many others in which the minimal diffuse interface width imposes a higher computational resolution.
It must be also noted that the size of the amplitude was chosen for consistency of the analytical model.
This highlights a limit case for the FSLBM where difficulties arise due to only small surface movement.
In this particular case where the surface oscillates back and forth around the same lattice cells, the amplitude of the oscillation can only be sufficiently resolved when cell conversions are triggered, i.e., when the surface movement extends beyond a single layer of cells.
In other setups where there is persistent directional movement of the surface, this is not a problem and the surface position is resolved well anywhere between lattice cells.

\subsubsection{Capillary wave} \label{subsubsec:cap-wave}
In contrast to the gravity wave, the fluid dynamics of the capillary wave are purely dominated by surface tension forces, while gravitational forces are neglected.

\paragraph{Simulation setup} \label{par:cap-wave-setup}
The simulation setup was equivalent to the one of the gravity wave in \Cref{par:grav-wave-setup}.
As for the gravity wave, a standing capillary wave evolves, oscillating around a liquid depth, $d$, because of surface tension forces.
The decay of the wave is again caused by energy dissipation due to viscous friction.
While the definition of $\mathrm{Re}$ in \Cref{eq:re-wave} is also used for the capillary wave, the angular frequency of the wave is given by,
\begin{equation}
\omega_{0} = \sqrt{\frac{\sigma k^{3}}{\rho_{H} + \rho_{L}}}.
\end{equation}
Here, it can be seen that it is now defined with the surface tension, $\sigma$, and the densities of the heavy, $\rho_{\text{H}}$, and light phase, $\rho_{\text{L}}$.
Except for hydrostatic pressure, which is not present due to the absence of gravity, the simulation parameters and evaluation procedure were identical to those presented in \Cref{par:grav-wave-setup}.\par

All simulations were performed with $\mathrm{Re}=10$ and $L \in \{50,100,200,400,800\}$ for the FSLBM and $L \in \{50,100,200,400\}$ for the PFLBM.
In the latter, the density ratio, $\tilde{\rho}=1000$, and kinematic viscosity ratio, $\tilde{\nu}=1$, mimic a liquid--gas system and conform with the capillary wave's analytical model.
The relaxation rate was set to $\omega = 1.8$ and $\omega_{\text{H}} = 1.99$, for the FSLBM and for the heavy phase in the PFLBM, respectively.
As in \Cref{par:grav-wave-results}, the mobility, $M=0.02$, and interface width, $\xi=5$, were used in the PFLBM.

\paragraph{Analytical model} \label{par:cap-wave-model}
Prosperetti~\cite{prosperetti1981MotionTwoSuperposed} presented an analytical model for small-amplitude capillary waves in viscous fluids.
The model assumes that there is either a single fluid with a free-surface ($\rho_{\text{L}} = 0$, $\mu_{\text{L}} = 0$) or two fluids with equal kinematic viscosity such that $\tilde{\nu} = 1$.
It is derived from the linearized Navier--Stokes equations and therefore only valid in the limit of infinitesimally small wave amplitudes.\par
Assuming no gravitational forces and no initial velocity, the capillary wave amplitude, $a(t)$, with respect to time, $t$, is described by,
\begin{equation}\label{eq:prosperetti-amplitude}
a\left( t \right)
= 
\frac{4\left(1-4\beta\right)\nu^{2}k^{4}}{8\left(1-4\beta\right)\nu^{2}k^{4}+\omega_{0}^{2}} 
a_{0}\mathrm{erfc}\sqrt{\nu k^2 t}
+
\sum_{i=1}^{4} \frac{z_{i}}{Z_{i}}
\left( \frac{w_{0}^2 a_{0}}{z_{i}^{2} - \nu k^2} \right)
\cdot
\exp\left( \left( z_{i}^{2} - \nu k^{2} \right)t \right) 
\cdot 
\mathrm{erfc}\left( z_{i} \sqrt{t} \right),
\end{equation}
where $z_{i}$ are the roots of the polynomial,
\begin{equation}
z^{4} - 4\beta\left( k^{2} \nu \right)^{\frac{1}{2}} z^{3} + 2\left( 1-6\beta \right)k^{2}\nu z^{2} + 4\left( 1-3\beta \right) \left(k^{2} \nu \right)^{\frac{3}{2}} z + \left(1-4\beta \right)\nu^{2} k^{4} + \omega_{0}^{2} = 0,
\end{equation}
and $Z_{i}$ are computed by circular permutation of the index $i$ in $z_{i}$,
\begin{equation}
Z_{i} =
\prod_{1 \leq j \leq 4, j \neq i}
\left( z_{j} - z_{i} \right).
\end{equation}
The expression $\mathrm{erfc}(x) = 1-\mathrm{erf}(x)$ is the complementary error function and $\beta$ is a dimensionless parameter defined by,
\begin{equation}
\beta = \frac{\rho_{\text{L}}\rho_{\text{H}}}{\left( \rho_{\text{L}} + \rho_{\text{H}} \right)^{2}}.
\end{equation}\par

The analytical model is only applicable for small amplitudes such that a correction factor was proposed extending the validity of the model to amplitudes of up to $a_{0}\lesssim 0.1L$~\cite{denner2017DispersionViscousAttenuation}.
For $a_{0}= 0.01L$ as chosen here, this correction factor is only $1.0023$ and it can be assumed that the original analytical model is valid to be used as a reference in this study.

\begin{figure}[htbp]
	\centering
	\setlength{\figureheight}{0.5\textwidth}
	\setlength{\figurewidth}{0.8\textwidth}
	\input{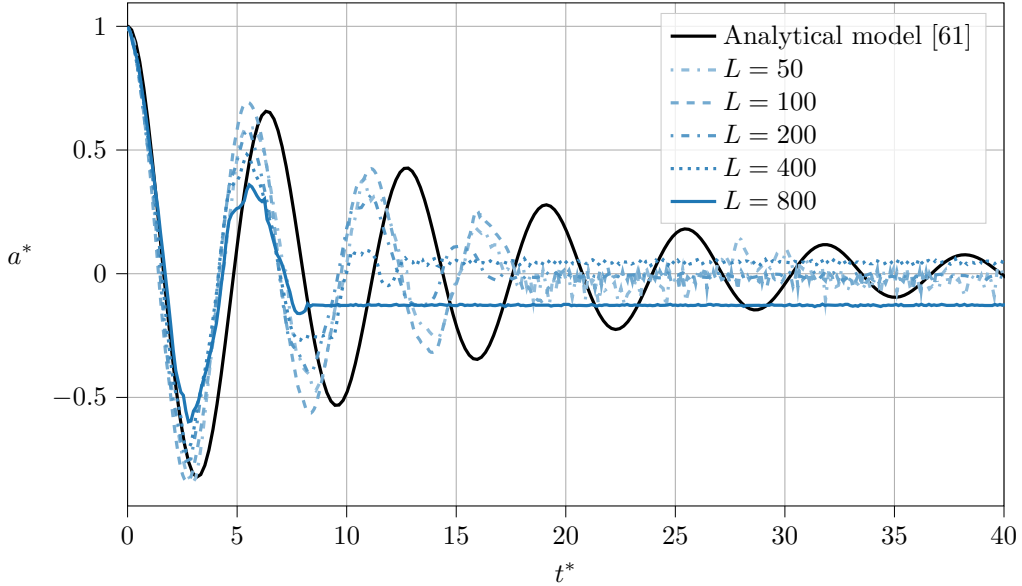}%
	\caption{
		Capillary wave as simulated by the FSLBM with $L \in \{50,100,200,400,800\}$ in terms of non-dimensional amplitude, $a^{*}(t^{*})$, and time, $t^{*}$.
		The FSLBM did not converge with higher resolution due to deficiencies in all investigated curvature computation models.
	}
	\label{fig:cap-wave-convergence-fslbm}
\end{figure}

\paragraph{Results and discussion} \label{par:cap-wave-results}
As illustrated in~\Cref{fig:cap-wave-convergence-fslbm}, the simulations of the FSLBM did not converge with increasing resolution of the computational grid.
The only difference between the gravity wave test case and the capillary wave test case is the driving force, which is a body force in the former and the surface tension in the latter.
In the gravity wave test case, the FSLBM simulation converged and the results agreed well with the analytical model.
This suggests the potential existence of errors in the surface tension model used within the FSLBM.
There, surface tension forces are incorporated by the Laplace pressure, $p^{\text{L}}$, from \Cref{eq:fslbm-laplace-pressure}, with the interface curvature $\kappa$ being the only non-constant parameter in the equation.
Therefore, it is apparent that the diverging behavior must be caused by a diverging interface curvature computation.\par

As described in \Cref{subsec:fslbm}, the simulations and results shown here are based on a curvature computation using the finite difference method (FDM)~\cite{bogner2016CurvatureEstimationVolumeoffluid}.
A similar result was obtained when computing the interface curvature using a local triangulation model and the algorithm from Taubin~\cite{taubin1995EstimatingTensorCurvature} in \walberla{}, as suggested by Reference~\cite{pohl2008HighPerformanceSimulation}.
This is in agreement with Reference~\cite{bogner2016CurvatureEstimationVolumeoffluid}, where both approaches were found to diverge with increasing resolution when computing the curvature of a resting spherical gas bubble.
On the other hand, a curvature model based on local triangulation and a least squares fit optimization (LSQR) was found to be second-order convergent in the same test case~\cite{bogner2016CurvatureEstimationVolumeoffluid}.
However, using a similar LSQR approach~\cite{lehmann2022AnalyticSolutionPiecewise} in FluidX3D, also no convergent behavior could be obtained in the capillary wave test case at the largest resolutions.\par

It is vital to remark that the absolute value of the curvature decreases when increasing the resolution.
With the parametrization chosen here, also the absolute numerical value of the surface tension decreases with increasing $L$.
Therefore, although the LSQR curvature model converges, the model's constant error in curvature has increasingly more influence at higher resolution.\par

The capillary wave has been previously simulated and compared to a different analytical model~\cite{mecke2010DynamicsNanoscopicCapillary}, where gravitational forces are also considered~\cite{donath2011VerificationSurfaceTension}.
There, simulations were performed with the curvature computation using the algorithm of Taubin but the authors did not present a convergence study.
Using the same capillary wave setup and resolution as in Reference~\cite{donath2011VerificationSurfaceTension}, a moderate agreement with the analytical model was observed with the FSLBM implementations used in this work.
However, in a convergence study, again the FSLBM did not converge to the analytical model regardless of the curvature computation model used.
In the work of Körner et al.~\cite{korner2005LatticeBoltzmannModel}, the FSLBM has also been used to simulate a capillary wave and found to agree well with the analytical solution.
The authors did not present a convergence study but verbally argued that the error decreases linearly with increasing resolution.
The curvature model used there is based on the two-dimensional template--sphere method~\cite{bullard1995NumericalMethodsComputing} that uses a neighborhood of 25 cells to compute the curvature.
However, the implementations presented here are explicitly targeted at parallel computing environments, in which such a calculation is not feasible.
To maintain reasonable parallel efficiency, only information from nearest neighbor cells is desired for curvature computation.\par

\begin{figure}[htbp]
	\centering
	\setlength{\figureheight}{0.5\textwidth}
	\setlength{\figurewidth}{0.8\textwidth}
	\input{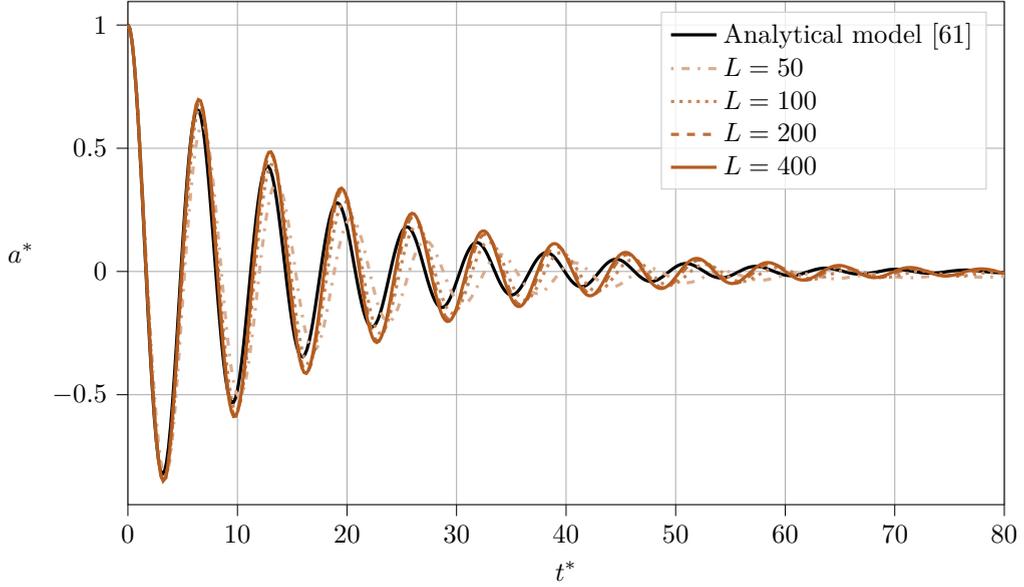}%
	\caption{
		Capillary wave as simulated by the PFLBM with $L \in \{50,100,200,400\}$ in terms of non-dimensional amplitude, $a^{*}(t^{*})$, and time, $t^{*}$.
		The model captured the wave's phase and damping with reasonable accuracy.
	}
	\label{fig:cap-wave-convergence-pf}
\end{figure}

In contrast, as illustrated in \Cref{fig:cap-wave-convergence-pf}, the PFLBM converged well towards the analytical solution but slightly underestimated the analytical model's damping with the parameters from this study.
A comparable capillary wave test case has been simulated with the PFLBM in Reference~\cite{Kumar2019}.
However, compared to the parameters chosen here, the initial amplitude and Reynolds numbers were significantly smaller in Reference~\cite{Kumar2019}, leading to a more accurate prediction of the damping.\par

It can be concluded that special attention must be paid when simulating surface tension dominated flows with very low curvature with the FSLBM.
While the capillary wave resembles an extreme case with small amplitudes leading to infinitesimal values of absolute curvature, other test cases with major surface tension influence have been simulated with good accuracy with the FSLBM~\cite{lehmann2021ejection, lehmann2022AnalyticSolutionPiecewise}.
On the other hand, the PFLBM accurately simulates this test case and as in \Cref{par:grav-wave-results}, it has to be pointed out explicitly that the PFLBM is capable of also simulating very small amplitudes.\par

\subsection{Buoyancy driven flows} \label{subsec:buoyancy-flows}
This section presents numerical simulations of buoyancy driven flows.
The first test case is an unconfined flow, where a single gas bubble rises in liquid.
In the second test case, a large gas bubble rises in liquid contained in a cylindrical tube.
This large gas bubble in the confined, buoyancy driven flow is commonly referred to as Taylor bubble.

\subsubsection{Rising bubble} \label{subsubsec:rising-bubble}
The more practically oriented third test case is an unconfined buoyancy driven flow, i.e., the rise of a single gas bubble in a liquid column.
In order to correctly simulate the bubble shape and rise velocity, the balance between buoyancy, viscous, and surface tension forces must be correct.
As there are no analytical models available predicting a rising bubble's shape and velocity, the comparison is drawn using experimental data from Bhaga and Weber~\cite{bhaga1981BubblesViscousLiquids}.

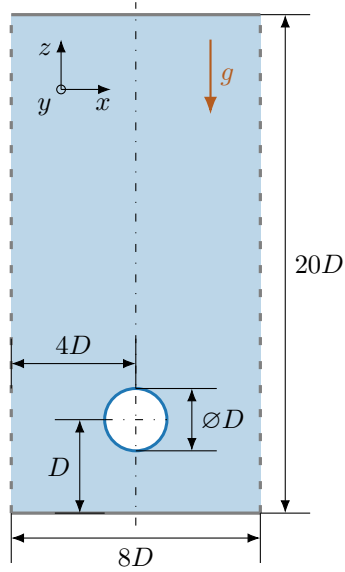
\begin{figure}[htbp]
	\centering
	\setlength{\figureheight}{0.4\textwidth}
	\setlength{\figurewidth}{0.2\textwidth}
	\setlength\mdist{0.02\textwidth}
	\begin{tikzpicture}
\definecolor{steelblue31119180}{RGB}{31,119,180}
\definecolor{sienna1809231}{RGB}{180,92,31}
\setlength\radius{0.125\figurewidth}

\draw [fill=steelblue31119180!30, draw=none] (0,0) rectangle (\figurewidth,\figureheight);

\draw[very thick, loosely dashed, black!50] (0,0)--(0,\figureheight);
\draw[very thick, loosely dashed, black!50] (\figurewidth,0)--(\figurewidth,\figureheight);

\draw[very thick, black!50] (0,\figureheight)--(\figurewidth,\figureheight);
\draw[very thick, black!50] (0,0)--(\figurewidth,0);

\draw[<->, >=Latex] (\figurewidth+\mdist,0)--(\figurewidth+\mdist,\figureheight) node [pos=0.5,right] {$20D$};
\draw[-] (\figurewidth,0)--(\figurewidth+2\mdist,0);
\draw[-] (\figurewidth,\figureheight)--(\figurewidth+2\mdist,\figureheight);

\draw[<->, >=Latex] (0,-\mdist)--(\figurewidth,-\mdist) node [pos=0.5,below] {$8D$};
\draw[-] (0,0)--(0,-2\mdist);
\draw[-] (\figurewidth,0)--(\figurewidth,-2\mdist);

\draw[thick, draw=steelblue31119180, fill=white, fill opacity=1, line width=0.4mm] (0.5\figurewidth,3\radius) circle [radius=\radius] node {};
\draw[loosely dashdotted] (0.5\figurewidth-\radius,3\radius)--(0.5\figurewidth+\radius,3\radius);

\draw[<->, >=Latex] (0.5\figurewidth-\radius-\mdist,0)--(0.5\figurewidth-\radius-\mdist,3\radius) node [pos=0.5,left] {$D$};
\draw[-] (0.5\figurewidth-\radius-2\mdist,3\radius)--(0.5\figurewidth-\radius,3\radius);
\draw[-] (0.5\figurewidth-\radius-2\mdist,0)--(0.5\figurewidth-\radius,0);

\draw[<->, >=Latex] (0,3\radius+\radius+\mdist)--(0.5\figurewidth,3\radius+\radius+\mdist) node [pos=0.5,above] {$4D$};
\draw[-] (0,3\radius+\radius+\mdist-\mdist)--(0,3\radius+\radius+\mdist+\mdist);
\draw[-] (0.5\figurewidth,3\radius+\radius+\mdist-\mdist)--(0.5\figurewidth,3\radius+\radius+\mdist+\mdist);

\draw[<->, >=Latex] (0.5\figurewidth+\radius+\mdist,3\radius-\radius)--(0.5\figurewidth+\radius+\mdist,3\radius+\radius) node [pos=0.5,right] {$\varnothing D$};
\draw[-] (0.5\figurewidth,3\radius-\radius)--(0.5\figurewidth+\radius+2\mdist,3\radius-\radius);
\draw[-] (0.5\figurewidth,3\radius+\radius)--(0.5\figurewidth+\radius+2\mdist,3\radius+\radius);

\draw[loosely dashdotted] (0.5\figurewidth,-0.025\figureheight)--(0.5\figurewidth,1.025\figureheight);

\draw[thick, ->, >=Latex, sienna1809231] (\figurewidth-2\mdist,\figureheight-\mdist)--(\figurewidth-2\mdist,\figureheight-4\mdist) node [pos=0.5,right] {$g$};

\draw[->, >=Latex] (2\mdist,\figureheight-3\mdist)--(4\mdist,\figureheight-3\mdist) node [pos=0.5,below right] {$x$};
\draw[->, >=Latex] (2\mdist,\figureheight-3\mdist)--(2\mdist,\figureheight-1\mdist) node [pos=0.5,above left] {$z$};
\draw[draw=black] (2\mdist,\figureheight-3\mdist) circle [radius=0.175\mdist] node[opacity=1, below left] {$y$};
\end{tikzpicture}%
	\caption{
		Simluation setup of the three-dimensional rising bubble test case with initial bubble diameter, $D$, and gravitational acceleration, $g$.
		The domain's side walls in $x$-direction are periodic, whereas at the top and bottom walls in $z$-direction, no-slip boundary conditions are applied.
	}
	\label{fig:rising-bubble-setup}
\end{figure}

\paragraph{Simulation setup} \label{par:rising-bubble-setup}
As shown in \Cref{fig:rising-bubble-setup}, a gas bubble was initialized as a sphere of diameter, $D$, centered at $(4D, 4D, 1D)$ in a computational domain of size $8D \times 8D \times 20D$ ($x$-, $y$-, $z$-direction).
Gravity was applied in the negative $z$-direction causing the bubble to rise due to buoyancy.
The top and bottom walls (in $z$-direction) were realized as no-slip boundaries, while the side walls of the domain were periodic.
The size of the domain was tested, and determined to be sufficiently large so as not to influence the results of the simulations.
Hydrostatic pressure was initialized such that the reference density, $\rho_{0}=1$, was positioned at $10D$ in the $z$-direction.\par

The rise of a single gas bubble in liquid is characterized by the Morton number,
\begin{equation}\label{eq:Mo}
\mathrm{Mo} = \frac{g \mu^{4}}{\rho \sigma^{3}},
\end{equation}
which describes the ratio of viscous to surface tension forces, and the Bond number,
\begin{equation}\label{eq:Bo}
\mathrm{Bo} = \frac{g D^{2} \rho}{\sigma},
\end{equation}
which describes the ratio of gravitational forces, i.e., buoyancy, to surface tension forces.
It is commonly also referred to as the Eötvös number ($\mathrm{Eo}$).
The definitions of these dimensionless numbers are taken from Reference~\cite{bhaga1981BubblesViscousLiquids} and the density, $\rho$, and dynamic viscosity, $\mu$, refer to the heavier fluid.\par

The bubble shape and position in terms of its center of mass, were monitored at every reference time interval,
\begin{equation}
t^{*} = t\sqrt{\frac{g}{D}}.
\end{equation}
From the bubble position in the $z$-direction at time, $t^{*}=5$, and $t^{*}=10$, the rise velocity $u$ and Reynolds number,
\begin{equation}\label{eq:Re}
\mathrm{Re} = \frac{\rho D u}{\mu},
\end{equation}
were evaluated.
The simulations were stopped at $t^{*}=10$. 
The bubble shape and the Reynolds number were then compared with experimental observations from Reference~\cite{bhaga1981BubblesViscousLiquids}.\par

The simulations were carried out with $D \in \{8,16,32,64\}$ for both models.
Additionally, as in Reference \cite{Mitchell2019}, a fixed mobility, $M=0.02$, and interface width, $\xi=5$, were used for the PFLBM.
Furthermore, to close the system parameters, the density of the liquid phase was specified as $\rho_{H} = 1$, and the density ratio, $\tilde{\rho}=1000$, and dynamic viscosity ratio, $\tilde{\mu} ~= \mu_{\text{H} / \mu_{\text{L}}}=100$, were chosen to mimic an air--water system.
For the FSLBM, the initial pressure of the bubble was set to the reference pressure, $p_{0}=\rho_{0} c_{s}^{2} = 1/3$, with reference density, $\rho_{0}=1$.
The dimensionless numbers that define the four cases tested, and the employed LBM relaxation rates are listed in \Cref{tab:rising-bubble-setups}.
Hydrostatic pressure was initialized in the domain such that the pressure is equivalent to the LBM reference density, $\rho_{0}=1$, in the center of the domain in $z$-direction.
\begin{table}[htbp]
	\centering
	\begin{tabular}{>{\raggedright}m{0.2\textwidth}
			>{\centering\arraybackslash}m{0.1\textwidth}
			>{\centering\arraybackslash}m{0.1\textwidth}
			>{\centering\arraybackslash}m{0.1\textwidth}
			>{\centering\arraybackslash}m{0.1\textwidth}}
		
		\toprule
		Case & $1$ & $2$ & $3$ & $4$ \\
		\midrule
		Bo & $32.2$ & $115$ & $243$ & $339$ \\
		
		Mo & $8.2\cdot 10^{-4}$ & $4.63\cdot 10^{-3}$ & $266$ & $43.1$ \\
		
		$\omega$ (FSLBM) & $1.95$ & $1.95$ & $1.65$ & $1.8$ \\
		
		$\omega_{\text{H}}$ (PFLBM) & $1.97$ & $1.98$ & $1.83$ & $1.92$ \\
		\bottomrule
	\end{tabular}
	\caption{
		The rising bubble test cases are defined by the Bond number, $\mathrm{Bo}$, and the Morton number, $\mathrm{Mo}$.
		The FSLBM's relaxation rate, $\omega$, and the PFLBM's relaxation rate in the heavy phase, $\omega_{\text{H}}$, are kept constant for all resolutions to achieve diffusive scaling.
	}
	\label{tab:rising-bubble-setups}
\end{table}

\paragraph{Results and discussion} \label{par:rising-bubble-results}
The simulated bubble shapes at $t^{*}=10$ are presented in \Cref{fig:bo-115-mo-4.63e-3} and in the Appendix in \Cref{fig:bo-32.2-mo-8.2e-4,fig:bo-243-mo-266,fig:bo-339-mo-43.1}.
It can be seen that both models converged to the Reynolds numbers reported in the experiments in Reference~\cite{bhaga1981BubblesViscousLiquids}.
The FSLBM simulated the rising bubble with reasonable accuracy for computational resolutions of $D\geq16$.
Although surface tension forces are not fully governing the rising bubble test cases, they still significantly determine the bubble shape and rise velocity here.
In contrast to the capillary wave test case in \Cref{subsubsec:cap-wave}, the FSLBM showed consistent convergence with increasing computational resolution.
This emphasizes that the FSLBM can still be applicable in problems where surface tension is non-negligible.
However, the FSLBM predicted the detachment of several satellite bubbles in cases 2 to 4 that can not be observed in the photographs of the experiments.
Qualitatively similar bubble shapes were also predicted by the FSLBM with the LSQR curvature model, as shown in the Appendix in \Cref{fig:fluidx3d-bo-32.2-mo-8.2e-4,fig:fluidx3d-bo-115-mo-4.63e-3,fig:fluidx3d-bo-243-mo-266,fig:fluidx3d-bo-339-mo-43.1}.\par

\begin{figure}[htbp]
	\centering
	\begin{tabular}{>{\centering\arraybackslash}m{0.05\textwidth}
			>{\centering\arraybackslash}m{0.2\textwidth}
			>{\centering\arraybackslash}m{0.2\textwidth}
			>{\centering\arraybackslash}m{0.2\textwidth}
			>{\centering\arraybackslash}m{0.2\textwidth}}
		
		\rotatebox[origin=l]{90}{Experiment}
		& \multicolumn{4}{c}{\makecell{
				\includegraphics[width=0.15\textwidth]{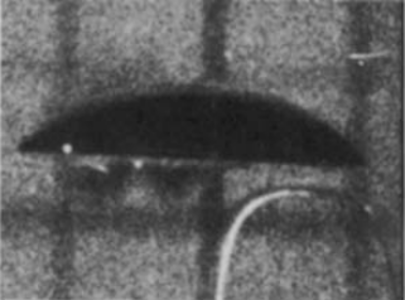} \\ 
				$\mathrm{Re}=94$~\cite{bhaga1981BubblesViscousLiquids}}} \\
		\rotatebox[origin=l]{90}{FSLBM} &
		\makecell{\includegraphics[width=0.2\textwidth]{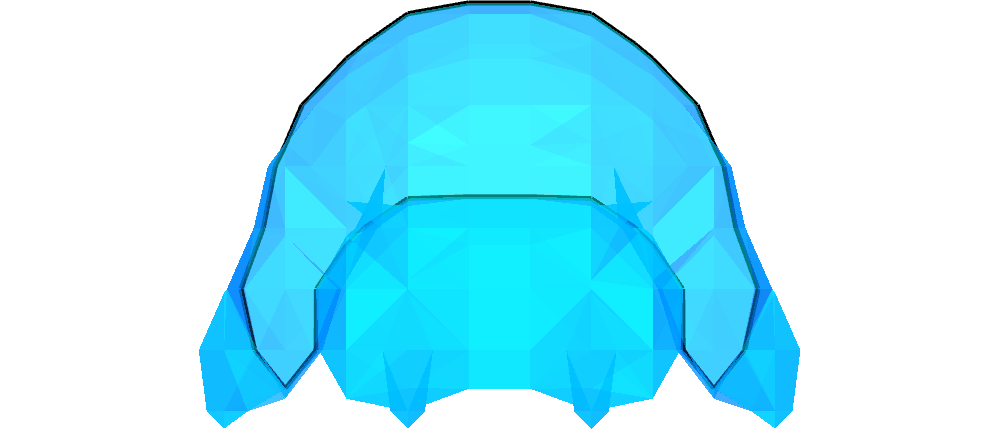} \\ (a) $\mathrm{Re}=51.0$} &   		\makecell{\includegraphics[width=0.2\textwidth]{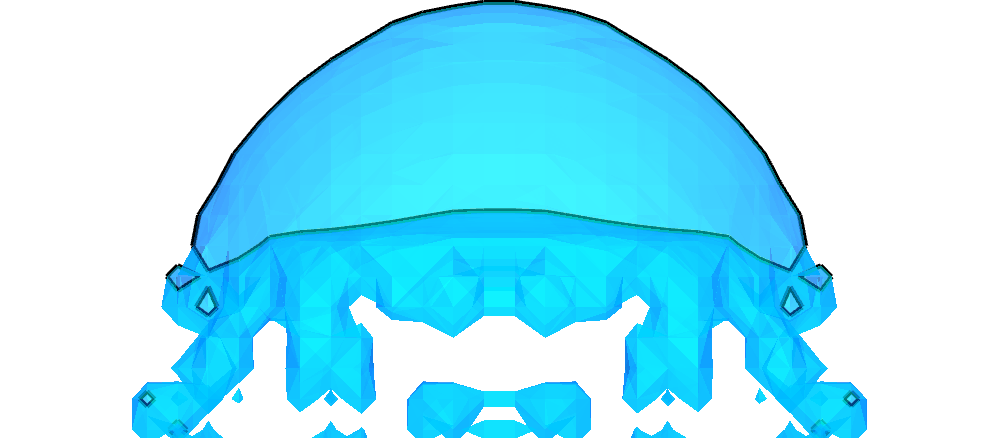} \\ (b) $\mathrm{Re}=78.8$} &	
		\makecell{\includegraphics[width=0.2\textwidth]{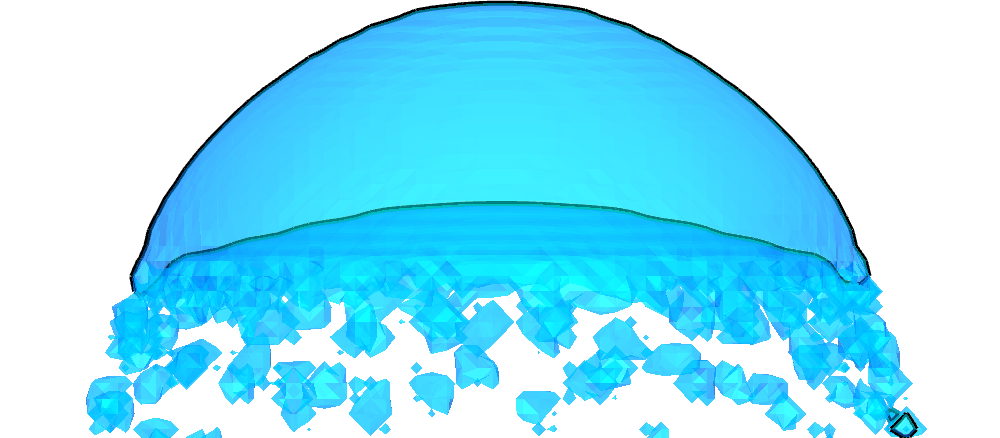} \\ (c) $\mathrm{Re}=89.4$} & \makecell{\includegraphics[width=0.2\textwidth]{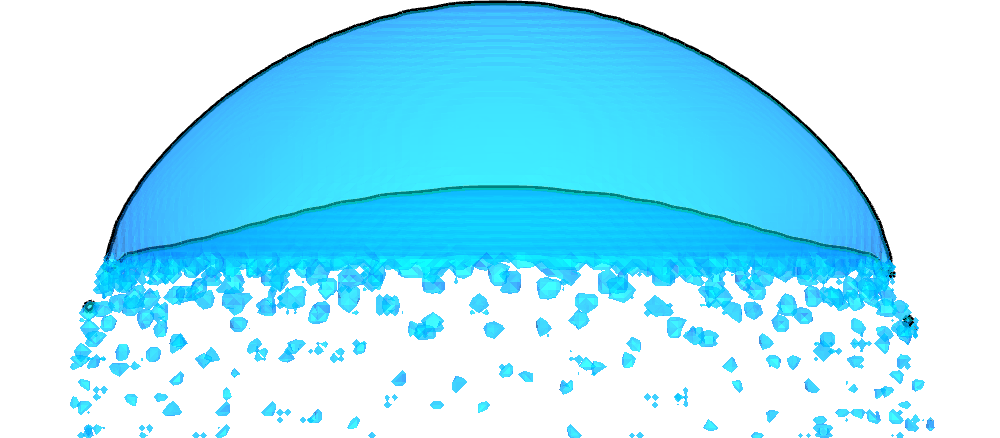} \\ (d) $\mathrm{Re}=91.3$} \\
		\rotatebox[origin=l]{90}{PFLBM} & unstable & unstable & 
		\makecell{\includegraphics[width=0.2\textwidth]{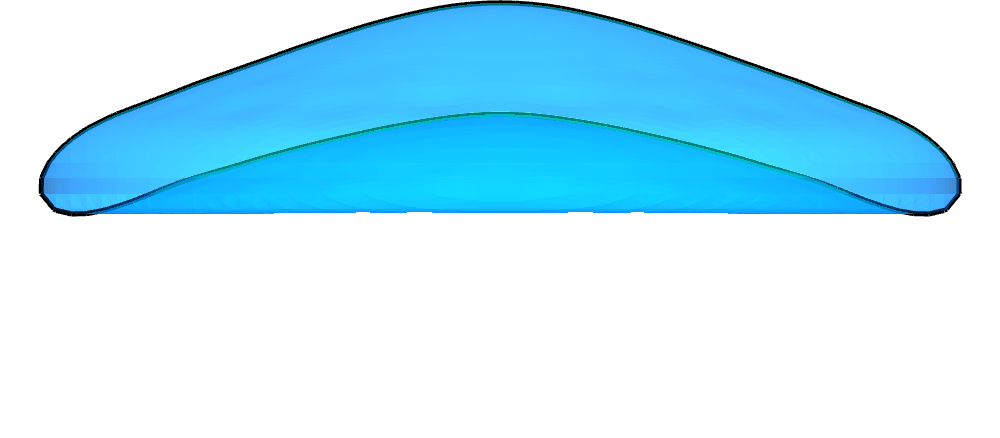} \\ (g) $\mathrm{Re}=65.6$} &   		\makecell{\includegraphics[width=0.2\textwidth]{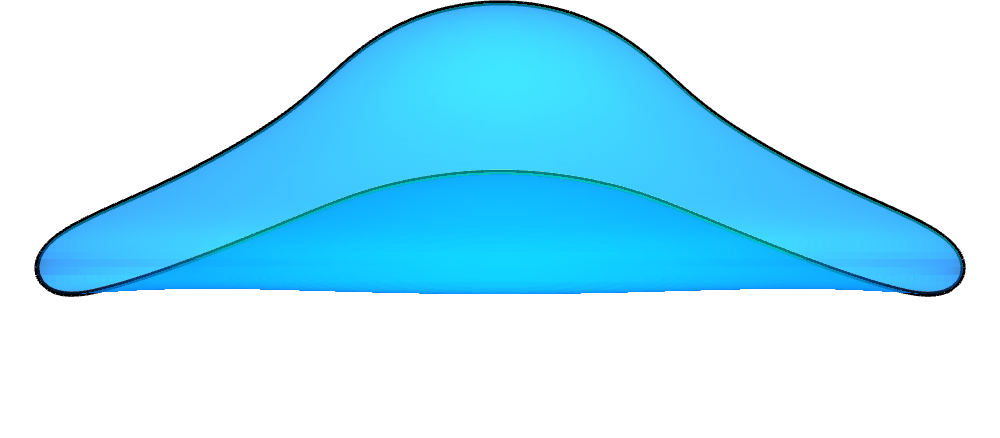} \\ (h) $\mathrm{Re}=77.7$} \\
		& $D=8$ & $D=16$ & $D=32$ & $D=64$\\
	\end{tabular}
	\caption{
		Simulated bubble shape and Reynolds number, Re, at time, $t^{*}=10$, for case 2 in \Cref{tab:rising-bubble-setups} with $\mathrm{Bo}=115$ and $\mathrm{Mo}=4.63\cdot10^{-3}$.
		Different computational resolutions according to the initial bubble diameter, $D$, are shown.
		The solid black lines illustrate the bubble's contour in the center cross-section with normal in the $x$-direction.
		The photograph of the laboratory experiment was reprinted from Reference~\cite{bhaga1981BubblesViscousLiquids} with the permission of Cambridge University Press.
	}
	\label{fig:bo-115-mo-4.63e-3}
\end{figure}

\begin{figure}[htbp]
	\centering
	\begin{subfigure}[b]{0.2\textwidth}
		\centering
		\vskip 0pt 
		\includegraphics[width=1\textwidth,trim={200 0 200 0}, clip]{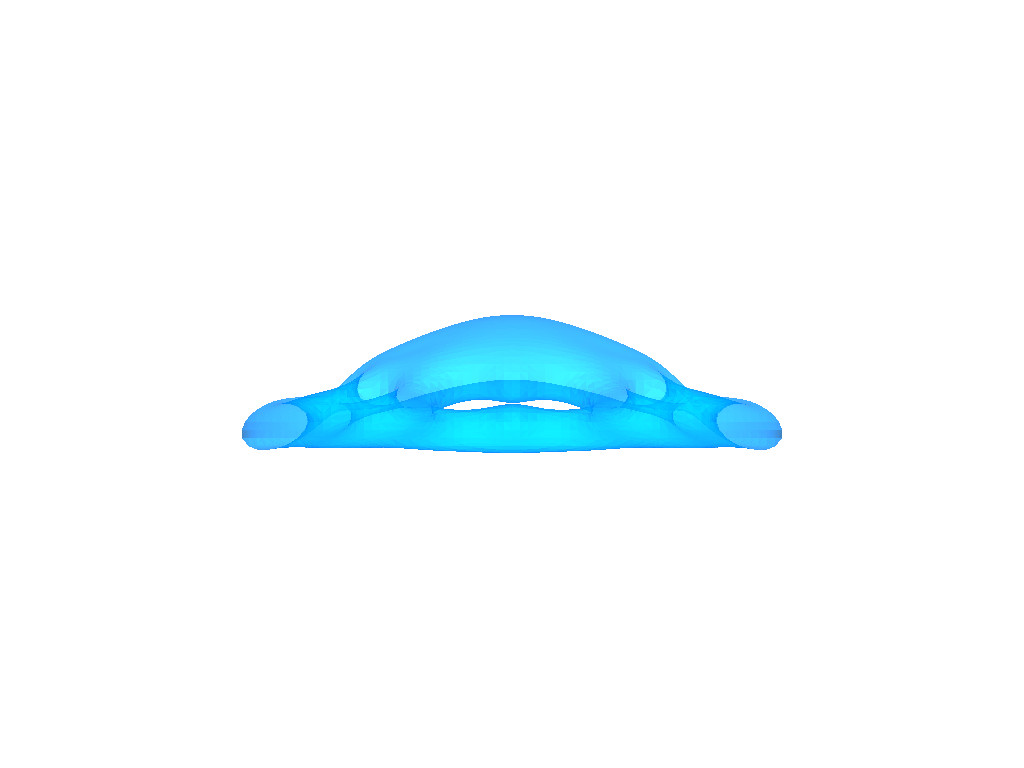}
		\caption{$M=0.02$}
		\label{fig:pflbm-rising-bubble-unstable}
	\end{subfigure}
	\hfill
	\begin{subfigure}[b]{0.2\textwidth}
		\centering
		\vskip 0pt 
		\includegraphics[width=1\textwidth,trim={200 0 200 0}, clip]{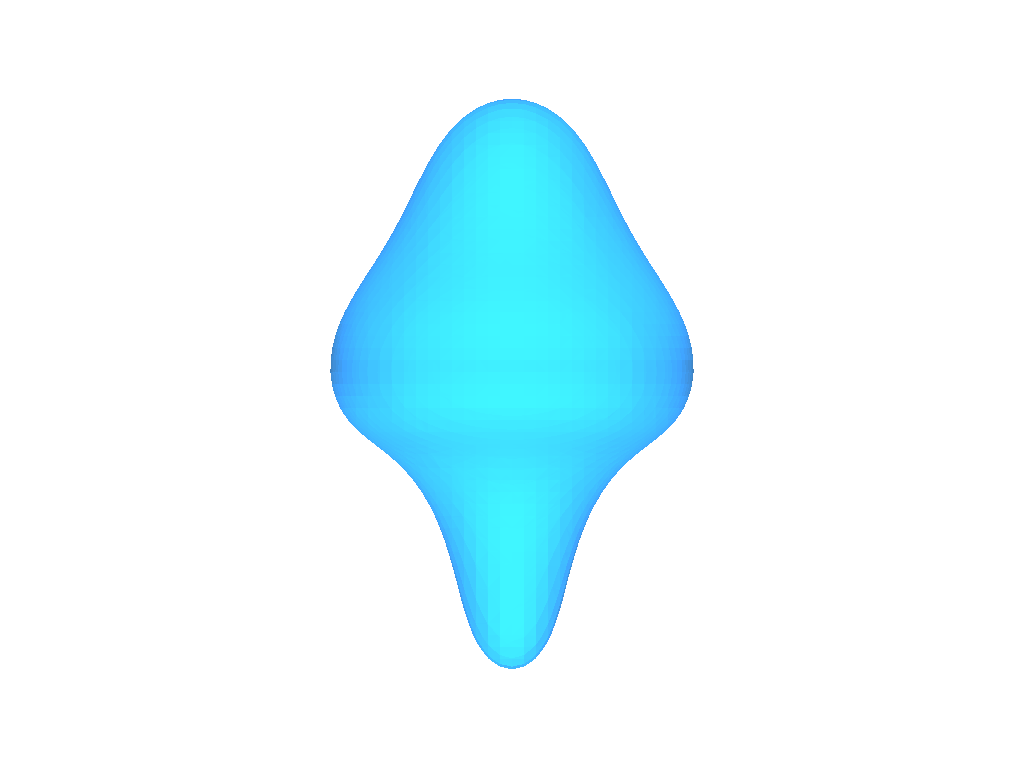}
		\caption{$M=0.05$}
	\end{subfigure}
	\hfill
	\begin{subfigure}[b]{0.2\textwidth}
		\centering
		\vskip 0pt 
		\includegraphics[width=1\textwidth,trim={200 0 200 0}, clip]{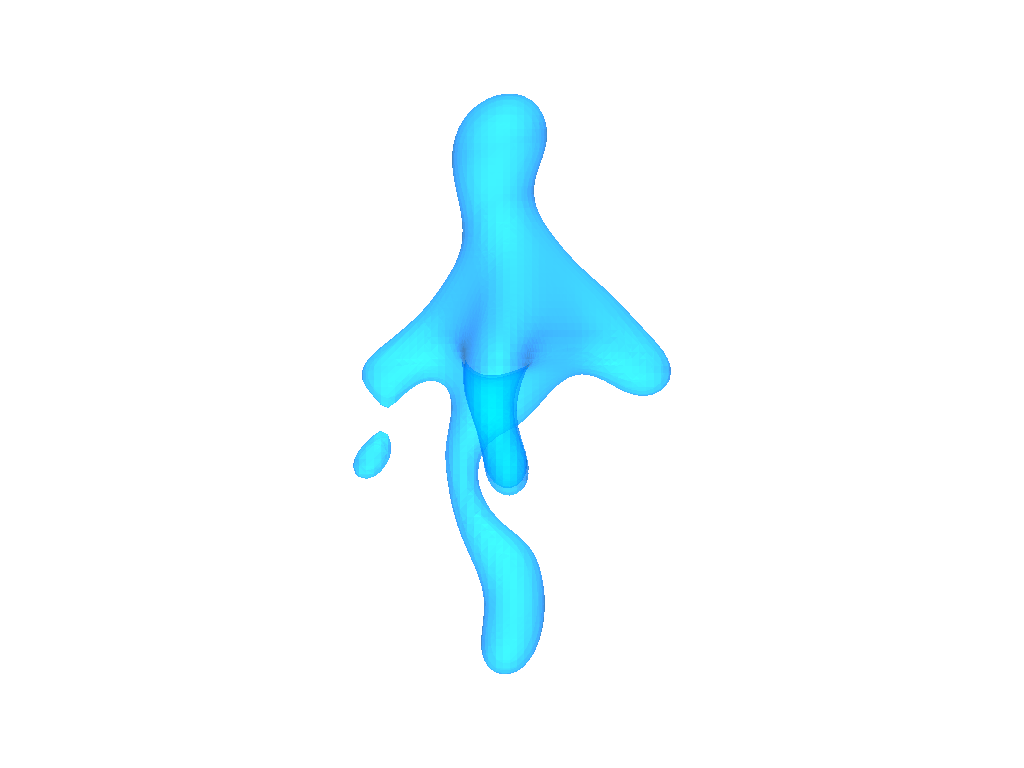}
		\caption{$M=0.1$}
	\end{subfigure}
	\hfill
	\begin{subfigure}[b]{0.2\textwidth}
		\centering
		\vskip 0pt 
		\includegraphics[width=1\textwidth,trim={200 0 200 0}, clip]{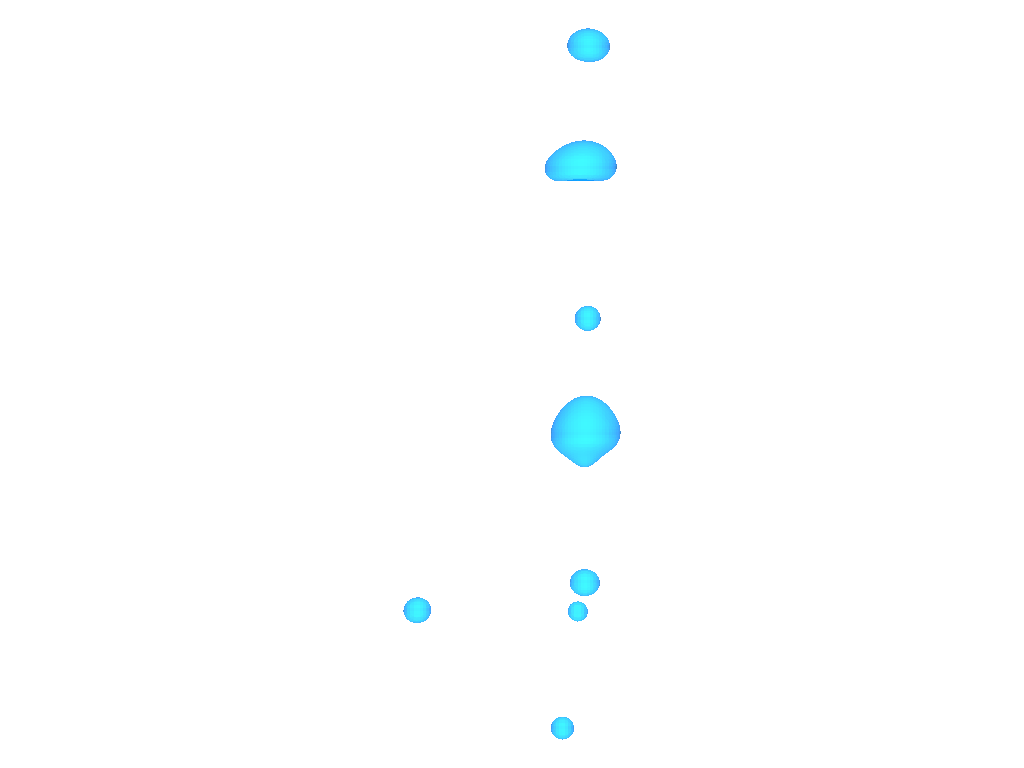}
		\caption{$M=0.15$}
	\end{subfigure}
	\caption{
		Shape of the rising bubble as simulated with the PFLBM at time, $t^{*}=12$, for case 2 in \Cref{tab:rising-bubble-setups} with $\mathrm{Bo}=115$ and $\mathrm{Mo}=4.63\cdot10^{-3}$.
		The computational resolution was set according to $D=32$.
		Larger values for the mobility, $M$, lead to non-physical bubble shapes, i.e., break-up.
	}
	\label{fig:rising-bubble-pflbm-mobility}
\end{figure}

In contrast, for the PFLBM, it was not possible to obtain results for resolutions of $D<32$.
Furthermore, for case 2, neither the bubble shape nor the Reynolds number was predicted reasonably well, regardless of the resolution, as illustrated in \Cref{fig:bo-115-mo-4.63e-3}.
Moreover, when increasing the simulation run time, non-physical bubble shapes and collapse were also observed with the PFLBM. \Cref{fig:pflbm-rising-bubble-unstable} shows this behavior for case 2 with $D=32$ in which the skirted bubble film ruptures at $t^{*}>10$.
With increased computational resolution, this effect occurred at later $t^{*}$.\par

It was shown in the literature, that phase-field models are sensitive to the choice of the mobility parameter, $M$~\cite{REN2016100}.
However, in general there appears to be no robust solution for how this parameter should be specified for arbitrary cases. 
In a study performed here, as depicted in \Cref{fig:rising-bubble-pflbm-mobility}, it was observed that larger values of $M$ seem to boost such non-physical effects. 
On the other hand, with $M<0.02$, instabilities were observed, as the relaxation time, $\tau_{\phi}$, in \Cref{eq:pflbm-tau} decreases and approaches its lower stability limits.
These instabilities occurred even when using the weighted MRT scheme, which is generally known for good stability properties~\cite{Mitchell2021}.
A rigorous study of this behavior is outside the scope of this work, but is proposed for future investigation.\par

The test cases used in this study have also been simulated in two dimensions by Kumar et al.~\cite{Kumar2019}, using the PFLBM.
To check the implementations' validity, these two-dimensional simulations were also performed here, agreeing with Reference~\cite{Kumar2019} and without becoming unstable or leading to implausible bubble shapes.\par

While the reason for these instabilities is not yet clear, it must be pointed out that the expected bubble shapes consist of only a thin film of gas.
In the literature, similar circular destabilization of films with the PFLBM could be observed in other test cases, however, often of a thin liquid rather than gas film~\cite{FAKHARI201722}.\par

\subsubsection{Taylor bubble} \label{subsubsec:taylor-bubble}
The fourth test case is a buoyancy driven confined flow, a large gas bubble rising through stagnant liquid in a cylindrical tube.
During the bubble's rise, it takes an elongated shape with a rounded leading edge.
Its length is several times the tube's diameter and it is commonly referred to as Taylor bubble~\cite{davies_taylor_1950,bretherton_1961}.

\begin{figure}[htbp]
	\centering
	\setlength{\figureheight}{0.5\textwidth}
	\setlength{\figurewidth}{0.15\textwidth}
	\setlength\mdist{0.02\textwidth}
	\begin{tikzpicture}
\definecolor{steelblue31119180}{RGB}{31,119,180}
\definecolor{sienna1809231}{RGB}{180,92,31}
\setlength\radius{0.375\figurewidth}

\draw [fill=steelblue31119180!30, draw=none] (0,0) rectangle (\figurewidth,\figureheight);

\draw[very thick, black!50] (0,0)--(0,\figureheight);
\draw[very thick, black!50] (\figurewidth,0)--(\figurewidth,\figureheight);

\draw[very thick, black!50] (0,\figureheight)--(\figurewidth,\figureheight);
\draw[very thick, black!50] (0,0)--(\figurewidth,0);

\draw[<->, >=Latex] (\figurewidth+\mdist,0)--(\figurewidth+\mdist,\figureheight) node [pos=0.5,right] {$10D$};
\draw[-] (\figurewidth,0)--(\figurewidth+2\mdist,0);
\draw[-] (\figurewidth,\figureheight)--(\figurewidth+2\mdist,\figureheight);

\draw[<->, >=Latex] (0,-\mdist)--(\figurewidth,-\mdist) node [pos=0.5,below] {$\varnothing D$};
\draw[-] (0,0)--(0,-2\mdist);
\draw[-] (\figurewidth,0)--(\figurewidth,-2\mdist);

\draw [fill=white, draw=steelblue31119180, thick] (0.5\figurewidth-\radius,0.1\figureheight) rectangle (0.5\figurewidth+\radius,0.4\figureheight);

\draw[<->, >=Latex] (0.5\figurewidth-\radius,0.4\figureheight+\mdist)--(0.5\figurewidth+\radius,0.4\figureheight+\mdist) node [pos=0.5,above] {$\varnothing 0.75 D$};
\draw[-] (0.5\figurewidth-\radius,0.4\figureheight)--(0.5\figurewidth-\radius,0.4\figureheight+2\mdist);
\draw[-] (0.5\figurewidth+\radius,0.4\figureheight)--(0.5\figurewidth+\radius,0.4\figureheight+2\mdist);

\draw[<->, >=Latex] (0.5\figurewidth,0.1\figureheight-\mdist)--(\figurewidth,0.1\figureheight-\mdist) node [pos=0.5,below] {$0.5 D$};
\draw[-] (0.5\figurewidth,0.1\figureheight)--(0.5\figurewidth,0.1\figureheight-2\mdist);
\draw[-] (\figurewidth,0.1\figureheight)--(\figurewidth,0.1\figureheight-2\mdist);

\draw[<->, >=Latex] (-\mdist,0)--(-\mdist,0.1\figureheight) node [pos=0.5,left] {$D$};
\draw[-] (-2\mdist,0.1\figureheight)--(0.5\figurewidth-\radius,0.1\figureheight);
\draw[-] (-2\mdist,0)--(0.5\figurewidth-\radius,0);

\draw[<->, >=Latex] (-\mdist,0.1\figureheight)--(-\mdist,0.4\figureheight) node [pos=0.5,left] {$3D$};
\draw[-] (-2\mdist,0.4\figureheight)--(0.5\figurewidth-\radius,0.4\figureheight);

\draw[loosely dashdotted] (0.5\figurewidth,-0.025\figureheight)--(0.5\figurewidth,1.025\figureheight);

\draw[thick, ->, >=Latex, sienna1809231] (\figurewidth-1.5\mdist,\figureheight-\mdist)--(\figurewidth-1.5\mdist,\figureheight-4\mdist) node [pos=0.5,right] {$g$};

\draw[->, >=Latex] (2\mdist,\figureheight-3\mdist)--(4\mdist,\figureheight-3\mdist) node [pos=0.5,below right] {$x$};
\draw[->, >=Latex] (2\mdist,\figureheight-3\mdist)--(2\mdist,\figureheight-1\mdist) node [pos=0.5,above left] {$z$};
\draw[draw=black] (2\mdist,\figureheight-3\mdist) circle [radius=0.175\mdist] node[opacity=1, below left] {$y$};
\end{tikzpicture}%
	\caption{
		Simulation setup of the three-dimensional Taylor bubble test case, with an initially cylindrical gas bubble in a cylindrical tube of diameter, $D$, and gravitational acceleration, $g$.
		No-slip boundary conditions are applied at the tube's walls and at the domain's top and bottom walls in $z$-direction.
	}
	\label{fig:taylor-bubble-setup}
\end{figure}
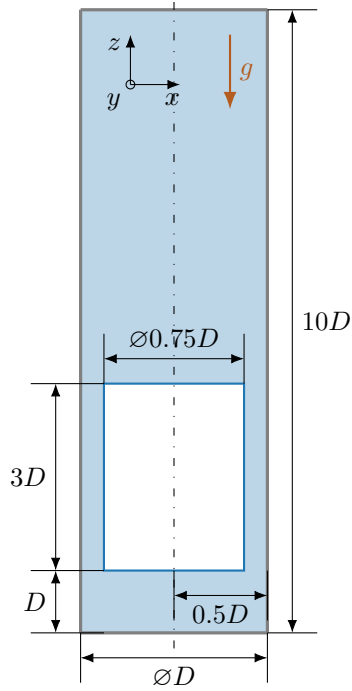

\paragraph{Setup} \label{par:taylor-bubble-setup}
The simulation setup chosen here is similar to the one in Reference~\cite{Mitchell2018}, conforming to the experiments in Reference~\cite{bugg2002VelocityFieldTaylor}.
As illustrated in~\Cref{fig:taylor-bubble-setup}, in a computational domain of size $1D \times 1D \times 10D$ ($x$-, $y$-, $z$-direction), the domain walls formed a cylindrical tube of diameter, $D$, pointing in $z$-direction.
A gas bubble was initialized as cylinder with diameter, $0.75D$, and length, $3D$, oriented concentrically to the boundary tube.
The gas bubble's bottom was located at $D$ in positive $z$-direction.
The rest of the domain was filled with a stagnant liquid.
According to the gravitational acceleration, $g$, the liquid was initialized with hydrostatic pressure such that the reference pressure, $p_{0}=\rho_{0} c_{s}^{2} = 1/3$, was set at $5D$ in $z$-direction.
As in \Cref{par:rising-bubble-setup}, the Morton number, Mo, Bond number, Bo, and the reference time, $t^{*}$, characterize the system.
Here, the tube diameter, $D$, was used as characteristic length~\cite{bugg2002VelocityFieldTaylor} in these non-dimensional numbers.\par

The experiments in Reference~\cite{bugg2002VelocityFieldTaylor} were conducted with  Bo$=100$, Mo=$0.015$, and olive oil. 
Following Reference~\cite{Mitchell2018}, it is assumed that the density and viscosity of the air injected into the oil was $\rho^{\text{SI}}=1.225$\,kg/m\textsuperscript{3} and $\mu^{\text{SI}}=1.983\cdot10^{5}$\,kg/(m$\cdot$s), respectively.
Therefore, the density ratio, $\tilde{\rho}=744$, and the dynamic viscosity ratio, $\tilde{\mu}=4236$, were used.
As in \Cref{subsubsec:rising-bubble}, for the FSLBM, the initial pressure of the bubble was set to the reference pressure, $p_{0}=\rho_{0} c_{s}^{2} = 1/3$.
The simulations were performed with computational resolutions according to the tube diameter, $D \in \{16, 32, 64, 128\}$.
However, in the PFLBM, simulating a tube diameter of $D\leq32$ was not possible, as the diffuse interface led to non-physical wall interactions with the interfacial region.
Based on the investigations from \Cref{subsubsec:rising-bubble}, the mobility parameter was set to the lowest value at which the simulations at any tested resolution were stable, namely $M=0.08$. 
The interface width was chosen as $\xi = 3$.
For all simulations, the relaxation rate was set to $\omega=1.8$ in the FSLBM, and $\omega_{\text{H}}=1.76$ in the heavier phase of the PFLBM's hydrodynamic LBM step.\par

\begin{figure}[htbp]
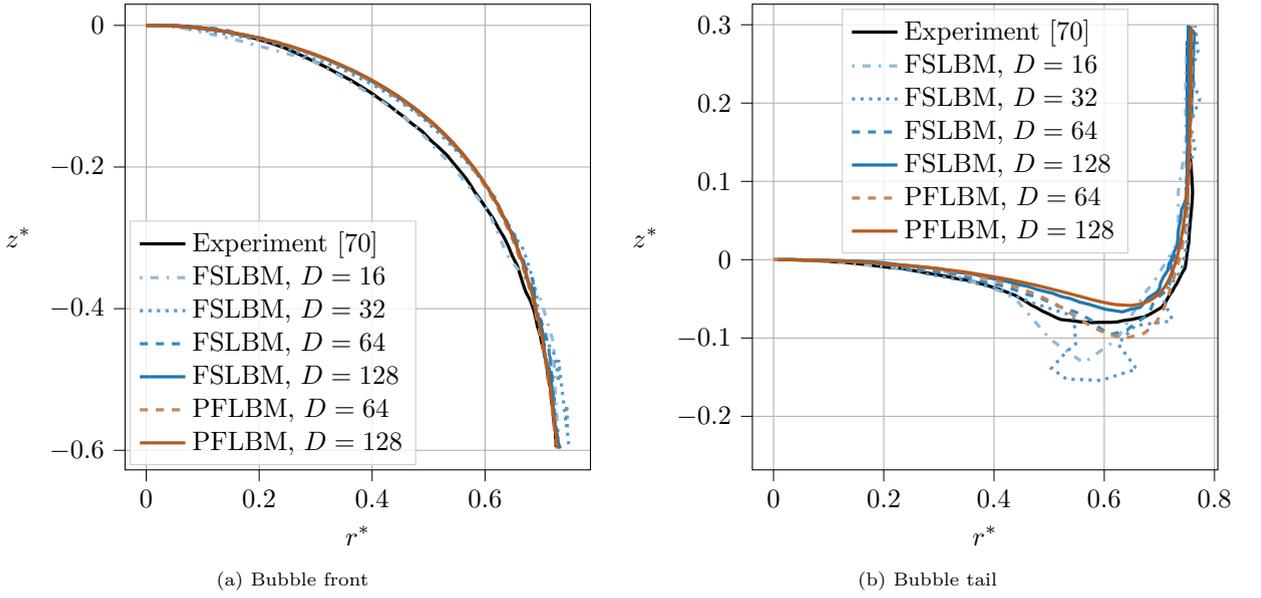

	\centering
	\setlength{\figureheight}{0.47\textwidth}
	\setlength{\figurewidth}{0.47\textwidth}
	\begin{subfigure}[t]{0.49\textwidth}
		\centering
		\input{figures/taylor-bubble/shape-front.tex}%
		\caption{Bubble front}
		\label{fig:taylor-bubble-shape-front}
	\end{subfigure}
	\hfill
	\begin{subfigure}[t]{0.49\textwidth}
		\centering
		\input{figures/taylor-bubble/shape-tail.tex}%
		\caption{Bubble tail}
		\label{fig:taylor-bubble-shape-tail}
	\end{subfigure}
	\caption{
		Shape of the front and tail of the simulated Taylor bubble at different computational resolutions, specified by the tube diameter, $D$.
		The comparison with experimental data~\cite{bugg2002VelocityFieldTaylor} is drawn in terms of the non-dimensionalized axial location, $z^{*}=z/D$, and radial location, $r^{*}=r/(0.5D)$ at time, $t^{*}=15$.
	}
	\label{fig:taylor-bubble-shape}
\end{figure}

\paragraph{Results and discussion} \label{par:taylor-bubble-results}
\Cref{fig:taylor-bubble-shape} compares the simulated Taylor bubble's shape at different computational resolutions at time $t^{*}=15$ with the experimental measurement~\cite{bugg2002VelocityFieldTaylor}.
To ease comparison, the axial location, $z^{*}=z/D$ and radial location, $r^{*}=r/(0.5D)$ are non-dimensionalized.
Additionally, an axial shift is employed as to set $z^{*}=0$ at $r^{*}=0$ for the bubble's front and tail individually.
Both models converged well, but showed minor deviations to the experimental data from Reference~\cite{bugg2002VelocityFieldTaylor}.
The shape of the front of the bubble was predicted with reasonable accuracy at all computational resolutions tested.
However, at the tail of the bubble, a resolution of $D\geq64$ was required for the FSLBM to capture the interface contour moderately well.
As also observed in \Cref{subsubsec:rising-bubble}, satellite bubbles separated from the main bubble in the case of the FSLBM, as shown in the Appendix in \Cref{fig:taylor-bubble-shape-mesh}.
In contrast to the observations for the rising bubble test, this effect vanished with increasing computational resolution.\par

In \Cref{tab:taylor-bubble-re}, the simulated Reynolds number, Re, as defined in \Cref{eq:Re}, is shown.
The tube diameter, $D$, and the Taylor bubble's rise velocity, $U$, are used as characteristic quantities to determine Re.
The rise velocity, $U$, was computed by the bubble's center of mass location in $z$-direction at time, $t^{*}=10$, and $t^{*}=15$.
In comparison to the PFLBM, which agreed well with the experimental measurement~\cite{bugg2002VelocityFieldTaylor}, the FSLBM showed larger deviations.
This was even more pronounced at lower computational resolutions, where it could capture the bubble's axial movement only moderately well.\par

\begin{table}[htbp]
	\centering
	\begin{tabular}{>{\raggedright}m{0.2\textwidth}
			>{\centering\arraybackslash}m{0.1\textwidth}
			>{\centering\arraybackslash}m{0.1\textwidth}
			>{\centering\arraybackslash}m{0.1\textwidth}
			>{\centering\arraybackslash}m{0.1\textwidth}}
		
		\toprule
		$D$ & $16$ & $32$ & $64$ & $128$ \\
		\midrule
		
		Re\textsubscript{FSLBM} & $22.15$ & $24.12$ & $25.35$ & $25.89$ \\
		
		Re\textsubscript{PFLBM} & unstable & unstable & $26.83$ & $27.12$ \\
		
		\midrule
		Re\textsubscript{Experiment}~\cite{bugg2002VelocityFieldTaylor} & \multicolumn{4}{c}{$27$} \\
		\bottomrule
	\end{tabular}
	\caption{
		Reynolds number, Re, of the simulated Taylor bubble for different computational resolutions as specified by the tube diameter, $D$.
		The bubble's rise velocity, as used in Re, was computed from the Taylor bubble's locations in axial direction at time $t^{*}=10$ and $t^{*}=15$.
	}
	\label{tab:taylor-bubble-re}
\end{table}

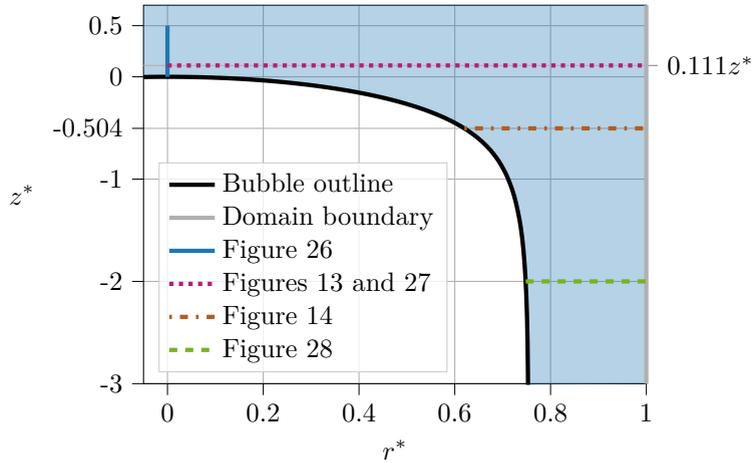
\begin{figure}[htbp]
	\centering
	\setlength{\figureheight}{0.4\textwidth}
	\setlength{\figurewidth}{0.5\textwidth}
\begin{tikzpicture}

\definecolor{darkgray176}{RGB}{176,176,176}
\definecolor{lightgray204}{RGB}{204,204,204}
\definecolor{mediumvioletred18031119}{RGB}{180,31,119}
\definecolor{olivedrab11918031}{RGB}{119,180,31}
\definecolor{sienna1809231}{RGB}{180,92,31}
\definecolor{steelblue31119180}{RGB}{31,119,180}

\begin{axis}[
height=\figureheight,
width=\figurewidth,
axis y line*=right,
axis x line=none,
ylabel style={rotate=-90.0},
xmin=0, xmax=0,
ymin=-3, ymax=0.7,
ytick={0.111},
yticklabels = {0.111$z^{*}$},
tick align=outside,
tick pos=right,
y grid style={darkgray176},
ymajorgrids,
]
\end{axis}

\begin{axis}[
height=\figureheight,
legend cell align={left},
legend style={
  fill opacity=0.8,
  draw opacity=1,
  text opacity=1,
  at={(0.03,0.03)},
  anchor=south west,
  draw=lightgray204
},
tick align=outside,
tick pos=left,
width=\figurewidth,
x grid style={darkgray176},
xlabel={\(\displaystyle r^{*}\)},
xmajorgrids,
xmin=-0.05, xmax=1,
xtick style={color=black},
y grid style={darkgray176},
ylabel style={rotate=-90.0},
ylabel={\(\displaystyle z^{*}\)},
ymajorgrids,
ymin=-3, ymax=0.7,
ytick style={color=black},
ytick={-3, -2, -1, -0.504, 0, 0.5},
yticklabels = {-3,-2,-1,-0.504, 0, 0.5},
yticklabel style={
	/pgf/number format/fixed,
	/pgf/number format/precision=3},
clip=false,
]
\addplot [ultra thick, black, name path=bubble]
table {%
-0.05 -0.00331425666809082
-0.0468618869781494 -0.0018768310546875
-0.0312368869781494 -0.000843405723571777	
-0.0156118869781494 -0.000216245651245117
-1.31130218505859e-05 0
0.0156118869781494 -0.000216245651245117
0.0312368869781494 -0.000843405723571777
0.0468618869781494 -0.0018768310546875
0.0624868869781494 -0.00331425666809082
0.0781118869781494 -0.00516211986541748
0.0937368869781494 -0.00744259357452393
0.11590051651001 -0.0115705728530884
0.124986886978149 -0.0134730339050293
0.140611886978149 -0.0170856714248657
0.156236886978149 -0.0210834741592407
0.176718354225159 -0.0271955728530884
0.203111886978149 -0.0361255407333374
0.2204270362854 -0.0428205728530884
0.234361886978149 -0.0486605167388916
0.255831003189087 -0.0584455728530884
0.265611886978149 -0.0632587671279907
0.286195278167725 -0.0740705728530884
0.296861886978149 -0.0800961256027222
0.312974691390991 -0.0896955728530884
0.336896896362305 -0.105320572853088
0.343736886978149 -0.110080242156982
0.359361886978149 -0.121415257453918
0.378740310668945 -0.136570572853088
0.397185802459717 -0.152195572853088
0.414325714111328 -0.167820572853088
0.430325031280518 -0.183445572853088
0.445314407348633 -0.199070572853088
0.459401726722717 -0.214695572853088
0.472667455673218 -0.230320572853088
0.485145807266235 -0.245945572853088
0.499986886978149 -0.265821099281311
0.515611886978149 -0.288233757019043
0.518711566925049 -0.292820572853088
0.531236886978149 -0.312572002410889
0.538240194320679 -0.324070572853088
0.547301888465881 -0.339695572853088
0.564130306243896 -0.370945572853088
0.579358339309692 -0.402195572853088
0.593736886978149 -0.434839367866516
0.599634408950806 -0.449070572853088
0.611665010452271 -0.480320572853088
0.62250554561615 -0.511570572853088
0.627724647521973 -0.527195572853088
0.637151956558228 -0.558445572853088
0.645975351333618 -0.589695572853088
0.657675385475159 -0.636570572853088
0.664606094360352 -0.667820572853088
0.674072861671448 -0.714695572853088
0.679645299911499 -0.745945572853088
0.687486886978149 -0.794471502304077
0.691862821578979 -0.824070572853088
0.69793701171875 -0.870945572853088
0.705211281776428 -0.933445572853088
0.709868550300598 -0.980320572853088
0.715261220932007 -1.04282057285309
0.720094203948975 -1.10532057285309
0.724194288253784 -1.16782057285309
0.727686405181885 -1.23032057285309
0.731468200683594 -1.30844557285309
0.735384106636047 -1.40219557285309
0.738070726394653 -1.48032057285309
0.740710735321045 -1.57407057285309
0.743194937705994 -1.68344557285309
0.745447278022766 -1.80844557285309
0.747605800628662 -1.96469557285309
0.749356508255005 -2.1365704536438
0.750810861587524 -2.3396954536438
0.75194787979126 -2.5896954536438
0.752784967422485 -2.9178204536438
0.752936363220215 -3.0115704536438
};
\addlegendentry{Bubble outline}
\addplot [ultra thick, darkgray176]
table {%
	1 -3
	1 0.7
};
\addlegendentry{Domain boundary}
\addplot [ultra thick, steelblue31119180]
table {%
	0 0
	0 0.5
};
\addlegendentry{\Cref{fig:taylor-bubble-axial}}
\addplot [ultra thick, mediumvioletred18031119, dotted]
table {%
0 0.111000061035156
1 0.111000061035156
};
\addlegendentry{\Cref{fig:taylor-bubble-axial-0.111,fig:taylor-bubble-radial-0.111}}
\addplot [ultra thick, sienna1809231, dash pattern=on 1pt off 3pt on 3pt off 3pt]
table {%
0.620000004768372 -0.503999948501587
1 -0.503999948501587
};
\addlegendentry{\Cref{fig:taylor-bubble-0.504}}
\addplot [ultra thick, olivedrab11918031, dashed]
table {%
0.748000025749207 -2
1 -2
};
\addlegendentry{\Cref{fig:taylor-bubble-axial-2}}

\path[name path=xaxis] (-0.05,0.7) -- (1,0.7) -- (1,-3);
\addplot [fill=steelblue31119180, fill opacity=0.33] fill between[of=bubble and xaxis];
\end{axis}
\end{tikzpicture}%
	\caption{
		Definition of the locations at the Taylor bubble's front, where the velocity profiles are evaluated in the subsequent figures.
		The monitored lines are expressed in terms of the non-dimensionalized axial location, $z^{*}=z/D$, and radial location, $r^{*}=r/(0.5D)$.
		The test case is radially symmetric such that the evaluation can be performed at an arbitrary cross-section.
	}
	\label{fig:taylor-bubble-schematic}
\end{figure}

\begin{figure}[htbp]
	\centering
	\setlength{\figureheight}{0.6\textwidth}
	\setlength{\figurewidth}{0.475\textwidth}
	\begin{subfigure}[t]{0.49\textwidth}
		\centering
\begin{tikzpicture}

\definecolor{darkgray176}{RGB}{176,176,176}
\definecolor{lightgray204}{RGB}{204,204,204}
\definecolor{peru18911259}{RGB}{189,112,59}
\definecolor{sienna1809231}{RGB}{180,92,31}
\definecolor{skyblue143187218}{RGB}{143,187,218}
\definecolor{steelblue31119180}{RGB}{31,119,180}
\definecolor{steelblue59136189}{RGB}{59,136,189}
\definecolor{steelblue87153199}{RGB}{87,153,199}

\begin{axis}[
height=\figureheight,
legend cell align={left},
legend style={
  fill opacity=0.8,
  draw opacity=1,
  text opacity=1,
  at={(0.03,0.97)},
  anchor=north west,
  draw=lightgray204
},
tick align=outside,
tick pos=left,
width=\figurewidth,
x grid style={darkgray176},
xlabel={\(\displaystyle r^{*}\)},
xmajorgrids,
xmin=0.671875, xmax=1.015625,
xtick style={color=black},
y grid style={darkgray176},
ylabel style={rotate=-90.0},
ylabel={\(\displaystyle U_{a}^{*}\)},
ymajorgrids,
ymin=-1.98010001478583, ymax=-0.0113356096839041,
ytick style={color=black}
]
\addplot [semithick, black, mark=*, mark size=3, mark options={solid}, only marks]
table {%
0.704999923706055 -1.52999997138977
0.720999956130981 -1.54999995231628
0.73799991607666 -1.61000001430511
0.754999995231628 -1.52999997138977
0.771000027656555 -1.5900000333786
0.787999987602234 -1.53999996185303
0.805000066757202 -1.42999994754791
0.820999979972839 -1.4099999666214
0.838000059127808 -1.29999995231628
0.853999972343445 -1.21000003814697
0.871000051498413 -1.08000004291534
0.888000011444092 -0.993000030517578
0.904000043869019 -0.777999997138977
0.921000003814697 -0.670000076293945
0.937999963760376 -0.593999981880188
0.953999996185303 -0.347000002861023
0.970999956130981 -0.185999989509583
0.98799991607666 -0.296000003814697
};
\addlegendentry{Experiment~\cite{bugg2002VelocityFieldTaylor}}
\addplot [semithick, skyblue143187218, mark=triangle, mark size=3, mark options={solid,rotate=180,fill opacity=0}, only marks]
table {%
0.8125 -1.70201539993286
1 -0.758584976196289
};
\addlegendentry{FSLBM, $D=16$}
\addplot [semithick, steelblue87153199, mark=triangle, mark size=3, mark options={solid,fill opacity=0}, only marks]
table {%
0.71875 -1.70847928524017
0.7890625 -1.64600050449371
0.859375 -1.44073224067688
0.9296875 -1.0386780500412
1 -0.397133827209473
};
\addlegendentry{FSLBM, $D=32$}
\addplot [semithick, steelblue59136189, mark=triangle, mark size=3, mark options={solid,rotate=270,fill opacity=0}, only marks]
table {%
0.734375 -1.5602867603302
0.767578125 -1.55284297466278
0.80078125 -1.50315022468567
0.833984375 -1.40884971618652
0.8671875 -1.27261090278625
0.900390625 -1.08971810340881
0.93359375 -0.853496551513672
0.966796875 -0.558734178543091
1 -0.198622703552246
};
\addlegendentry{FSLBM, $D=64$}
\addplot [semithick, steelblue31119180, mark=triangle, mark size=3, mark options={solid,rotate=90,fill opacity=0}, only marks]
table {%
0.7109375 -1.53462839126587
0.726996541023254 -1.54563581943512
0.743055582046509 -1.54960572719574
0.759114742279053 -1.54152166843414
0.775173664093018 -1.52435517311096
0.791232585906982 -1.4988135099411
0.807291507720947 -1.46304142475128
0.823350667953491 -1.41723906993866
0.839409708976746 -1.36088848114014
0.85546875 -1.29299116134644
0.871528148651123 -1.21380603313446
0.887587547302246 -1.12266314029694
0.903645277023315 -1.01910316944122
0.919704675674438 -0.902506828308105
0.935764074325562 -0.772350668907166
0.951823472976685 -0.627924680709839
0.967881202697754 -0.46860682964325
0.983940601348877 -0.293307781219482
1 -0.100941777229309
};
\addlegendentry{FSLBM, $D=128$}
\addplot [semithick, peru18911259, mark=square, mark size=3, mark options={solid,fill opacity=0}, only marks]
table {%
0.703125 -1.70039939880371
0.736111164093018 -1.67887675762177
0.769097208976746 -1.64160799980164
0.802083492279053 -1.57835602760315
0.835069417953491 -1.48082387447357
0.868055582046509 -1.34166276454926
0.901041507720947 -1.15543067455292
0.934027791023254 -0.917268872261047
0.967013835906982 -0.62147319316864
1 -0.232735514640808
};
\addlegendentry{PFLBM, $D=64$}
\addplot [semithick, sienna1809231, mark=diamond, mark size=3, mark options={solid,fill opacity=0}, only marks]
table {%
0.7109375 -1.63943660259247
0.726996541023254 -1.63412284851074
0.743055582046509 -1.62766265869141
0.759114742279053 -1.61490142345428
0.775173664093018 -1.59478402137756
0.791232585906982 -1.56584393978119
0.807291507720947 -1.52728748321533
0.823350667953491 -1.47855257987976
0.839409708976746 -1.41916191577911
0.85546875 -1.34865319728851
0.871528148651123 -1.26657545566559
0.887587547302246 -1.17248177528381
0.903645277023315 -1.06593418121338
0.919704675674438 -0.946483612060547
0.935764074325562 -0.813661336898804
0.951823472976685 -0.666911840438843
0.967881202697754 -0.505552530288696
0.983940601348877 -0.328648805618286
1 -0.120051145553589
};
\addlegendentry{PFLBM, $D=128$}
\end{axis}

\end{tikzpicture}%
		\label{fig:taylor-bubble-axial-0.504}
	\end{subfigure}
	\hfill
	\begin{subfigure}[t]{0.49\textwidth}
		\centering
\begin{tikzpicture}

\definecolor{darkgray176}{RGB}{176,176,176}
\definecolor{lightgray204}{RGB}{204,204,204}
\definecolor{peru18911259}{RGB}{189,112,59}
\definecolor{sienna1809231}{RGB}{180,92,31}
\definecolor{skyblue143187218}{RGB}{143,187,218}
\definecolor{steelblue31119180}{RGB}{31,119,180}
\definecolor{steelblue59136189}{RGB}{59,136,189}
\definecolor{steelblue87153199}{RGB}{87,153,199}

\begin{axis}[
height=\figureheight,
legend cell align={left},
legend style={fill opacity=0.8, draw opacity=1, text opacity=1, draw=lightgray204},
tick align=outside,
tick pos=left,
width=\figurewidth,
x grid style={darkgray176},
xlabel={\(\displaystyle r^{*}\)},
xmajorgrids,
xmin=0.671875, xmax=1.015625,
xtick style={color=black},
y grid style={darkgray176},
ylabel style={rotate=-90.0},
ylabel={\(\displaystyle U_{r}^{*}\)},
ymajorgrids,
ymin=-0.0191229025709116, ymax=0.336900953989142,
ytick style={color=black},
yticklabel style={%
	/pgf/number format/.cd,
	fixed,
},
]
\addplot [semithick, black, mark=*, mark size=3, mark options={solid}, only marks]
table {%
0.704999923706055 0.184000015258789
0.720999956130981 0.177999973297119
0.73799991607666 0.162999987602234
0.754999995231628 0.148000001907349
0.771000027656555 0.138999938964844
0.787999987602234 0.136000037193298
0.805000066757202 0.101999998092651
0.820999979972839 0.0559999942779541
0.838000059127808 0.0486999750137329
0.853999972343445 0.0228999853134155
0.871000051498413 0.0448000431060791
0.888000011444092 0.0616999864578247
0.904000043869019 0.0169999599456787
0.921000003814697 0.0118000507354736
0.937999963760376 0.00337004661560059
0.953999996185303 0.00521004199981689
0.970999956130981 0.00581002235412598
0.98799991607666 -0.00294005870819092
};
\addlegendentry{Experiment~\cite{bugg2002VelocityFieldTaylor}}
\addplot [semithick, skyblue143187218, mark=triangle, mark size=3, mark options={solid,rotate=180,fill opacity=0}, only marks]
table {%
0.8125 0.170094609260559
1 0.0250577926635742
};
\addlegendentry{FSLBM, $D=16$}
\addplot [semithick, steelblue87153199, mark=triangle, mark size=3, mark options={solid,fill opacity=0}, only marks]
table {%
0.71875 0.265585660934448
0.7890625 0.176230430603027
0.859375 0.0954520702362061
0.9296875 0.0362814664840698
1 0.0052635669708252
};
\addlegendentry{FSLBM, $D=32$}
\addplot [semithick, steelblue59136189, mark=triangle, mark size=3, mark options={solid,rotate=270,fill opacity=0}, only marks]
table {%
0.734375 0.202325463294983
0.767578125 0.161665916442871
0.80078125 0.124759316444397
0.833984375 0.0924372673034668
0.8671875 0.0643014907836914
0.900390625 0.0401526689529419
0.93359375 0.0210850238800049
0.966796875 0.0078740119934082
1 0.00111377239227295
};
\addlegendentry{FSLBM, $D=64$}
\addplot [semithick, steelblue31119180, mark=triangle, mark size=3, mark options={solid,rotate=90,fill opacity=0}, only marks]
table {%
0.7109375 0.196348190307617
0.726996541023254 0.176600813865662
0.743055582046509 0.16409707069397
0.759114742279053 0.152887225151062
0.775173664093018 0.13964581489563
0.791232585906982 0.124281883239746
0.807291507720947 0.108263254165649
0.823350667953491 0.0929234027862549
0.839409708976746 0.0782536268234253
0.85546875 0.0648690462112427
0.871528148651123 0.0526858568191528
0.887587547302246 0.0414456129074097
0.903645277023315 0.0312329530715942
0.919704675674438 0.0223314762115479
0.935764074325562 0.0149909257888794
0.951823472976685 0.00912237167358398
0.967881202697754 0.00459420680999756
0.983940601348877 0.00159800052642822
1 0.000191688537597656
};
\addlegendentry{FSLBM, $D=128$}
\addplot [semithick, peru18911259, mark=square, mark size=3, mark options={solid,fill opacity=0}, only marks]
table {%
0.703125 0.266327500343323
0.736111164093018 0.217748761177063
0.769097208976746 0.16700553894043
0.802083492279053 0.131625890731812
0.835069417953491 0.0920437574386597
0.868055582046509 0.0670467615127563
0.901041507720947 0.0376332998275757
0.934027791023254 0.0234395265579224
0.967013835906982 0.00550651550292969
1 0.003517746925354
};
\addlegendentry{PFLBM, $D=64$}
\addplot [semithick, sienna1809231, mark=diamond, mark size=3, mark options={solid,fill opacity=0}, only marks]
table {%
0.7109375 0.241501450538635
0.726996541023254 0.216018557548523
0.743055582046509 0.194280862808228
0.759114742279053 0.17170786857605
0.775173664093018 0.15239417552948
0.791232585906982 0.132040619850159
0.807291507720947 0.114966869354248
0.823350667953491 0.0969133377075195
0.839409708976746 0.0822553634643555
0.85546875 0.0667173862457275
0.871528148651123 0.0546815395355225
0.887587547302246 0.0418459177017212
0.903645277023315 0.0325902700424194
0.919704675674438 0.0225880146026611
0.935764074325562 0.0162062644958496
0.951823472976685 0.00909674167633057
0.967881202697754 0.00561225414276123
0.983940601348877 0.00139486789703369
1 0.000830769538879395
};
\addlegendentry{PFLBM, $D=128$}
\end{axis}

\end{tikzpicture}%
		\label{fig:taylor-bubble-radial-0.504}
	\end{subfigure}
	\caption{
		Simulated non-dimensionalized axial velocity, $U^{*}_{a}$, and radial velocity, $U^{*}_{r}$, along a radial line positioned at $0.504D$ behind the Taylor bubble's front (see \Cref{fig:taylor-bubble-schematic}), with tube diameter $D$.
		The comparison with experimental data~\cite{bugg2002VelocityFieldTaylor} is drawn in terms of the non-dimensionalized radial location, $r^{*}=r/(0.5D)$, at time, $t^{*}=15$.
	}
	\label{fig:taylor-bubble-0.504}
\end{figure}
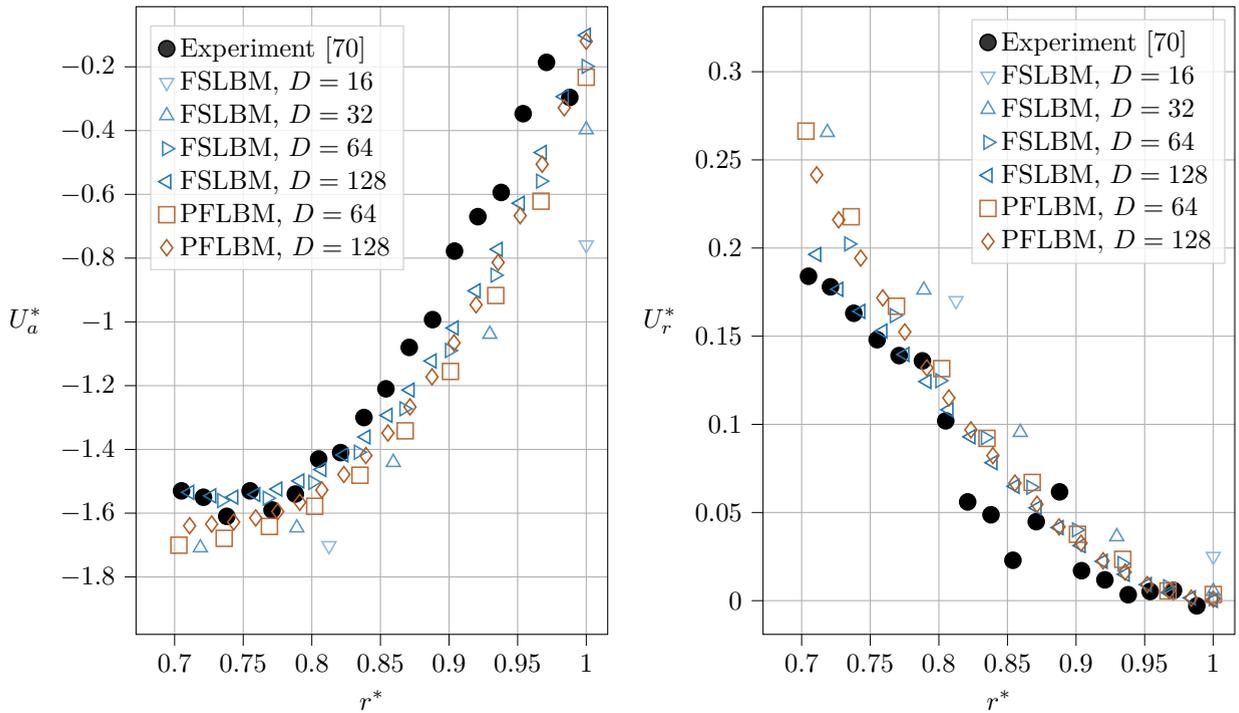

At the locations specified in \Cref{fig:taylor-bubble-schematic}, the flow field around the bubble was evaluated.
The non-dimensionalized axial fluid velocity, $U_a^{*}=U_{a}/U$, along a central axial line of length $0.5D$ in front of the bubble is presented in the Appendix in \Cref{fig:taylor-bubble-axial}.
Both models converged and agreed well with the experimental data~\cite{bugg2002VelocityFieldTaylor}.
On the other hand, at a radial line situated at $0.111D$ in front of the bubble, the non-dimensionalized radial fluid velocity $U_r^{*}=U_{r}/U$ (see \Cref{fig:taylor-bubble-radial-0.111}), and axial velocity (see Appendix, \Cref{fig:taylor-bubble-axial-0.111}) showed larger deviations when using the PFLBM, favoring the FSLBM at higher computational resolution.
A similar observation could be made at a radial line at $0.504D$ behind the bubble's front, as visualized in \Cref{fig:taylor-bubble-0.504}.
\Cref{fig:taylor-bubble-axial-2} illustrates that at a radial line located $2D$ behind the front of the bubble, the predicted axial velocity by both models agreed reasonably well with the experimental data.\par

\begin{figure}[htbp]
	\centering
	\setlength{\figureheight}{0.5\textwidth}
	\setlength{\figurewidth}{0.8\textwidth}
	\input{figures/taylor-bubble/radial-0.111.tex}%
	\caption{
		Simulated non-dimensionalized radial velocity, $U^{*}_{r}$, along a radial line positioned at $0.111D$ in front of the Taylor bubble (see \Cref{fig:taylor-bubble-schematic}), with tube diameter, $D$.
		The comparison with experimental data~\cite{bugg2002VelocityFieldTaylor} is drawn in terms of the non-dimensionalized radial location, $r^{*}=r/(0.5D)$, at time, $t^{*}=15$.
	}
	\label{fig:taylor-bubble-radial-0.111}
\end{figure}

\subsection{Dynamic coalescence -- crown splash} \label{subsec:dynamic-coalescence}
Understanding the dynamics of splashing during liquid drop impacts has many implications, including aerosol production~\cite{joung2015aerosol}, erosion processes~\cite{lu2016effects} and microplastic transfer in the environment~\cite{lehmann2021ejection}.
In this study, two drop impact test cases are simulated for which photographs of the laboratory experiments are available in the literature~\cite{wang2000SplashingImpactSingle}.
Both test cases have already been simulated using the FSLBM with LSQR curvature computation model~\cite{lehmann2021ejection}.
Due to the absence of quantitative experimental data, the comparisons with reference data can only be made qualitatively.

\subsubsection{Vertical drop impact} \label{subsubsec:vertical-drop}
In the fifth test case, a vertical drop impacting on a thin film of liquid was simulated and compared with experimental data~\cite{wang2000SplashingImpactSingle}.

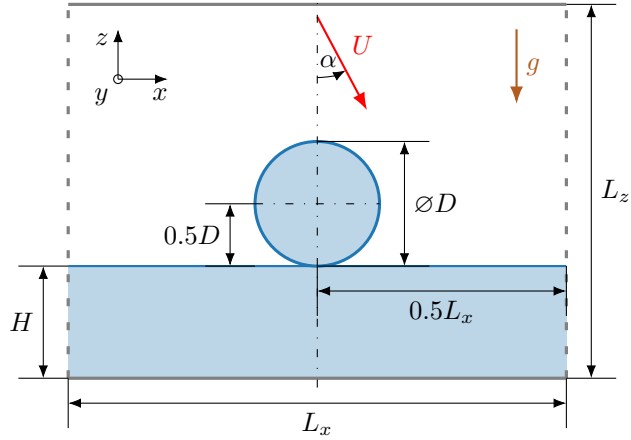
\begin{figure}[htbp]
	\centering
	\setlength{\figureheight}{0.3\textwidth}
	\setlength{\figurewidth}{0.4\textwidth}
	\setlength\mdist{0.02\textwidth}
	\begin{tikzpicture}
\definecolor{steelblue31119180}{RGB}{31,119,180}
\definecolor{sienna1809231}{RGB}{180,92,31}
\setlength\radius{0.125\figurewidth}

\coordinate (intersectPoint) at (0.5\figurewidth,0.3\figureheight+4\radius);
\coordinate (obliquePoint) at (0.6\figurewidth,0.3\figureheight+2.5\radius);
\coordinate (verticalPoint) at (0.5\figurewidth,0.3\figureheight);

\draw [thick, fill=steelblue31119180!30, draw=none] (0,0) rectangle (\figurewidth,0.3\figureheight);
\draw[thick, steelblue31119180] (0,0.3\figureheight)--(\figurewidth,0.3\figureheight);

\draw[very thick, loosely dashed, black!50] (0,0)--(0,\figureheight);
\draw[very thick, loosely dashed, black!50] (\figurewidth,0)--(\figurewidth,\figureheight);

\draw[very thick, black!50] (0,\figureheight)--(\figurewidth,\figureheight);
\draw[very thick, black!50] (0,0)--(\figurewidth,0);

\draw[<->, >=Latex] (\figurewidth+\mdist,0)--(\figurewidth+\mdist,\figureheight) node [pos=0.5,right] {$L_{z}$};
\draw[-] (\figurewidth,0)--(\figurewidth+2\mdist,0);
\draw[-] (\figurewidth,\figureheight)--(\figurewidth+2\mdist,\figureheight);

\draw[<->, >=Latex] (0,-\mdist)--(\figurewidth,-\mdist) node [pos=0.5,below] {$L_{x}$};
\draw[-] (0,0)--(0,-2\mdist);
\draw[-] (\figurewidth,0)--(\figurewidth,-2\mdist);

\draw[<->, >=Latex] (-\mdist,0)--(-\mdist,0.3\figureheight) node [pos=0.5,left] {$H$};
\draw[-] (-2\mdist,0.3\figureheight)--(0,0.3\figureheight);
\draw[-] (-2\mdist,0)--(0,0);

\draw[thick, draw=steelblue31119180, fill=steelblue31119180!30, fill opacity=1, line width=0.4mm] (0.5\figurewidth,0.3\figureheight+\radius) circle [radius=\radius] node {};
\draw[loosely dashdotted] (0.5\figurewidth-\radius,0.3\figureheight+\radius)--(0.5\figurewidth+\radius,0.3\figureheight+\radius);

\draw[<->, >=Latex] (0.5\figurewidth-\radius-\mdist,0.3\figureheight)--(0.5\figurewidth-\radius-\mdist,0.3\figureheight+\radius) node [pos=0.5,left] {$0.5D$};
\draw[-] (0.5\figurewidth-\radius-2\mdist,0.3\figureheight+\radius)--(0.5\figurewidth-\radius,0.3\figureheight+\radius);
\draw[-] (0.5\figurewidth-\radius-2\mdist,0.3\figureheight)--(0.5\figurewidth-\radius,0.3\figureheight);

\draw[<->, >=Latex] (0.5\figurewidth,0.3\figureheight-\mdist)--(\figurewidth,0.3\figureheight-\mdist) node [pos=0.5,below] {$0.5L_{x}$};
\draw[-] (0.5\figurewidth,0.3\figureheight)--(0.5\figurewidth,0.3\figureheight-2\mdist);
\draw[-] (\figurewidth,0.3\figureheight)--(\figurewidth,0.3\figureheight-2\mdist);

\draw[<->, >=Latex] (0.5\figurewidth+\radius+\mdist,0.3\figureheight)--(0.5\figurewidth+\radius+\mdist,0.3\figureheight+2\radius) node [pos=0.5,right] {$\varnothing D$};
\draw[-] (0.5\figurewidth,0.3\figureheight+2\radius)--(0.5\figurewidth+\radius+2\mdist,0.3\figureheight+2\radius);
\draw[-] (0.5\figurewidth,0.3\figureheight)--(0.5\figurewidth+\radius+2\mdist,0.3\figureheight);

\draw[thick, ->, >=Latex, red] (intersectPoint) -- (obliquePoint) node [pos=0.5,above right] {$U$};
\pic[draw, ->, >=Latex, "$\alpha$", angle eccentricity=0.75, angle radius=0.8cm] {angle = verticalPoint--intersectPoint--obliquePoint};

\draw[loosely dashdotted] (0.5\figurewidth,-0.025\figureheight)--(0.5\figurewidth,1.025\figureheight);

\draw[thick, ->, >=Latex, sienna1809231] (\figurewidth-2\mdist,\figureheight-\mdist)--(\figurewidth-2\mdist,\figureheight-4\mdist) node [pos=0.5,right] {$g$};

\draw[->, >=Latex] (2\mdist,\figureheight-3\mdist)--(4\mdist,\figureheight-3\mdist) node [pos=0.5,below right] {$x$};
\draw[->, >=Latex] (2\mdist,\figureheight-3\mdist)--(2\mdist,\figureheight-1\mdist) node [pos=0.5,above left] {$z$};
\draw[draw=black] (2\mdist,\figureheight-3\mdist) circle [radius=0.175\mdist] node[opacity=1, below left] {$y$};
\end{tikzpicture}%
	\caption{
		Simulation setup of the drop impact test cases.
		In a domain of size $L_{x} \times L_{y} \times L_{z}$, a spherical drop of liquid with diameter, $D$, is initialized right above the surface of a liquid pool of height, $H$.
		While the gravitational acceleration, $g$, points in negative $z$-direction, the droplet is initialized with impact velocity, $U$, acting at an angle, $\alpha$.
		The domain's side walls in $x$- and $y$-direction are periodic, whereas the domain's top and bottom walls in $z$-direction are set to no-slip.
	}
	\label{fig:drop-impact-setup}
\end{figure}

\paragraph{Setup} \label{par:vertical-drop-setup}
As shown in \Cref{fig:drop-impact-setup}, a thin liquid film of height, $H=0.5D$, was initialized in a computational domain of size, $L_{x} \times L_{y} \times L_{z}$ ($x$-, $y$-, $z$-direction) with $L_{x}=L_{y}=10D$ and $L_{z}=5D$.
At the pool's surface, a spherical droplet with diameter, $D$, was initialized with an impact velocity, $U$, in the negative $z$-direction with $\alpha=0^{\circ}$, leading to a vertical impact.
The domain's side walls in the $x$- and $y$-direction were periodic, whereas there were no-slip boundary conditions at the domain's top- and bottom walls.
Conforming with the gravitational acceleration, $g$, hydrostatic pressure was initialized such that the reference density was $\rho_{0}=1$ at the surface of the pool.\par

The drop impact is described by the Weber number
\begin{equation}
\mathrm{We} = \frac{\rho U^{2} D}{\sigma},
\end{equation}
which relates inertial and surface tension forces, and by the Ohnesorge number,
\begin{equation}
\mathrm{Oh} = \frac{\mu}{\sqrt{\sigma \rho D}},
\end{equation}
which is defined by the relation of viscous to inertial and surface tension forces.
The drop diameter, $D$, the Bond number, Bo (see \Cref{eq:Bo}), and reference time, $t^{*}=t \, U/D$, close the definition of the system.
As found in Reference~\cite{lehmann2021ejection}, the simulation results must be offset by $t^{*}=0.16$ to synchronize the first photograph of the laboratory experiment with the simulation setup chosen in this study.\par

In the experiments of Reference~\cite{wang2000SplashingImpactSingle}, a 70\,\% glycerol--water mixture at 23\,\textdegree C was used with $\rho^{\text{SI}}=1200$\,kg/m\textsuperscript{3} and $\mu^{\text{SI}}=0.022$\,kg/(m$\cdot$s).
The experiment obeyed the non-dimensional numbers, We=$2010$, and, Oh=$0.0384$.
Assuming $g^{\text{SI}}=9.81$\,m/s\textsuperscript{2}, the system is closed by Bo$=3.18$.
As in \Cref{par:oblique-drop-results}, the density ratio is set to $\tilde{\rho}=1000$ and the dynamic viscosity ratio is set to $\tilde{\mu} = 100$.\par

The simulations were performed with computational resolutions according to $D \in \{20, 40, 80\}$.
The FSLBM's relaxation rate was chosen at $\omega=1.989$ and the PFLBM's hydrodynamic relaxation rate in the heavy phase was set to $\omega_{\text{H}}=1.988$.
In agreement with the findings from \Cref{subsec:buoyancy-flows}, lower values for the PFLBM's interface width, $\xi$, and mobility, $M$, tended to give more physically realistic results, as shown in the Appendix in \Cref{fig:drop-vertical-pflbm-width-interface}.
Therefore, $\xi=4$ and $M=0.03$ were chosen as they are the lowest values that allowed stable simulations for all tested computational resolutions.\par

\begin{figure}
	\centering
	\begin{tabular}{>{\centering\arraybackslash}m{0.025\textwidth}
			>{\centering\arraybackslash}m{0.3\textwidth}
			>{\centering\arraybackslash}m{0.3\textwidth}
			>{\centering\arraybackslash}m{0.3\textwidth}}
		\rotatebox[origin=l]{90}{Experiment~\cite{wang2000SplashingImpactSingle}}
		& \multicolumn{3}{c}{\makecell{\includegraphics[width=0.25\textwidth]{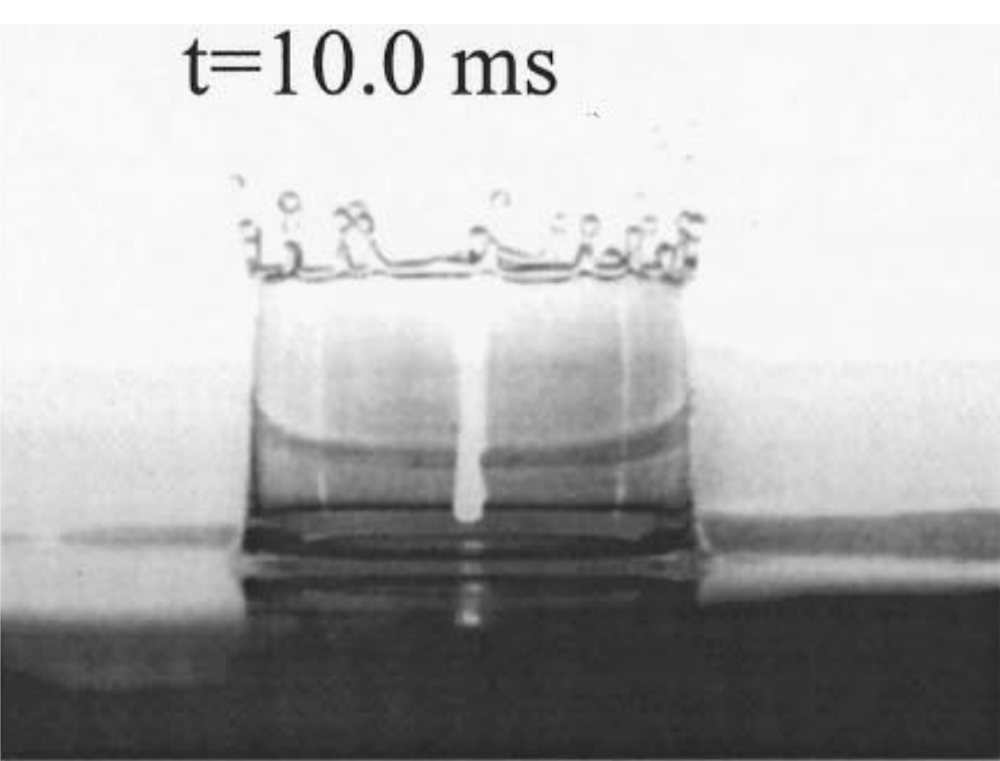}}} \\
		
		\rotatebox[origin=l]{90}{FSLBM} &		
		\includegraphics[width=0.2\textwidth]{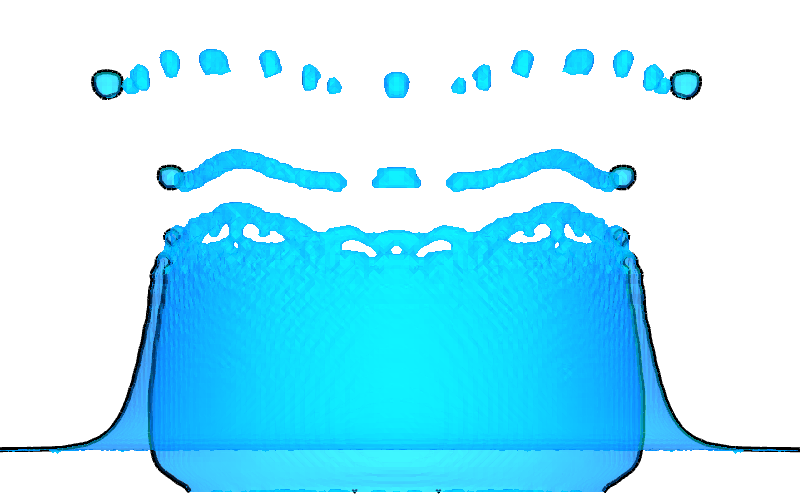} &
		\includegraphics[width=0.2\textwidth]{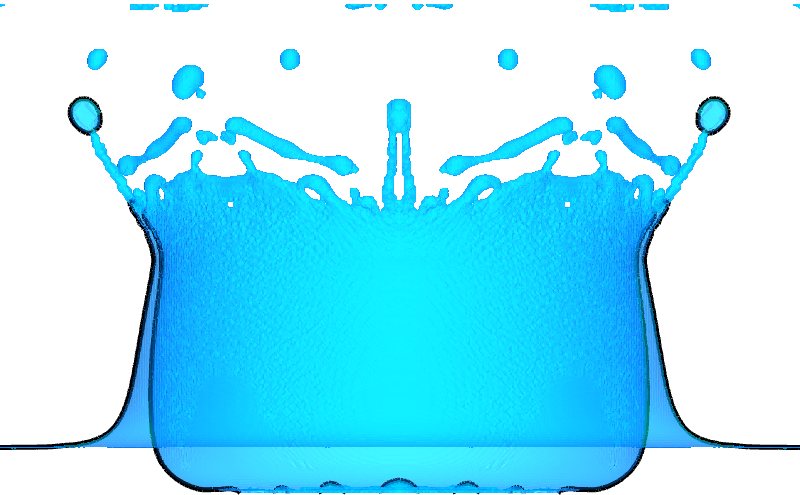} & 
		\includegraphics[width=0.2\textwidth]{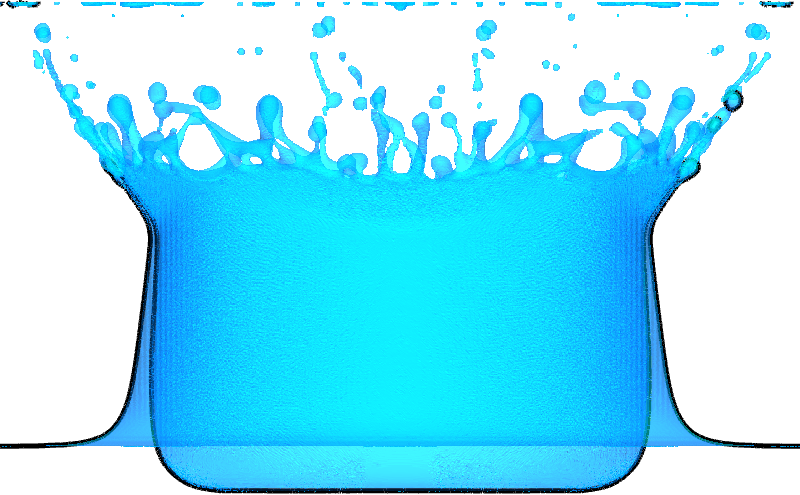} \\
		
		\rotatebox[origin=l]{90}{PFLBM} &
		\includegraphics[width=0.2\textwidth]{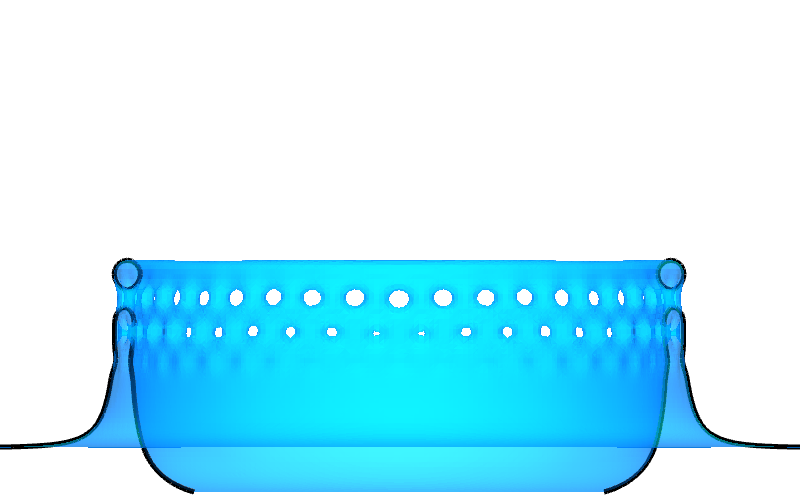} &
		\includegraphics[width=0.2\textwidth]{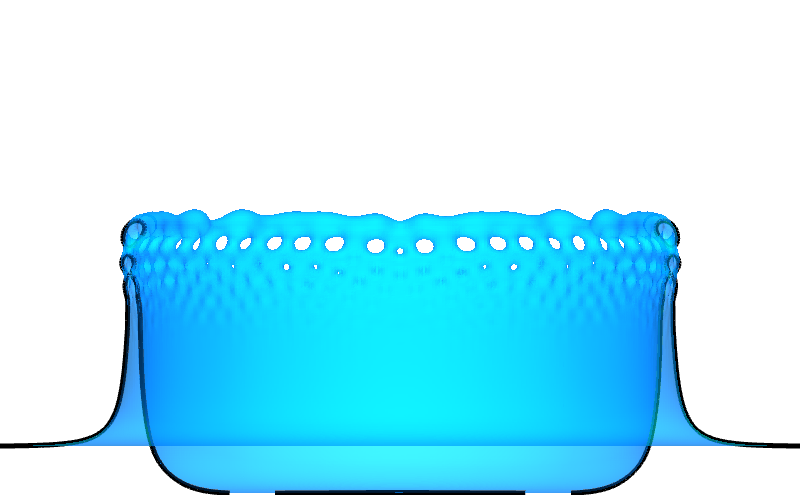} & 
		\includegraphics[width=0.2\textwidth]{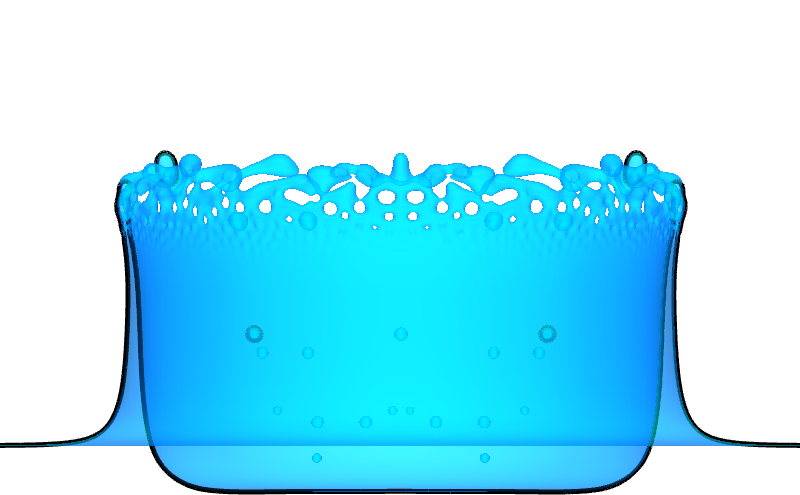} \\
		
		& $D=20$ & $D=40$ & $D=80$ \\
	\end{tabular}
	\caption{
		Simulated vertical drop impact at time, $t^{*}=12$, at different computational resolutions defined by the initial drop diameter, $D$.
		While the simulation results are true to scale, no scale bar is available for the photograph of the experiment~\cite{wang2000SplashingImpactSingle}.
		Therefore, the splash crown's dimension can only be compared between simulations rather than with the experiment.
		The solid black line illustrates the crown's contour in the center cross-section with normal in the $x$-direction.
		The photograph of the laboratory experiment was reprinted from Reference~\cite{wang2000SplashingImpactSingle} with the permission of AIP Publishing.
	}
	\label{fig:drop-vertical-t-12}
\end{figure}

\paragraph{Results and discussion} \label{par:vertical-drop-results}
In \Cref{fig:drop-vertical-t-12}, the crown formation at time, $t^{*}=12$, is shown for both models at various computational resolutions.
In the Appendix, \Cref{fig:drop-vertical-fslbm,fig:drop-vertical-pflbm} compare the simulated and experimental drop impact dynamically, i.e., with respect to time.
While no scale bars for the photograph of the laboratory experiments are available, it can be noted that all simulations converged well with increasing resolution, and the dimensions of the simulated splash crowns agreed with each other.
The measured simulated cavity depths and splash crowns' inner diameters are presented in the Appendix in \Cref{tab:drop-vertical-cavity,tab:drop-vertical-crown}.
The FSLBM captured the droplets ejected from the crown qualitatively well, even at low computational resolution.
Similar results have also been obtained with the FSLBM and LSQR curvature computation~\cite{lehmann2021ejection}.
In contrast, the PFLBM with the parameters chosen here, could not sufficiently predict these droplets.\par
It must be emphasized, that the PFLBM is sensitive to the choice of the interface width and mobility parameter (see Appendix, \Cref{fig:drop-vertical-pflbm-width-interface}).
That is, for consistency reasons, these values were chosen as to be stable with the lowest computational resolution, $D=20$, and kept constant for higher resolutions.
A rigorous study of the individual lower limits of these parameters at each resolution might improve the quality of the results.\par

\subsubsection{Oblique drop impact} \label{subsubsec:oblique-drop}
In the final test case, an oblique drop impact is simulated as in the experiments of References~\cite{gielen2017ObliqueDropImpact,reijers2019ObliqueDropletImpact}.

\paragraph{Setup} \label{par:oblique-drop-setup}
The setup is similar to \Cref{subsubsec:vertical-drop} and presented in \Cref{fig:drop-impact-setup}.
However, the computational domain is cubical with $L_{x}=L_{y}=L_{z}=10D$, and the liquid pool is of height, $H=5D$.
The droplet's impact velocity, $U$, is oriented in an angle, $\alpha=28.5^{\circ}$, from negative $z$-direction.\par

The experimental investigations were performed with We$=416.5$, $D^{\text{SI}}=1.15\cdot 10^{-4}$\,m, and liquid water with $\rho^{\text{SI}}=1000$\,kg/m\textsuperscript{3} and $\sigma^{\text{SI}}=0.072$\,kg/(s\textsuperscript{2}).
Assuming $\mu^{\text{SI}}=10^{-3}$\,kg/(m$\cdot$s) for water at 20\,\textdegree C and $g^{\text{SI}}=9.81$\,m/s\textsuperscript{2}, the setup is defined by, Oh$=0.011$, and, Bo$=0.0018$.
The density ratio, $\tilde{\rho}=1000$, and dynamic viscosity ratio, $\tilde{\mu}=100$, are chosen as to mimic an air--water system~\cite{reijers2019ObliqueDropletImpact}.
The computational resolution, relaxation rates, and hydrostatic pressure were set as for the vertical drop impact in \Cref{subsubsec:vertical-drop}.
Here, the lowest interface width and mobility that allowed stable simulations in the PFLBM for all tested resolutions were, $\xi=4$, and, $M=0.09$, respectively.

\begin{figure}[htbp]
	\centering
	\begin{tabular}{>{\centering\arraybackslash}m{0.025\textwidth}
			>{\centering\arraybackslash}m{0.3\textwidth}
			>{\centering\arraybackslash}m{0.3\textwidth}
			>{\centering\arraybackslash}m{0.3\textwidth}}
		\rotatebox[origin=l]{90}{Experiment~\cite{reijers2019ObliqueDropletImpact}}
		& \multicolumn{3}{c}{\makecell{\includegraphics[width=0.15\textwidth]{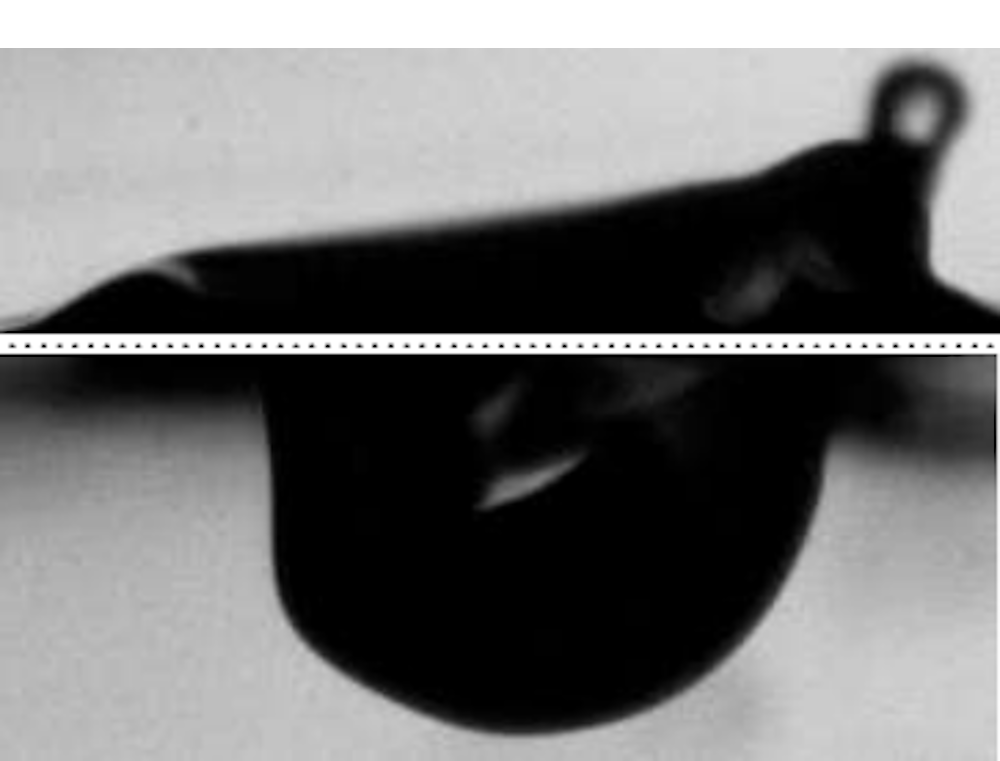}}} \\
		
		\rotatebox[origin=l]{90}{FSLBM} &		
		\includegraphics[width=0.2\textwidth]{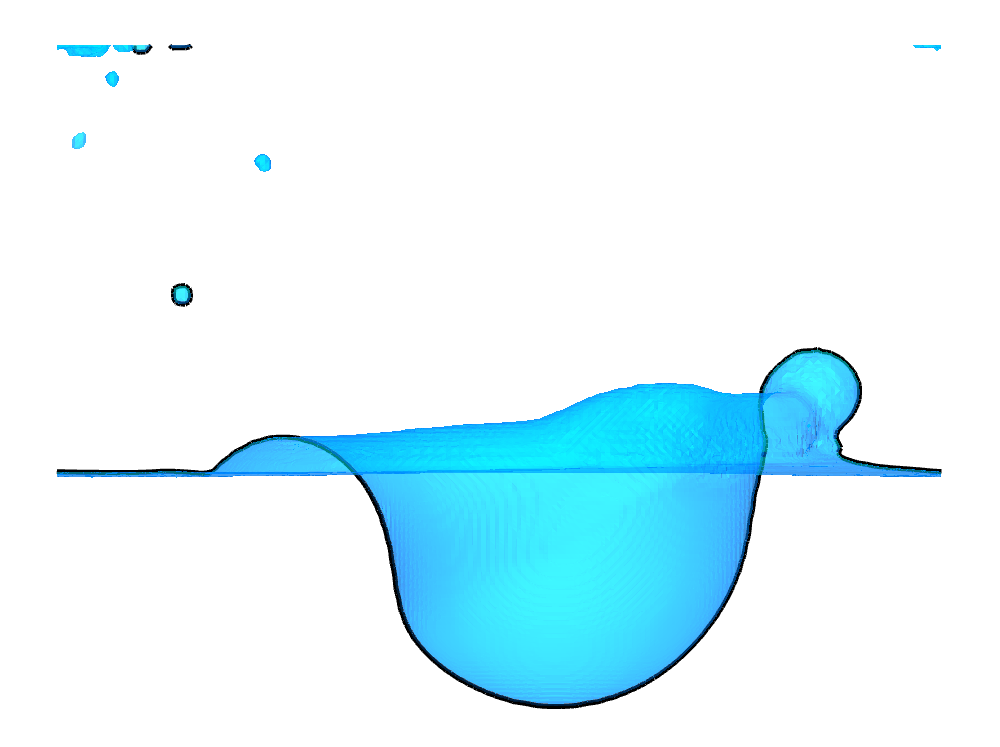} &
		\includegraphics[width=0.2\textwidth]{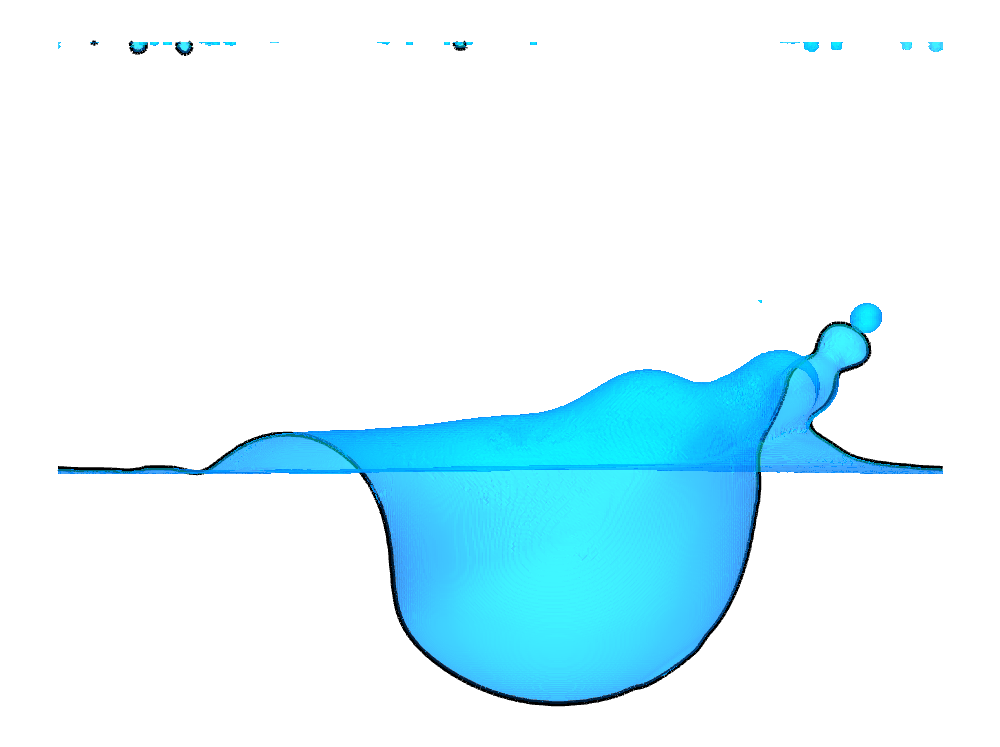} & 
		\includegraphics[width=0.2\textwidth]{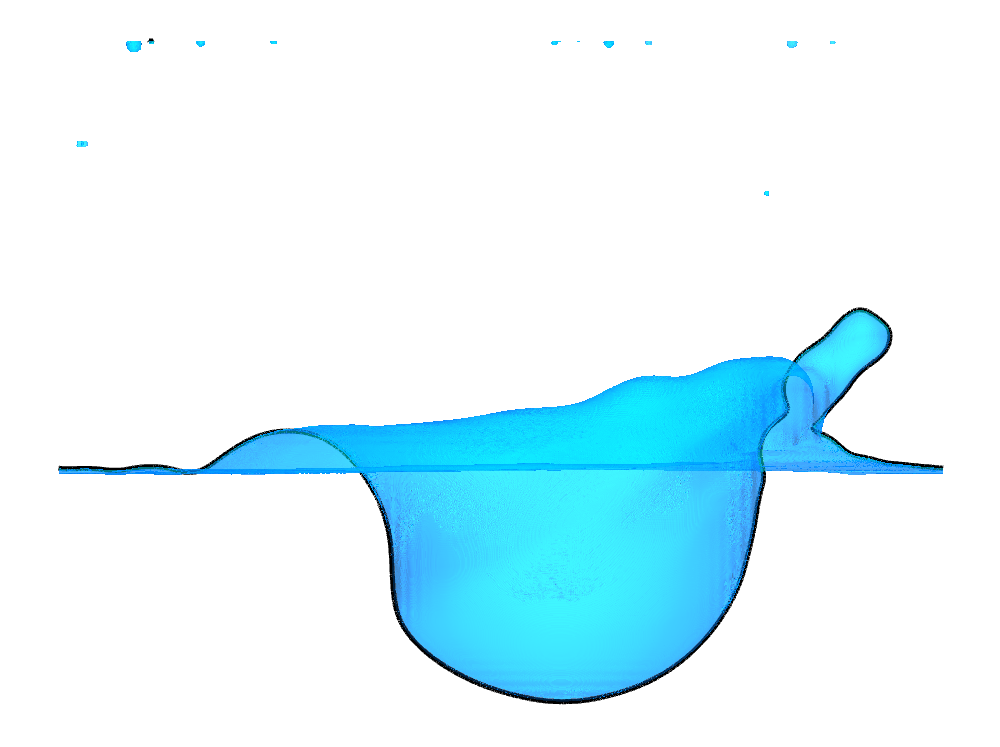} \\
		
		\rotatebox[origin=l]{90}{PFLBM} &
		\includegraphics[width=0.2\textwidth]{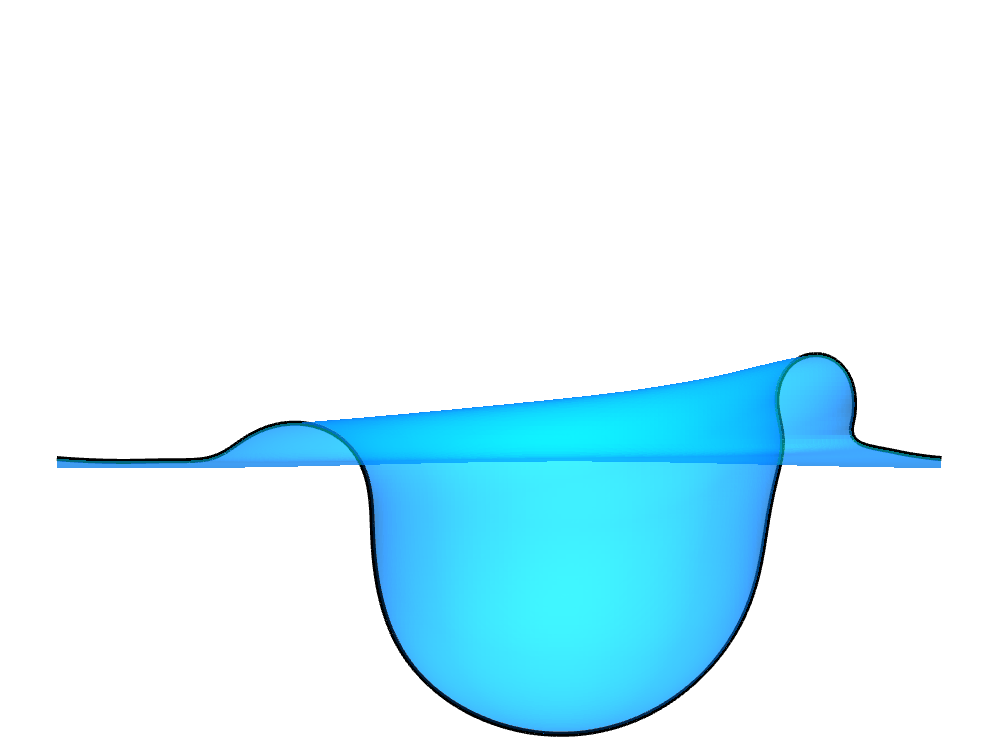} &
		\includegraphics[width=0.2\textwidth]{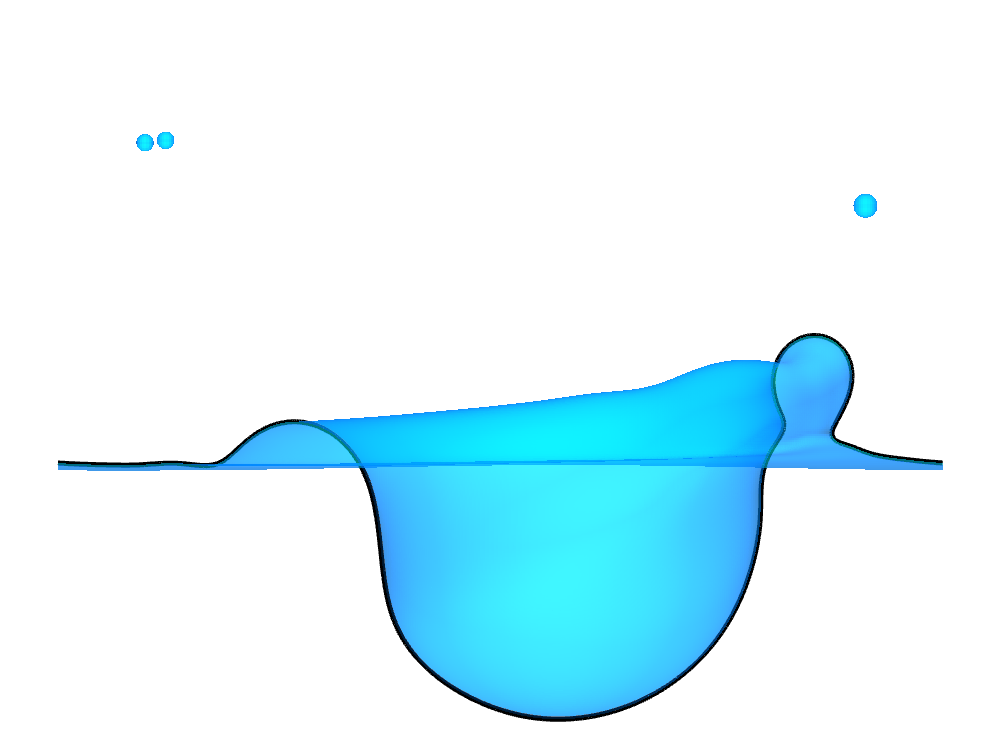} & 
		\includegraphics[width=0.2\textwidth]{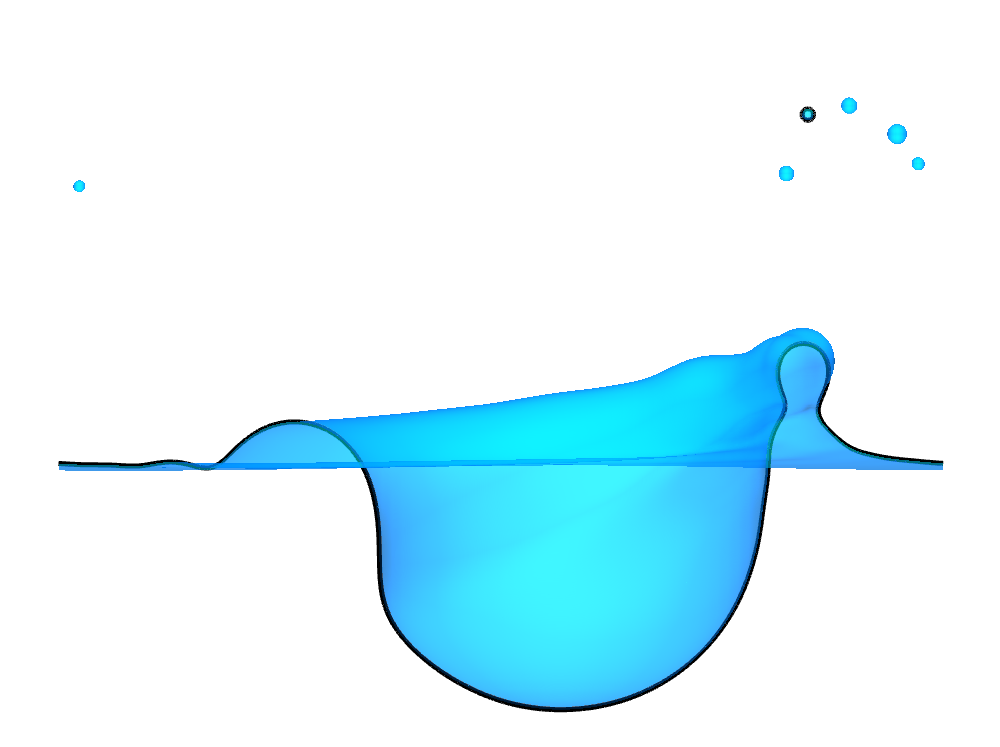} \\
		
		& $D=20$ & $D=40$ & $D=80$ \\
	\end{tabular}
	\caption{
		Simulated oblique drop impact at time, $t^{*}=18$, at different computational resolutions defined by the initial drop diameter, $D$.
		While the simulation results are true to scale, no scale bar is available for the photograph of the experiment~\cite{reijers2019ObliqueDropletImpact}.
		Therefore, the splash crown's dimension can only be compared between simulations rather than with the experiment.
		The solid black line illustrates the crown's contour in the center cross-section with normal in the $x$-direction.
		The photograph of the laboratory experiment was reprinted from Reference~\cite{reijers2019ObliqueDropletImpact} with the permission of the original authors.
	}
	\label{fig:drop-oblique-t-18}
\end{figure}

\paragraph{Results and discussion} \label{par:oblique-drop-results}
In \Cref{fig:drop-oblique-t-18}, the crown formation at time, $t=18t^{*}$, is shown for both models at various computational resolutions and are compared with photographs of the laboratory experiments~\cite{reijers2019ObliqueDropletImpact}.
Additionally, \Cref{fig:drop-oblique-fslbm,fig:drop-oblique-pflbm} in the Appendix show the drop impact as simulated by the FSLBM and PFLBM over time.
The FSLBM and PFLBM converged well and the dimensions of the simulated splash crowns agreed well with each other.
As for the vertical impact, scale bars for the photograph of the laboratory experiments are missing and no quantitative comparison with reference data could be drawn.
Nevertheless, the measured simulated cavity depths and splash crowns' inner diameters are presented in the Appendix in \Cref{tab:drop-oblique-cavity,tab:drop-oblique-crown}.
In contrast to the vertical impact, less droplets were ejected from the crown and the PFLBM agreed qualitatively well at high computational resolution.
The FSLBM captured the shape of the drop cavity and splash crown qualitatively well, even with low computational resolution.
Similar results have been obtained with the FSLBM and LSQR curvature computation~\cite{lehmann2021ejection}.\par

\section{Conclusion}
This study has compared two different LBM approaches for simulating flows in which the dynamics of the lighter phase are assumed negligible.
After an introduction of the numerical foundation of the FSLBM and PFLBM, both models were applied to a series of benchmark cases and their performance was discussed in terms of their numerical properties and implementation aspects.\par

The FSLBM ignores fluid flow in the secondary phase and requires much less memory, making it efficient and more applicable to limited-memory hardware. 
On the other hand, the PFLBM simulates flow in both phases but is well suited for massively parallel computing and can be implemented more easily in a flexible way using code generation technology.
A very distinct difference between the two models is their sharp and diffuse interface representation in the FSLBM and PFLBM, respectively.
Therefore, six numerical experiments were shown in which the models' accuracy is compared at different resolutions of the computational grid.\par

While the standing gravity wave was simulated more accurately by the FSLBM, a much higher resolution was required than in the PFLBM to capture the motion of the interface at low amplitude.
However, it has to be remarked that this test setup represents a limit case of the FSLBM.
In consistency with the analytical solution, the interface motion is limited to only a few LBM cells, even for highly resolved grids.\par

In the capillary wave test case, the FSLBM diverged with increasing resolution due to deficiencies in all tested approaches for the computation of infinitesimal interface curvature.
In contrast, the PFLBM could simulate the capillary wave with reasonable accuracy.\par

The third and fourth test case featured buoyancy driven flows.
That is, an unconfined single rising gas bubble in liquid in four different characteristic parameter sets, and a confined Taylor bubble traversing a cylindrical tube, were simulated.
The FSLBM was able to capture the bubble shape and Reynolds number with reasonable accuracy, even with moderate resolution.
On the other hand, with the parameters used in this study, the PFLBM required higher resolutions for simulations to be stable.
While it predicted bubble shape and accuracy well in the initial phase of the single gas bubble rise, the bubbles tended to evolve into non-physical shapes leading to eventually collapse of the simulation.
This observation was made even with the highest computational resolution used in this study.
A sensitivity of the chosen mobility on the phase-field was observed. However, the mobility can only be chosen in a certain range to obtain stable simulations. In this range no generally suitable values could be found.\par

In the fifth and sixth test case, the models' ability to capture dynamic coalescence was validated.
To do this, a vertical and oblique drop impact into a pool of liquid were simulated.
The FSLBM predicted the shape of the splash crown reasonably well, even with low computational resolution.
With sufficiently high computational resolution, the PFLBM was also able to simulate the oblique drop impact with satisfying accuracy.
However, for the vertical drop impact, only the FSLBM was able to capture the droplets ejected from the crown formation sufficiently well.
As for the rising bubble, the PFLBM was observed to be sensitive to the choice of the mobility and interface width, but no generally applicable choice could be identified.\par

The investigation of the optimal choice of mobility and interface width in the PFLBM remains future work.
Extending the implementations of both models with adaptive refinement of the computational grid is expected to significantly enhance the issues observed, and the efficiency of the implementations.
Additionally, the applicability of the FSLBM to code generation should be explored leading to a flexible and portable code basis.

\FloatBarrier
\newpage

\renewcommand{\thesection}{A}

\section{Appendix} \label{sec:appendix}
The appendix presents additional simulation results that were not shown in the main part of the manuscript for reasons of brevity.

\subsection{Rising bubble} \label{subsec:app-rising-bubble}
\begin{figure}[h!]
	\centering
	\begin{tabular}{>{\centering\arraybackslash}m{0.025\textwidth}
					>{\centering\arraybackslash}m{0.2\textwidth}
					>{\centering\arraybackslash}m{0.2\textwidth}
					>{\centering\arraybackslash}m{0.2\textwidth}
					>{\centering\arraybackslash}m{0.2\textwidth}}
		
		\rotatebox[origin=l]{90}{Experiment}
		& \multicolumn{4}{c}{\makecell{
			\includegraphics[width=0.25\textwidth]{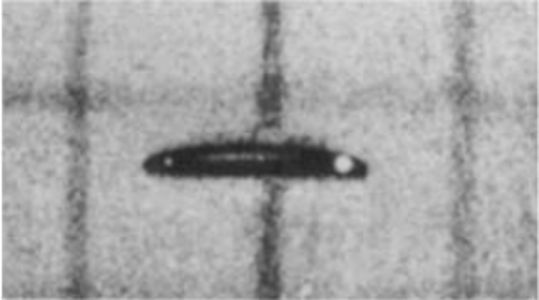} \\ 
			$\mathrm{Re}=55.3$~\cite{bhaga1981BubblesViscousLiquids}}} \\
			
		\rotatebox[origin=l]{90}{FSLBM} &
		\makecell{\includegraphics[width=0.2\textwidth]{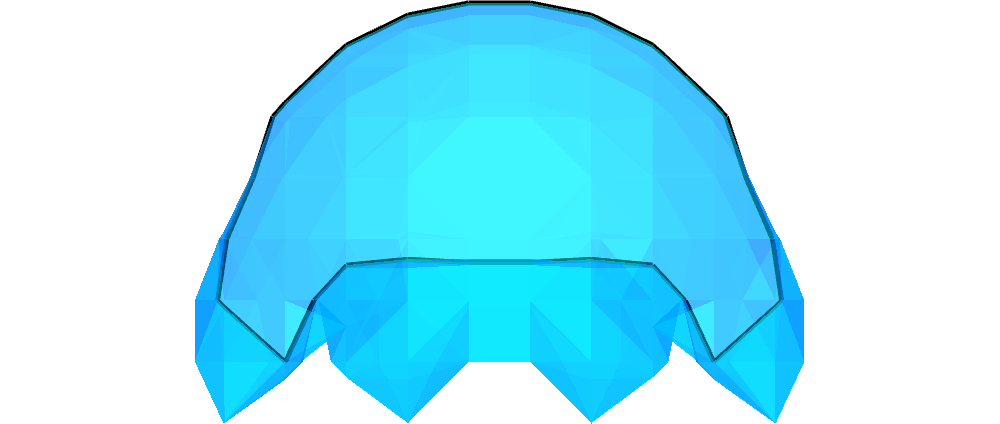} \\ (a) $\mathrm{Re}=33.9$} &
		\makecell{\includegraphics[width=0.2\textwidth]{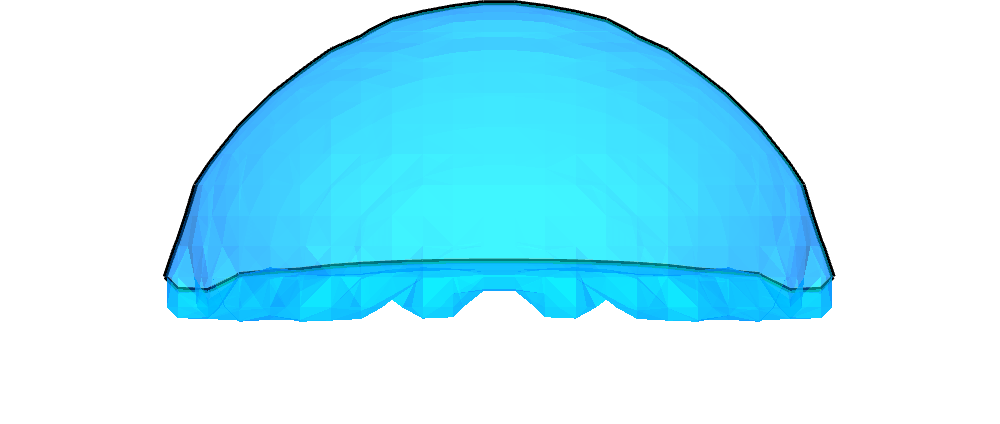} \\ (b) $\mathrm{Re}=45.5$} &	
		\makecell{\includegraphics[width=0.2\textwidth]{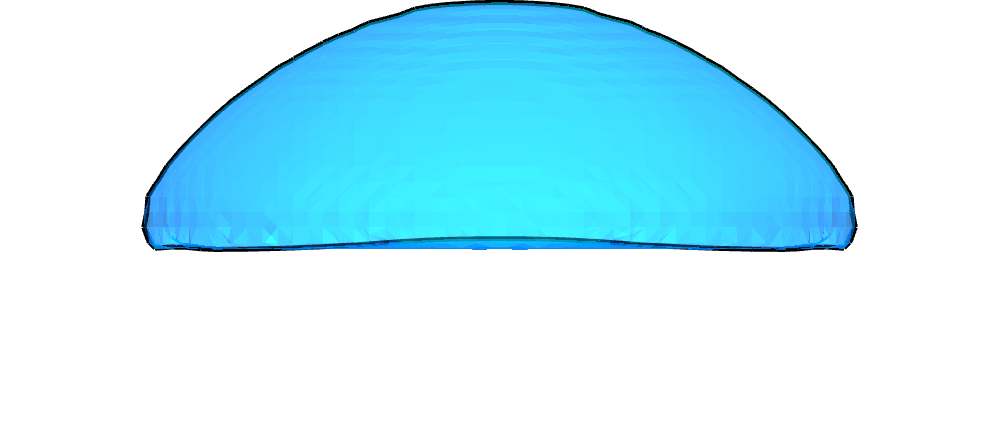} \\ (c) $\mathrm{Re}=50.7$} &
		\makecell{\includegraphics[width=0.2\textwidth]{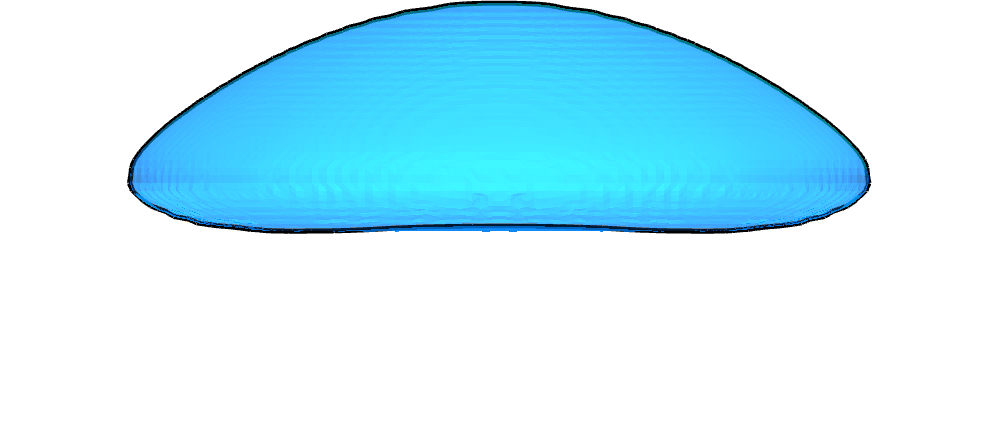} \\ (d) $\mathrm{Re}=52.4$} \\
		
		\rotatebox[origin=l]{90}{PFLBM} & unstable & unstable & 
		\makecell{\includegraphics[width=0.2\textwidth]{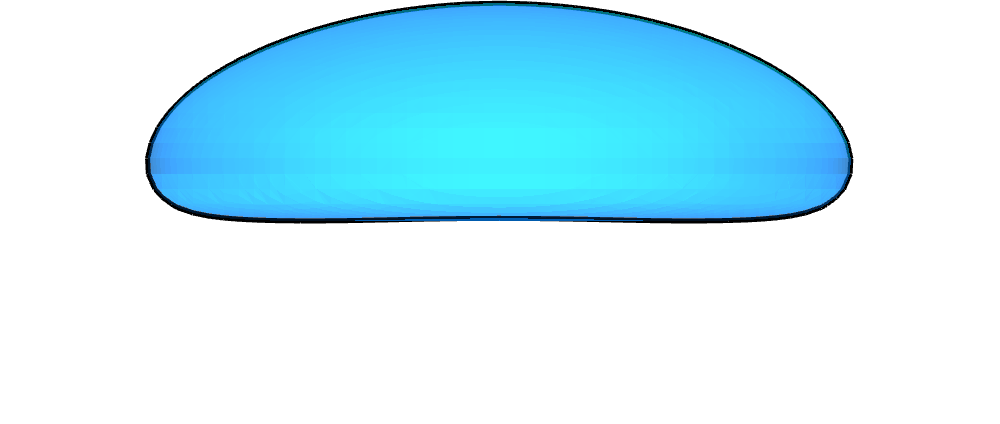} \\ (g) $\mathrm{Re}=51.4$} &
		\makecell{\includegraphics[width=0.2\textwidth]{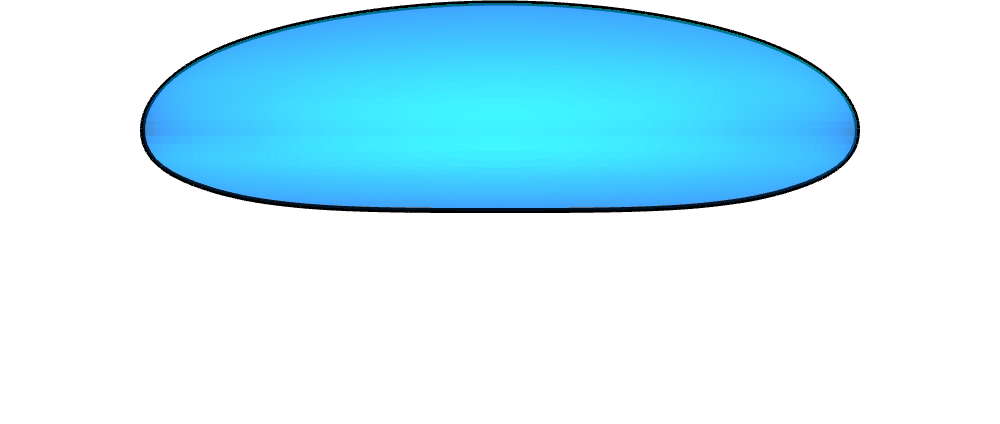} \\ (h) $\mathrm{Re}=55.2$} \\
		
		& $D=8$ & $D=16$ & $D=32$ & $D=64$\\
	\end{tabular}
	\caption{
		Simulated bubble shape and Reynolds number, Re, at time, $t^{*}=10$, for case 1 in \Cref{tab:rising-bubble-setups} with $\mathrm{Bo}=32.2$ and $\mathrm{Mo}=8.2\cdot10^{-4}$.
		Different computational resolutions according to the initial bubble diameter, $D$, are shown.
		The solid black lines illustrate the bubble's contour in the center cross-section with normal in the $x$-direction.
		The photograph of the laboratory experiment was reprinted from Reference~\cite{bhaga1981BubblesViscousLiquids} with the permission of Cambridge University Press.
	}
	\label{fig:bo-32.2-mo-8.2e-4}
\end{figure}

\begin{figure}[h!]
	\centering
	\begin{tabular}{>{\centering\arraybackslash}m{0.05\textwidth}
					>{\centering\arraybackslash}m{0.2\textwidth}
					>{\centering\arraybackslash}m{0.2\textwidth}
					>{\centering\arraybackslash}m{0.2\textwidth}
					>{\centering\arraybackslash}m{0.2\textwidth}}
		
		\rotatebox[origin=l]{90}{Experiment}
		& \multicolumn{4}{c}{\makecell{
			\includegraphics[width=0.15\textwidth]{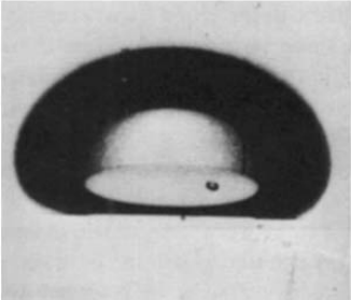} \\ 
			$\mathrm{Re}=7.77$~\cite{bhaga1981BubblesViscousLiquids}}} \\

		\rotatebox[origin=l]{90}{FSLBM} &
		\makecell{\includegraphics[width=0.2\textwidth]{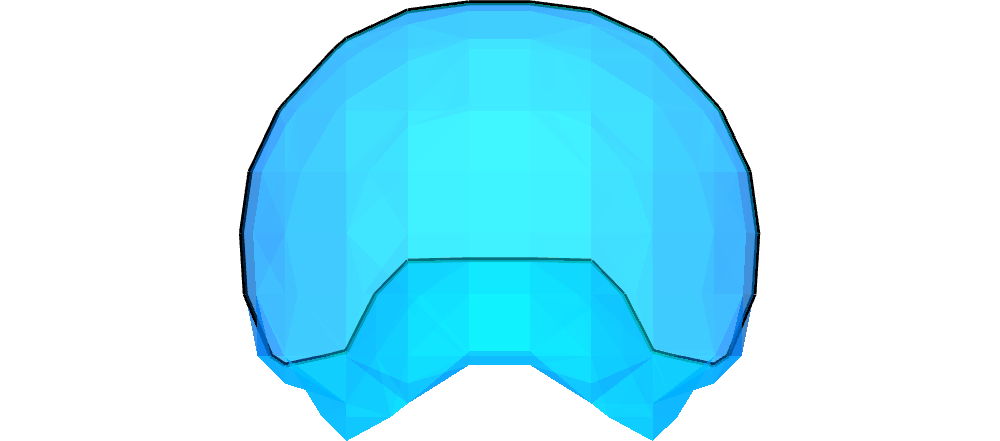} \\ (a) $\mathrm{Re}=4.4$} &
		\makecell{\includegraphics[width=0.2\textwidth]{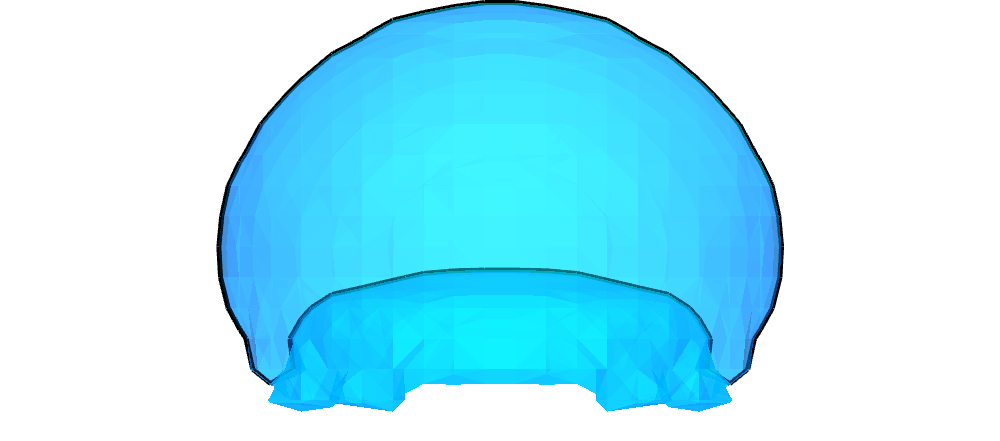} \\ (b) $\mathrm{Re}=6.4$} &
		\makecell{\includegraphics[width=0.2\textwidth]{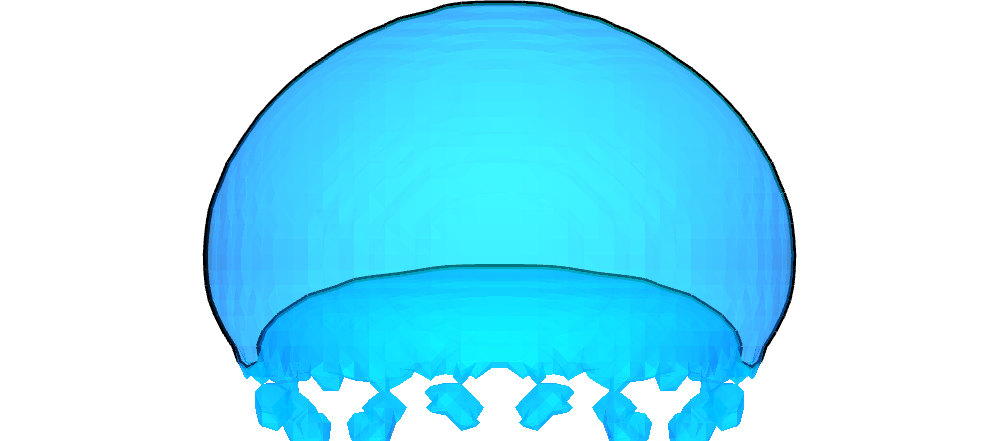} \\ (c) $\mathrm{Re}=7.3$} &
		\makecell{\includegraphics[width=0.2\textwidth]{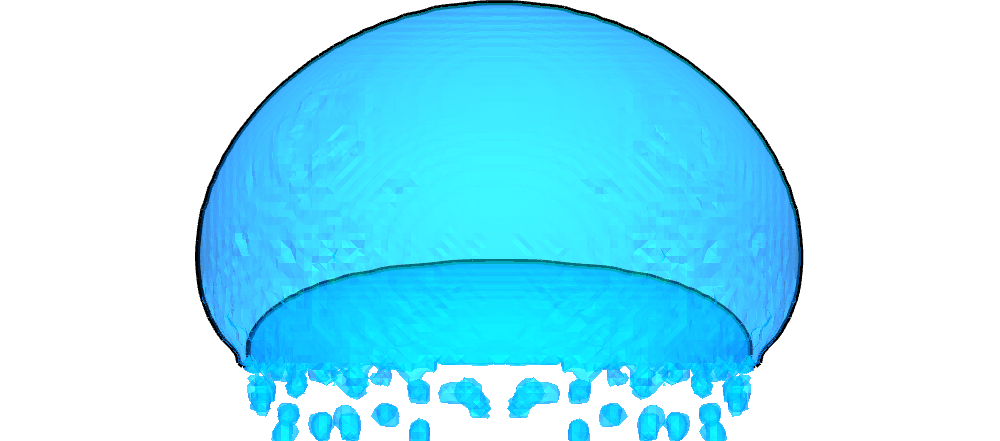} \\ (d) $\mathrm{Re}=7.6$} \\
		
		\rotatebox[origin=l]{90}{PFLBM} & unstable & unstable & 
		\makecell{\includegraphics[width=0.2\textwidth]{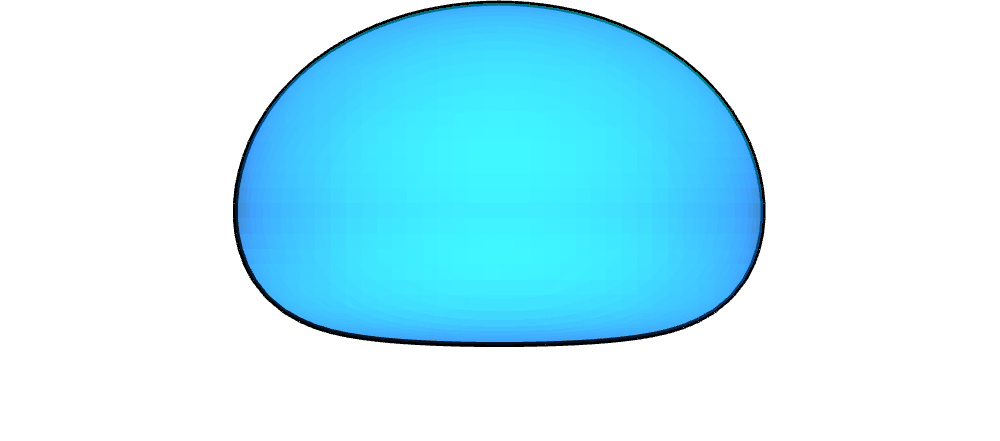} \\ (g) $\mathrm{Re}=6.1$} &
		\makecell{\includegraphics[width=0.2\textwidth]{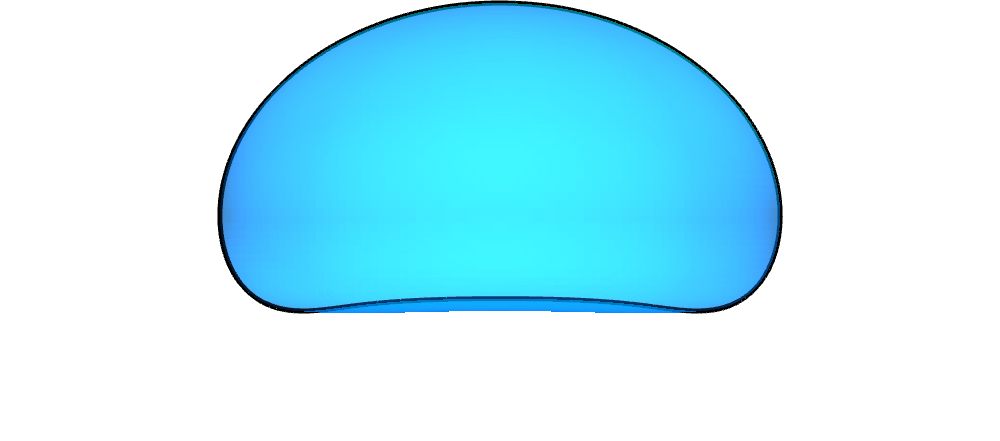} \\ (h) $\mathrm{Re}=7.1$} \\
		
		& $D=8$ & $D=16$ & $D=32$ & $D=64$\\
	\end{tabular}
	\caption{
		Simulated bubble shape and Reynolds number, Re, at time, $t^{*}=10$, for case 3 in \Cref{tab:rising-bubble-setups} with $\mathrm{Bo}=243$ and $\mathrm{Mo}=266$.
		Different computational resolutions according to the initial bubble diameter, $D$, are shown.
		The solid black lines illustrate the bubble's contour in the center cross-section with normal in the $x$-direction.
		The photograph of the laboratory experiment was reprinted from Reference~\cite{bhaga1981BubblesViscousLiquids} with the permission of Cambridge University Press.
	}
	\label{fig:bo-243-mo-266}
\end{figure}

\begin{figure}[h!]
	\centering
	\begin{tabular}{>{\centering\arraybackslash}m{0.05\textwidth}
					>{\centering\arraybackslash}m{0.2\textwidth}
					>{\centering\arraybackslash}m{0.2\textwidth}
					>{\centering\arraybackslash}m{0.2\textwidth}
					>{\centering\arraybackslash}m{0.2\textwidth}}
		
		\rotatebox[origin=l]{90}{Experiment}
		& \multicolumn{4}{c}{\makecell{
			\includegraphics[width=0.15\textwidth]{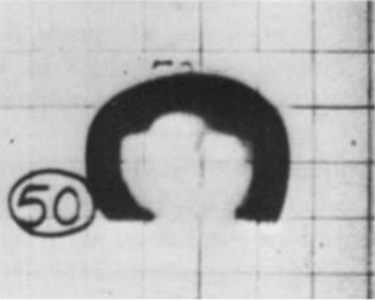} \\ 
			$\mathrm{Re}=18.3$~\cite{bhaga1981BubblesViscousLiquids}}} \\
			
		\rotatebox[origin=l]{90}{FSLBM} &
		\makecell{\includegraphics[width=0.2\textwidth]{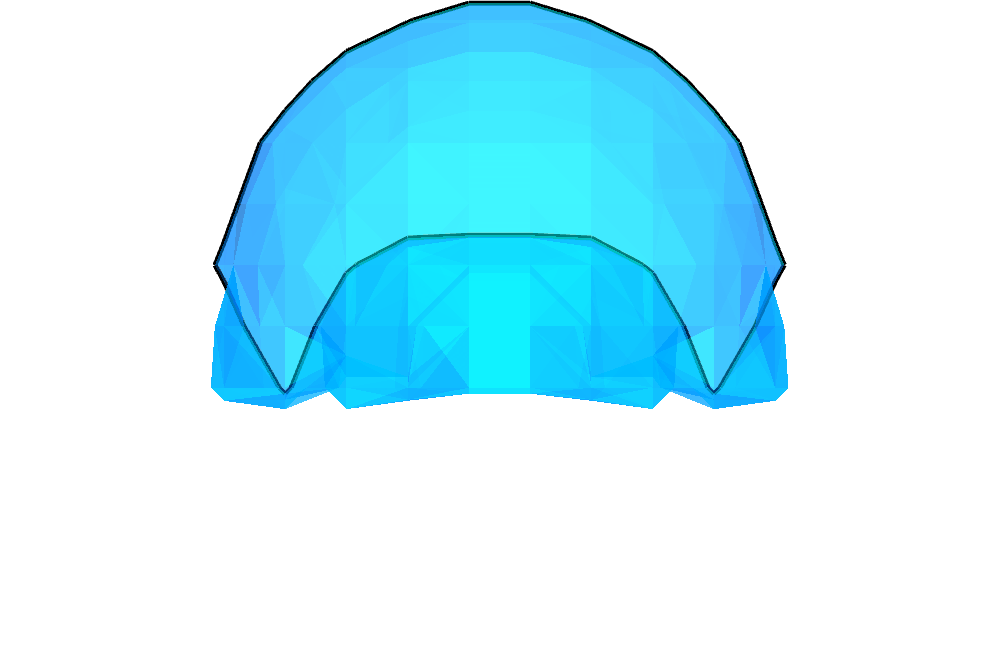} \\ (a) $\mathrm{Re}=10.2$} &   		
		\makecell{\includegraphics[width=0.2\textwidth]{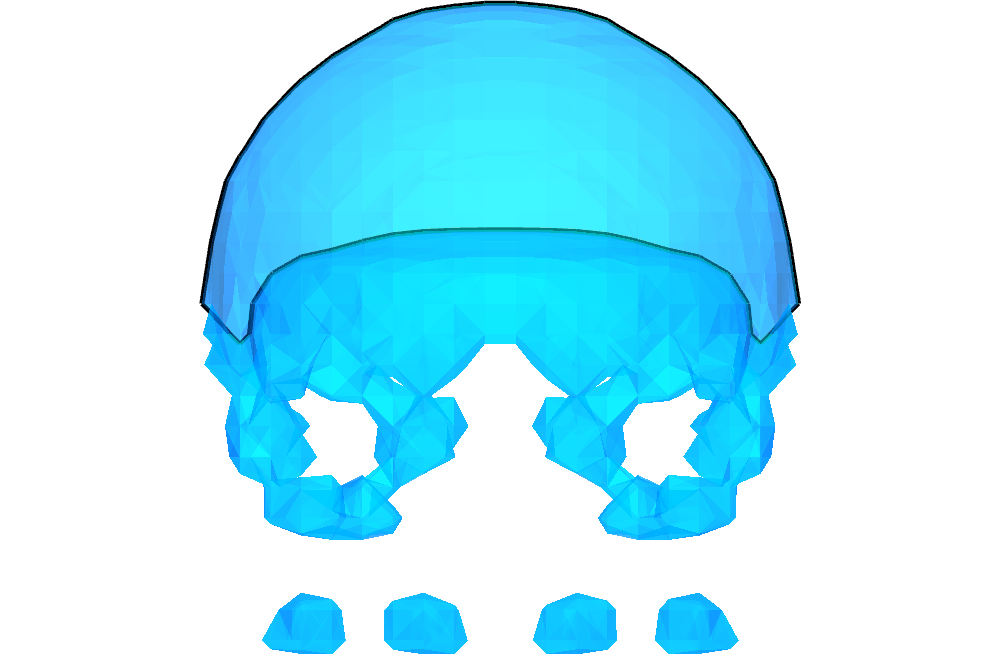} \\ (b) $\mathrm{Re}=15.1$} &	
		\makecell{\includegraphics[width=0.2\textwidth]{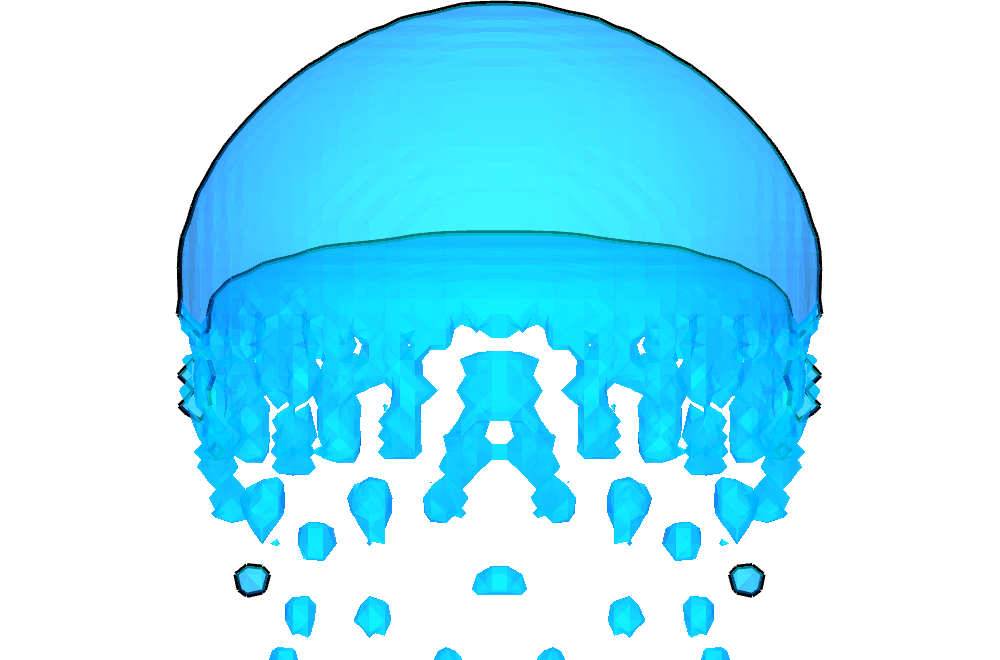} \\ (c) $\mathrm{Re}=17.4$} &
		 \makecell{\includegraphics[width=0.2\textwidth]{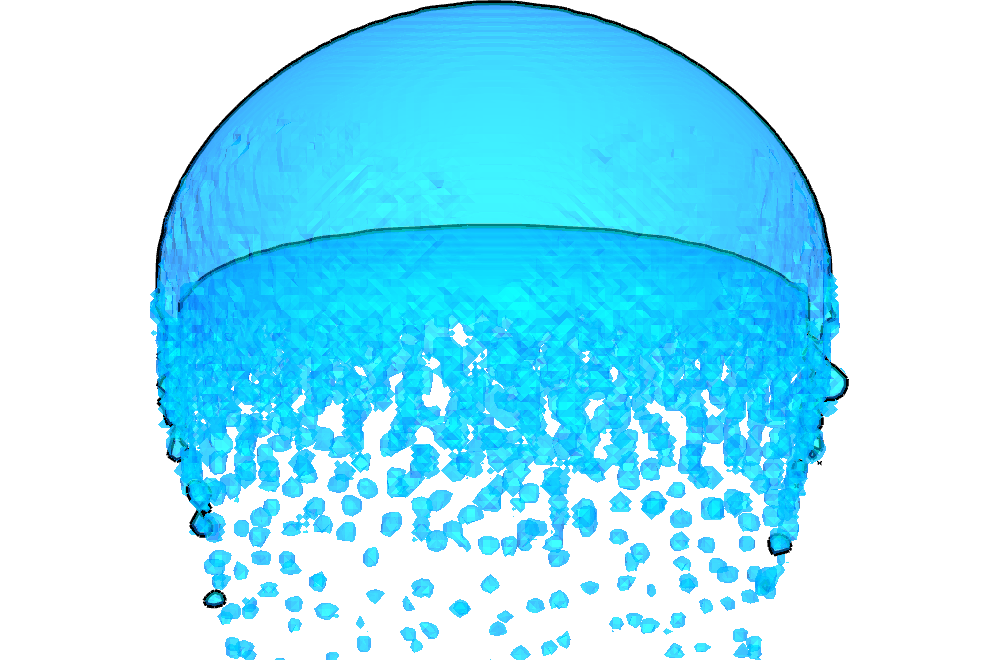} \\ (d) $\mathrm{Re}=18.2$} \\
		
		\rotatebox[origin=l]{90}{PFLBM} & unstable & unstable & 
		\makecell{\includegraphics[width=0.2\textwidth]{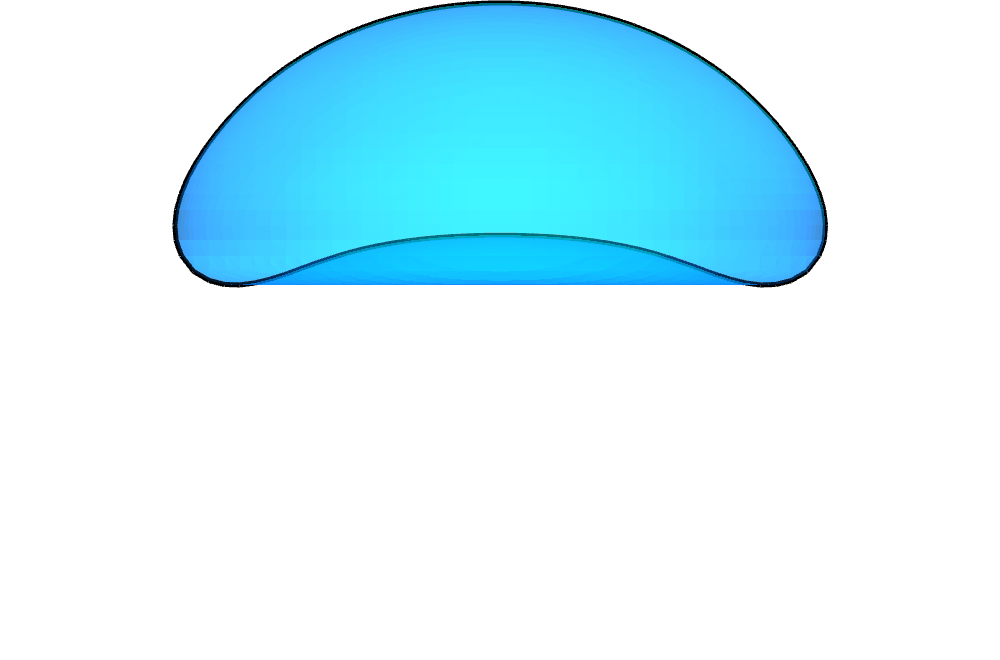} \\ (g) $\mathrm{Re}=16.3$} &
		\makecell{\includegraphics[width=0.2\textwidth]{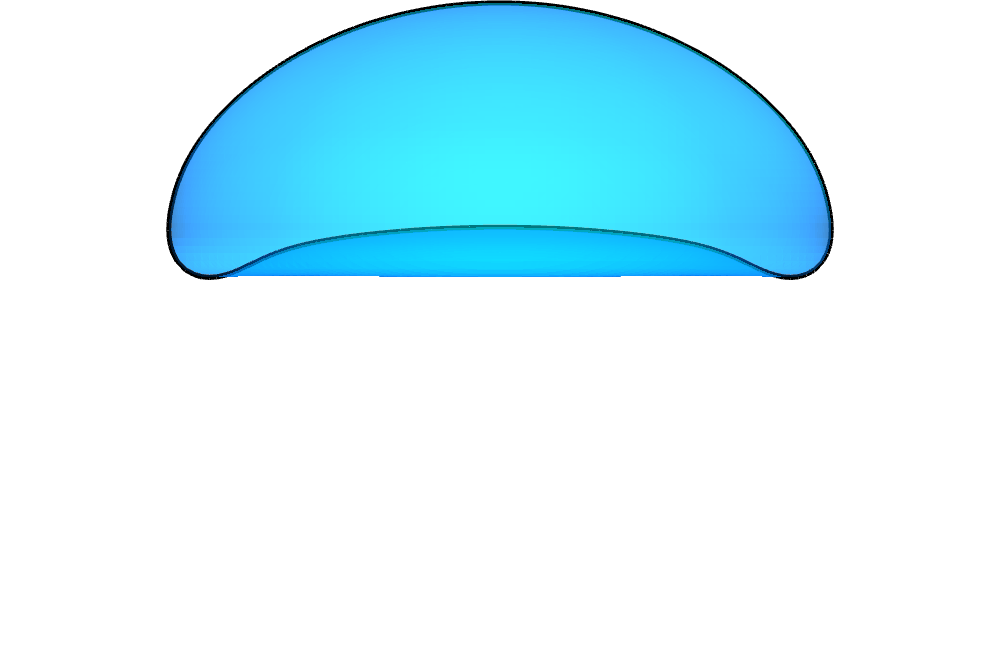} \\ (h) $\mathrm{Re}=18.5$} \\
		
		& $D=8$ & $D=16$ & $D=32$ & $D=64$\\
	\end{tabular}
	\caption{
		Simulated bubble shape and Reynolds number, Re, at time, $t^{*}=10$, for case 4 in \Cref{tab:rising-bubble-setups} with $\mathrm{Bo}=339$ and $\mathrm{Mo}=43.1$.
		Different computational resolutions according to the initial bubble diameter, $D$, are shown. 
		The solid black lines illustrate the bubble's contour in the center cross-section with normal in the $x$-direction.
		The photograph of the laboratory experiment was reprinted from Reference~\cite{bhaga1981BubblesViscousLiquids} with the permission of Cambridge University Press.
	}
	\label{fig:bo-339-mo-43.1}
\end{figure}

\Cref{fig:bo-32.2-mo-8.2e-4,fig:bo-243-mo-266,fig:bo-339-mo-43.1} extend \Cref{par:rising-bubble-results} with simulation results for the rising bubble test cases 1, 3, and 4 from \Cref{tab:rising-bubble-setups}.\par

\begin{figure}[h!]
	\centering
	\begin{tabular}{>{\centering\arraybackslash}m{0.025\textwidth}
					>{\centering\arraybackslash}m{0.2\textwidth}
					>{\centering\arraybackslash}m{0.2\textwidth}
					>{\centering\arraybackslash}m{0.2\textwidth}
					>{\centering\arraybackslash}m{0.2\textwidth}}
		
		\rotatebox[origin=l]{90}{Experiment~\cite{bhaga1981BubblesViscousLiquids}}
		& \multicolumn{4}{c}{\makecell{\includegraphics[width=0.25\textwidth]{figures/rising-bubble/experiments/bo-32.2-mo-8.2e-4.png}}} \\
		
		\rotatebox[origin=l]{90}{FSLBM/LSQR} &
		\includegraphics[width=0.2\textwidth]{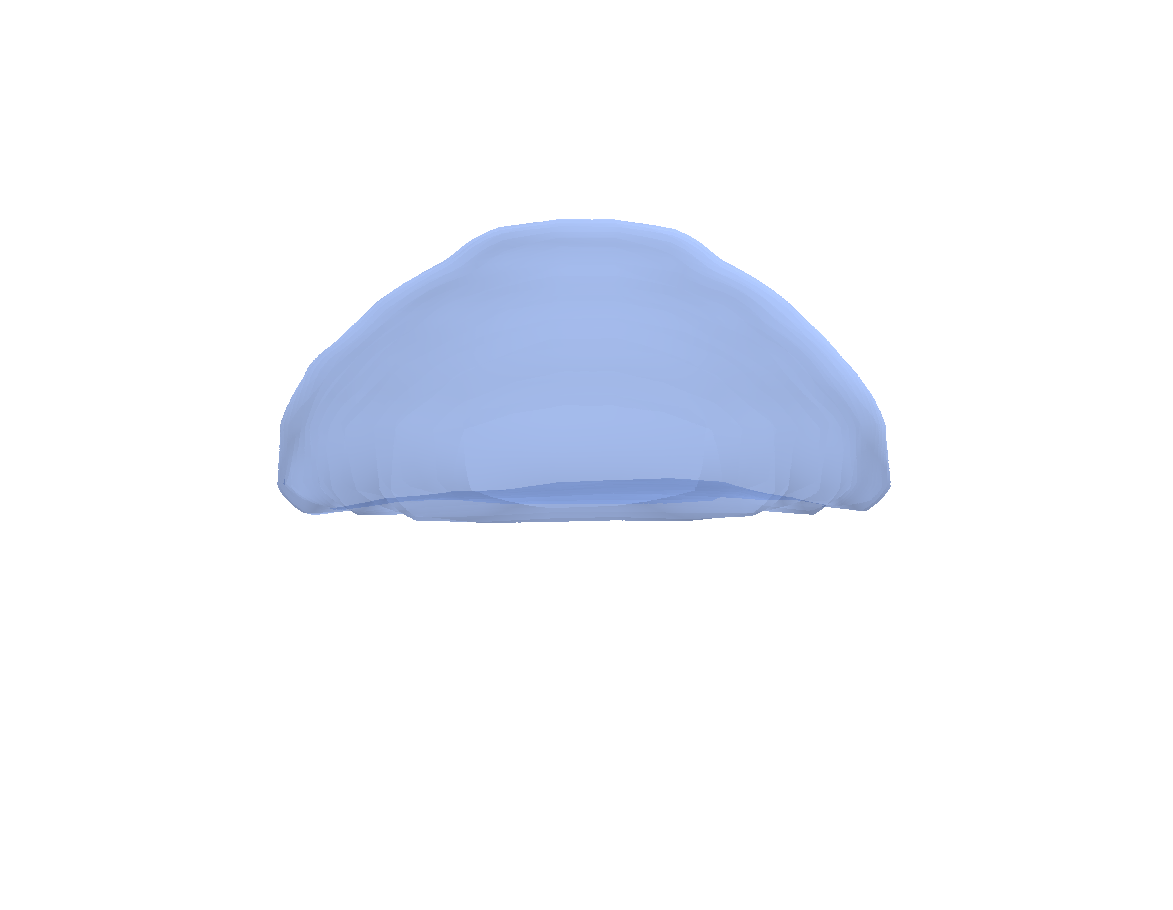} &
		\includegraphics[width=0.2\textwidth]{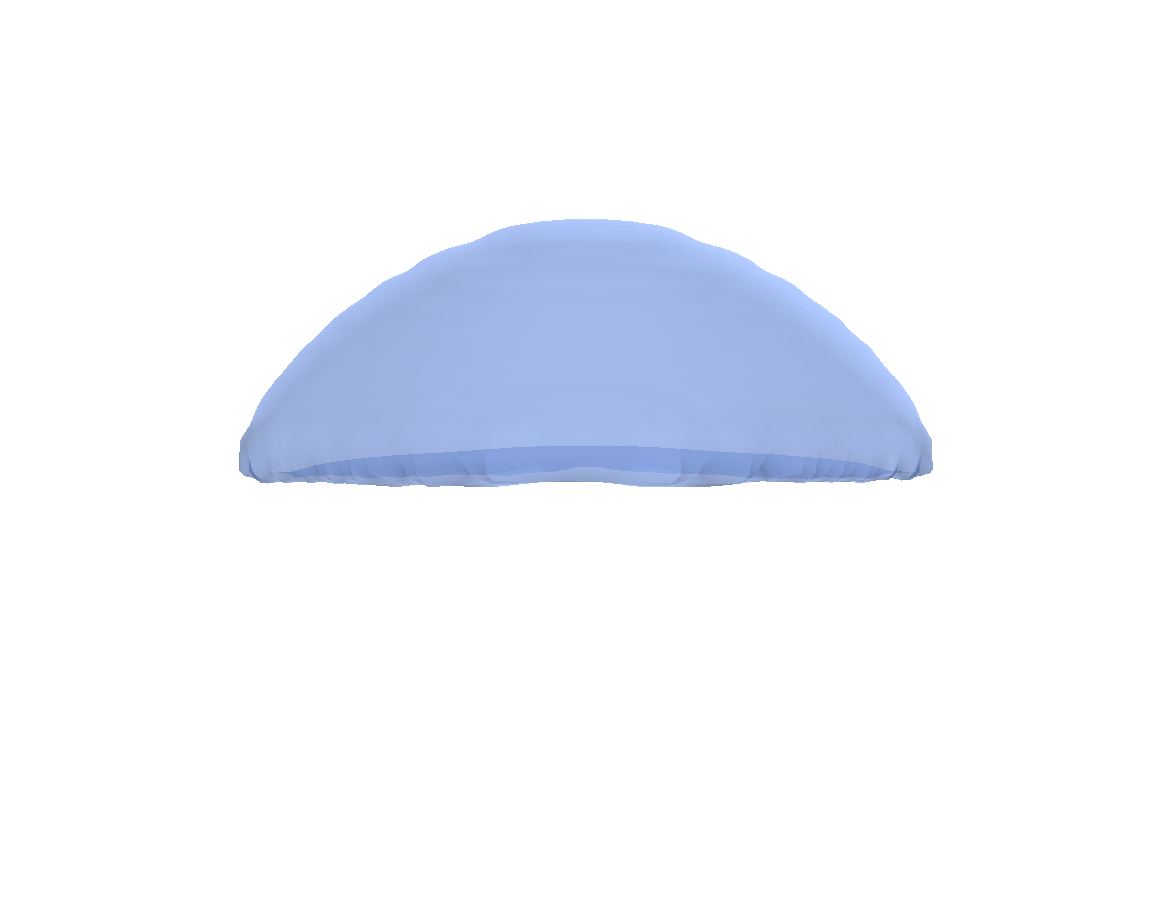} &
		\includegraphics[width=0.2\textwidth]{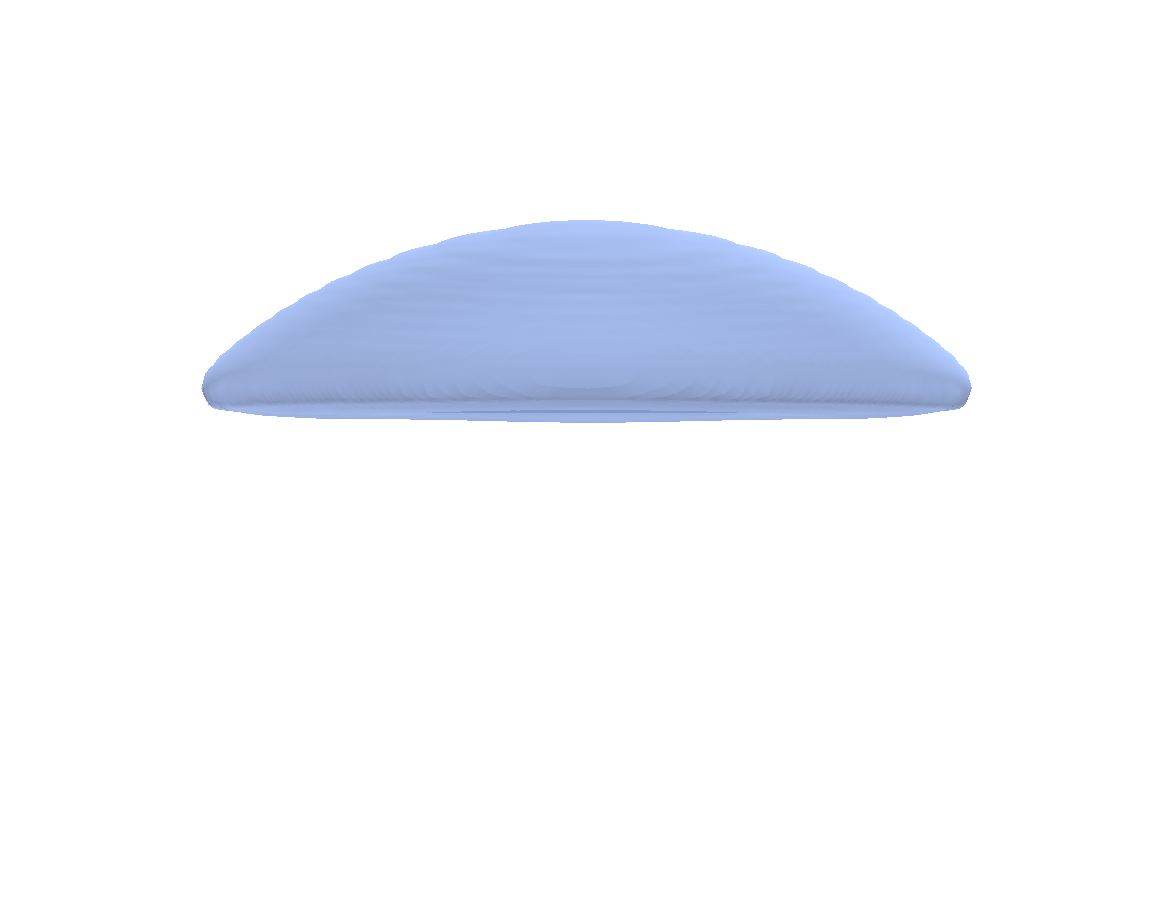} &
		\includegraphics[width=0.2\textwidth]{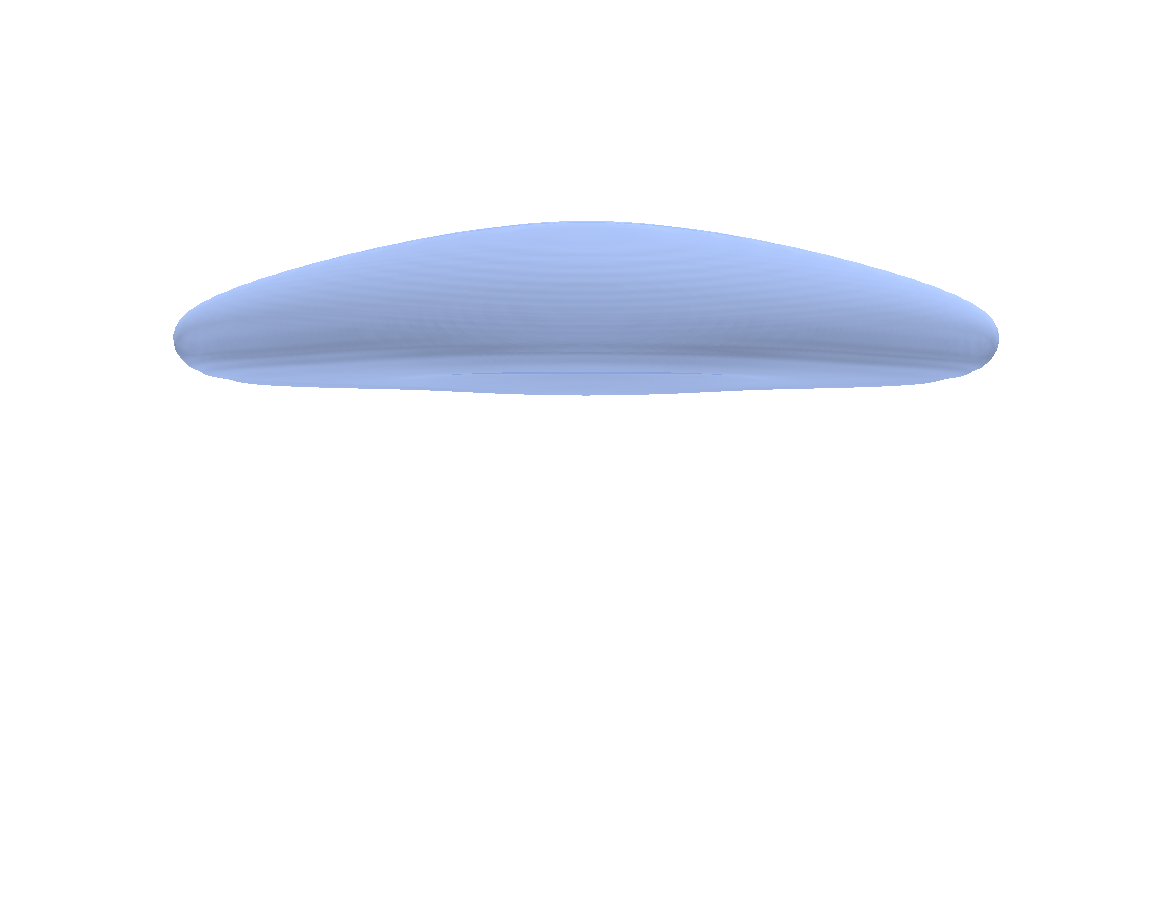} \\
		
		& $t^{*}=19.6$ & $t^{*}=16.5$ & $t^{*}=15.8$ & $t^{*}=8$\\
		& $D=8$ & $D=16$ & $D=32$ & $D=64$\\
	\end{tabular}
	\caption{
		Simulated bubble shape for case 1 in \Cref{tab:rising-bubble-setups} with $\mathrm{Bo}=32.2$ and $\mathrm{Mo}=8.2\cdot10^{-4}$.
		Different computational resolutions according to the initial bubble diameter, $D$, are shown.
		The simulations were performed with the FSLBM with LSQR curvature computation model implemented in FluidX3D~\cite{lehmann2021ejection, lehmann2022AnalyticSolutionPiecewise}.
		The photograph of the laboratory experiment was reprinted from Reference~\cite{bhaga1981BubblesViscousLiquids} with the permission of Cambridge University Press.
	}
	\label{fig:fluidx3d-bo-32.2-mo-8.2e-4}
\end{figure}

\begin{figure}[h!]
	\centering
	\begin{tabular}{>{\centering\arraybackslash}m{0.025\textwidth}
					>{\centering\arraybackslash}m{0.2\textwidth}
					>{\centering\arraybackslash}m{0.2\textwidth}
					>{\centering\arraybackslash}m{0.2\textwidth}
					>{\centering\arraybackslash}m{0.2\textwidth}}
			
		\rotatebox[origin=l]{90}{Experiment~\cite{bhaga1981BubblesViscousLiquids}}
		& \multicolumn{4}{c}{\makecell{\includegraphics[width=0.15\textwidth]{figures/rising-bubble/experiments/bo-115-mo-4.63e-3.png}}} \\
			
		\rotatebox[origin=l]{90}{FSLBM/LSQR} &
		\includegraphics[width=0.2\textwidth]{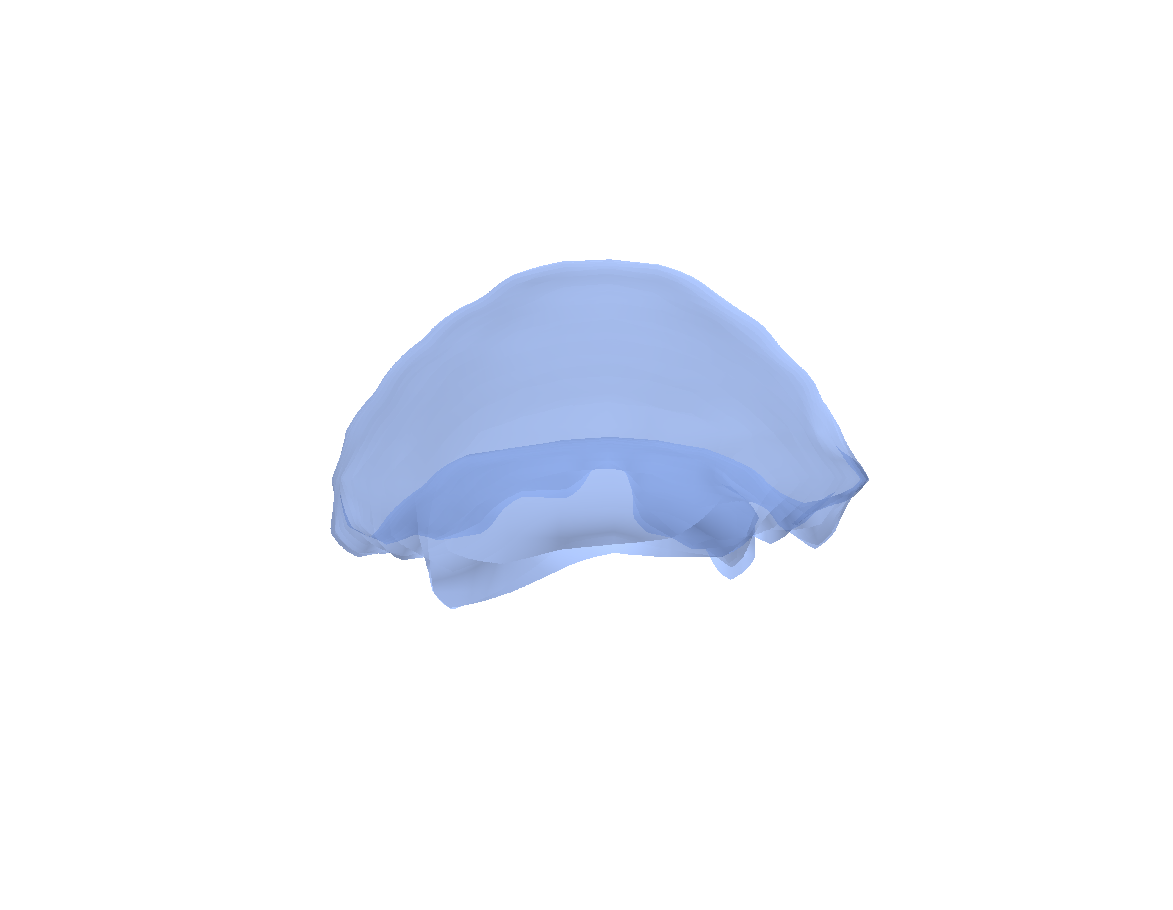} &
		\includegraphics[width=0.2\textwidth]{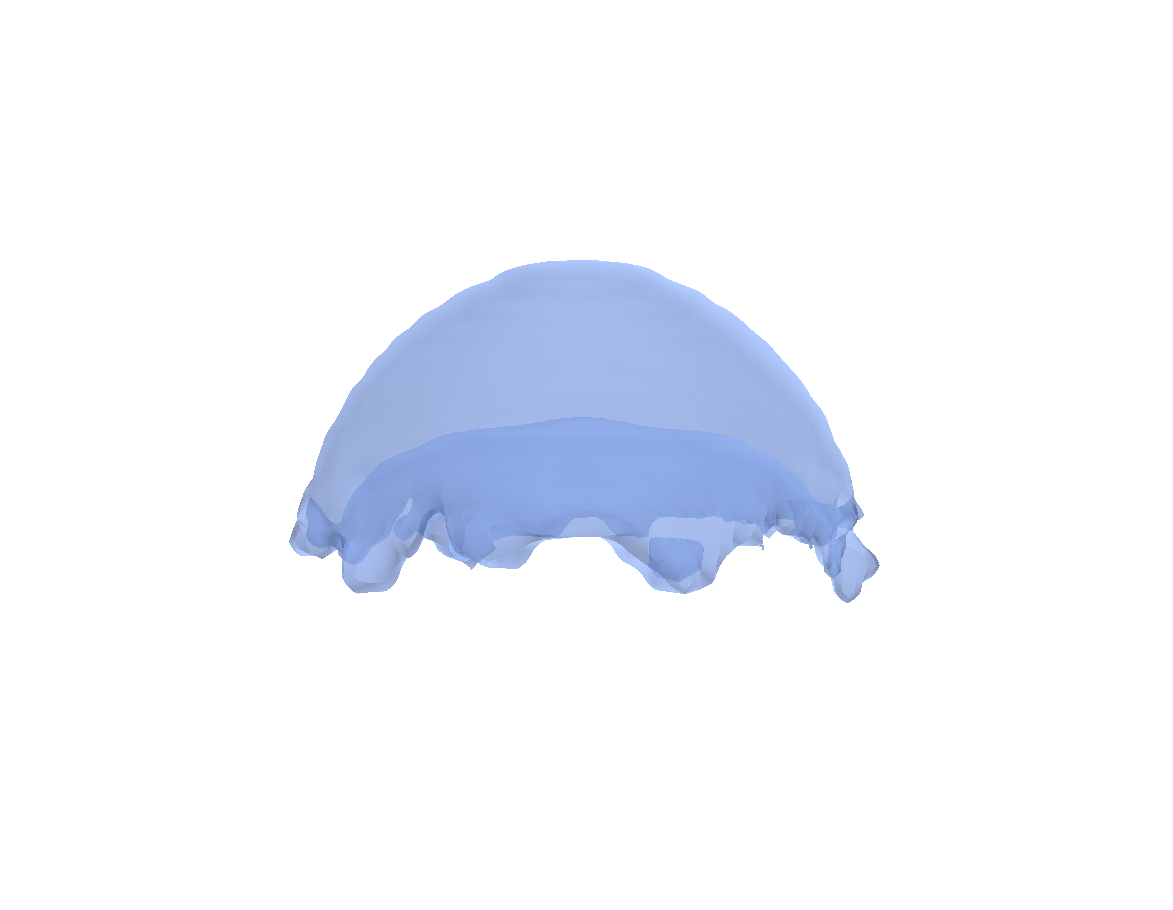} &
		\includegraphics[width=0.2\textwidth]{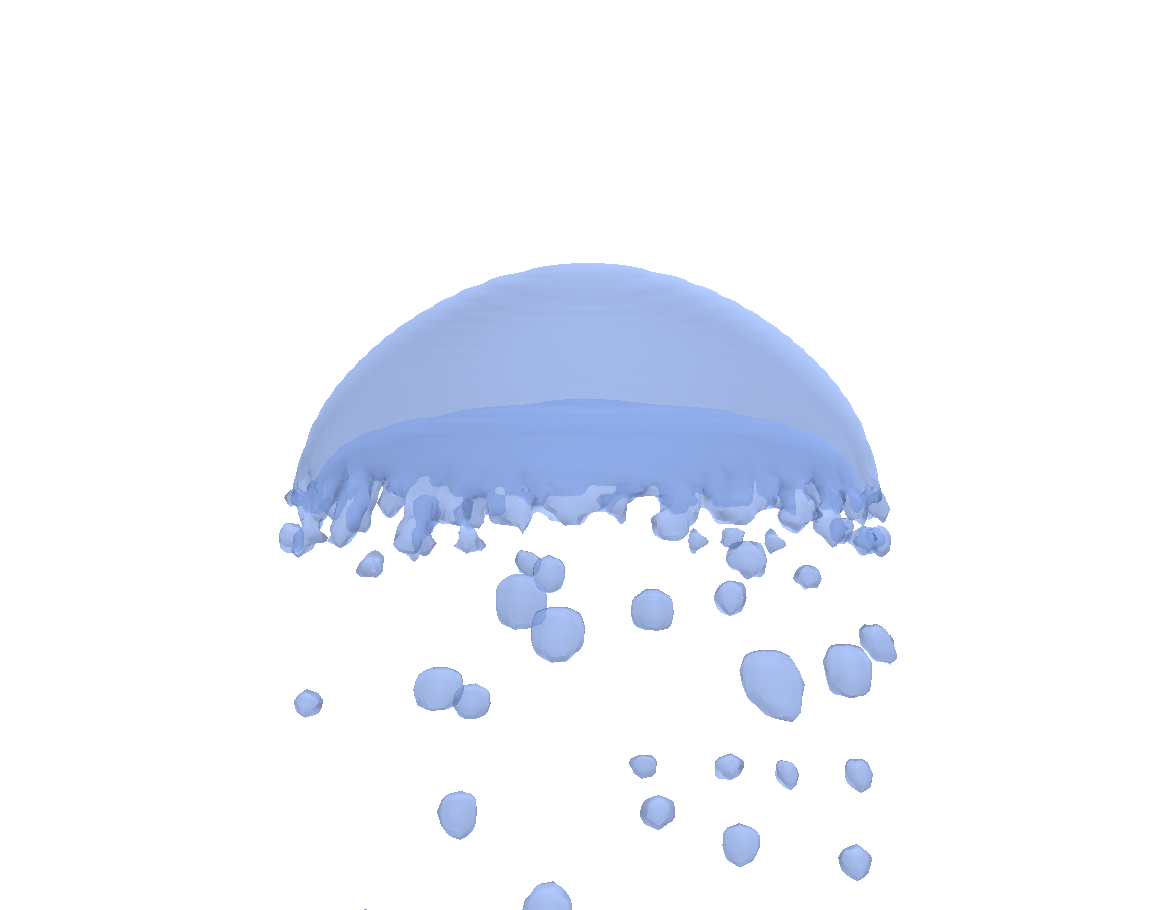} &
		\includegraphics[width=0.2\textwidth]{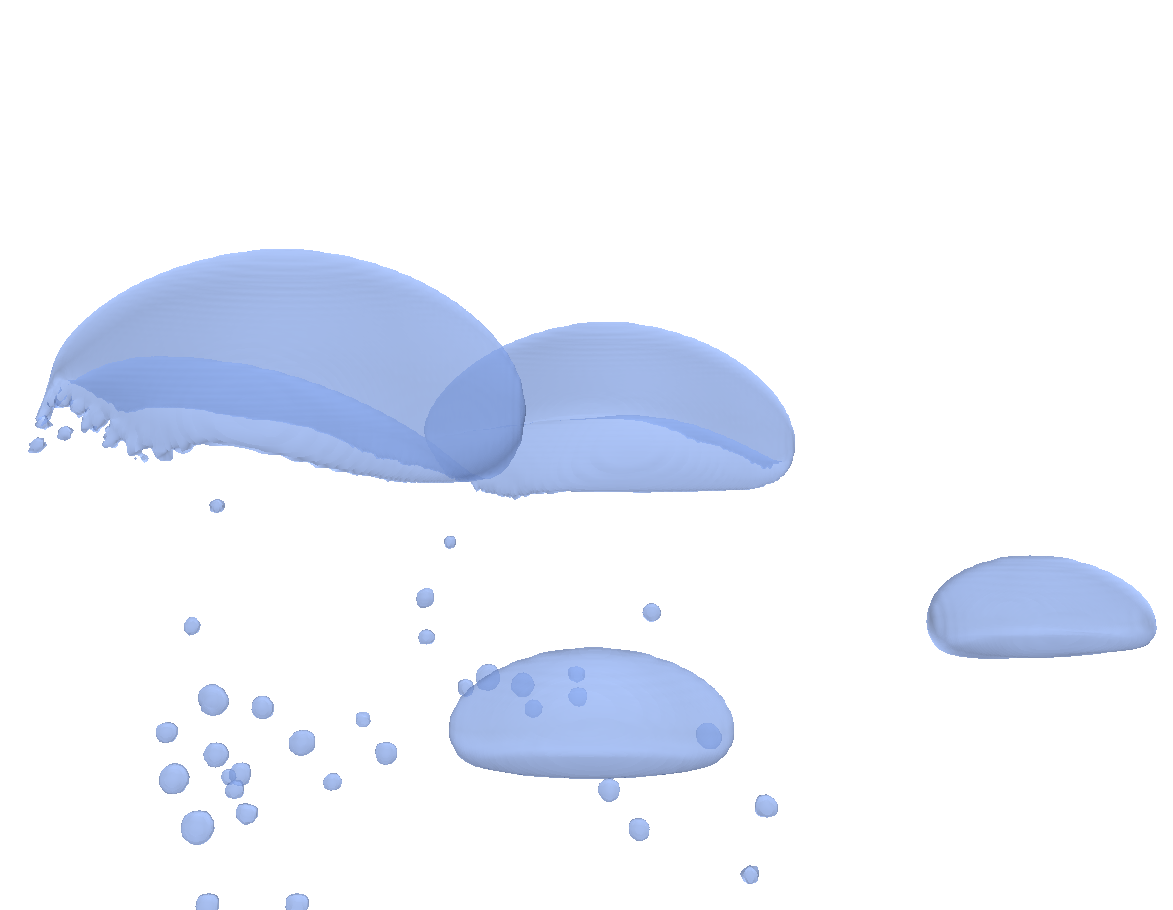} \\
		
		& $t^{*}=19.4$ & $t^{*}=15.8$ & $t^{*}=15.1$ & $t^{*}=8.6$\\
		& $D=8$ & $D=16$ & $D=32$ & $D=64$\\
	\end{tabular}
	\caption{
		Simulated bubble shape for case 2 in \Cref{tab:rising-bubble-setups} with $\mathrm{Bo}=115$ and $\mathrm{Mo}=4.63\cdot10^{-3}$.
		Different computational resolutions according to the initial bubble diameter, $D$, are shown.
		The simulations were performed with the FSLBM with LSQR curvature computation model implemented in FluidX3D~\cite{lehmann2021ejection, lehmann2022AnalyticSolutionPiecewise}.
		In contrast to the experiment, the bubble broke apart into several smaller bubbles in the simulation at a resolution of $D=64$.
		The photograph of the laboratory experiment was reprinted from Reference~\cite{bhaga1981BubblesViscousLiquids} with the permission of Cambridge University Press.
	}
	\label{fig:fluidx3d-bo-115-mo-4.63e-3}
\end{figure}

\begin{figure}[h!]
	\centering
	\begin{tabular}{>{\centering\arraybackslash}m{0.025\textwidth}
					>{\centering\arraybackslash}m{0.2\textwidth}
					>{\centering\arraybackslash}m{0.2\textwidth}
					>{\centering\arraybackslash}m{0.2\textwidth}
					>{\centering\arraybackslash}m{0.2\textwidth}}
			
		\rotatebox[origin=l]{90}{Experiment~\cite{bhaga1981BubblesViscousLiquids}}
		& \multicolumn{4}{c}{\makecell{\includegraphics[width=0.15\textwidth]{figures/rising-bubble/experiments/bo-243-mo-266.png}}} \\
		
		\rotatebox[origin=l]{90}{FSLBM/LSQR} &
		\includegraphics[width=0.2\textwidth]{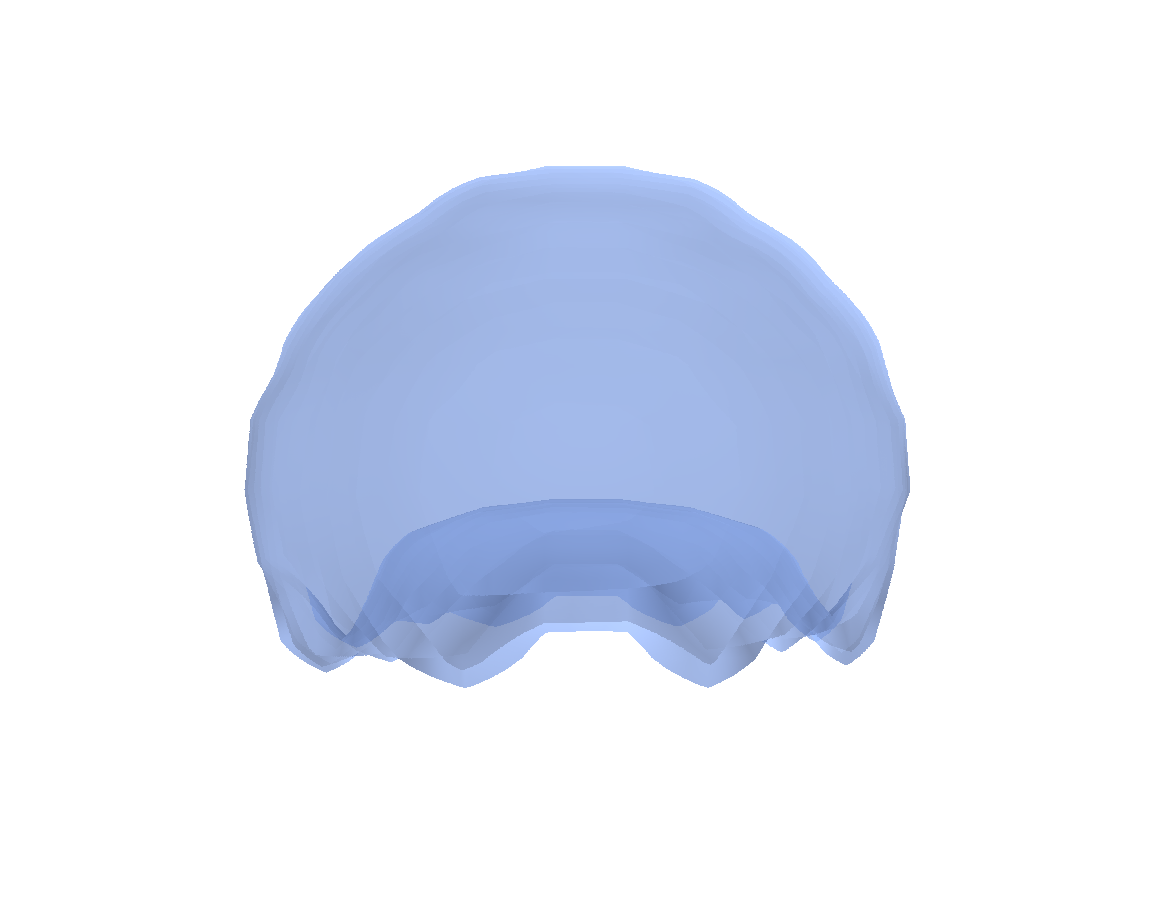} &
		\includegraphics[width=0.2\textwidth]{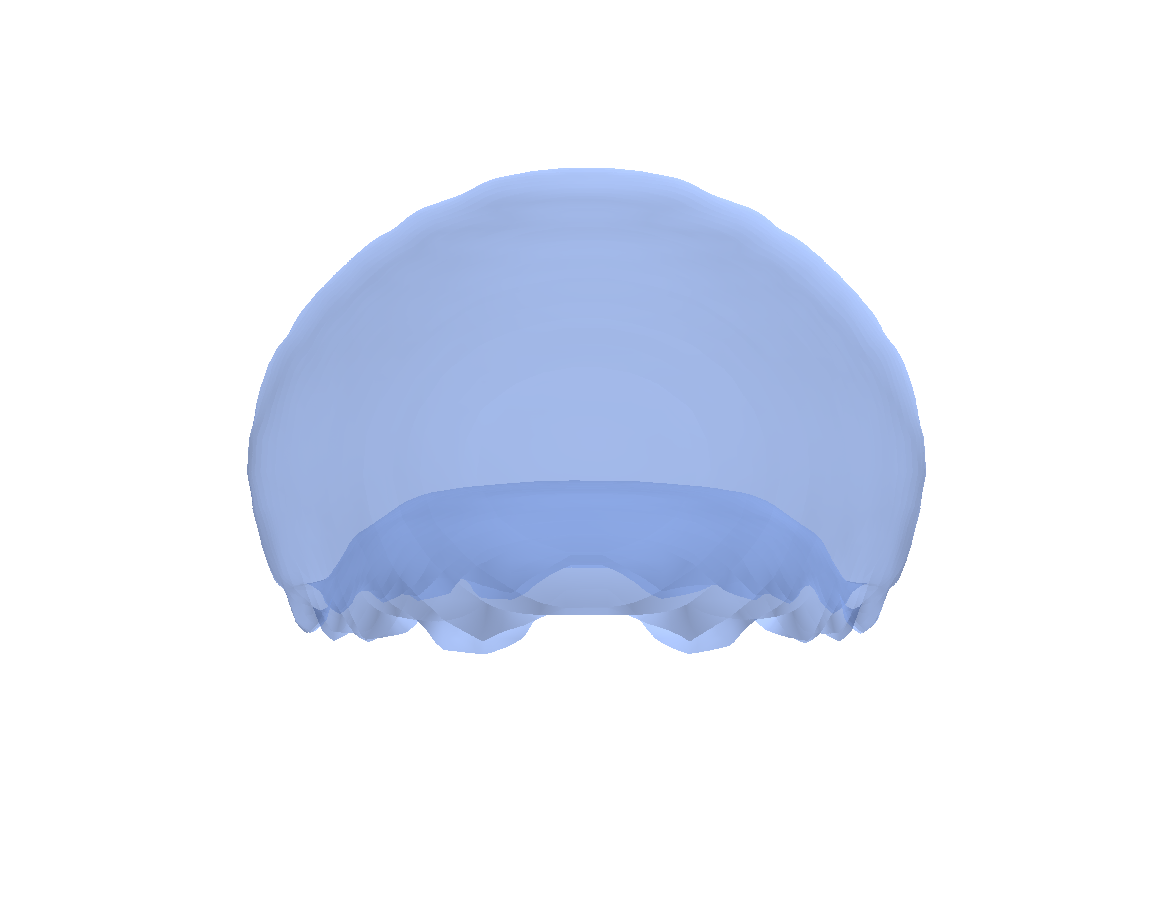} &
		\includegraphics[width=0.2\textwidth]{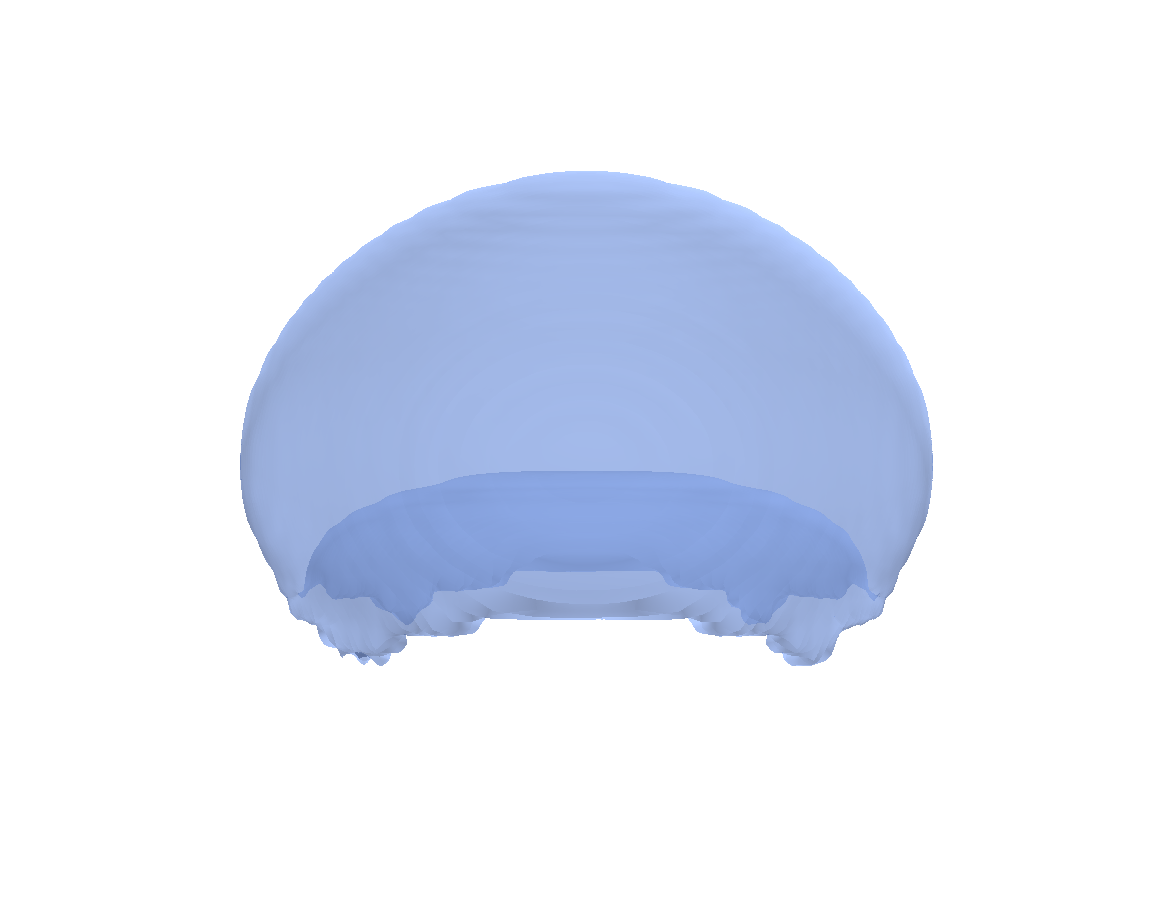} &
		\includegraphics[width=0.2\textwidth]{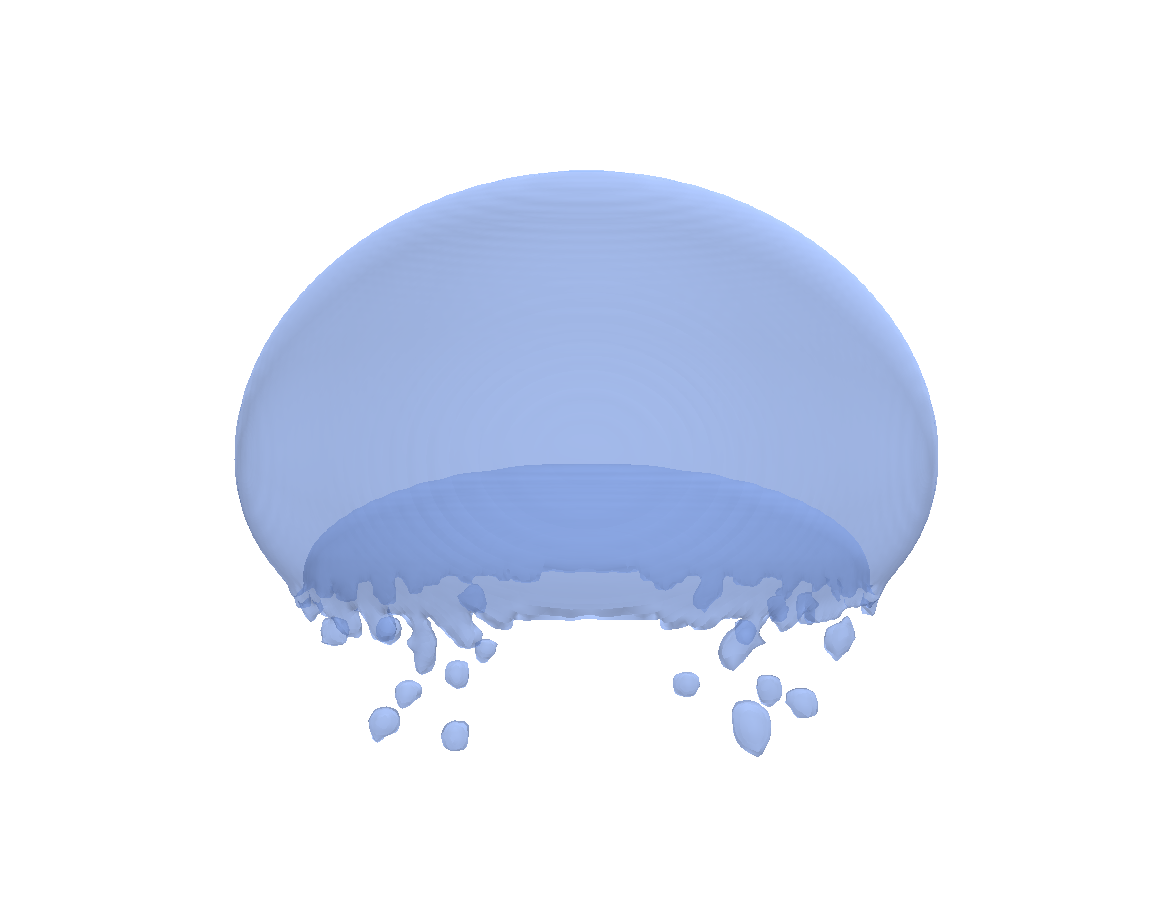} \\
		
		& $t^{*}=26.6$ & $t^{*}=21.4$ & $t^{*}=19.5$ & $t^{*}=9$\\
		& $D=8$ & $D=16$ & $D=32$ & $D=64$\\
	\end{tabular}
	\caption{
		Simulated bubble shape for case 3 in \Cref{tab:rising-bubble-setups} with $\mathrm{Bo}=243$ and $\mathrm{Mo}=266$.
		Different computational resolutions according to the initial bubble diameter, $D$, are shown.
		The simulations were performed with the FSLBM with LSQR curvature computation model implemented in FluidX3D~\cite{lehmann2021ejection, lehmann2022AnalyticSolutionPiecewise}.
		The photograph of the laboratory experiment was reprinted from Reference~\cite{bhaga1981BubblesViscousLiquids} with the permission of Cambridge University Press.
	}
	\label{fig:fluidx3d-bo-243-mo-266}
\end{figure}

\begin{figure}[h!]
	\centering
	\begin{tabular}{>{\centering\arraybackslash}m{0.025\textwidth}
					>{\centering\arraybackslash}m{0.2\textwidth}
					>{\centering\arraybackslash}m{0.2\textwidth}
					>{\centering\arraybackslash}m{0.2\textwidth}
					>{\centering\arraybackslash}m{0.2\textwidth}}
			
		\rotatebox[origin=l]{90}{Experiment~\cite{bhaga1981BubblesViscousLiquids}}
		& \multicolumn{4}{c}{\makecell{\includegraphics[width=0.15\textwidth]{figures/rising-bubble/experiments/bo-339-mo-43.1.png}}} \\	
			
		\rotatebox[origin=l]{90}{FSLBM/LSQR} &
		\includegraphics[width=0.2\textwidth]{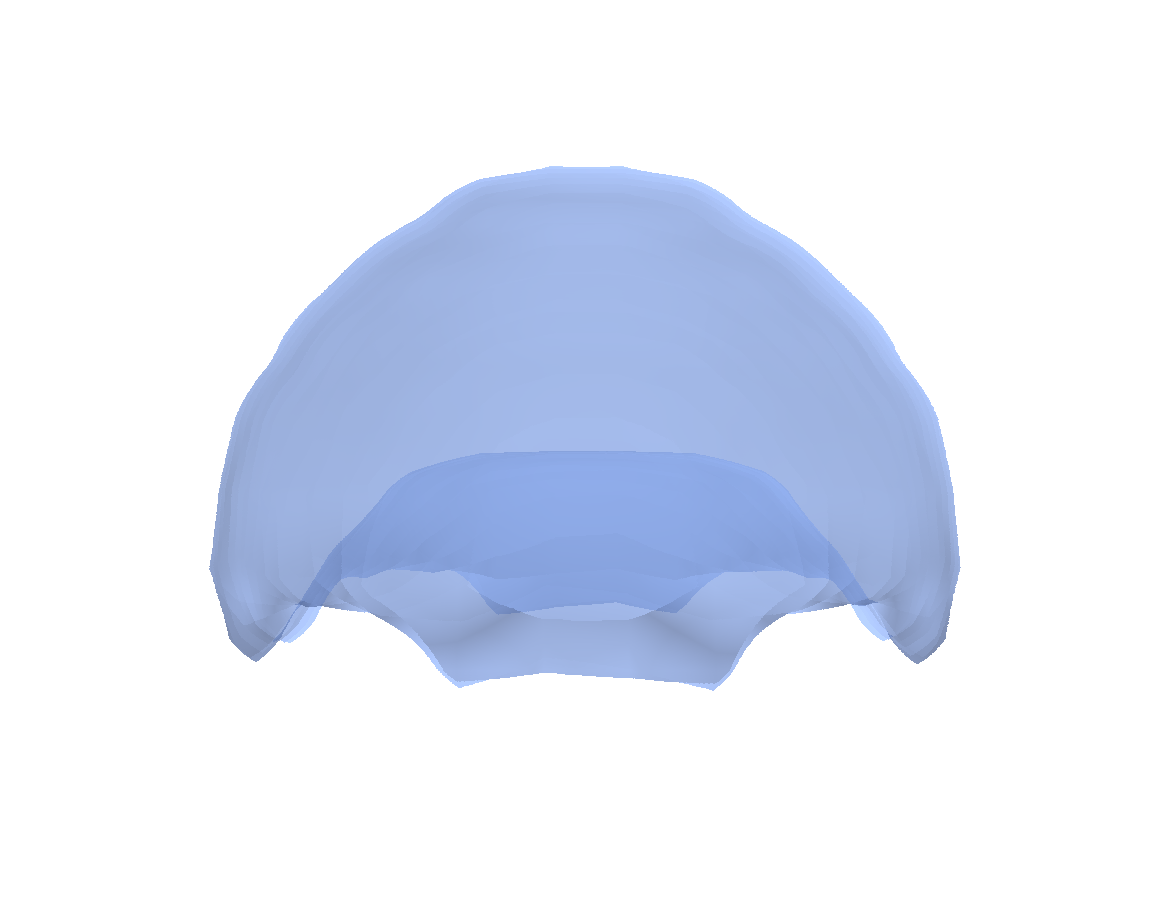} &
		\includegraphics[width=0.2\textwidth]{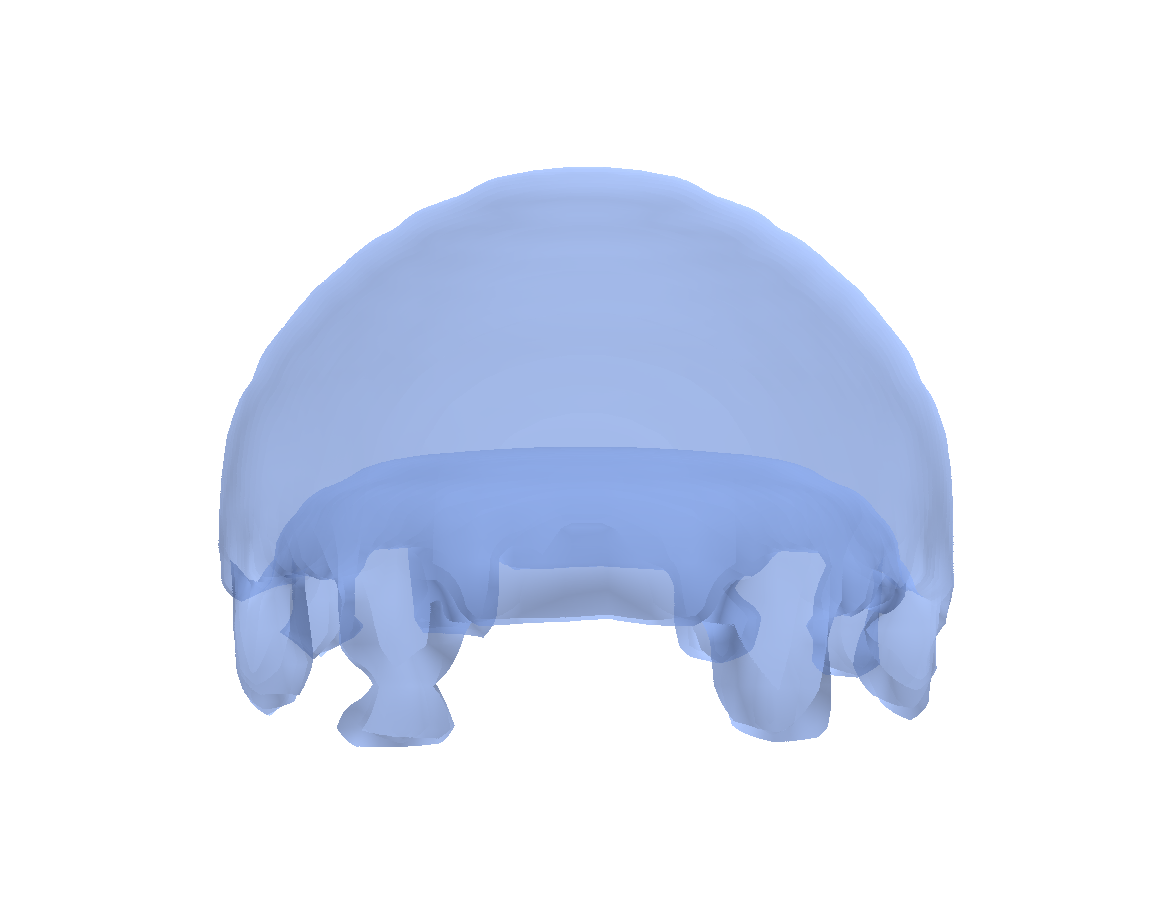} &
		\includegraphics[width=0.2\textwidth]{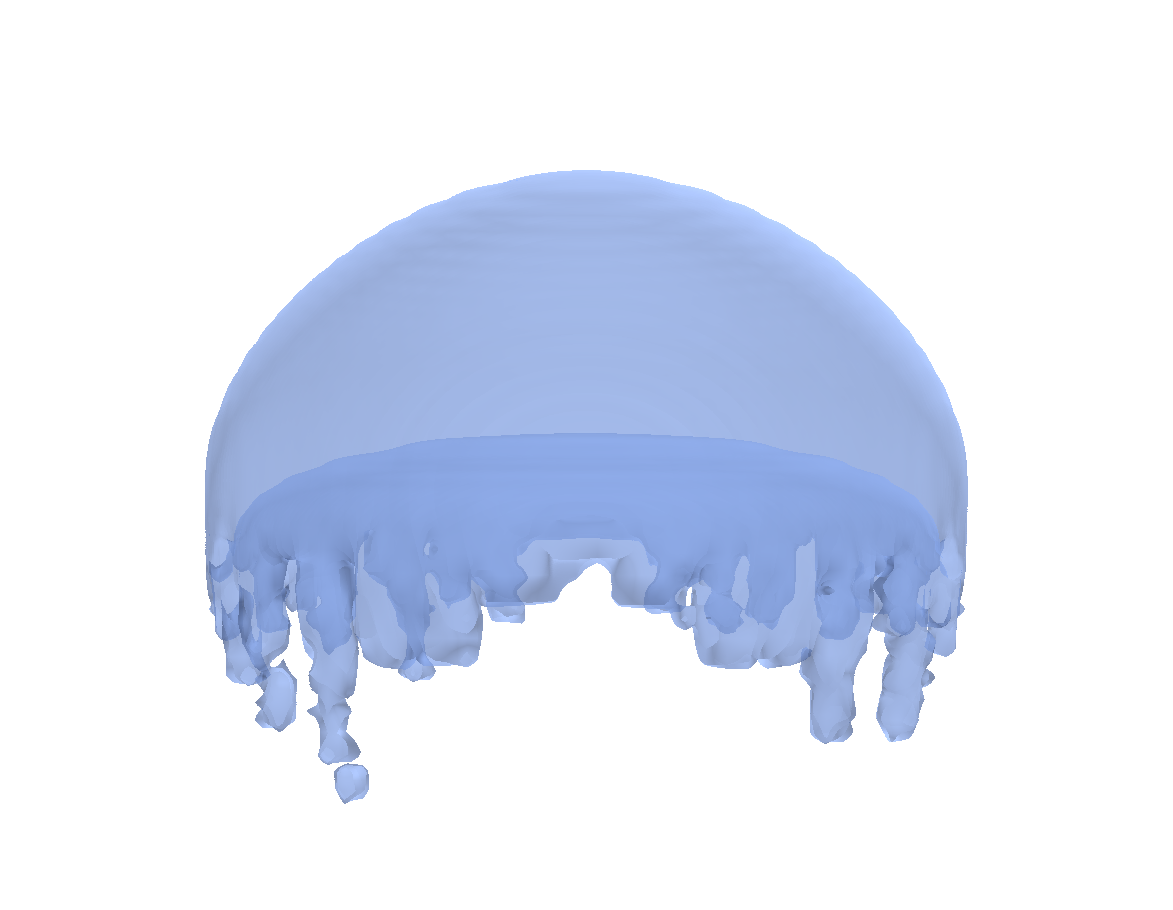} &
		\includegraphics[width=0.2\textwidth]{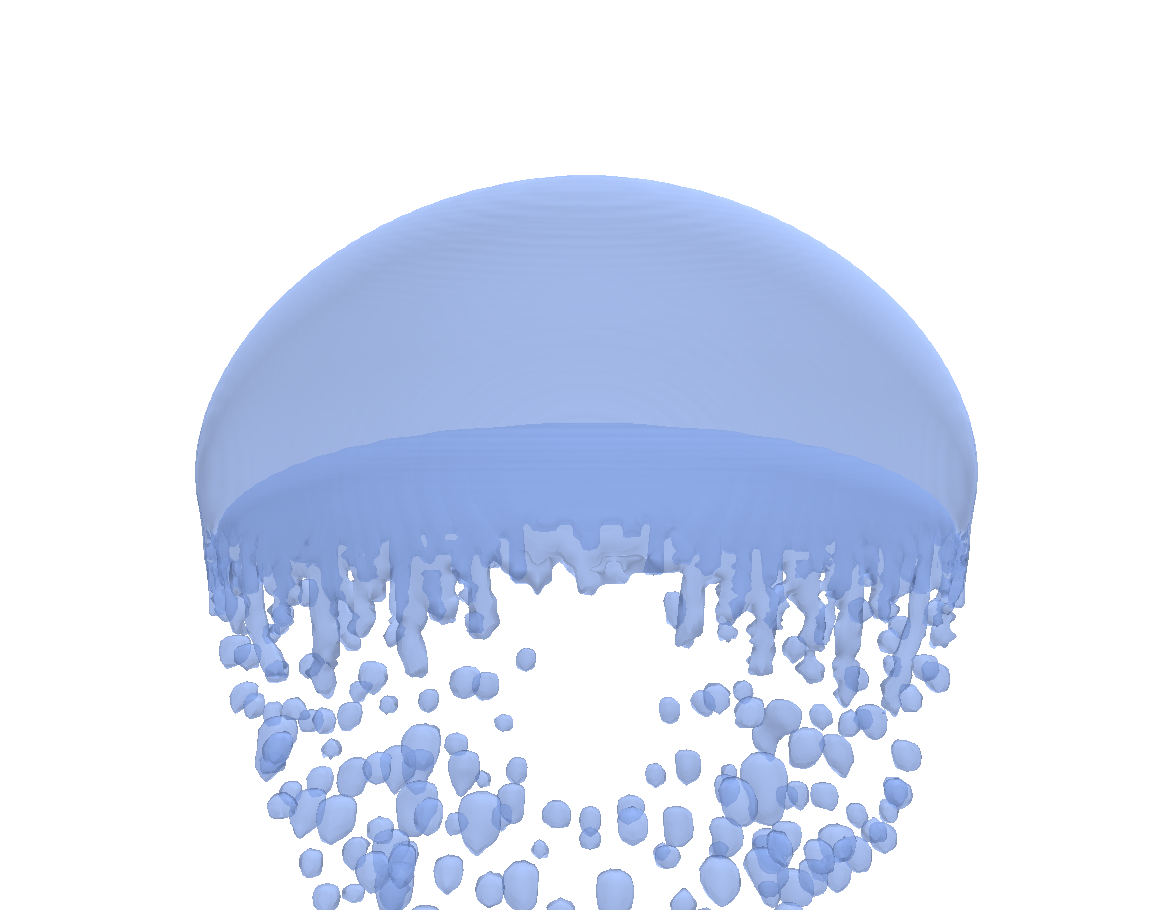} \\
		
		& $t^{*}=21.7$ & $t^{*}=18$ & $t^{*}=16.4$ & $t^{*}=7.7$\\
		& $D=8$ & $D=16$ & $D=32$ & $D=64$\\
	\end{tabular}
	\caption{
		Simulated bubble shape for case 4 in \Cref{tab:rising-bubble-setups} with $\mathrm{Bo}=339$ and $\mathrm{Mo}=43.1$.
		The simulations were performed with the FSLBM with LSQR curvature computation model implemented in FluidX3D~\cite{lehmann2021ejection, lehmann2022AnalyticSolutionPiecewise}.
		Different computational resolutions according to the initial bubble diameter, $D$, are shown.
		The photograph of the laboratory experiment was reprinted from Reference~\cite{bhaga1981BubblesViscousLiquids} with the permission of Cambridge University Press.
	}
	\label{fig:fluidx3d-bo-339-mo-43.1}
\end{figure}

Additionally, simulation results for the FSLBM with the LSQR curvature computation model as described in \Cref{par:cap-wave-results}, are presented in \Cref{fig:fluidx3d-bo-32.2-mo-8.2e-4,fig:fluidx3d-bo-115-mo-4.63e-3,fig:fluidx3d-bo-243-mo-266,fig:fluidx3d-bo-339-mo-43.1}.
The simulations were conducted with parameters as in \Cref{tab:rising-bubble-setups} using the software FluidX3D~\cite{lehmann2021ejection,lehmann2022AnalyticSolutionPiecewise}.
The results agree reasonably well with those of the FSLBM with FDM curvature computation model as presented in \Cref{fig:bo-32.2-mo-8.2e-4,fig:bo-115-mo-4.63e-3,fig:bo-243-mo-266,fig:bo-339-mo-43.1}, and therefore also with the experimental results.
However, as shown in \Cref{fig:fluidx3d-bo-115-mo-4.63e-3}, with the LSQR curvature model, the bubble broke apart into several smaller bubbles at resolution $D=64$ in case 2 with $\mathrm{Bo}=115$ and $\mathrm{Mo}=4.63\cdot10^{-3}$.

\FloatBarrier

\subsection{Taylor bubble} \label{subsec:app-taylor-bubble}
\begin{figure}[h!]
	\centering
	\begin{subfigure}[b]{0.16\textwidth}
		\centering
		\vskip 0pt 
		\includegraphics[width=0.6\textwidth]{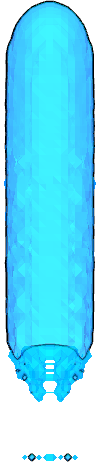}
		\caption*{FSLBM,\\ $D=16$}
	\end{subfigure}
	\hfill
	\begin{subfigure}[b]{0.16\textwidth}
		\centering
		\vskip 0pt 
		\includegraphics[width=0.6\textwidth]{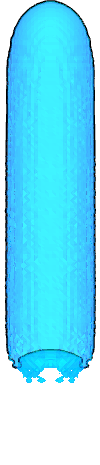}
		\caption*{FSLBM,\\ $D=32$}
	\end{subfigure}
	\hfill
	\begin{subfigure}[b]{0.16\textwidth}
		\centering
		\vskip 0pt 
		\includegraphics[width=0.6\textwidth]{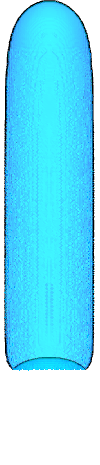}
		\caption*{FSLBM,\\ $D=64$}
	\end{subfigure}
	\hfill
	\begin{subfigure}[b]{0.16\textwidth}
		\centering
		\vskip 0pt 
		\includegraphics[width=0.6\textwidth]{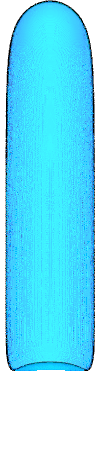}
		\caption*{FSLBM,\\ $D=128$}
	\end{subfigure}
	\hfill
	\begin{subfigure}[b]{0.16\textwidth}
		\centering
		\vskip 0pt 
		\includegraphics[width=0.6\textwidth]{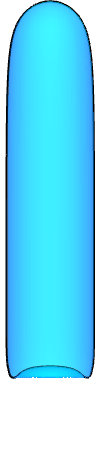}
		\caption*{PFLBM,\\ $D=64$}
	\end{subfigure}
	\hfill
	\begin{subfigure}[b]{0.16\textwidth}
		\centering
		\vskip 0pt 
		\includegraphics[width=0.6\textwidth]{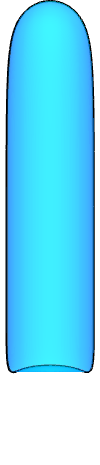}
		\caption*{PFLBM,\\ $D=128$}
	\end{subfigure}
	\caption{
		Shape of the Taylor bubble at time, $t^{*}=15$, as simulated with the FSLBM and PFLBM at different computational resolutions, defined by tube diameter, $D$.
		The solid black line illustrates the bubble's contour in the center cross-section with normal in the $x$-direction.
	}
	\label{fig:taylor-bubble-shape-mesh}
\end{figure}

\begin{figure}[h!]
	\centering
	\setlength{\figureheight}{0.5\textwidth}
	\setlength{\figurewidth}{0.8\textwidth}
	\input{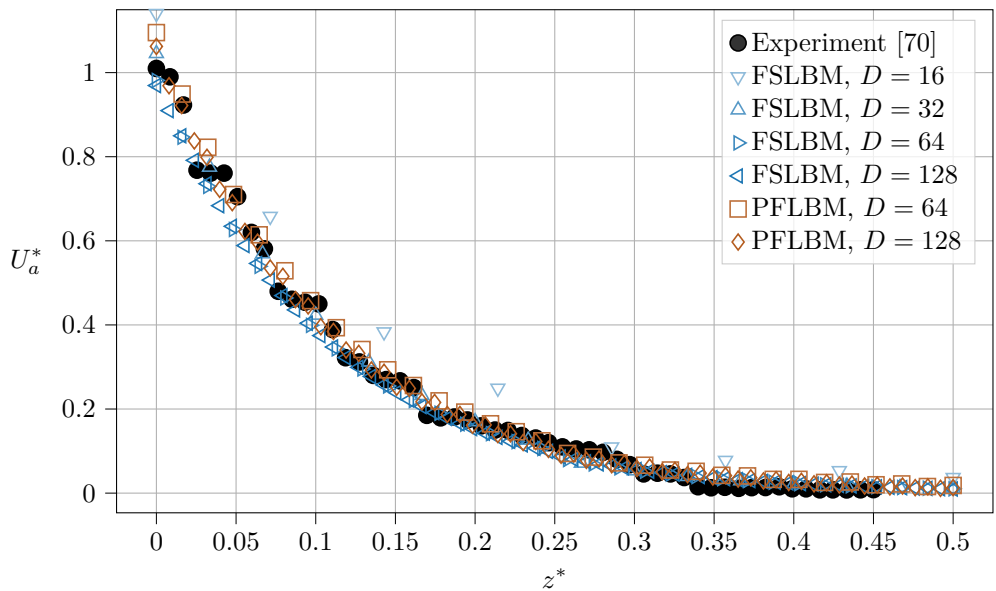}%
	\caption{
		Simulated non-dimensionalized axial velocity, $U^{*}_{a}$, along an axial line of length $0.5D$ in front of the Taylor bubble (cf.\ \Cref{fig:taylor-bubble-schematic}).
		The line is located in the center of the boundary tube with diameter, $D$.
		The comparison with experimental data~\cite{bugg2002VelocityFieldTaylor} is drawn in terms of the non-dimensionalized axial location, $z^{*}=z/D$, at time, $t^{*}=15$.
	}
	\label{fig:taylor-bubble-axial}
\end{figure}

\begin{figure}[h!]
	\centering
	\setlength{\figureheight}{0.5\textwidth}
	\setlength{\figurewidth}{0.8\textwidth}
\begin{tikzpicture}

\definecolor{darkgray176}{RGB}{176,176,176}
\definecolor{lightgray204}{RGB}{204,204,204}
\definecolor{peru18911259}{RGB}{189,112,59}
\definecolor{sienna1809231}{RGB}{180,92,31}
\definecolor{skyblue143187218}{RGB}{143,187,218}
\definecolor{steelblue31119180}{RGB}{31,119,180}
\definecolor{steelblue59136189}{RGB}{59,136,189}
\definecolor{steelblue87153199}{RGB}{87,153,199}

\begin{axis}[
height=\figureheight,
legend cell align={left},
legend style={
  fill opacity=0.8,
  draw opacity=1,
  text opacity=1,
  at={(0.03,0.97)},
  anchor=north west,
  draw=lightgray204
},
tick align=outside,
tick pos=left,
width=\figurewidth,
x grid style={darkgray176},
xlabel={\(\displaystyle r^{*}\)},
xmajorgrids,
xmin=0.75, xmax=1.01,
xtick style={color=black},
y grid style={darkgray176},
ylabel style={rotate=-90.0},
ylabel={\(\displaystyle U_{a}^{*}\)},
ymajorgrids,
ymin=-2.68132084561161, ymax=-0.0110321746373764,
ytick style={color=black}
]
\addplot [semithick, black, mark=*, mark size=3, mark options={solid}, only marks]
table {%
0.779999971389771 -2.16319990158081
0.796999931335449 -2.12409996986389
0.812999963760376 -2.0890998840332
0.829999923706055 -1.99909996986389
0.845999956130981 -1.81369996070862
0.86299991607666 -1.7849999666214
0.879999995231628 -1.64610004425049
0.896000027656555 -1.40359997749329
0.912999987602234 -1.15820002555847
0.929000020027161 -1.04789996147156
0.945999979972839 -0.782799959182739
0.963000059127808 -0.411499977111816
0.978999972343445 -0.405900001525879
};
\addlegendentry{Experiment~\cite{bugg2002VelocityFieldTaylor}}
\addplot [semithick, skyblue143187218, mark=triangle, mark size=3, mark options={solid,rotate=180,fill opacity=0}, only marks]
table {%
0.8125 -2.42596220970154
1 -1.05700659751892
};
\addlegendentry{FSLBM, $D=16$}
\addplot [semithick, steelblue87153199, mark=triangle, mark size=3, mark options={solid,fill opacity=0}, only marks]
table {%
0.78125 -2.39767408370972
0.854166865348816 -2.0916485786438
0.927083134651184 -1.47393107414246
1 -0.548947811126709
};
\addlegendentry{FSLBM, $D=32$}
\addplot [semithick, steelblue59136189, mark=triangle, mark size=3, mark options={solid,rotate=270,fill opacity=0}, only marks]
table {%
0.765625 -2.28672909736633
0.799107193946838 -2.21283745765686
0.832589387893677 -2.06480860710144
0.866071581840515 -1.8455194234848
0.899553418159485 -1.55630588531494
0.933035612106323 -1.19833755493164
0.966517806053162 -0.770538806915283
1 -0.270428538322449
};
\addlegendentry{FSLBM, $D=64$}
\addplot [semithick, steelblue31119180, mark=triangle, mark size=3, mark options={solid,rotate=90,fill opacity=0}, only marks]
table {%
0.7578125 -2.20958638191223
0.773958206176758 -2.19142127037048
0.790104150772095 -2.15509653091431
0.806249976158142 -2.10124754905701
0.822395801544189 -2.02944684028625
0.838541507720947 -1.94004201889038
0.854687452316284 -1.83373749256134
0.870833277702332 -1.71063494682312
0.886979222297668 -1.57067549228668
0.903125047683716 -1.41432905197144
0.919270277023315 -1.24187326431274
0.935417175292969 -1.05325746536255
0.951562523841858 -0.848390817642212
0.967707872390747 -0.627115726470947
0.9838547706604 -0.388885140419006
1 -0.132942318916321
};
\addlegendentry{FSLBM, $D=128$}
\addplot [semithick, peru18911259, mark=square, mark size=3, mark options={solid,fill opacity=0}, only marks]
table {%
0.765625 -2.55994415283203
0.799107193946838 -2.37838459014893
0.832589387893677 -2.17705321311951
0.866071581840515 -1.93294417858124
0.899553418159485 -1.63267552852631
0.933035612106323 -1.27110648155212
0.966517806053162 -0.845204830169678
1 -0.312628984451294
};
\addlegendentry{PFLBM, $D=64$}
\addplot [semithick, sienna1809231, mark=diamond, mark size=3, mark options={solid,fill opacity=0}, only marks]
table {%
0.7578125 -2.31654214859009
0.773958206176758 -2.26850485801697
0.790104150772095 -2.2168710231781
0.806249976158142 -2.15477323532104
0.822395801544189 -2.07805061340332
0.838541507720947 -1.98505771160126
0.854687452316284 -1.8754608631134
0.870833277702332 -1.74939966201782
0.886979222297668 -1.60717916488647
0.903125047683716 -1.44911336898804
0.919270277023315 -1.27547240257263
0.935417175292969 -1.08645689487457
0.951562523841858 -0.882105946540833
0.967707872390747 -0.662240982055664
0.9838547706604 -0.426299571990967
1 -0.153952598571777
};
\addlegendentry{PFLBM, $D=128$}
\end{axis}

\end{tikzpicture}%
	\caption{
		Simulated non-dimensionalized axial velocity, $U^{*}_{a}$, along a radial line positioned at $2D$ behind the Taylor bubble's front (cf.\ \Cref{fig:taylor-bubble-schematic}), with tube diameter, $D$.
		The comparison with experimental data~\cite{bugg2002VelocityFieldTaylor} is drawn in terms of the non-dimensionalized radial location, $r^{*}=r/(0.5D)$, at time, $t^{*}=15$.
	}
	\label{fig:taylor-bubble-axial-2}
\end{figure}
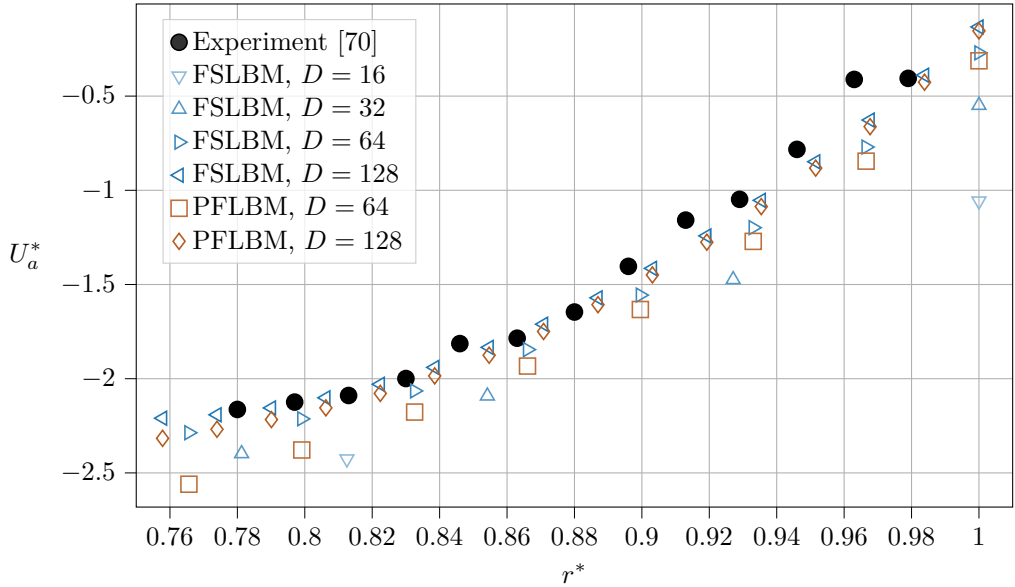

\Cref{fig:taylor-bubble-shape-mesh,fig:taylor-bubble-axial,fig:taylor-bubble-axial-0.111,fig:taylor-bubble-axial-2} extend \Cref{par:taylor-bubble-results} with the shape of the Taylor bubble and additional evaluations of the flow field in the surrounding fluid according to \Cref{fig:taylor-bubble-schematic}.

\begin{figure}[h!]
	\centering
	\setlength{\figureheight}{0.5\textwidth}
	\setlength{\figurewidth}{0.8\textwidth}
	\input{figures/taylor-bubble/axial-0.111.tex}%
	\caption{
		Simulated non-dimensionalized axial velocity, $U^{*}_{a}$, along a radial line positioned at $0.111D$ in front of the Taylor bubble (cf.\ \Cref{fig:taylor-bubble-schematic}), with tube diameter, $D$.
		The comparison with experimental data~\cite{bugg2002VelocityFieldTaylor} is drawn in terms of the non-dimensionalized radial location, $r^{*}=r/(0.5D)$, at time, $t^{*}=15$.
	}
	\label{fig:taylor-bubble-axial-0.111}
\end{figure}

\FloatBarrier

\subsection{Vertical drop impact} \label{subsec:app-drop-vertical}
\begin{figure}[h!]
	\centering
	\begin{tabular}{>{\centering\arraybackslash}m{0.025\textwidth}
			>{\centering\arraybackslash}m{0.3\textwidth}
			>{\centering\arraybackslash}m{0.3\textwidth}
			>{\centering\arraybackslash}m{0.3\textwidth}}
		\rotatebox[origin=l]{90}{Experiment~\cite{wang2000SplashingImpactSingle}}
		& \multicolumn{3}{c}{\makecell{\includegraphics[width=0.25\textwidth]{figures/drop-impact/vertical/experiment/t-12.png}}} \\
		
		\rotatebox[origin=l]{90}{$M=0.1$} &		
		\includegraphics[width=0.3\textwidth]{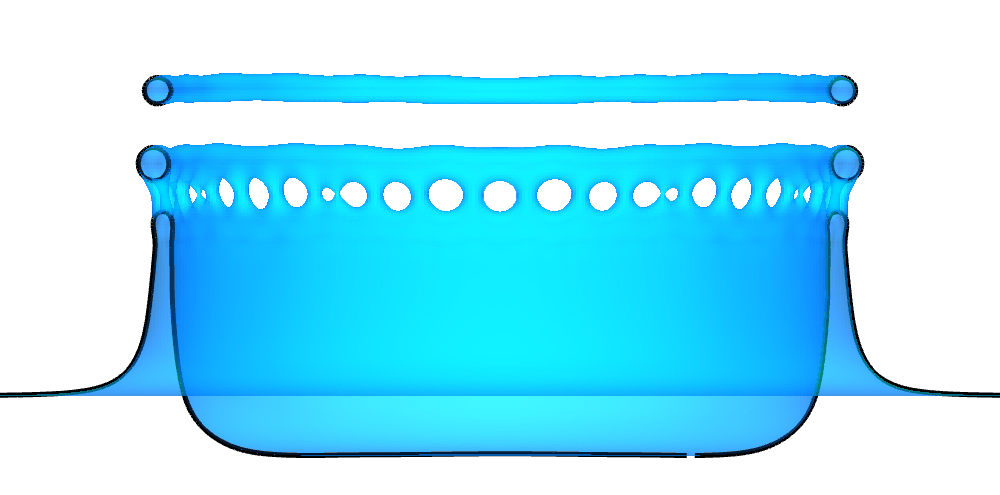} &
		\includegraphics[width=0.3\textwidth]{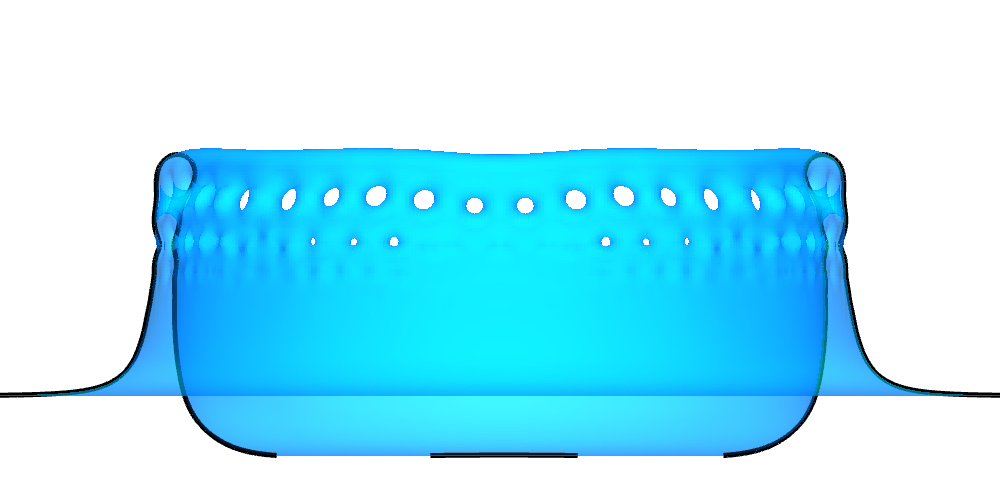} & 
		\includegraphics[width=0.3\textwidth]{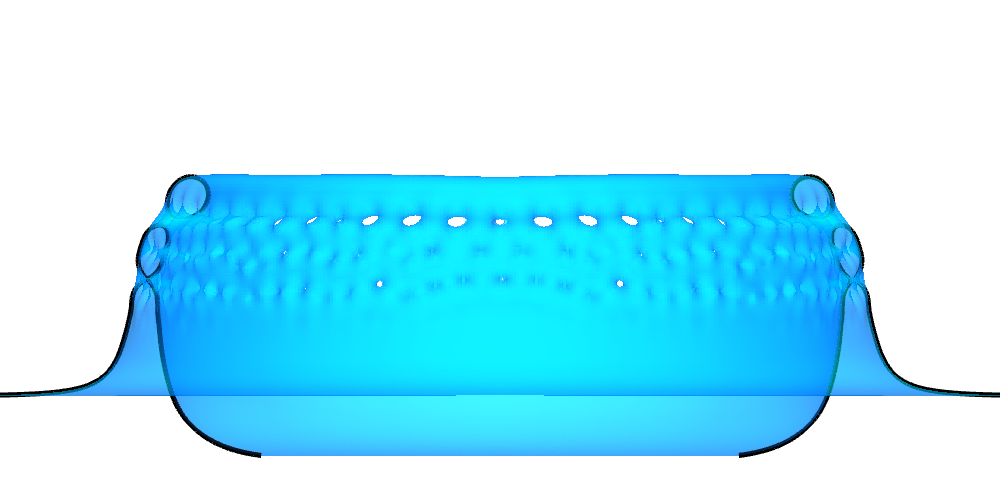} \\
		
		\rotatebox[origin=l]{90}{$M=0.05$} &	
		\includegraphics[width=0.3\textwidth]{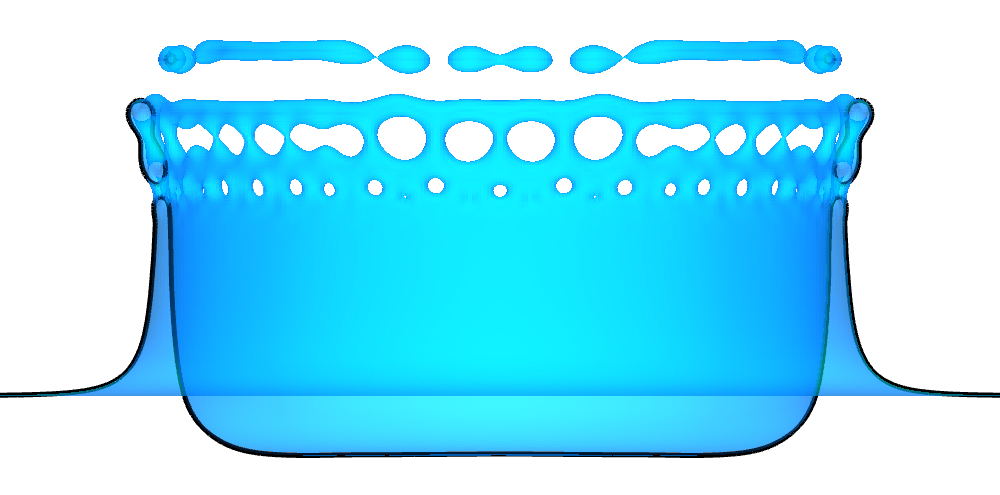} &
		\includegraphics[width=0.3\textwidth]{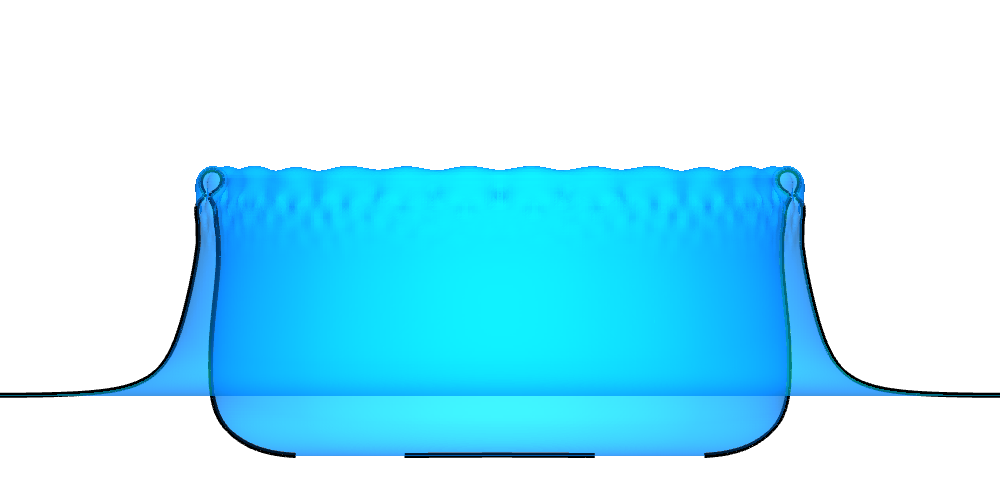} & 
		\includegraphics[width=0.3\textwidth]{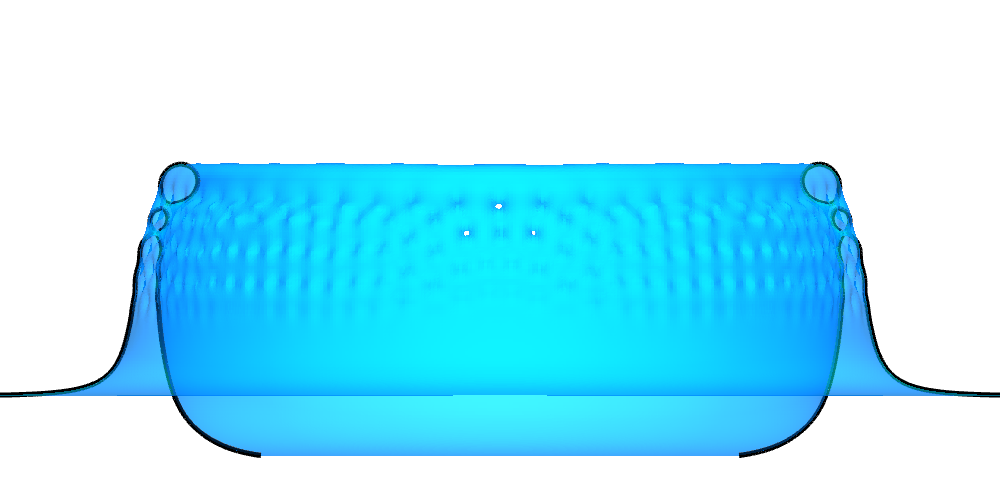} \\
		
		\rotatebox[origin=l]{90}{$M=0.03$} &	
		unstable &
		\includegraphics[width=0.3\textwidth]{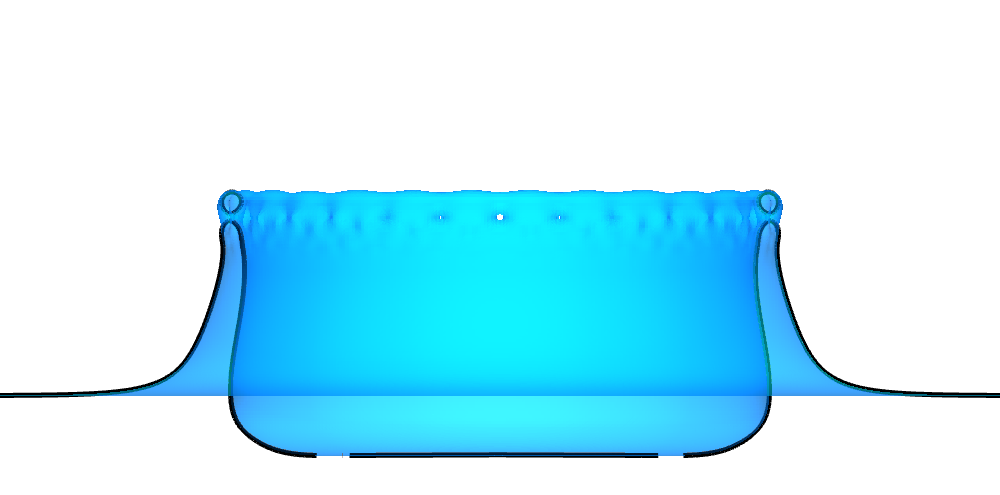} & 
		\includegraphics[width=0.3\textwidth]{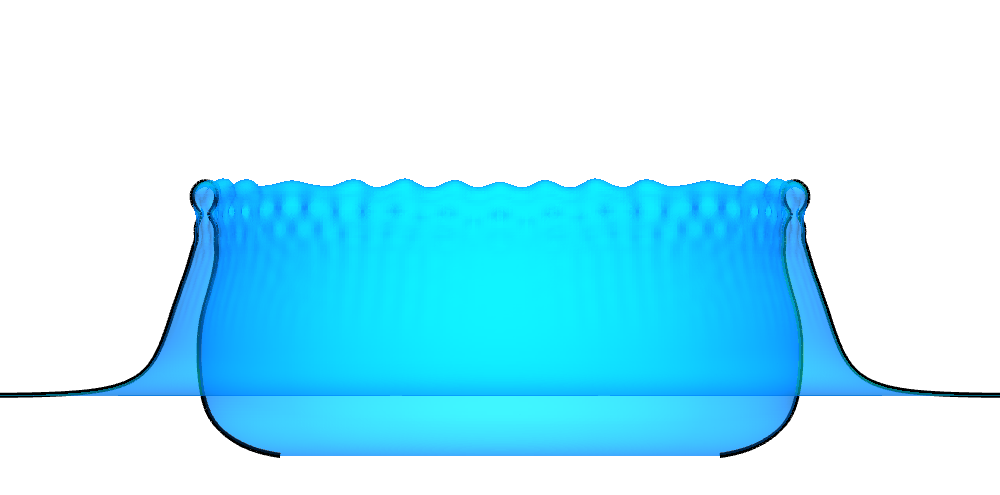} \\
		
		& $\xi=3$ & $\xi=5$ & $\xi=8$ \\
	\end{tabular}
	\caption{
		Vertical drop impact at reference time $t^{*}=12$, as simulated with the PFLBM with initial drop diameter, $D=40$.
		The influence of the mobility, $M$, and interface width, $\xi$, are shown.
		While the simulation results are true to scale, no scale bar is available for the photograph of the experiment~\cite{wang2000SplashingImpactSingle}.
		Therefore, the splash crown's dimension can only be compared between simulations rather than with the experiment.
		The solid black line illustrates the crown's contour in the center cross-section with normal in the $x$-direction.
		The photograph of the laboratory experiment was reprinted from Reference~\cite{wang2000SplashingImpactSingle} with the permission of AIP Publishing.
	}
	\label{fig:drop-vertical-pflbm-width-interface}
\end{figure}

\begin{figure}[h!]
	\begin{tabular}{>{\centering\arraybackslash}m{0.1\textwidth}
					>{\centering\arraybackslash}m{0.2\textwidth}
					>{\centering\arraybackslash}m{0.2\textwidth}
					>{\centering\arraybackslash}m{0.2\textwidth}
					>{\centering\arraybackslash}m{0.2\textwidth}
					>{\centering\arraybackslash}m{0.2\textwidth}
					>{\centering\arraybackslash}m{0.2\textwidth}}
	$t^{*}=0$ &
	\includegraphics[width=0.2\textwidth]{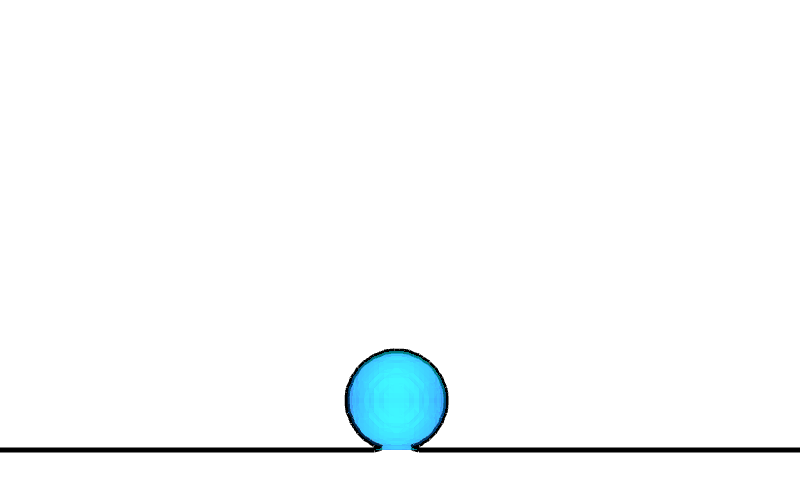} &
	\includegraphics[width=0.2\textwidth]{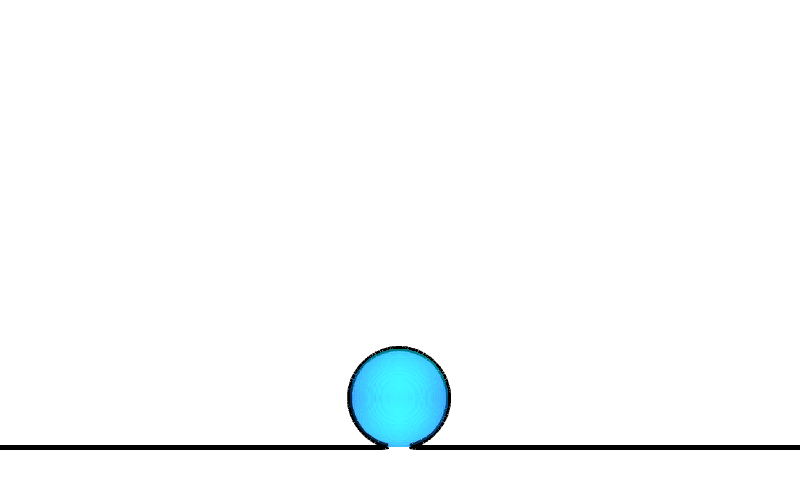} &
	\includegraphics[width=0.2\textwidth]{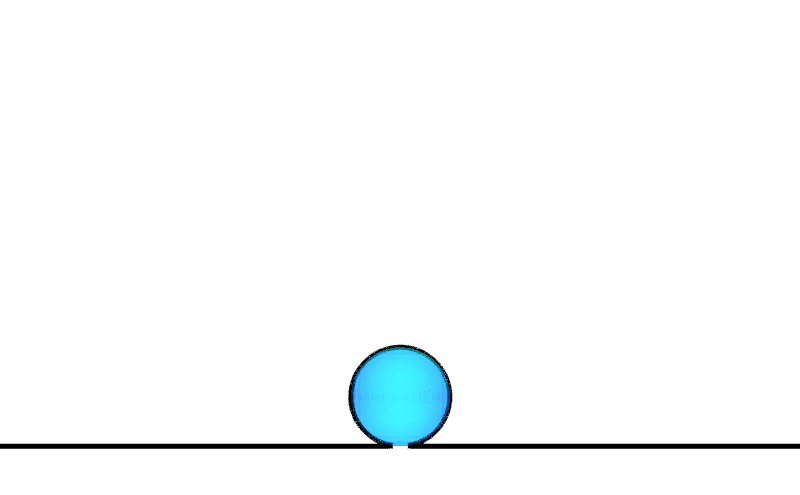} &
	\\
	
	$t^{*}=0.15$ &
	\includegraphics[width=0.2\textwidth]{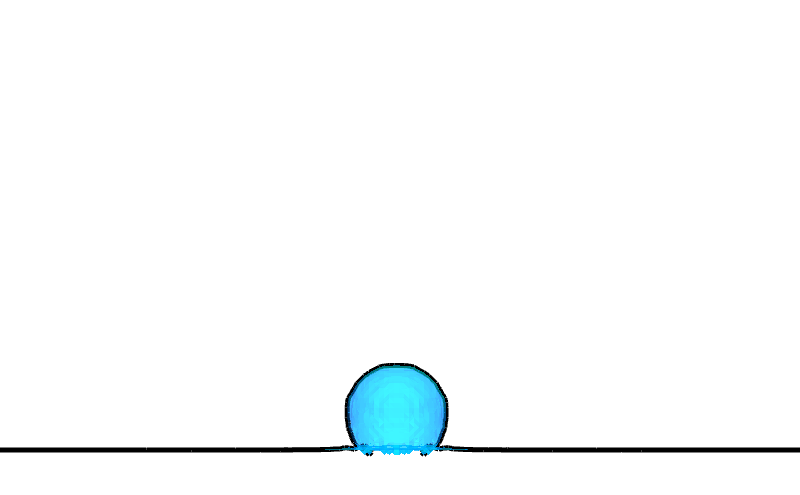} &
	\includegraphics[width=0.2\textwidth]{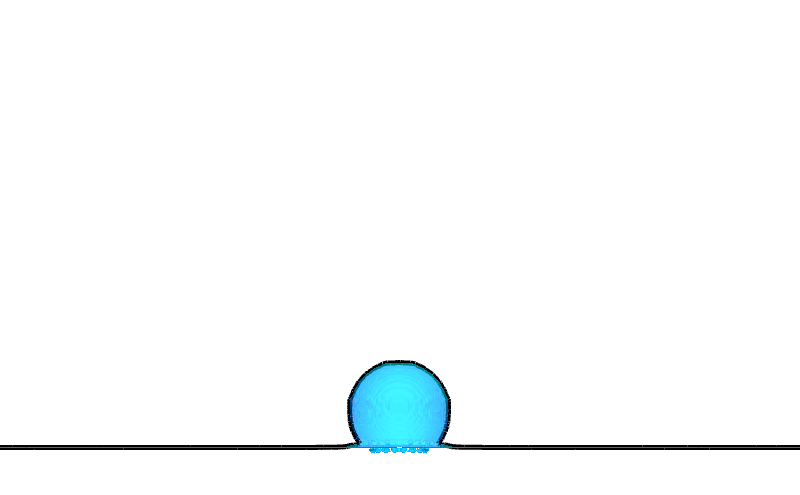} &
	\includegraphics[width=0.2\textwidth]{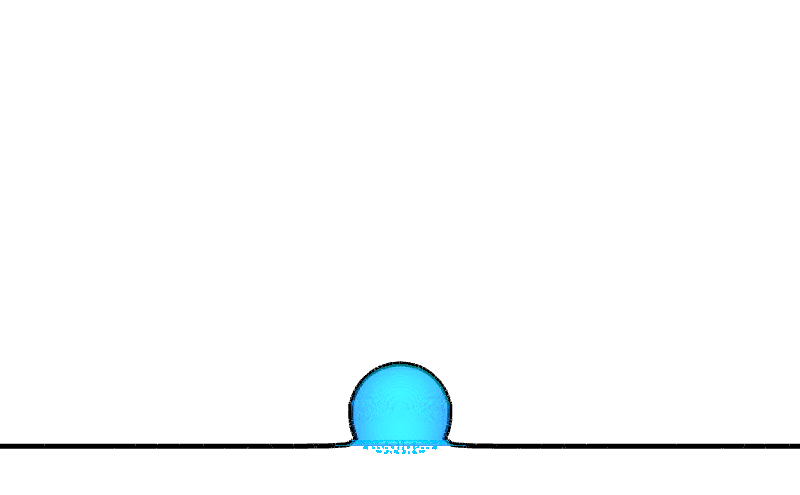} &
	\includegraphics[width=0.2\textwidth]{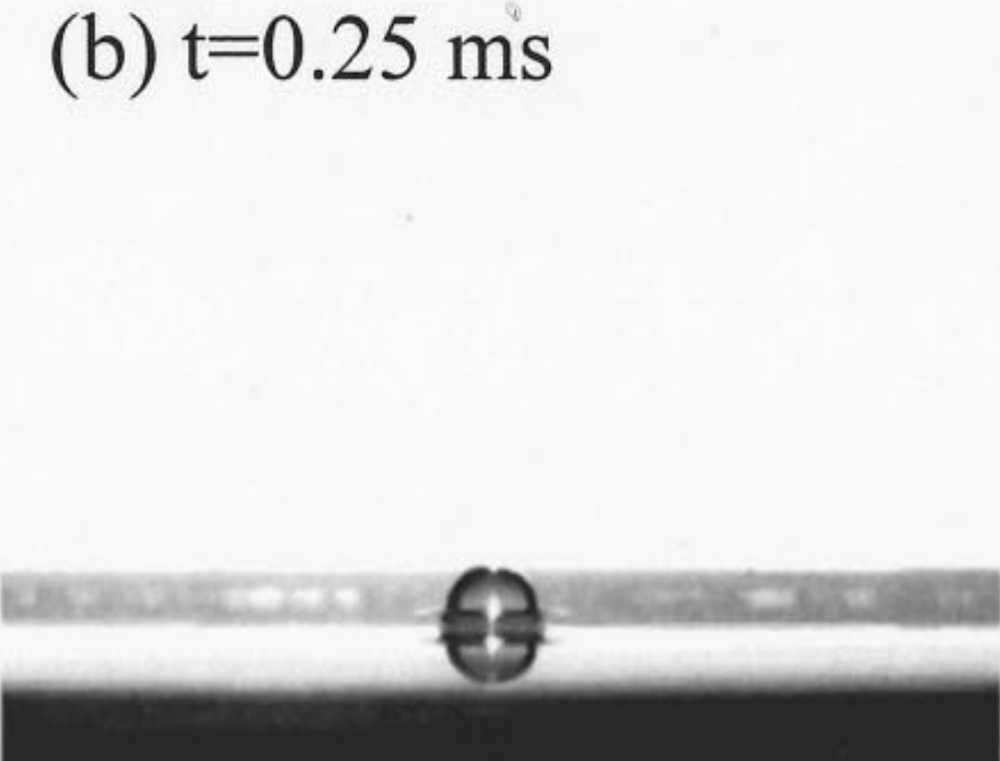} \\
	
	$t^{*}=1.1$ &
	\includegraphics[width=0.2\textwidth]{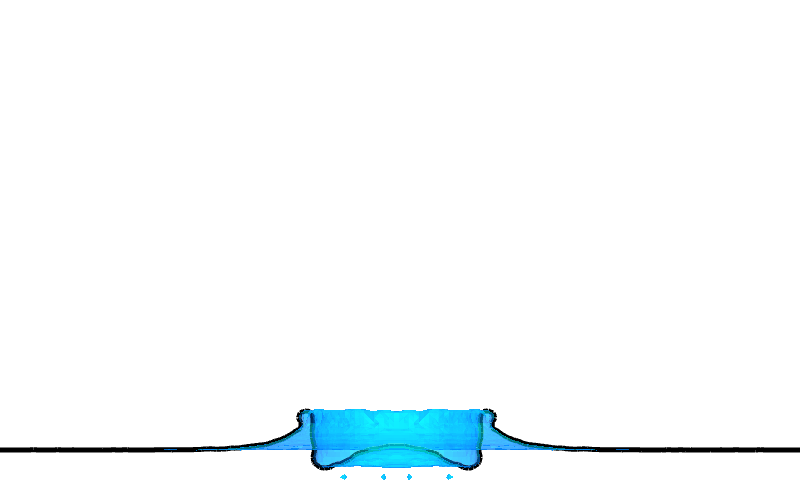} &
	\includegraphics[width=0.2\textwidth]{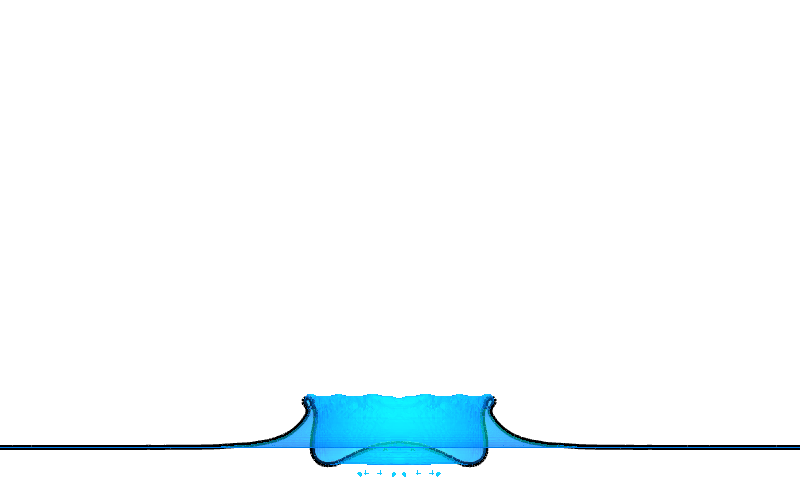} &
	\includegraphics[width=0.2\textwidth]{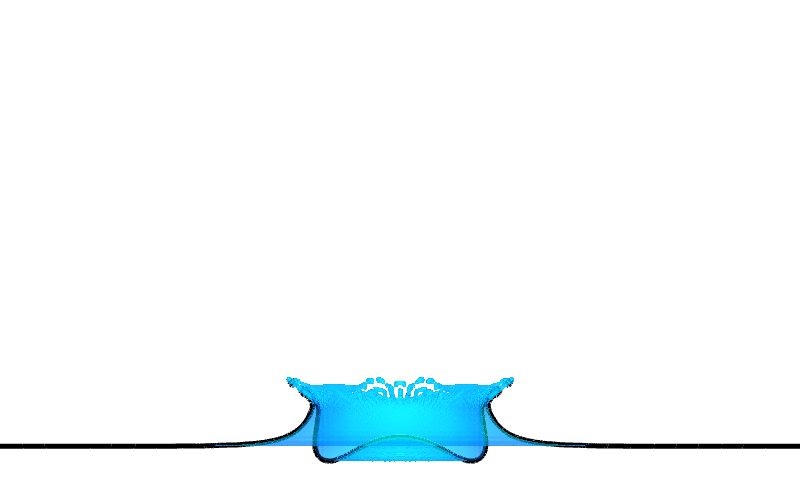} &
	\includegraphics[width=0.2\textwidth]{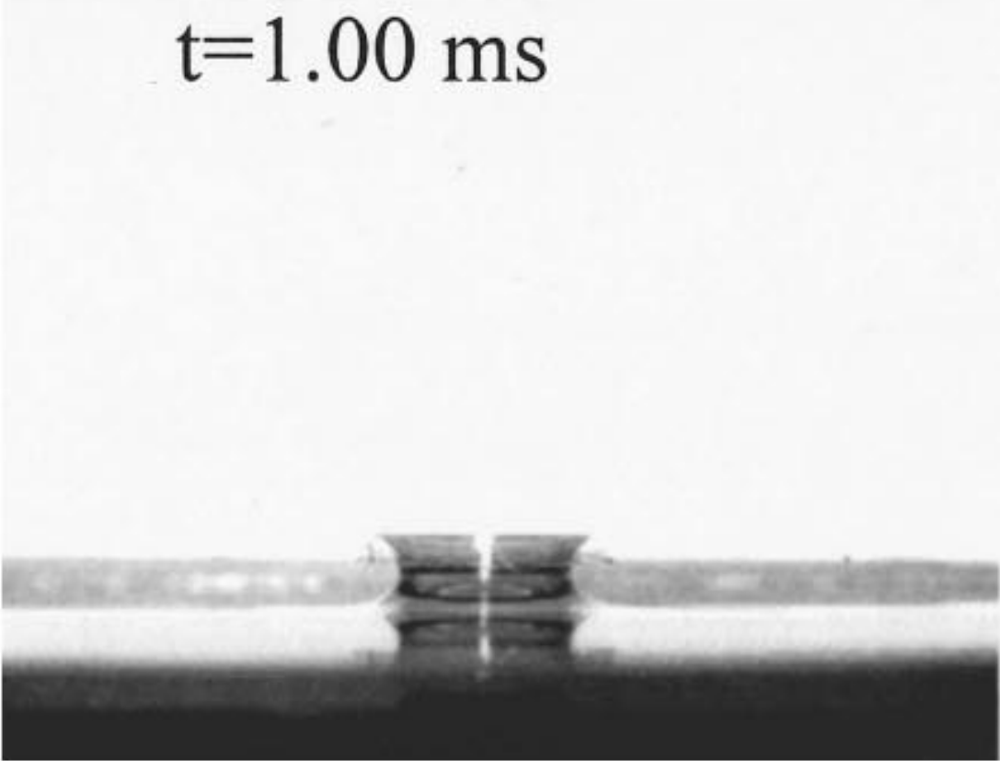} \\
	
	$t^{*}=3.5$ &
	\includegraphics[width=0.2\textwidth]{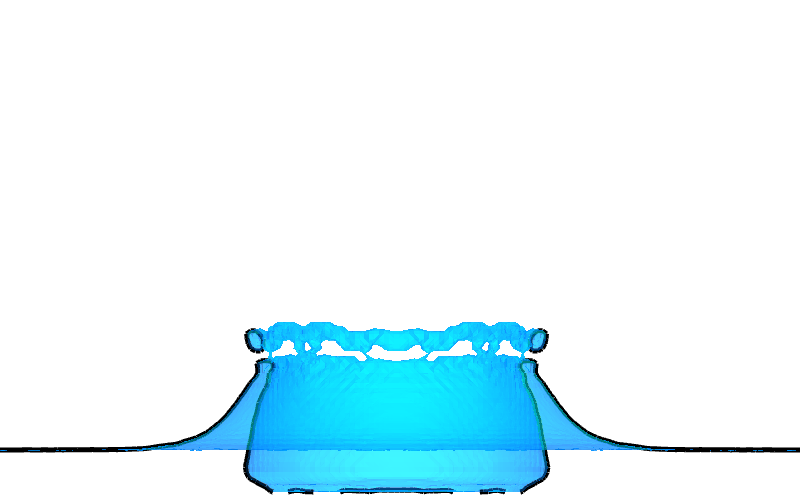} &
	\includegraphics[width=0.2\textwidth]{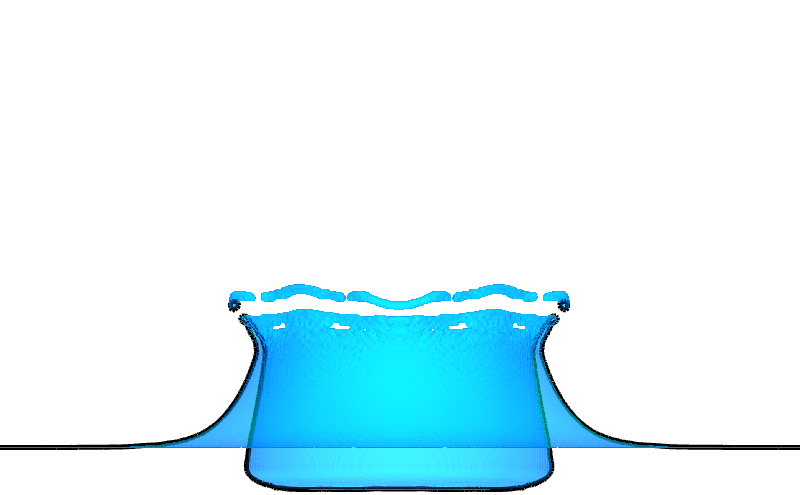} &
	\includegraphics[width=0.2\textwidth]{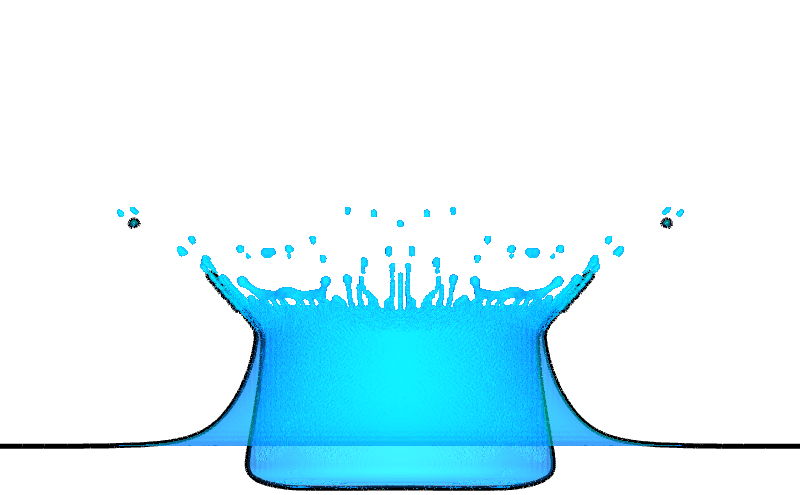} &
	\includegraphics[width=0.2\textwidth]{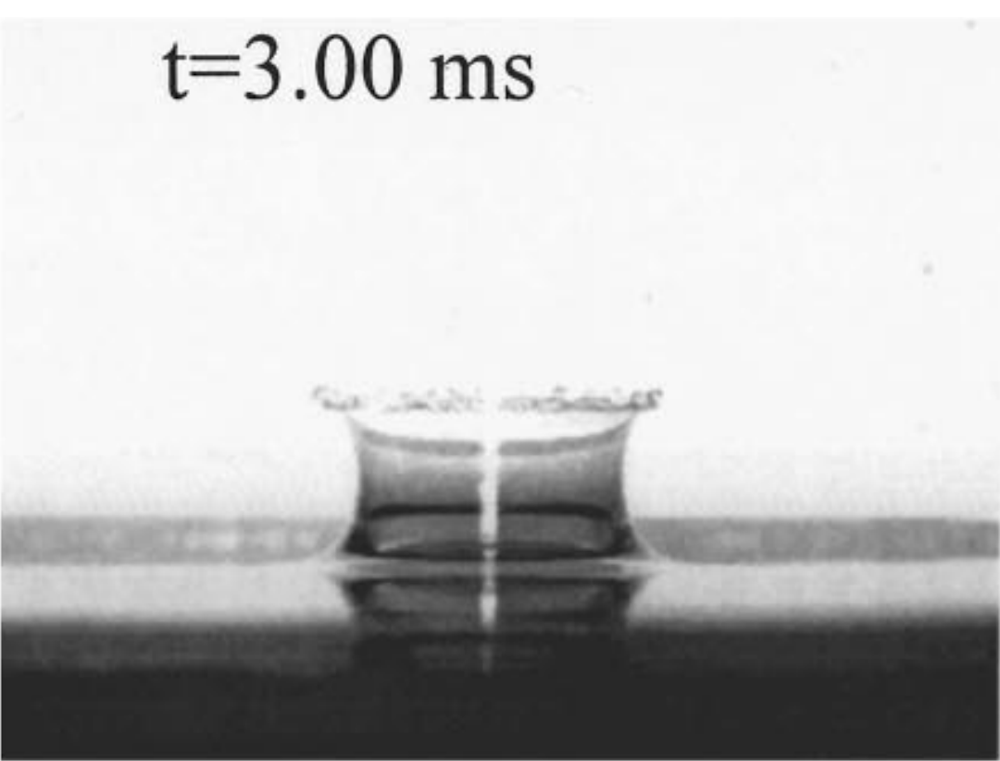} \\
	
	$t^{*}=9$ &
	\includegraphics[width=0.2\textwidth]{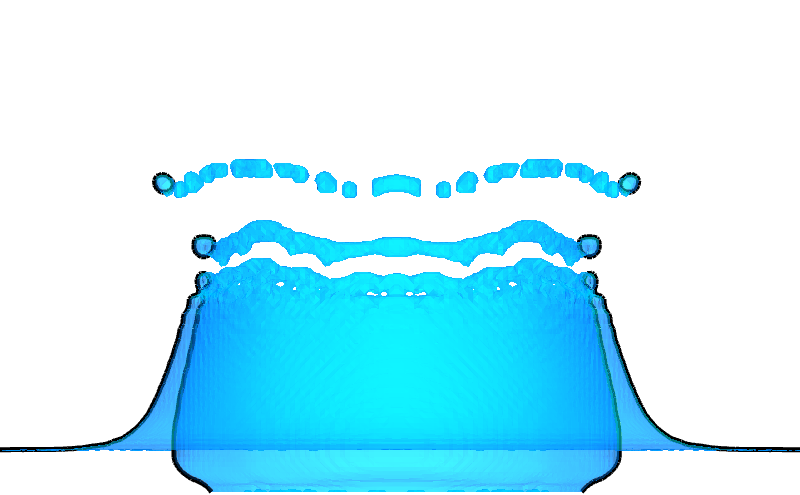} &
	\includegraphics[width=0.2\textwidth]{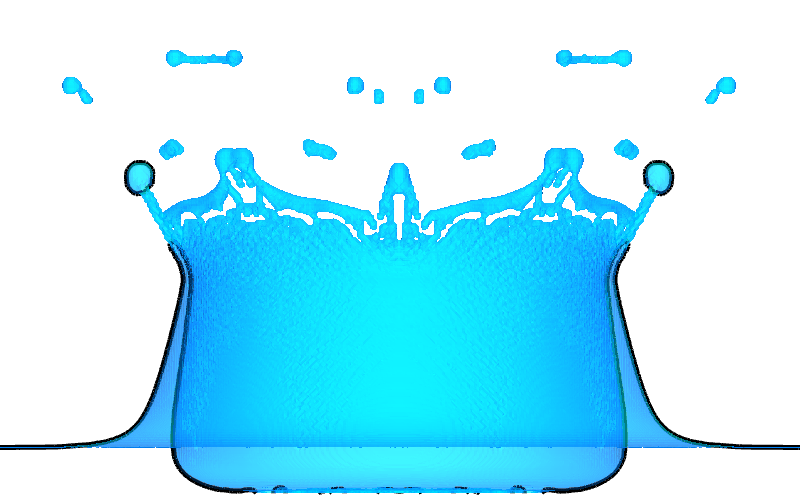} &
	\includegraphics[width=0.2\textwidth]{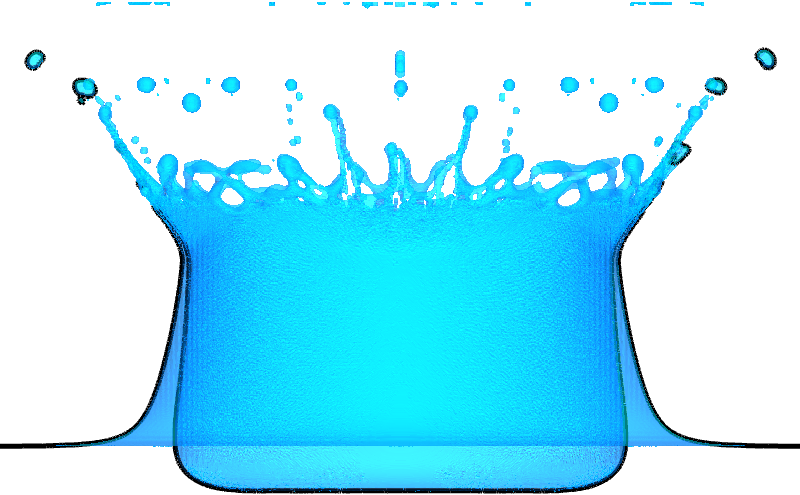} &
	\includegraphics[width=0.2\textwidth]{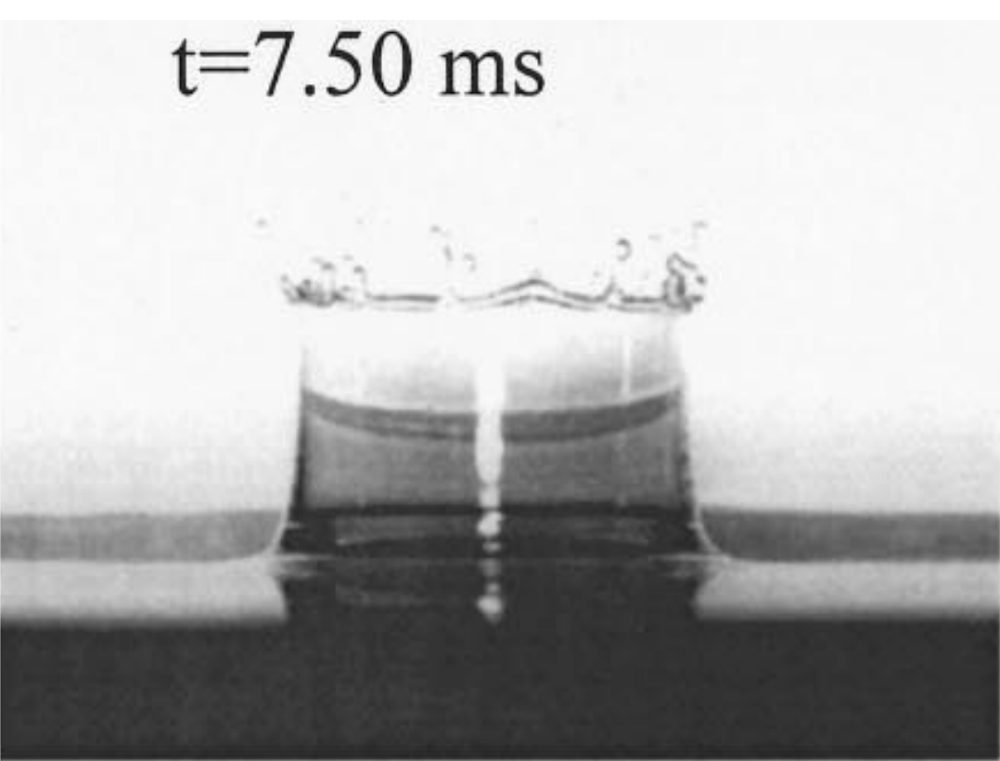} \\
	
	$t^{*}=12$ &
	\includegraphics[width=0.2\textwidth]{figures/drop-impact/vertical/fslbm/d-20/t-12.png} &
	\includegraphics[width=0.2\textwidth]{figures/drop-impact/vertical/fslbm/d-40/t-12.png} &
	\includegraphics[width=0.2\textwidth]{figures/drop-impact/vertical/fslbm/d-80/t-12.png} &
	\includegraphics[width=0.2\textwidth]{figures/drop-impact/vertical/experiment/t-12.png} \\
	
	& $D=20$ & $D=40$ & $D=80$ & Experiment~\cite{wang2000SplashingImpactSingle}
	\end{tabular}
	\caption{
		Vertical drop impact over non-dimensionalized time, $t^{*}$, as simulated with the FSLBM. The computational resolution is defined by the initial drop diameter, $D$.
		While the simulation results are true to scale, no scale bar is available for the photographs of the experiment~\cite{wang2000SplashingImpactSingle}.
		Therefore, the splash crown's dimension can only be compared between simulations rather than with the experiment.
		The solid black line illustrates the crown's contour in the center cross-section with normal in the $x$-direction.
		The photographs of the laboratory experiment were reprinted from Reference~\cite{wang2000SplashingImpactSingle} with the permission of AIP Publishing.
	}
	\label{fig:drop-vertical-fslbm}
\end{figure}

\begin{figure}[h!]
	\begin{tabular}{>{\centering\arraybackslash}m{0.1\textwidth}
					>{\centering\arraybackslash}m{0.2\textwidth}
					>{\centering\arraybackslash}m{0.2\textwidth}
					>{\centering\arraybackslash}m{0.2\textwidth}
					>{\centering\arraybackslash}m{0.2\textwidth}
					>{\centering\arraybackslash}m{0.2\textwidth}
					>{\centering\arraybackslash}m{0.2\textwidth}}
	
		$t^{*}=0$ &
		\includegraphics[width=0.2\textwidth]{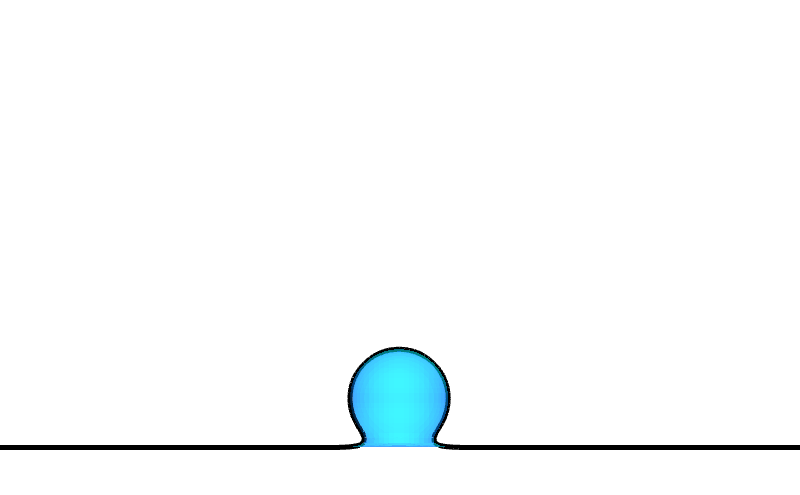} &
		\includegraphics[width=0.2\textwidth]{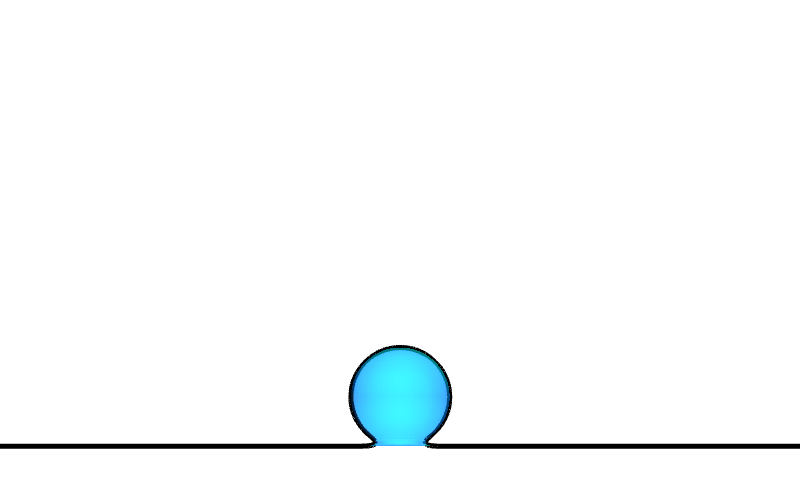} &
		\includegraphics[width=0.2\textwidth]{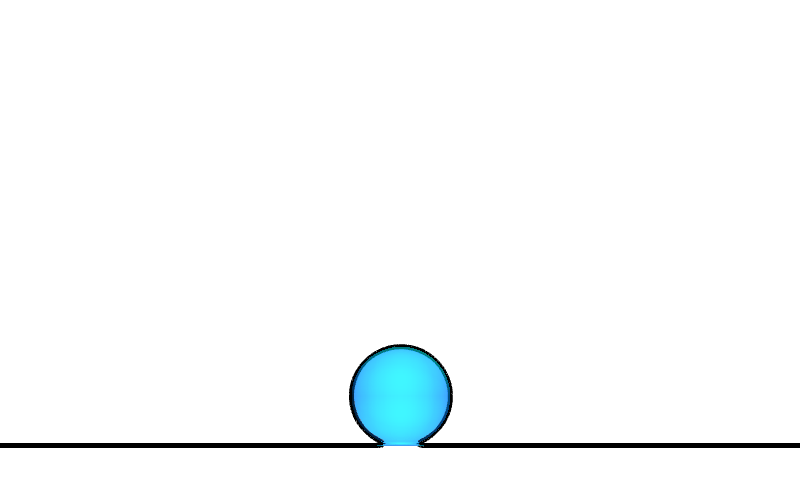} &
		\\
		
		$t^{*}=0.15$ &
		\includegraphics[width=0.2\textwidth]{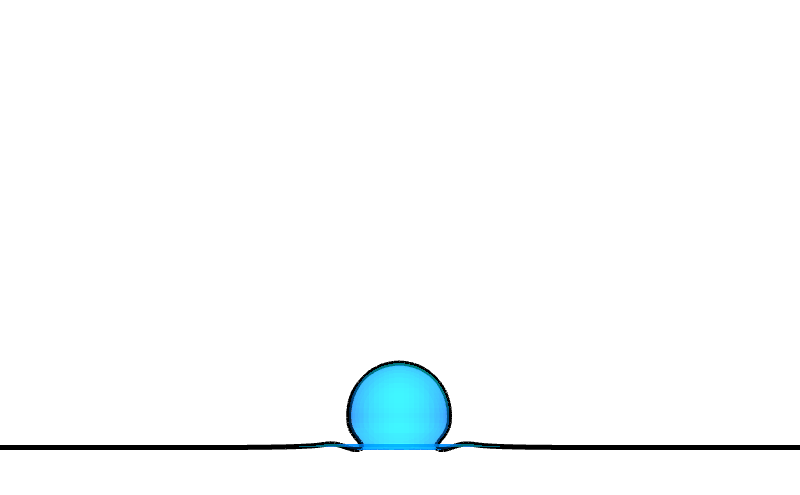} &
		\includegraphics[width=0.2\textwidth]{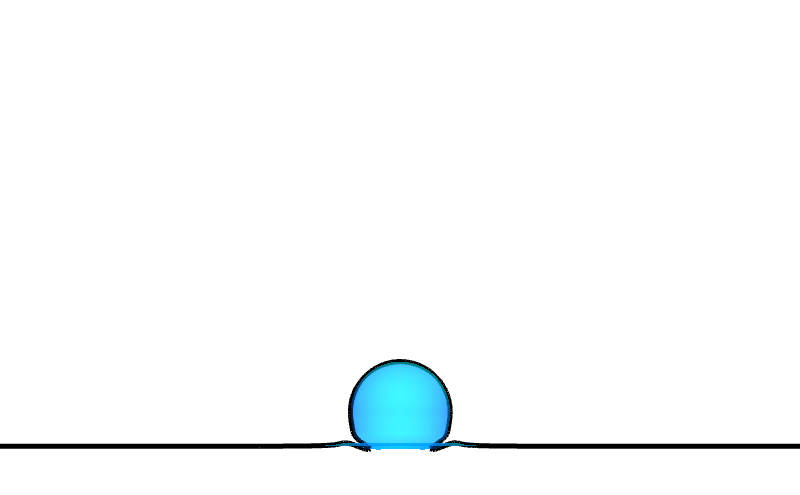} &
		\includegraphics[width=0.2\textwidth]{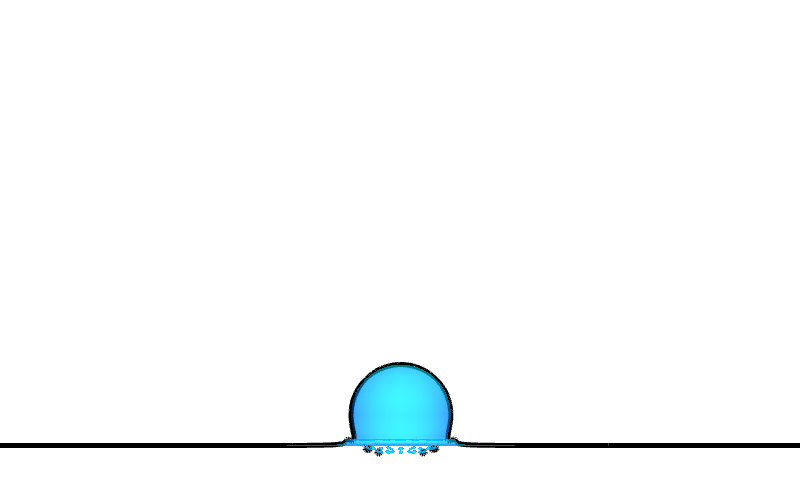} &
		\includegraphics[width=0.2\textwidth]{figures/drop-impact/vertical/experiment/t-0.15.png} \\
		
		$t^{*}=1.1$ &
		\includegraphics[width=0.2\textwidth]{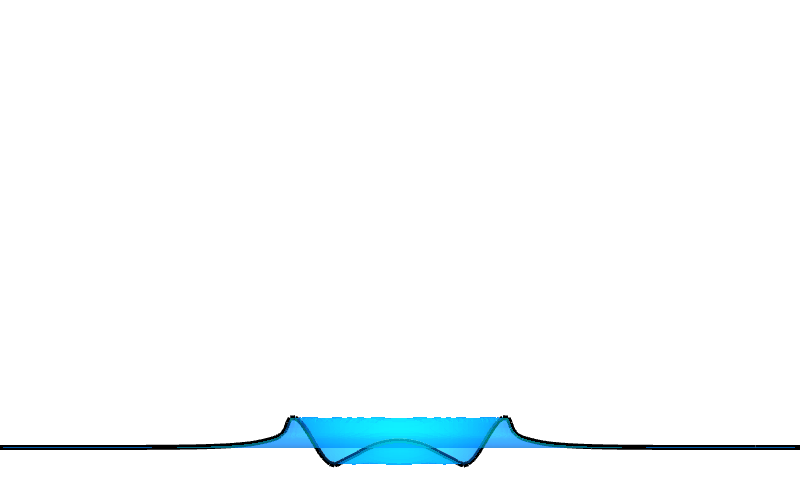} &
		\includegraphics[width=0.2\textwidth]{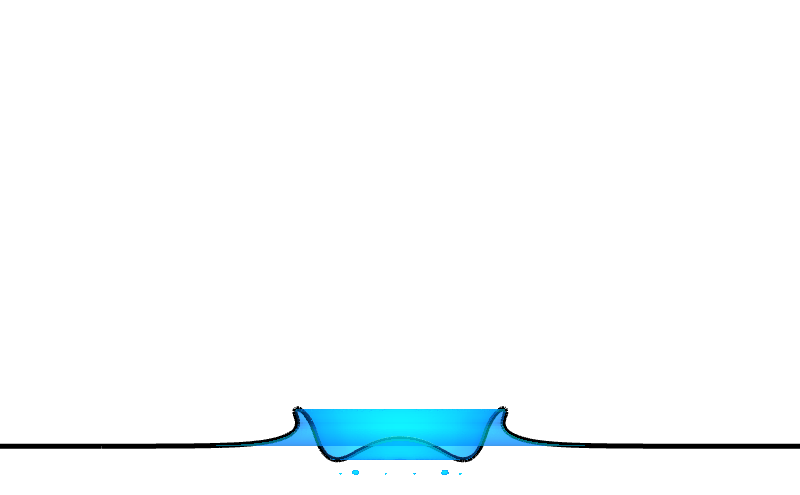} &
		\includegraphics[width=0.2\textwidth]{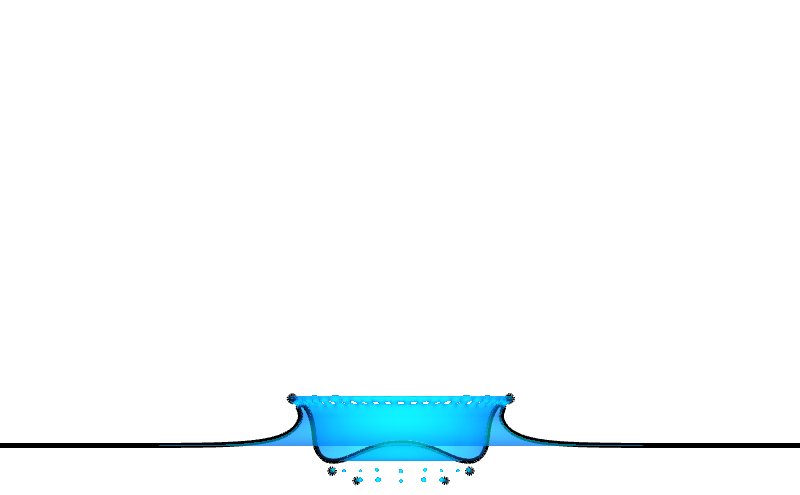} &
		\includegraphics[width=0.2\textwidth]{figures/drop-impact/vertical/experiment/t-1.1.png} \\
		
		$t^{*}=3.5$ &
		\includegraphics[width=0.2\textwidth]{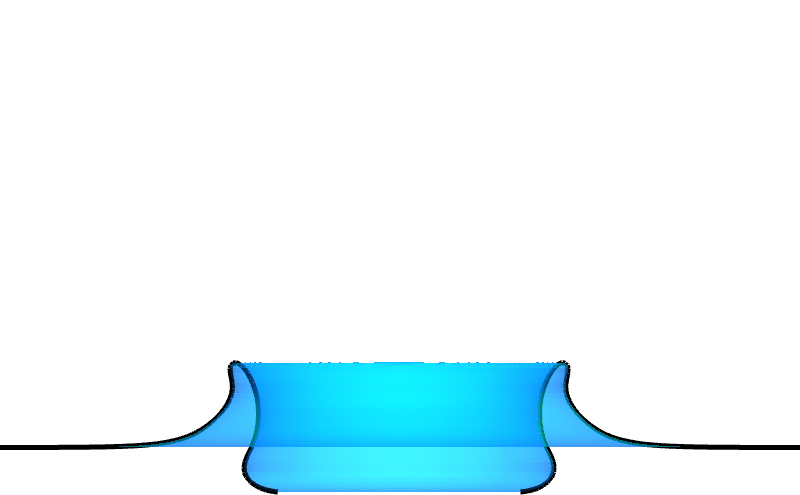} &
		\includegraphics[width=0.2\textwidth]{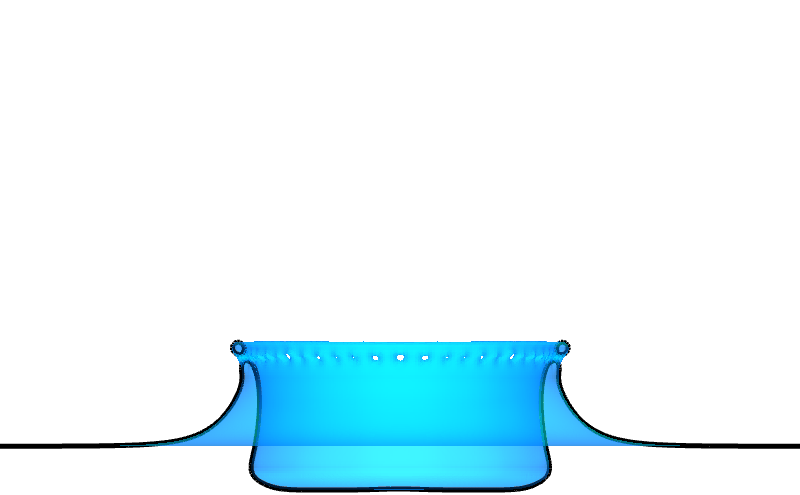} &
		\includegraphics[width=0.2\textwidth]{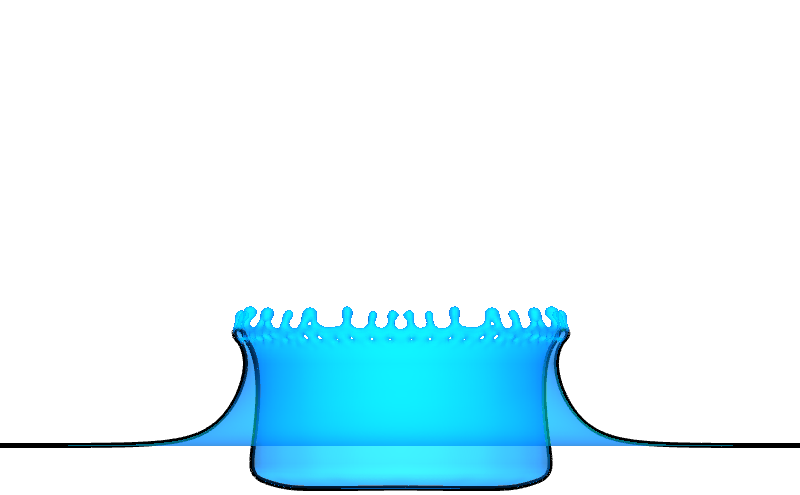} &
		\includegraphics[width=0.2\textwidth]{figures/drop-impact/vertical/experiment/t-3.5.png} \\
		
		$t^{*}=9$ &
		\includegraphics[width=0.2\textwidth]{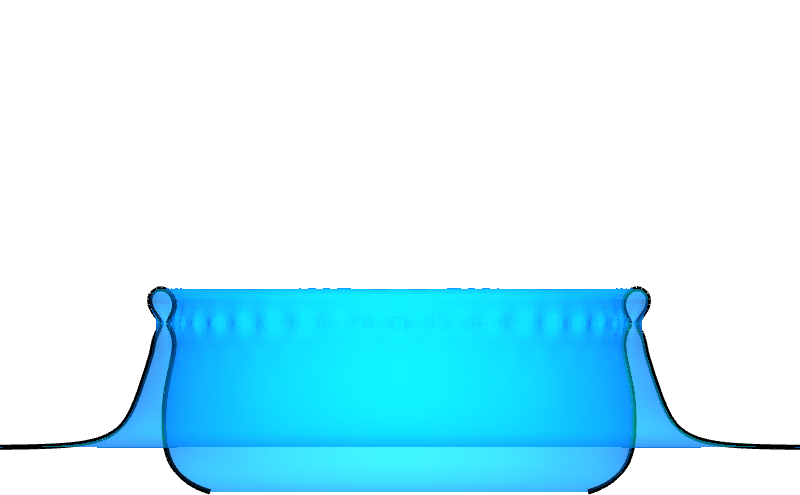} &
		\includegraphics[width=0.2\textwidth]{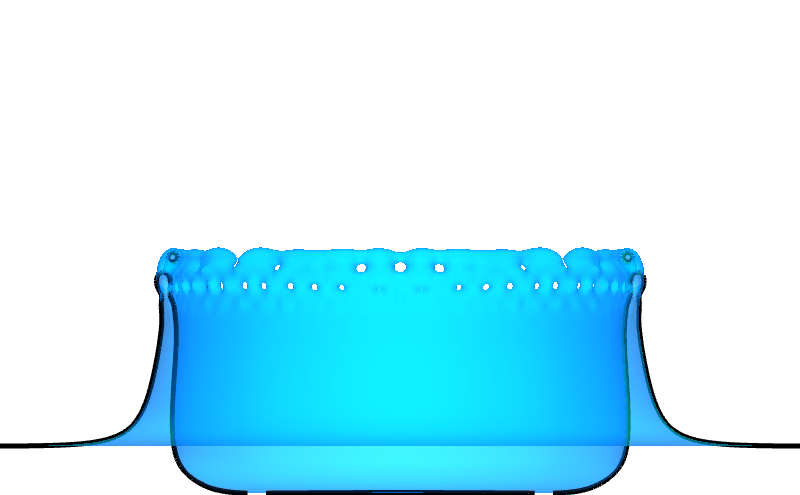} &
		\includegraphics[width=0.2\textwidth]{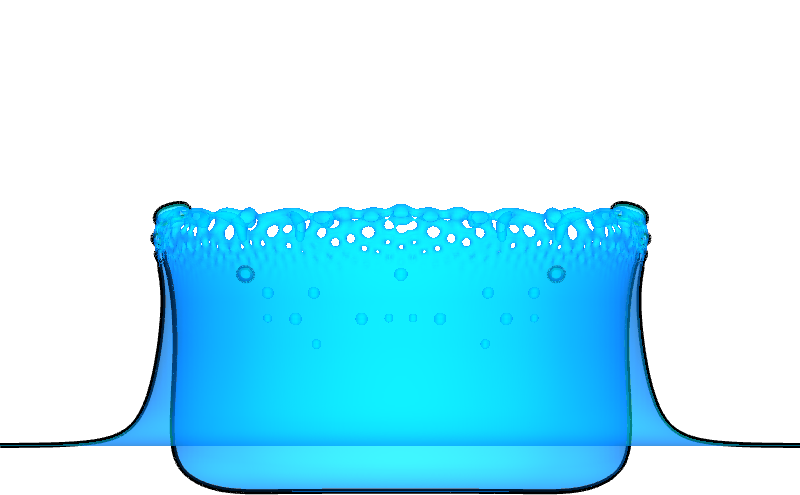} &
		\includegraphics[width=0.2\textwidth]{figures/drop-impact/vertical/experiment/t-9.png} \\
		
		$t^{*}=12$ &
		\includegraphics[width=0.2\textwidth]{figures/drop-impact/vertical/pflbm/d-20/t-12.png} &
		\includegraphics[width=0.2\textwidth]{figures/drop-impact/vertical/pflbm/d-40/t-12.png} &
		\includegraphics[width=0.2\textwidth]{figures/drop-impact/vertical/pflbm/d-80/t-12.png} &
		\includegraphics[width=0.2\textwidth]{figures/drop-impact/vertical/experiment/t-12.png} \\
		
		& $D=20$ & $D=40$ & $D=80$ & Experiment~\cite{wang2000SplashingImpactSingle}
	\end{tabular}
	\caption{
		Vertical drop impact over non-dimensionalized time, $t^{*}$, as simulated with the PFLBM. The computational resolution is defined by the initial drop diameter, $D$.
		While the simulation results are true to scale, no scale bar is available for the photographs of the experiment~\cite{wang2000SplashingImpactSingle}.
		Therefore, the splash crown's dimension can only be compared between simulations rather than with the experiment.
		The solid black line illustrates the crown's contour in the center cross-section with normal in the $x$-direction.
		The photographs of the laboratory experiment were reprinted from Reference~\cite{wang2000SplashingImpactSingle} with the permission of AIP Publishing.
	}
	\label{fig:drop-vertical-pflbm}
\end{figure}

Extending \Cref{par:vertical-drop-results}, \Cref{fig:drop-vertical-pflbm-width-interface} illustrates the PFLBM's sensitivity to the mobility parameter, $M$, and interface width, $\xi$, in the vertical drop impact test case.
\Cref{fig:drop-vertical-fslbm,fig:drop-vertical-pflbm} qualitatively compare the simulated vertical drop impact with experimental data at different points in time.

\FloatBarrier

\begin{table}[htbp]
	\centering
	\begin{tabular}{>{\raggedright}m{0.2\textwidth}
			>{\centering\arraybackslash}m{0.1\textwidth}
			>{\centering\arraybackslash}m{0.1\textwidth}
			>{\centering\arraybackslash}m{0.1\textwidth}
			>{\centering\arraybackslash}m{0.1\textwidth}
			>{\centering\arraybackslash}m{0.1\textwidth}}
		
		\toprule
		& $t^{*}$ & $1.1$ & $3.5$ & $9$ & $12$ \\
		\midrule
		FSLBM, $D=20$ & \multirow{6}{*}{$h_{\text{cav}}^{*}$} & $-0.18$ & $-0.42$ & $-0.42$ & $-0.42$ \\
		FSLBM, $D=40$ & & $-0.17$ & $-0.43$ & $-0.46$ & $-0.46$ \\
		FSLBM, $D=80$ & & $-0.15$ & $-0.43$ & $-0.45$ & $-0.45$ \\
		
		PFLBM, $D=20$ & & $-0.18$ & $-0.45$ & $-0.45$ & $-0.45$ \\
		PFLBM, $D=40$ & & $-0.14$ & $-0.44$ & $-0.48$ & $-0.48$ \\
		PFLBM, $D=80$ & & $-0.38$ & $-0.43$ & $-0.47$ & $-0.47$ \\
		\bottomrule
	\end{tabular}
	\caption{
		Simulated non-dimensionalized cavity depth, $h_{\text{ca}}^{*}(t^{*}) = h_{\text{ca}}(t^{*}) / D$, of the vertical drop impact.
		The cavity depth, $h_{\text{ca}}(t^{*})$, is the maximum distance of the cavity bottom to the initial position of the liquid surface at time, $t^{*}=0$, measured in the center cross-section with normal in the $x$-direction.
		The results are presented for different dimensionless times, $t^{*}$, and computational resolutions as defined by the initial drop diameter, $D$.
	}
	\label{tab:drop-vertical-cavity}
\end{table}

\begin{table}[htbp]
	\centering
	\begin{tabular}{>{\raggedright}m{0.2\textwidth}
			>{\centering\arraybackslash}m{0.1\textwidth}
			>{\centering\arraybackslash}m{0.1\textwidth}
			>{\centering\arraybackslash}m{0.1\textwidth}
			>{\centering\arraybackslash}m{0.1\textwidth}
			>{\centering\arraybackslash}m{0.1\textwidth}}
		
		\toprule
		& $t^{*}$ & $1.1$ & $3.5$ & $9$ & $12$ \\
		\midrule
		FSLBM, $D=20$ & \multirow{6}{*}{$d_{\text{cr}}^{*}$} & $1.66$ & $2.99$ & $4.50$ & $4.93$ \\
		FSLBM, $D=40$ & & $1.76$ & $3.03$ & $4.57$ & $4.93$ \\
		FSLBM, $D=80$ & & $1.73$ & $3.05$ & $4.55$ & $4.92$ \\
		
		PFLBM, $D=20$ & & $1.66$ & $2.99$ & $4.71$ & $5.17$ \\
		PFLBM, $D=40$ & & $1.62$ & $2.94$ & $4.58$ & $5.02$ \\
		PFLBM, $D=80$ & & $1.72$ & $2.98$ & $4.60$ & $5.04$ \\
		\bottomrule
	\end{tabular}
	\caption{
		Simulated non-dimensionalized splash crown diameter, $d_{\text{cr}}^{*}(t^{*}) = d_{\text{cr}}(t^{*}) / D$, of the vertical drop impact. 
		The splash crown diameter, $d_{\text{cr}}(t^{*})$, is the crown's inner diameter at the position of the initial liquid surface at time, $t^{*}=0$, measured in a center cross-section with normal in the $x$-direction.
		The results are presented for different dimensionless times, $t^{*}$, and computational resolutions as defined by the initial drop diameter, $D$.
	}
	\label{tab:drop-vertical-crown}
\end{table}

\FloatBarrier

\subsection{Oblique drop impact} \label{subsec:app-drop-oblique}
\begin{figure}[h!]
	\begin{tabular}{>{\centering\arraybackslash}m{0.1\textwidth}
					>{\centering\arraybackslash}m{0.2\textwidth}
					>{\centering\arraybackslash}m{0.2\textwidth}
					>{\centering\arraybackslash}m{0.2\textwidth}
					>{\centering\arraybackslash}m{0.2\textwidth}
					>{\centering\arraybackslash}m{0.2\textwidth}
					>{\centering\arraybackslash}m{0.2\textwidth}}
	
		$t^{*}=0$ &
		\includegraphics[width=0.2\textwidth]{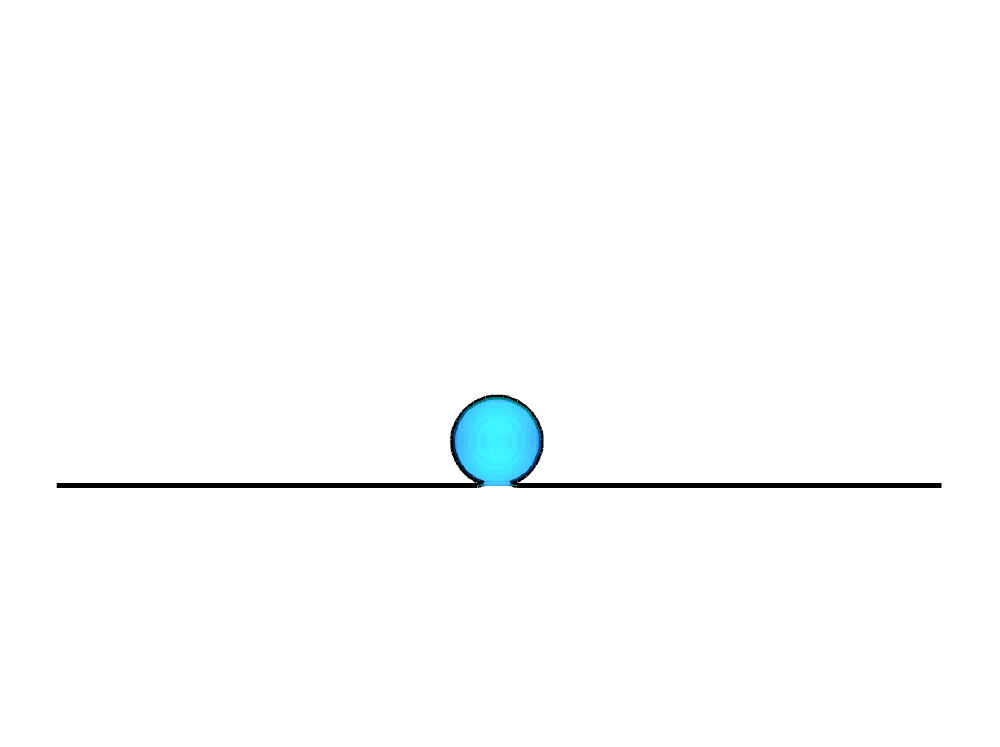} &
		\includegraphics[width=0.2\textwidth]{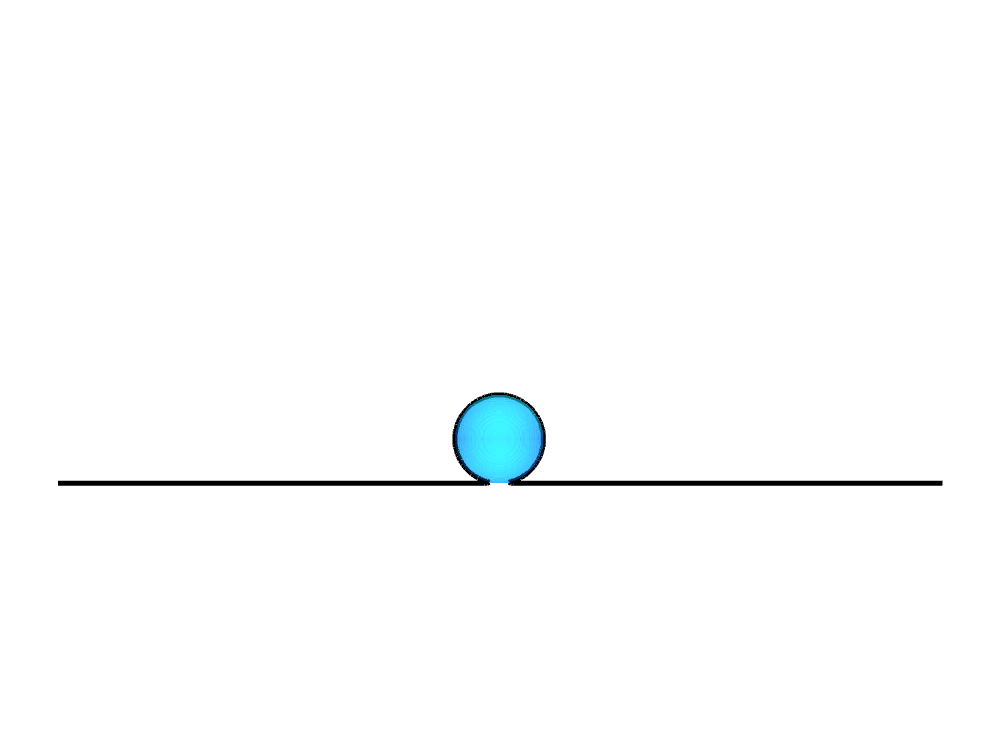} &
		\includegraphics[width=0.2\textwidth]{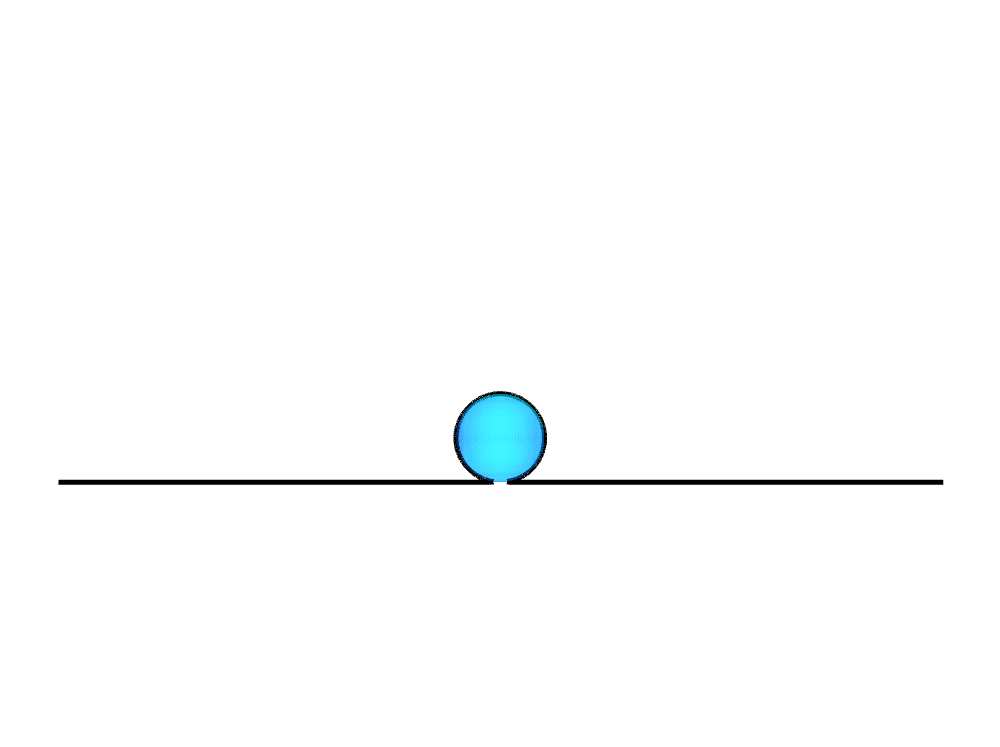} &
		\\
		
		$t^{*}=0.46$ &
		\includegraphics[width=0.2\textwidth]{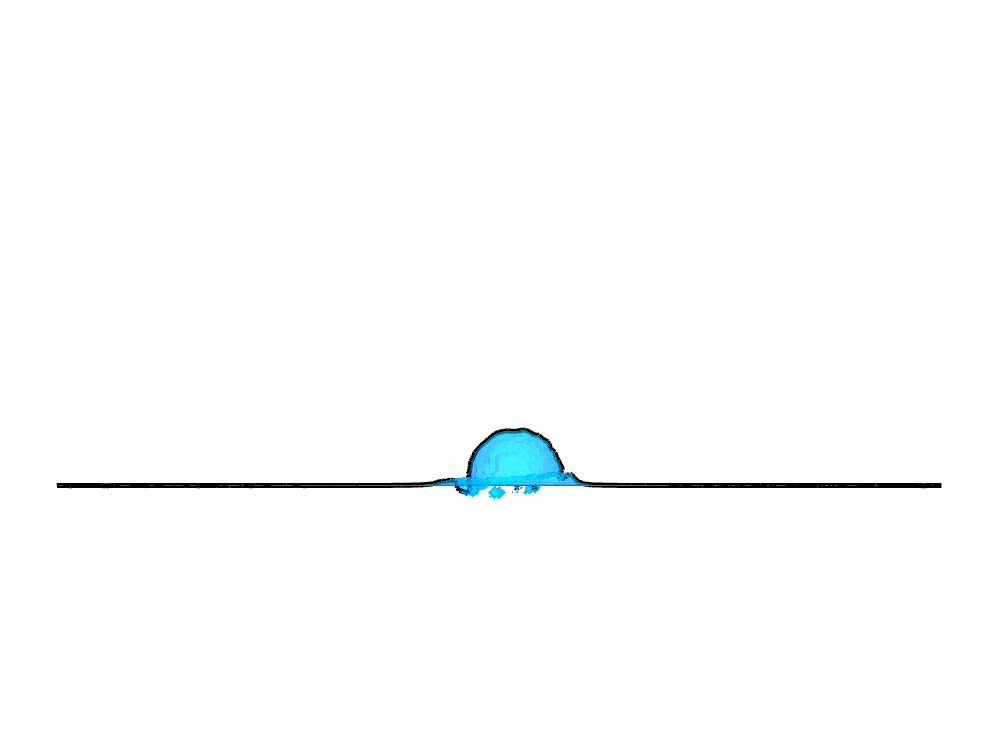} &
		\includegraphics[width=0.2\textwidth]{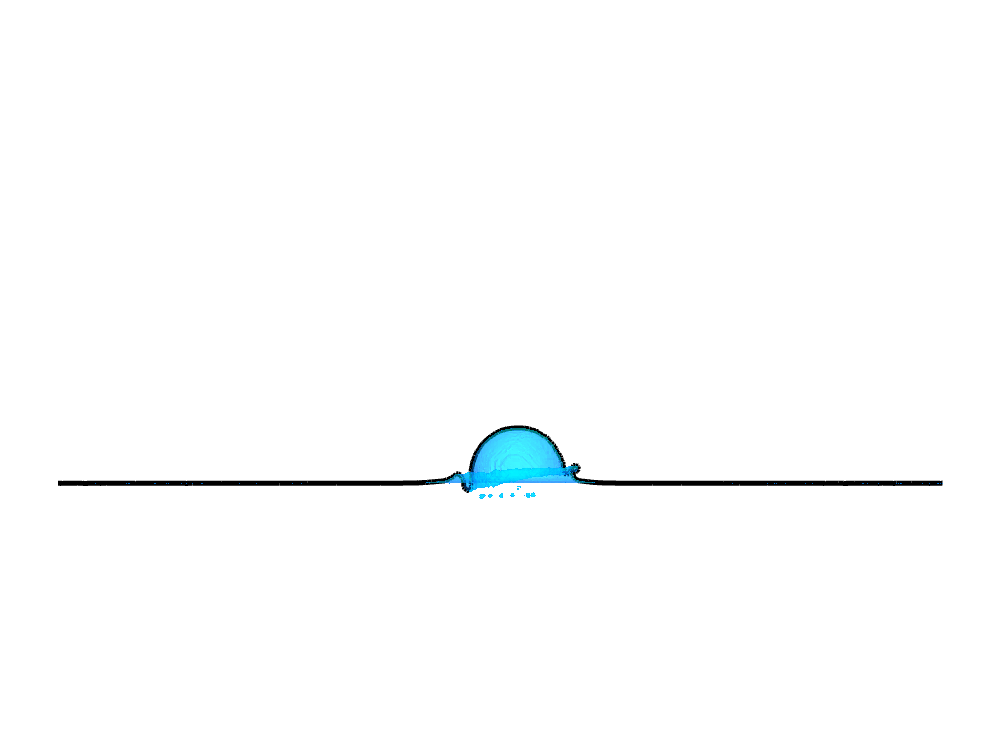} &
		\includegraphics[width=0.2\textwidth]{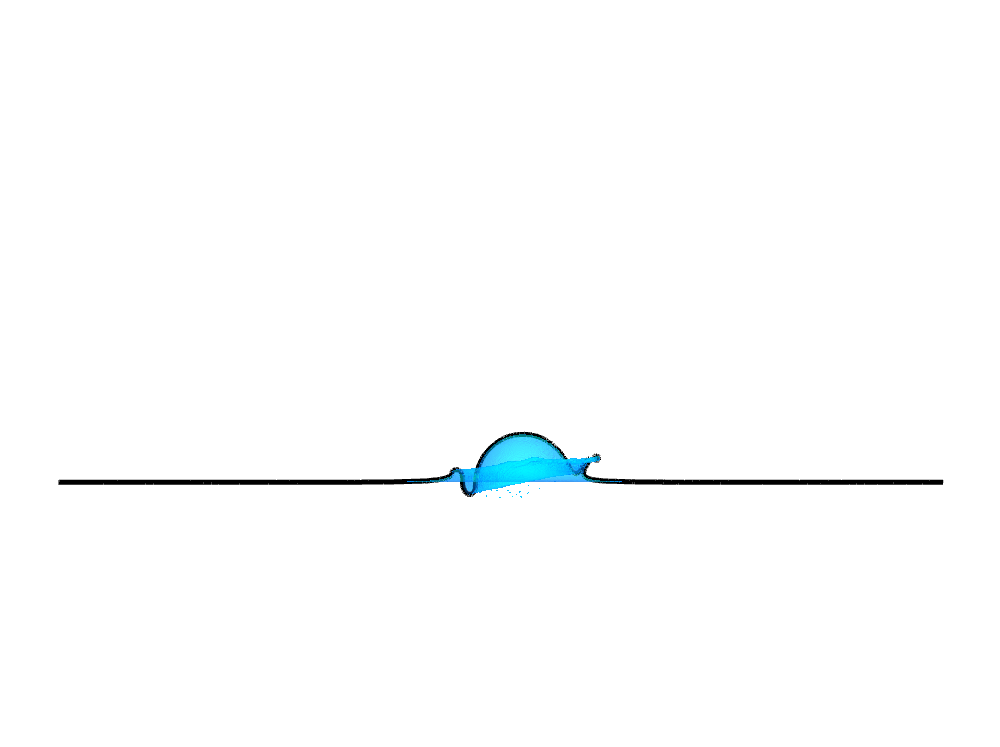} &
		\includegraphics[width=0.15\textwidth]{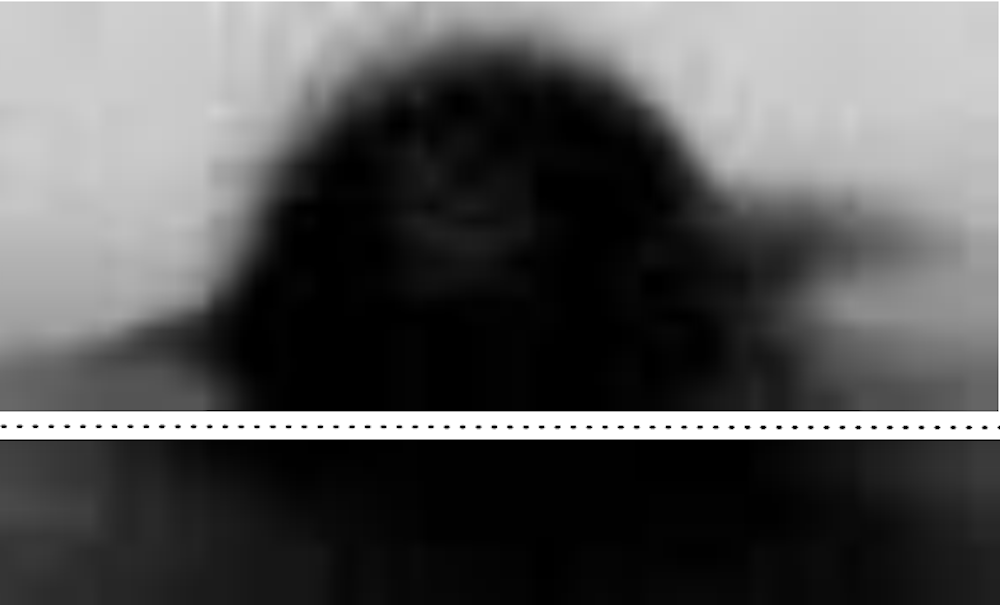} \\
		
		$t^{*}=2.33$ &
		\includegraphics[width=0.2\textwidth]{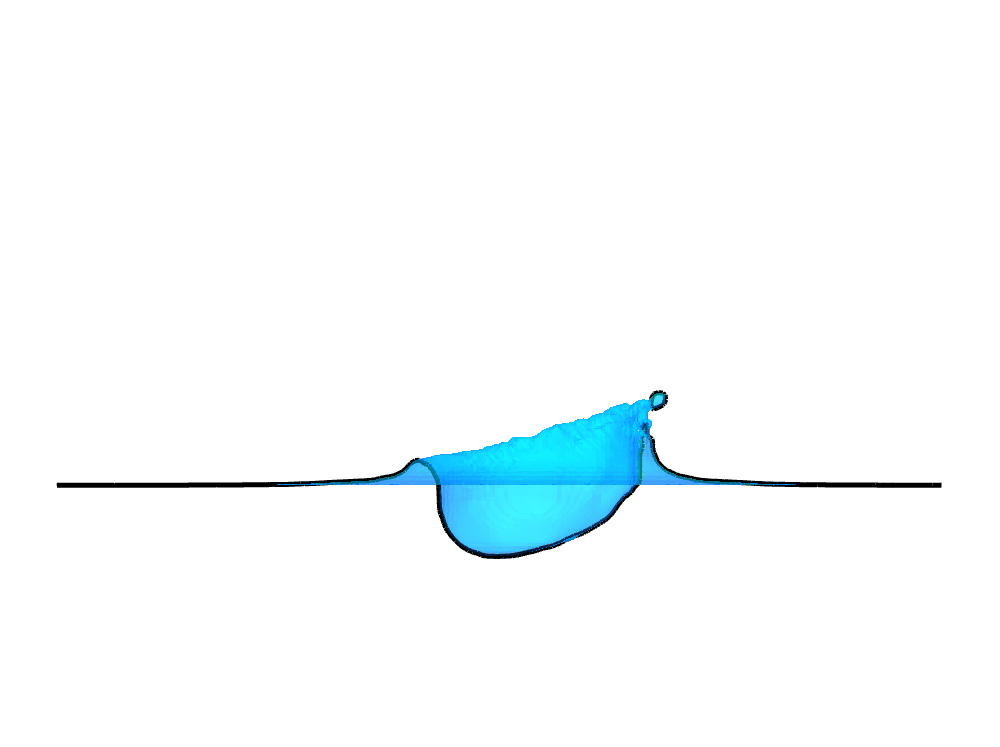} &
		\includegraphics[width=0.2\textwidth]{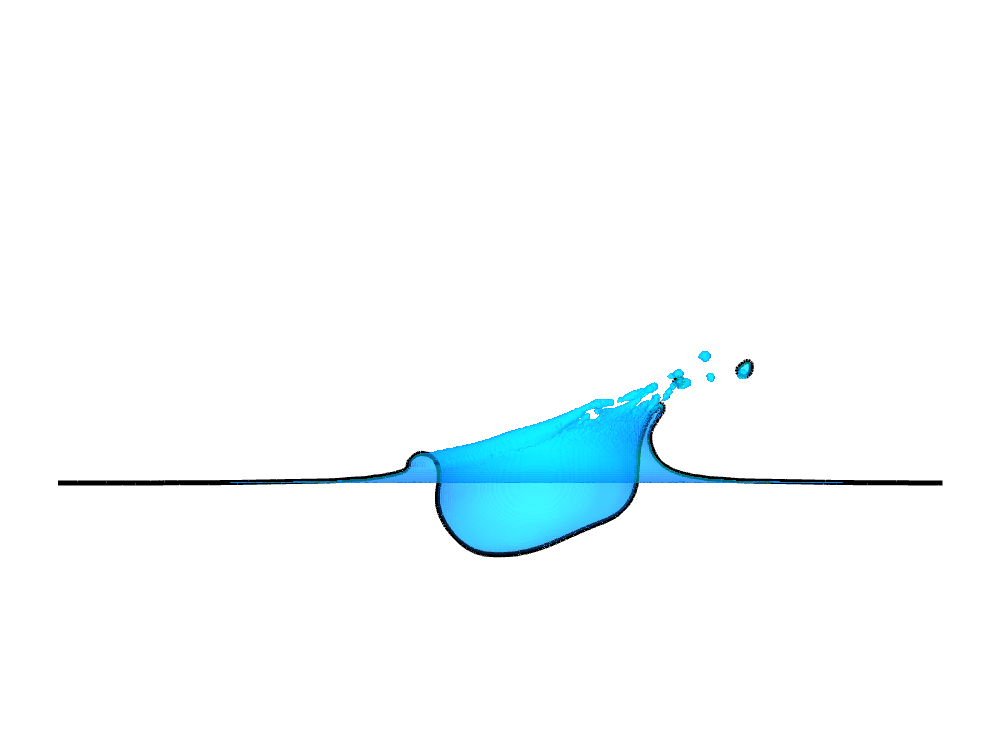} &
		\includegraphics[width=0.2\textwidth]{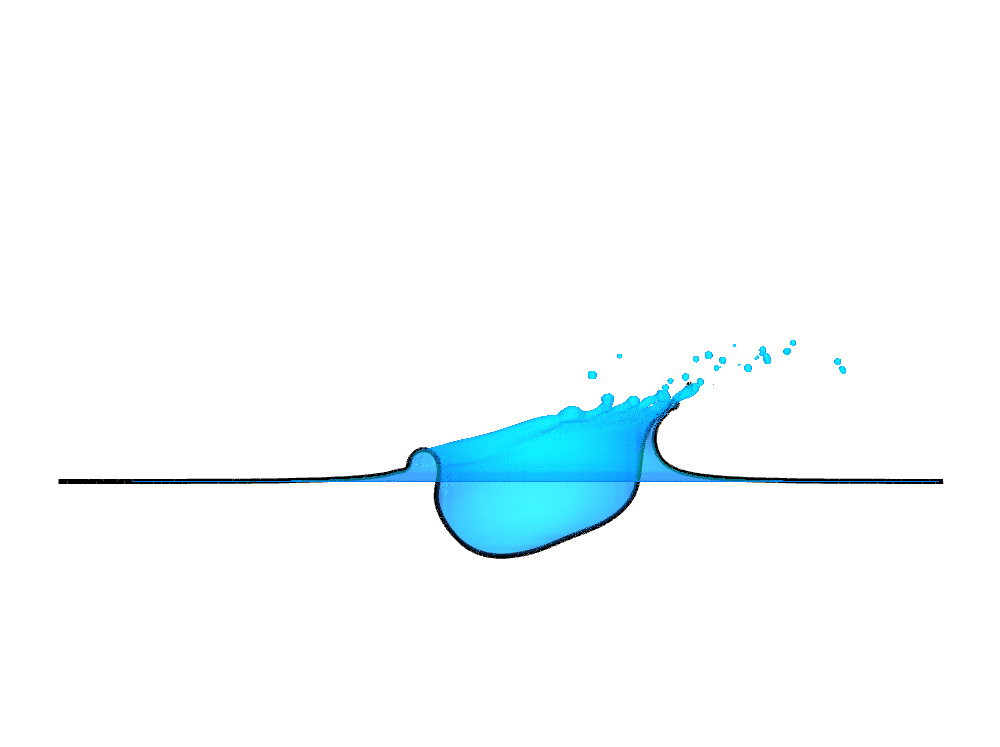} &
		\includegraphics[width=0.15\textwidth]{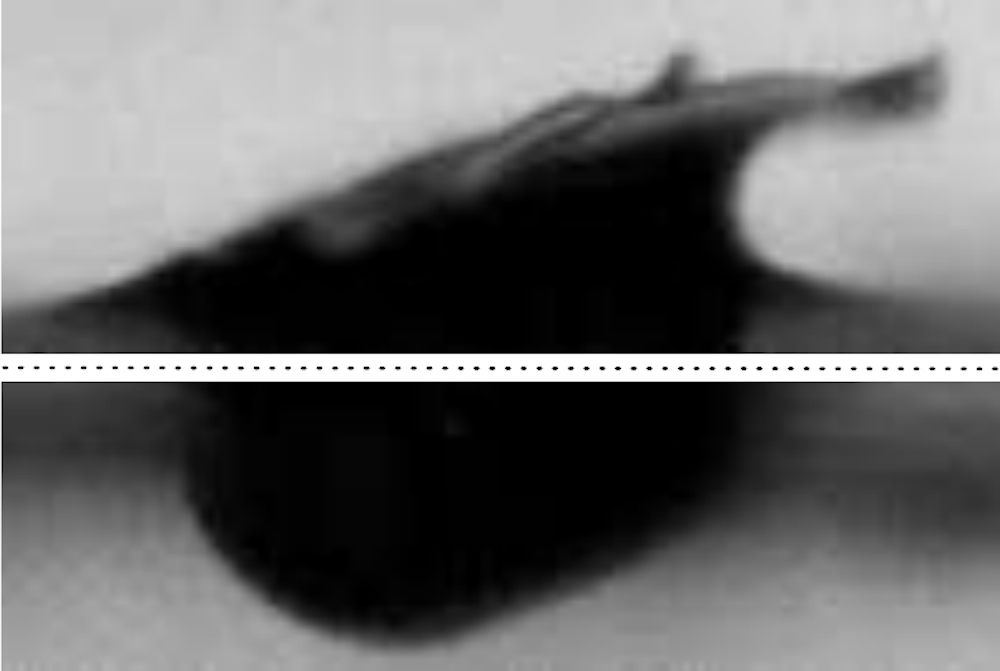} \\
		
		$t^{*}=8.22$ &
		\includegraphics[width=0.2\textwidth]{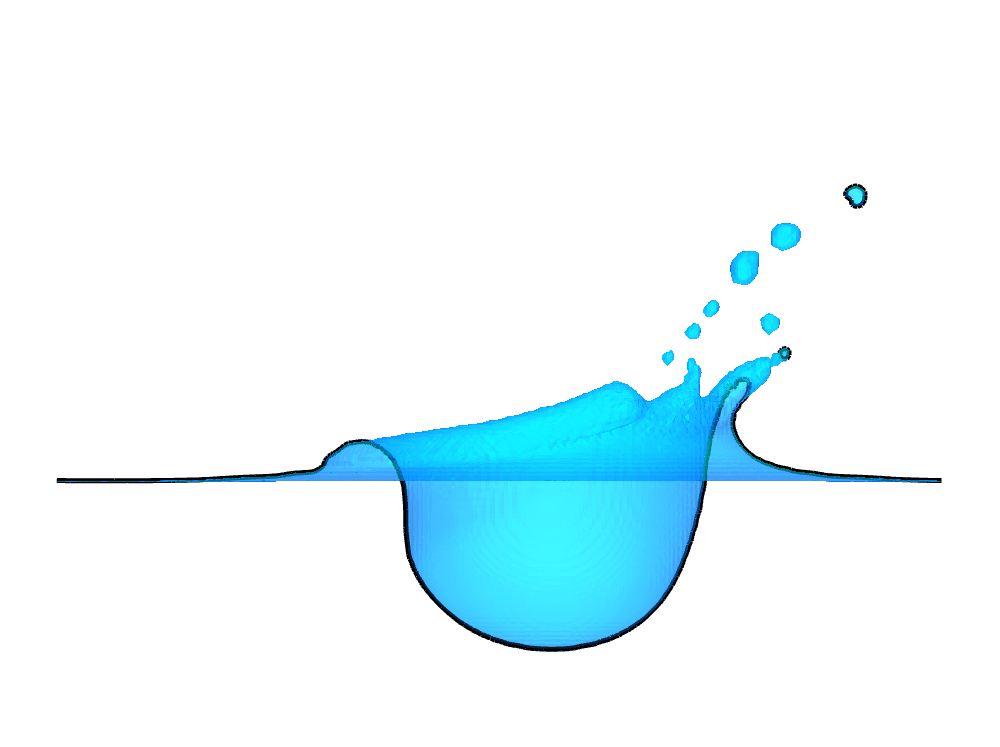} &
		\includegraphics[width=0.2\textwidth]{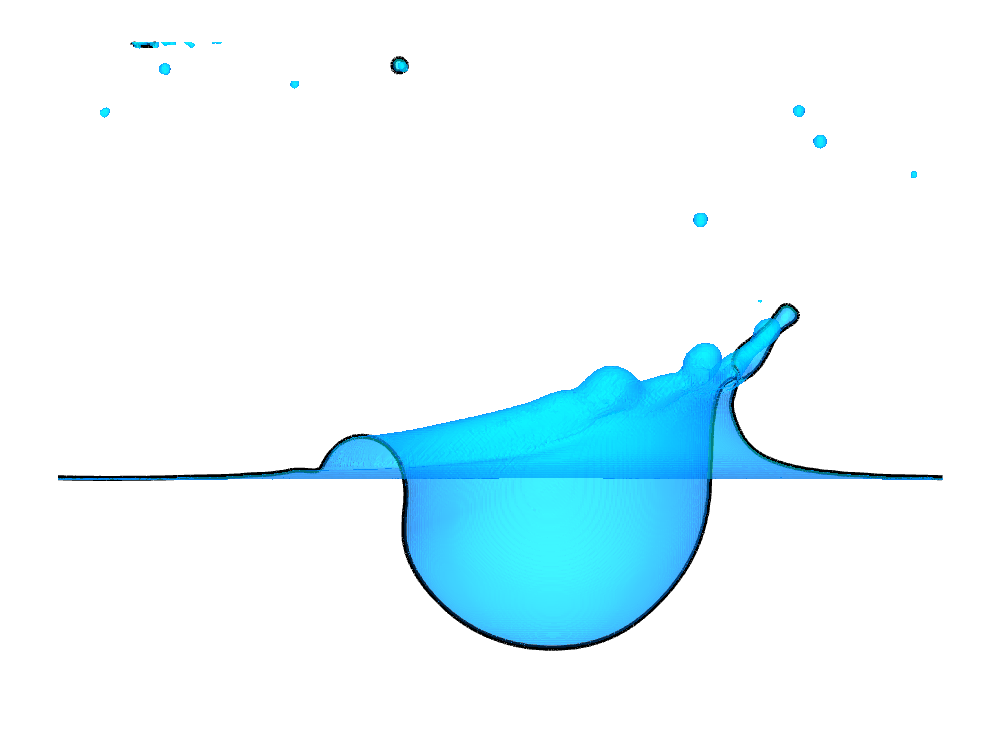} &
		\includegraphics[width=0.2\textwidth]{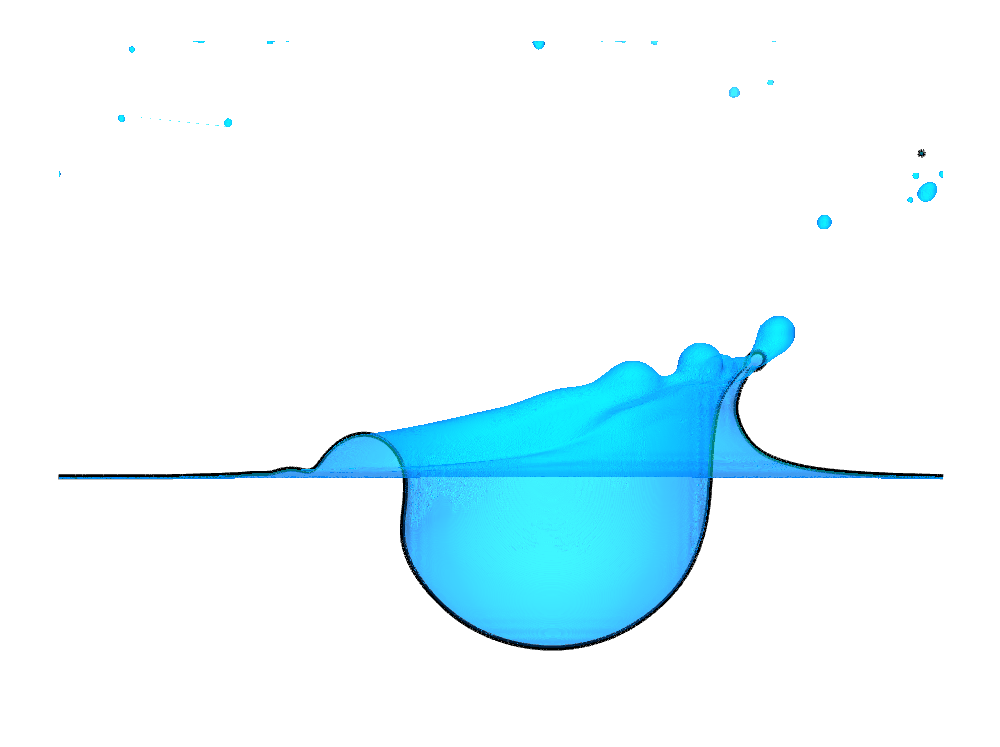} &
		\includegraphics[width=0.15\textwidth]{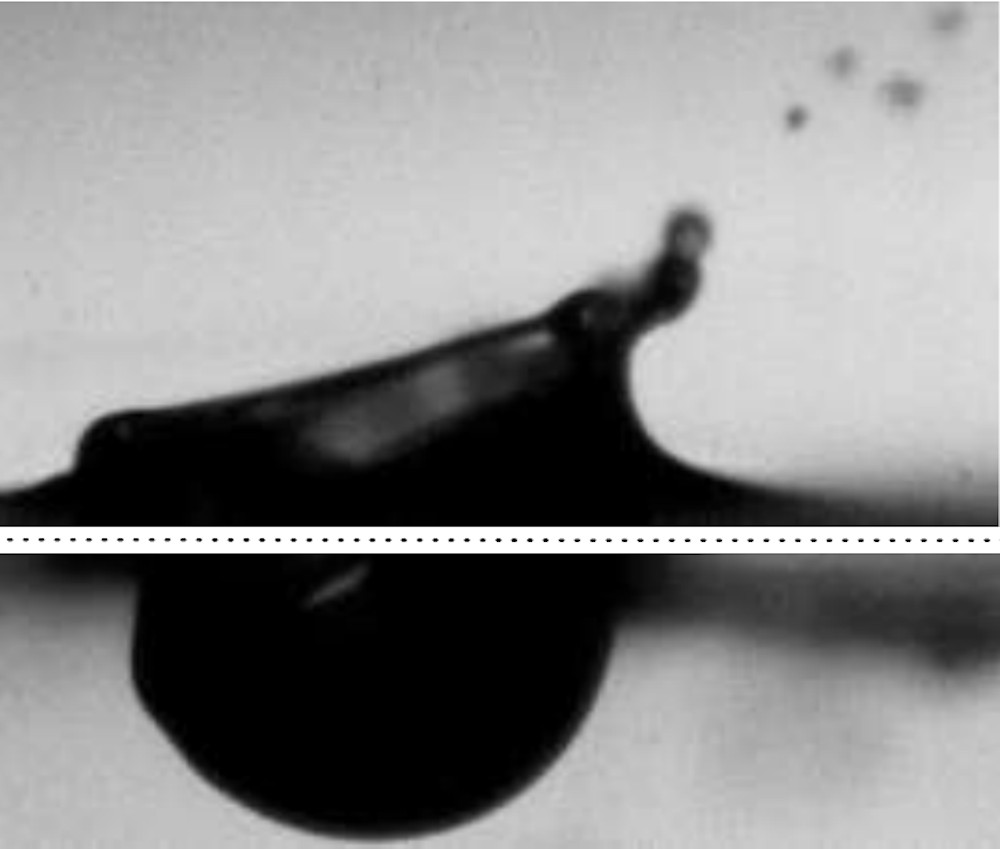} \\
		
		$t^{*}=12.15$ &
		\includegraphics[width=0.2\textwidth]{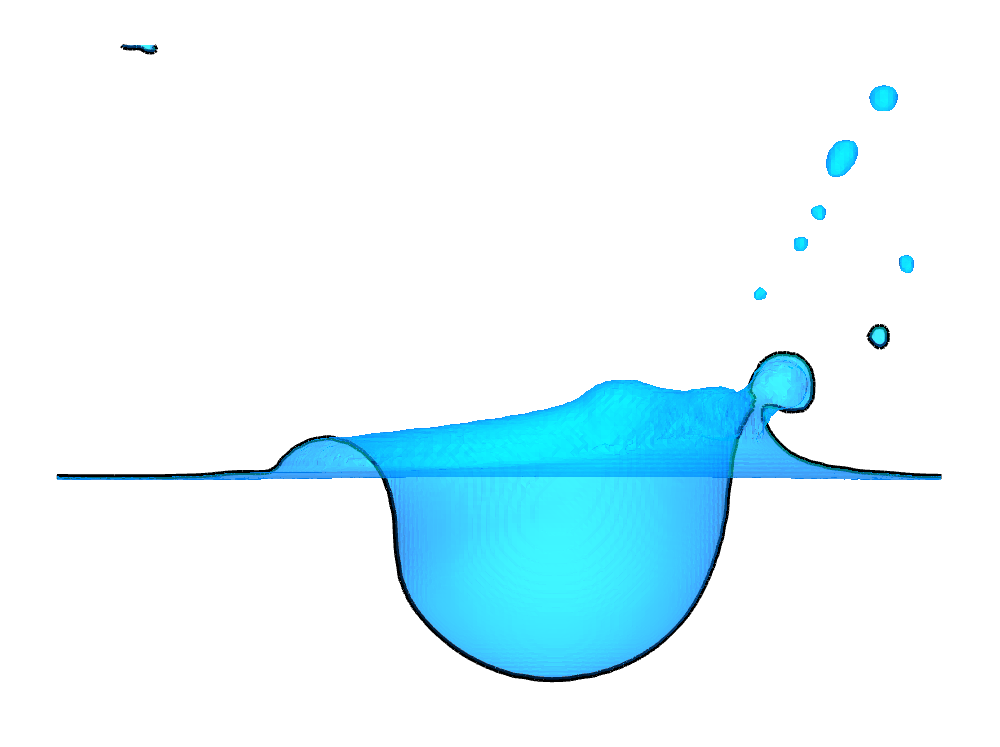} &
		\includegraphics[width=0.2\textwidth]{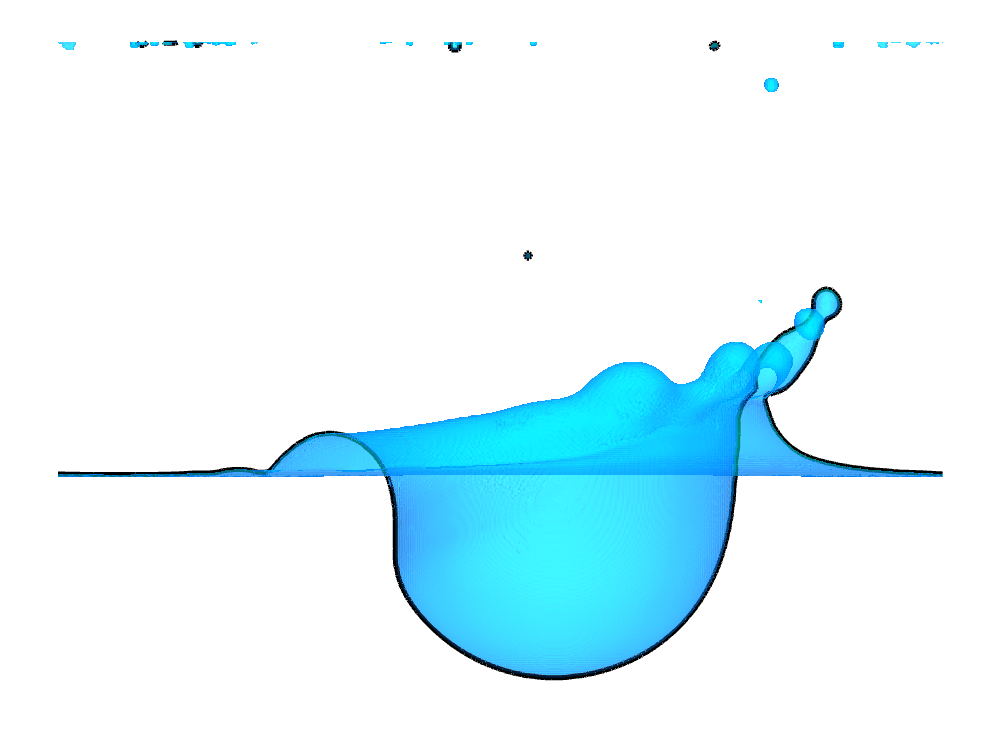} &
		\includegraphics[width=0.2\textwidth]{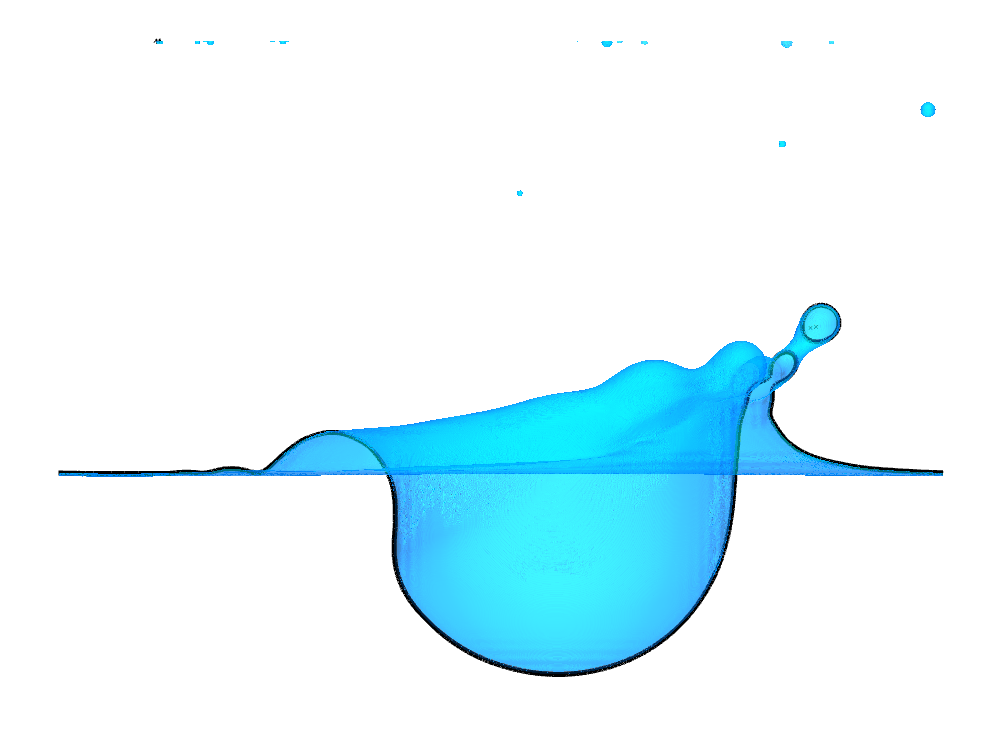} &
		\includegraphics[width=0.15\textwidth]{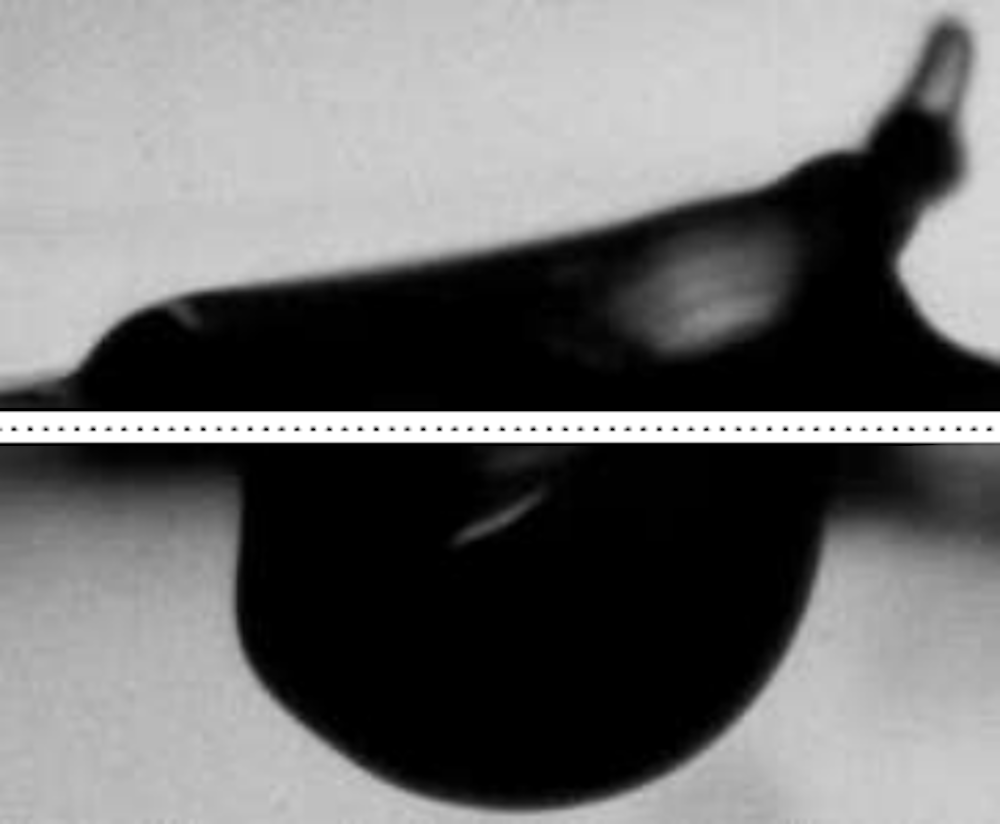} \\
		
		$t^{*}=18$ &
		\includegraphics[width=0.2\textwidth]{figures/drop-impact/oblique/fslbm/d-20/t-18.png} &
		\includegraphics[width=0.2\textwidth]{figures/drop-impact/oblique/fslbm/d-40/t-18.png} &
		\includegraphics[width=0.2\textwidth]{figures/drop-impact/oblique/fslbm/d-80/t-18.png} &
		\includegraphics[width=0.15\textwidth]{figures/drop-impact/oblique/experiment/t-18.png} \\
		
		& $D=20$ & $D=40$ & $D=80$ & Experiment~\cite{reijers2019ObliqueDropletImpact}
	\end{tabular}
	\caption{
		Oblique drop impact over non-dimensionalized time, $t^{*}$, as simulated with the FSLBM.
		The computational resolution is defined by the initial drop diameter, $D$.
		While the simulation results are true to scale, no scale bar is available for the photographs of the experiment~\cite{reijers2019ObliqueDropletImpact}.
		Therefore, the splash crown's dimension can only be compared between simulations rather than with the experiment.
		The solid black line illustrates the crown's contour in the center cross-section with normal in the $x$-direction.
		The photographs of the laboratory experiment were reprinted from Reference~\cite{reijers2019ObliqueDropletImpact} with the permission of the original authors.
	}
	\label{fig:drop-oblique-fslbm}
\end{figure}

\begin{figure}[h!]
	\begin{tabular}{>{\centering\arraybackslash}m{0.1\textwidth}
					>{\centering\arraybackslash}m{0.2\textwidth}
					>{\centering\arraybackslash}m{0.2\textwidth}
					>{\centering\arraybackslash}m{0.2\textwidth}
					>{\centering\arraybackslash}m{0.2\textwidth}
					>{\centering\arraybackslash}m{0.2\textwidth}
					>{\centering\arraybackslash}m{0.2\textwidth}}
	
		$t^{*}=0$ &
		\includegraphics[width=0.2\textwidth]{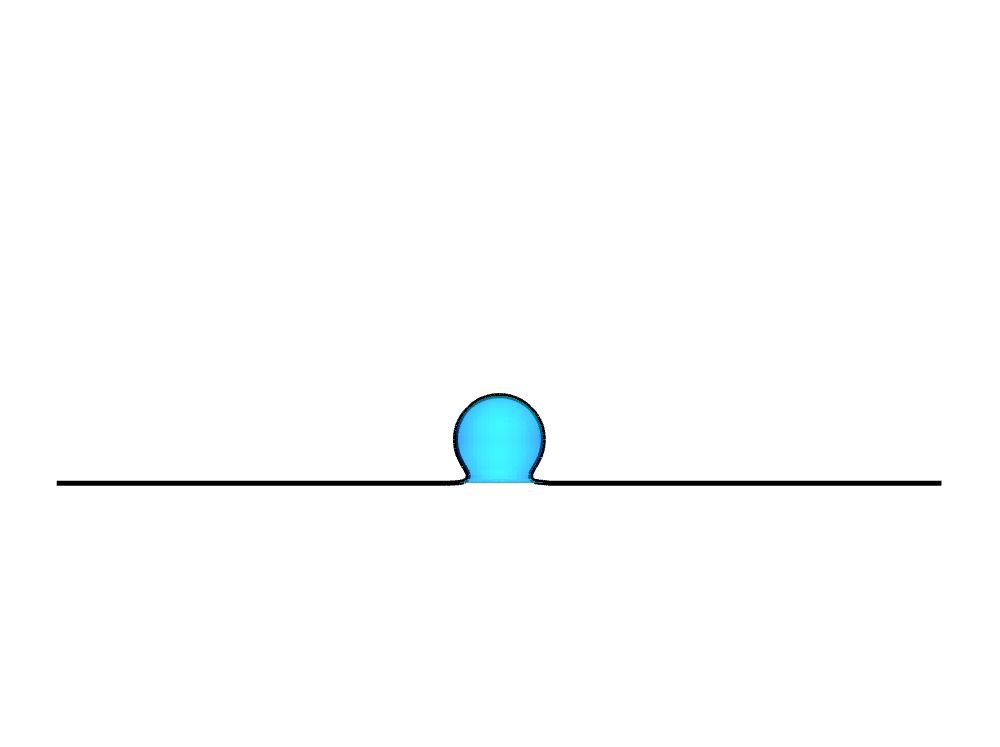} &
		\includegraphics[width=0.2\textwidth]{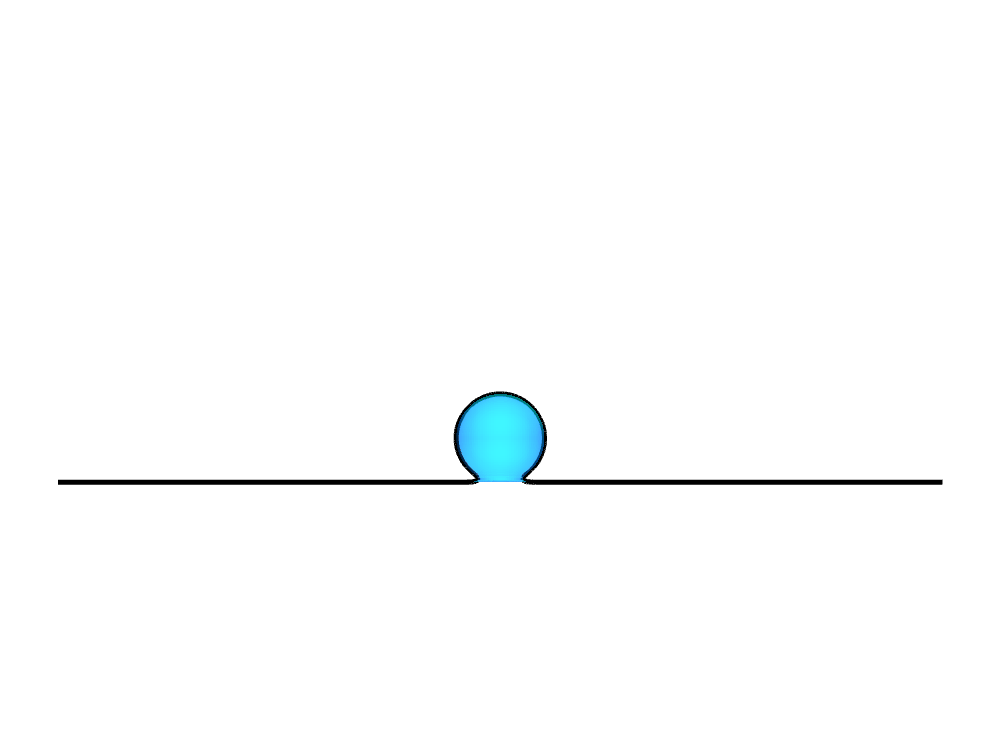} &
		\includegraphics[width=0.2\textwidth]{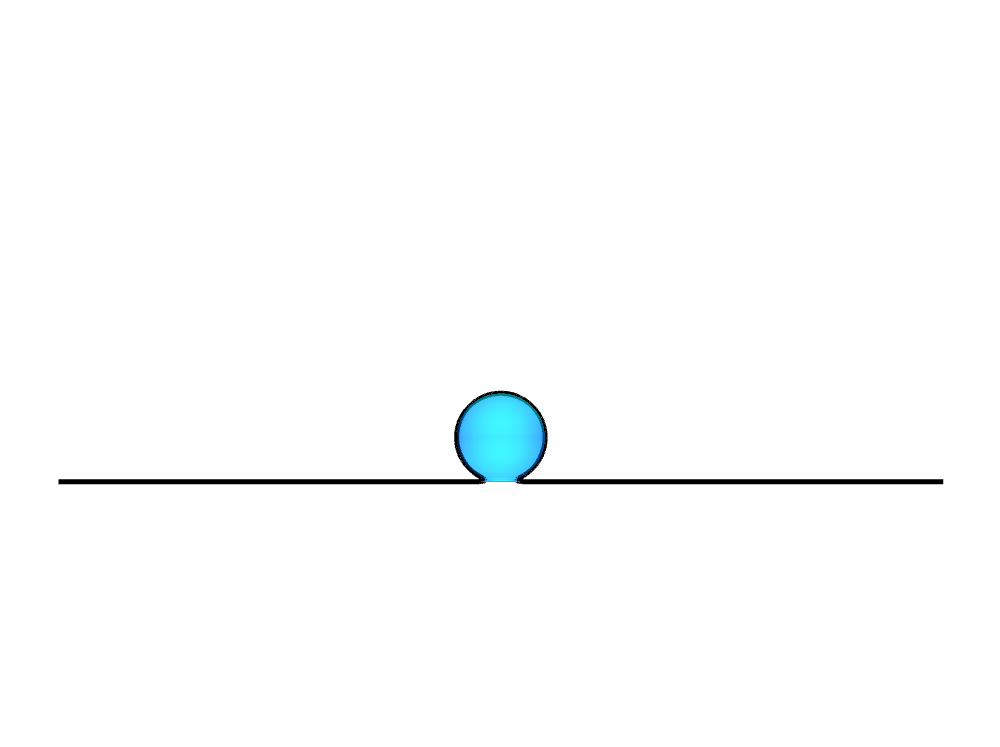} &
		 \\
		
		$t^{*} = 0.46$ &
		\includegraphics[width=0.2\textwidth]{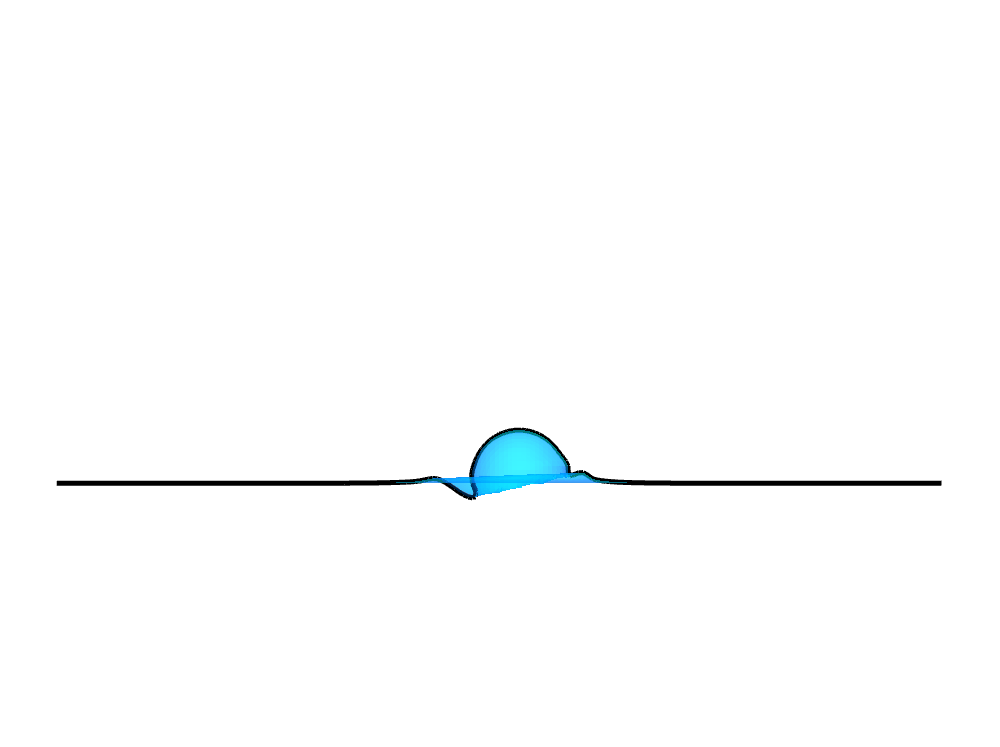} &
		\includegraphics[width=0.2\textwidth]{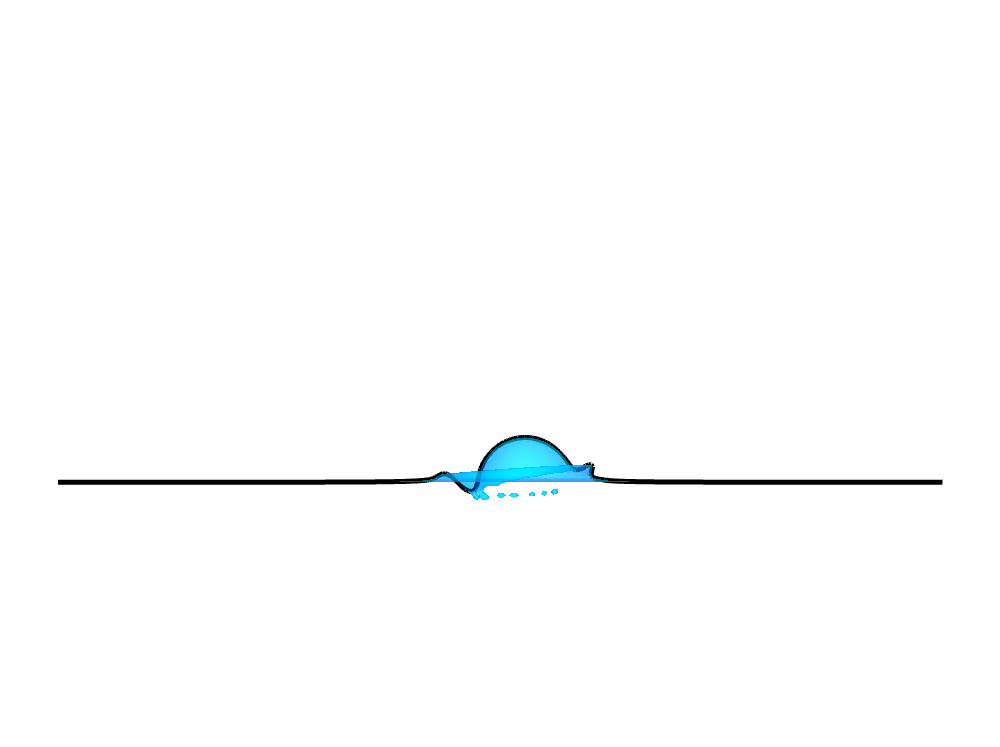} &
		\includegraphics[width=0.2\textwidth]{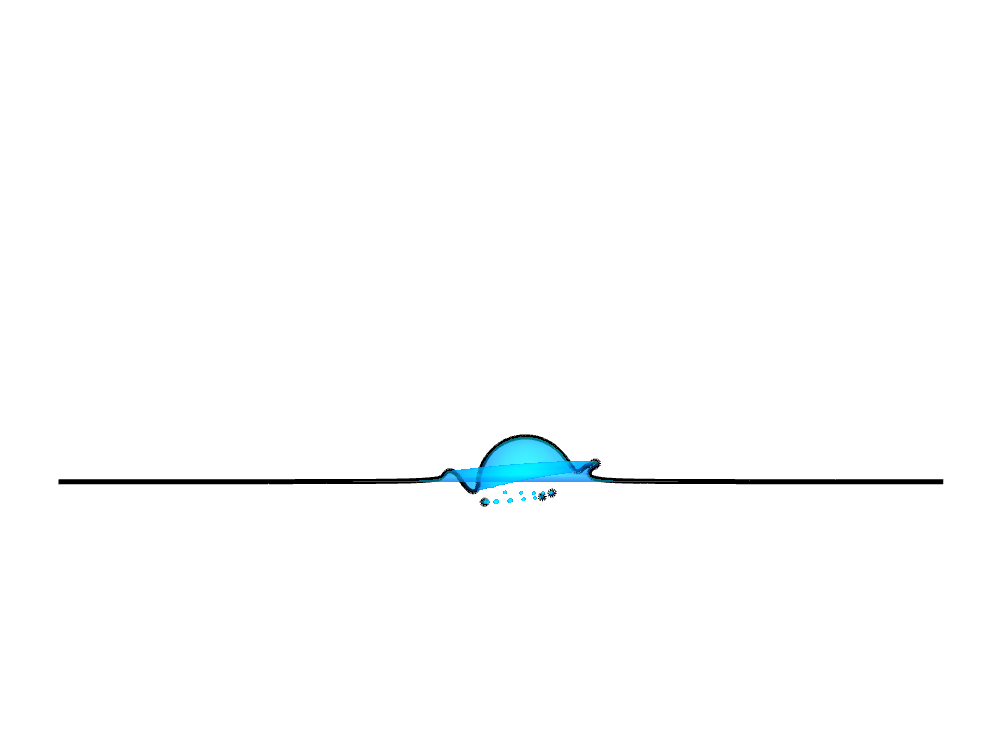} &
		\includegraphics[width=0.15\textwidth]{figures/drop-impact/oblique/experiment/t-0.46.png} \\
		
		$t^{*} = 2.33$ &
		\includegraphics[width=0.2\textwidth]{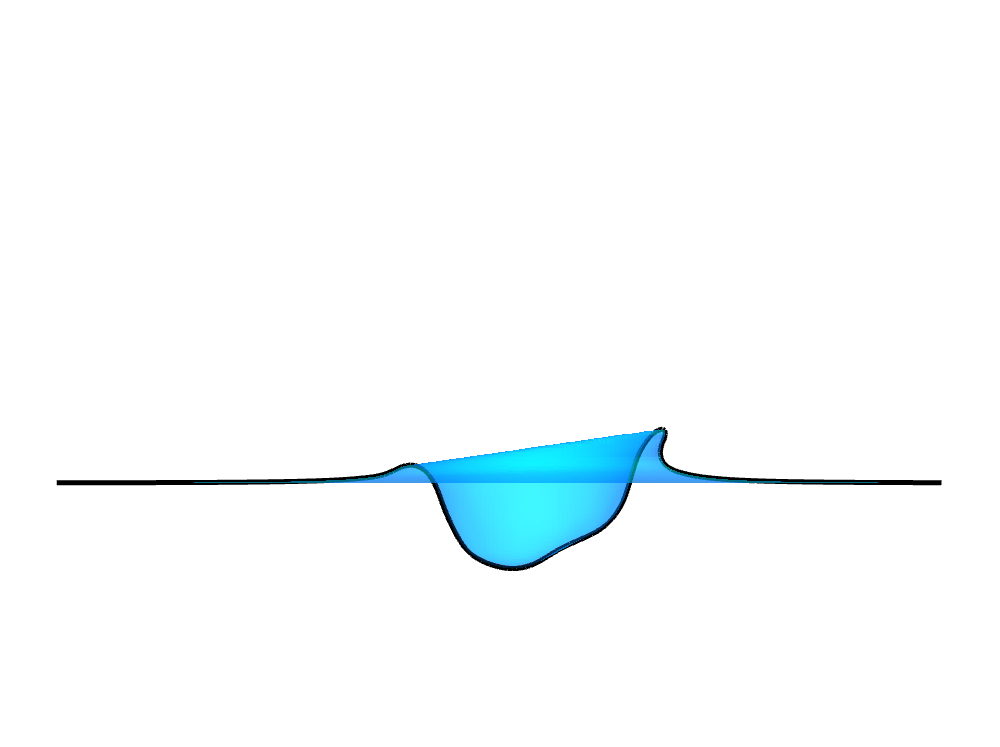} &
		\includegraphics[width=0.2\textwidth]{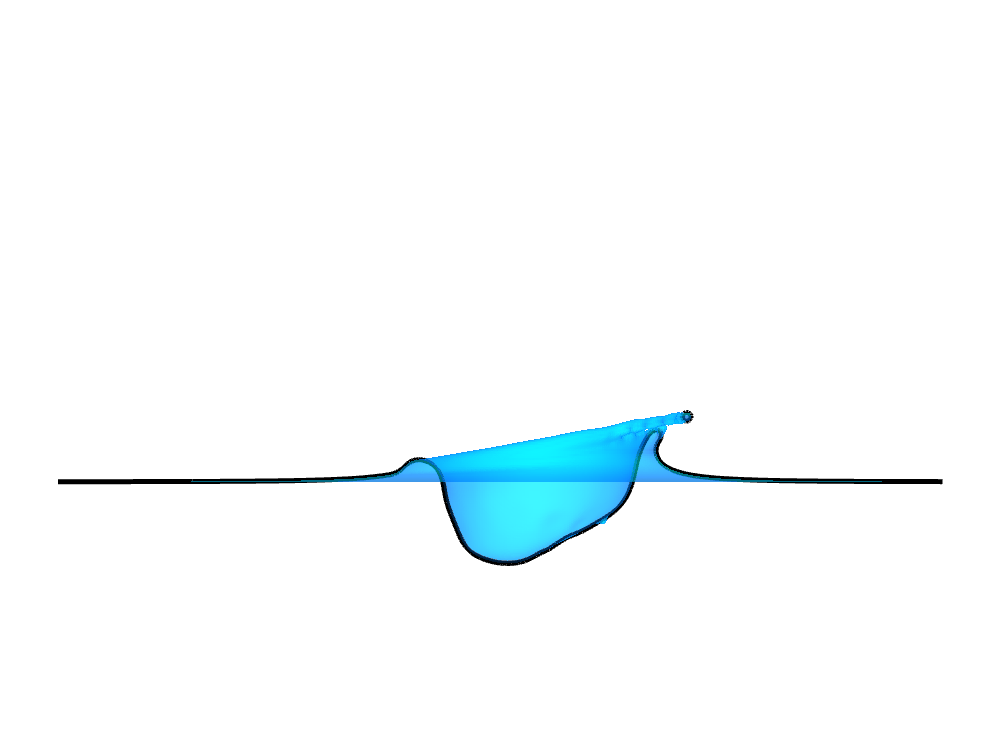} &
		\includegraphics[width=0.2\textwidth]{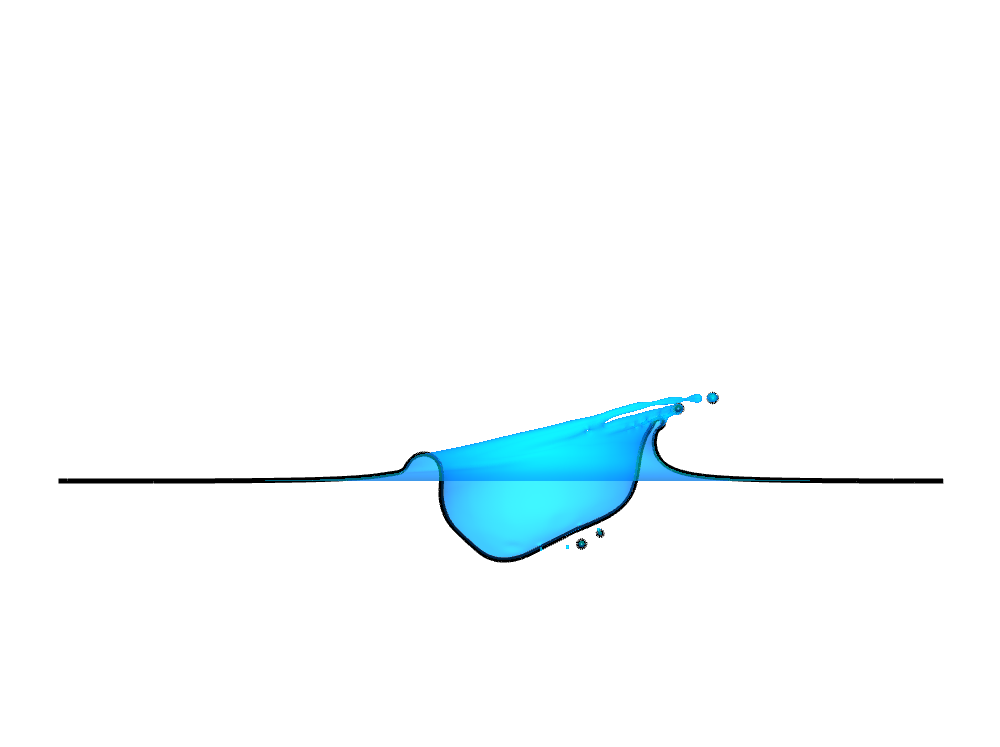} &
		\includegraphics[width=0.15\textwidth]{figures/drop-impact/oblique/experiment/t-2.33.png} \\
		
		$t^{*} = 8.22$ &
		\includegraphics[width=0.2\textwidth]{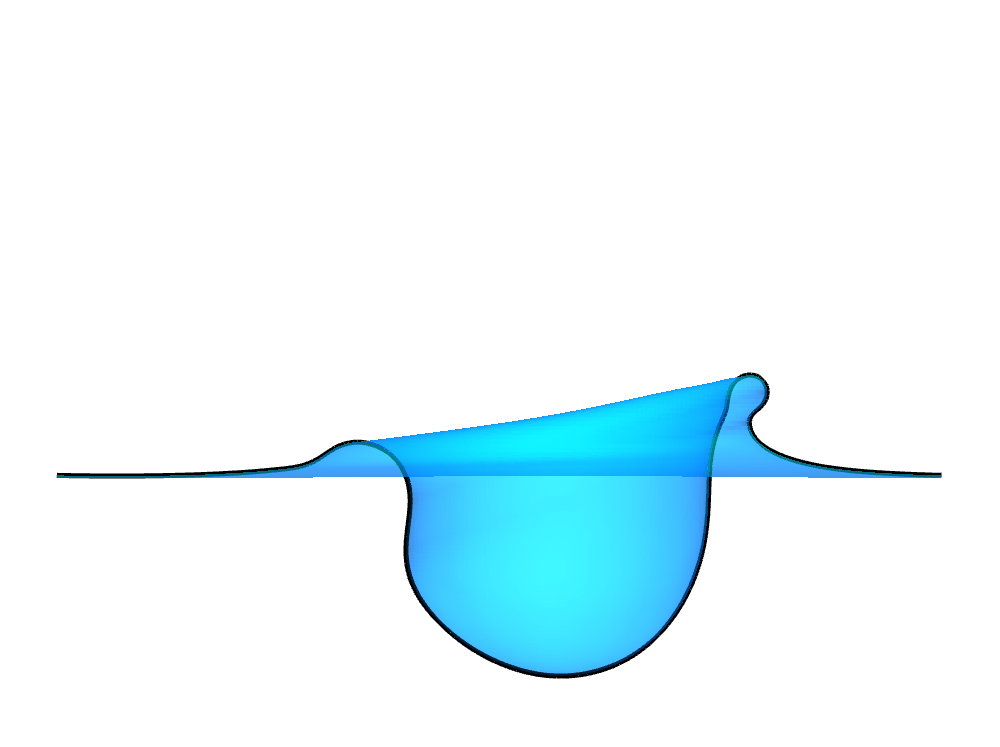} &
		\includegraphics[width=0.2\textwidth]{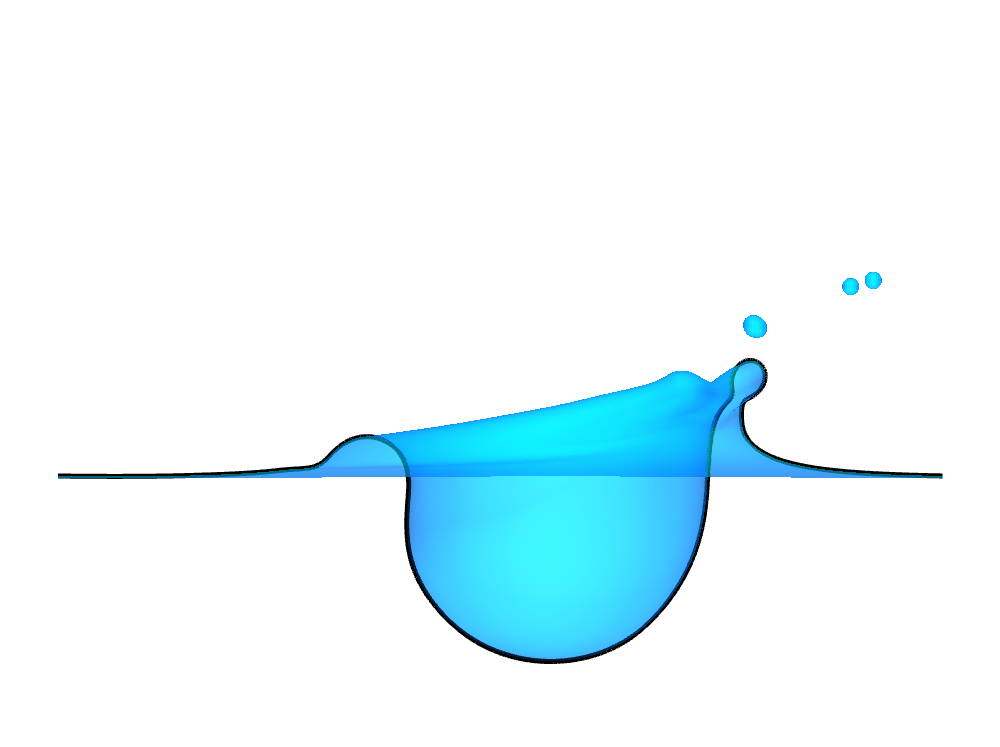} &
		\includegraphics[width=0.2\textwidth]{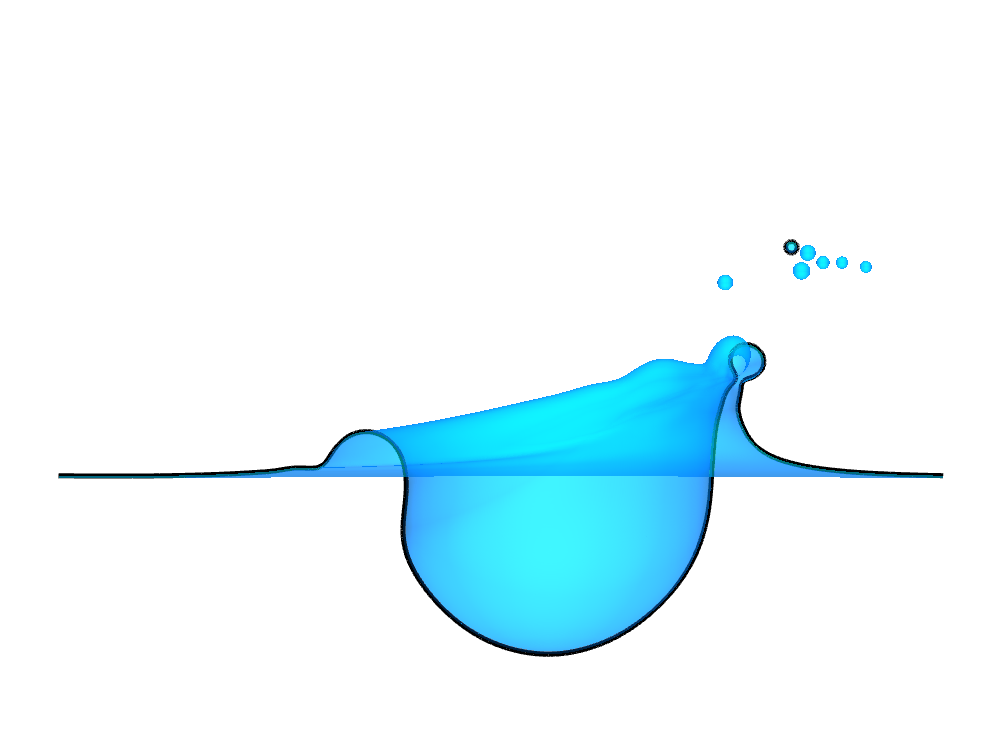} &
		\includegraphics[width=0.15\textwidth]{figures/drop-impact/oblique/experiment/t-8.22.png} \\
		
		$t^{*} = 12.15$ &
		\includegraphics[width=0.2\textwidth]{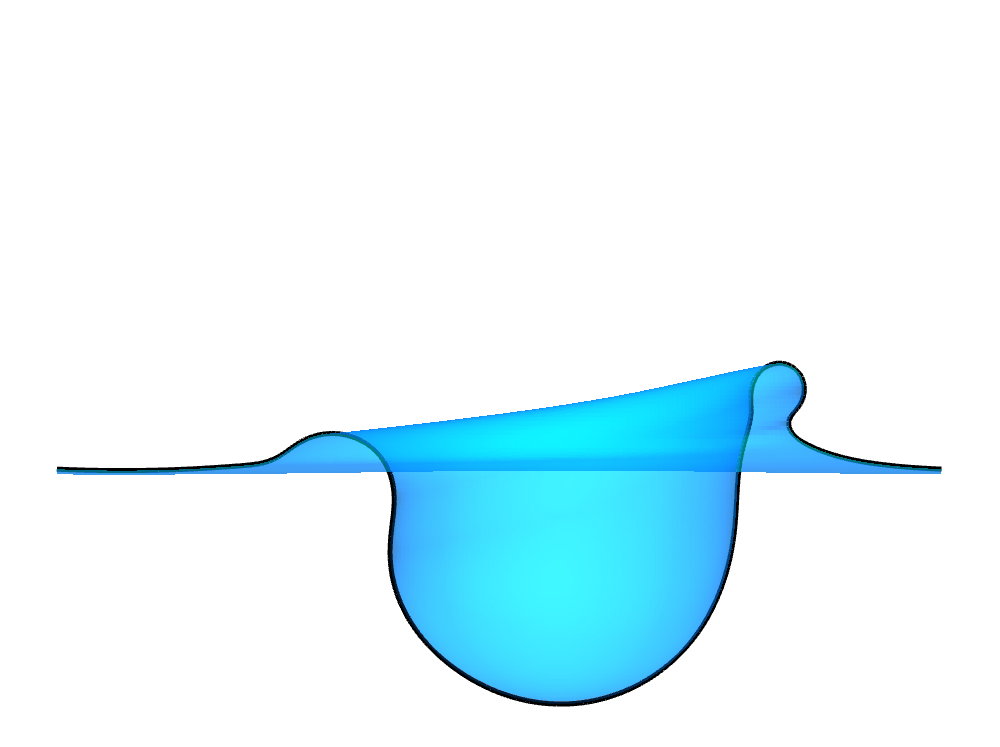} &
		\includegraphics[width=0.2\textwidth]{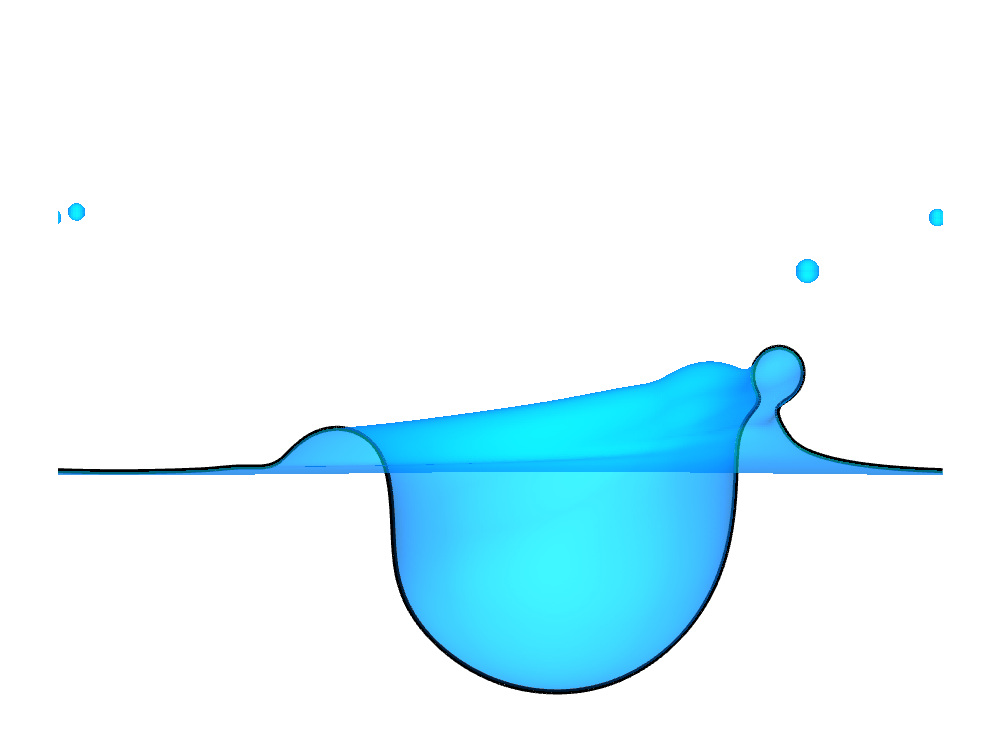} &
		\includegraphics[width=0.2\textwidth]{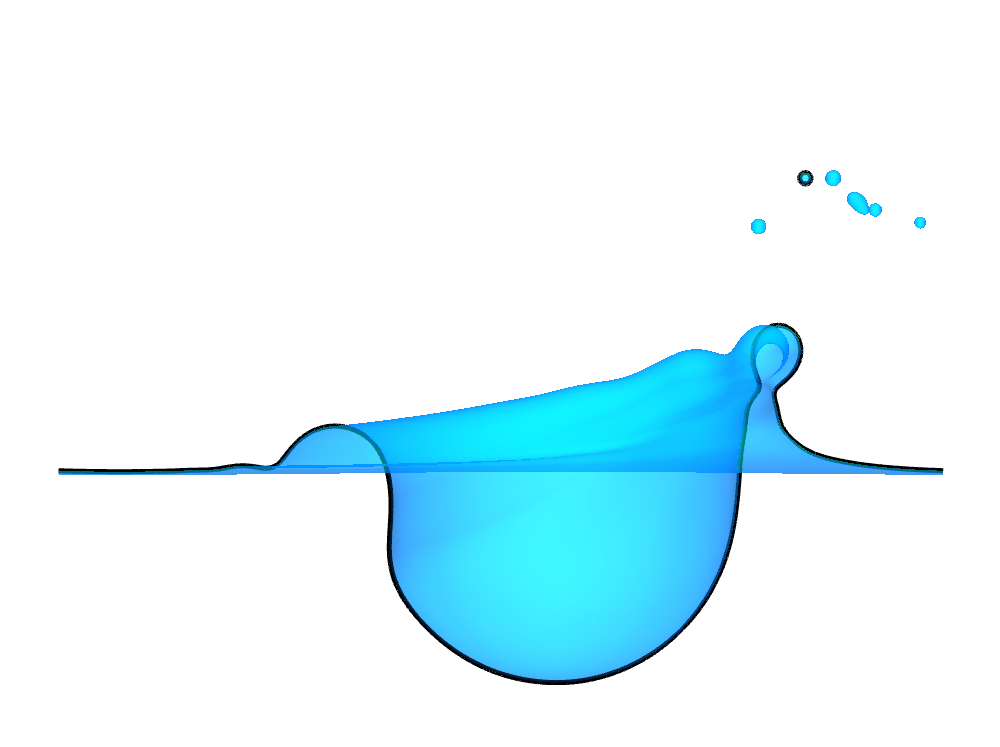} &
		\includegraphics[width=0.15\textwidth]{figures/drop-impact/oblique/experiment/t-12.15.png} \\
		
		$t^{*} = 18$ &
		\includegraphics[width=0.2\textwidth]{figures/drop-impact/oblique/pflbm/d-20/t-18.png} &
		\includegraphics[width=0.2\textwidth]{figures/drop-impact/oblique/pflbm/d-40/t-18.png} &
		\includegraphics[width=0.2\textwidth]{figures/drop-impact/oblique/pflbm/d-80/t-18.png} &
		\includegraphics[width=0.15\textwidth]{figures/drop-impact/oblique/experiment/t-18.png} \\
		
		& $D=20$ & $D=40$ & $D=80$ & Experiment~\cite{reijers2019ObliqueDropletImpact}
	\end{tabular}
	\caption{
		Oblique drop impact over non-dimensionalized time, $t^{*}$, as simulated with the PFLBM.
		The computational resolution is defined by the initial drop diameter, $D$.
		While the simulation results are true to scale, no scale bar is available for the photographs of the experiment~\cite{reijers2019ObliqueDropletImpact}.
		Therefore, the splash crown's dimension can only be compared between simulations rather than with the experiment.
		The solid black line illustrates the crown's contour in the center cross-section with normal in the $x$-direction.
		The photographs of the laboratory experiment were reprinted from Reference~\cite{reijers2019ObliqueDropletImpact} with the permission of the original authors.
	}
	\label{fig:drop-oblique-pflbm}
\end{figure}

\Cref{fig:drop-oblique-fslbm,fig:drop-oblique-pflbm} extend \Cref{par:oblique-drop-results} and qualitatively compare the simulated oblique drop impact with experimental data at different points in time.
\Cref{tab:drop-oblique-cavity,tab:drop-oblique-crown} present the temporal evolution of the quantified simulated cavity depth and inner crown diameter.

\FloatBarrier

\begin{table}[htbp]
	\centering
	\begin{tabular}{>{\raggedright}m{0.2\textwidth}
			>{\centering\arraybackslash}m{0.1\textwidth}
			>{\centering\arraybackslash}m{0.1\textwidth}
			>{\centering\arraybackslash}m{0.1\textwidth}
			>{\centering\arraybackslash}m{0.1\textwidth}
			>{\centering\arraybackslash}m{0.1\textwidth}}
		
		\toprule
		& $t^{*}$ & $2.33$ & $8.22$ & $12.15$ & $18$ \\
		\midrule
		FSLBM, $D=20$ & \multirow{6}{*}{$h_{\text{ca}}^{*}$} & $-0.80$ & $-1.85$ & $-2.19$ & $-2.5$ \\
		FSLBM, $D=40$ & & $-0.81$ & $-1.86$ & $-2.2$ & $-2.49$ \\
		FSLBM, $D=80$ & & $-0.84$ & $-1.87$ & $-2.13$ & $-2.48$ \\
		
		PFLBM, $D=20$ & & $-0.96$ & $-2.18$ & $-2.5$ & $-2.84$ \\
		PFLBM, $D=40$ & & $-0.92$ & $-2.03$ & $-2.37$ & $-2.68$ \\
		PFLBM, $D=80$ & & $-0.88$ & $-1.95$ & $-2.27$ & $-2.58$ \\
		\bottomrule
	\end{tabular}
	\caption{
		Simulated non-dimensionalized cavity depth, $h_{\text{ca}}^{*}(t^{*}) = h_{\text{ca}}(t^{*}) / D$, of the oblique drop impact.
		The cavity depth, $h_{\text{ca}}(t^{*})$, is the maximum distance of the cavity bottom to the initial position of the liquid surface at time, $t^{*}=0$, measured in the center cross-section with normal in the $x$-direction.
		The results are presented for different dimensionless times, $t^{*}$, and computational resolutions as defined by the initial drop diameter, $D$.
	}
	\label{tab:drop-oblique-crown}
\end{table}

\begin{table}[htbp]
	\centering
	\begin{tabular}{>{\raggedright}m{0.2\textwidth}
			>{\centering\arraybackslash}m{0.1\textwidth}
			>{\centering\arraybackslash}m{0.1\textwidth}
			>{\centering\arraybackslash}m{0.1\textwidth}
			>{\centering\arraybackslash}m{0.1\textwidth}
			>{\centering\arraybackslash}m{0.1\textwidth}}
		
		\toprule
		& $t^{*}$ & $2.33$ & $8.22$ & $12.15$ & $18$ \\
		\midrule
		FSLBM, $D=20$ & \multirow{6}{*}{$d_{\text{cr}}^{*}$} & $2.23$ & $3.39$ & $3.87$ & $4.43$ \\
		FSLBM, $D=40$ & & $2.23$ & $3.44$ & $3.9$ & $4.40$ \\
		FSLBM, $D=80$ & & $2.25$ & $3.44$ & $3.88$ & $4.46$ \\
		
		PFLBM, $D=20$ & & $2.20$ & $3.4$ & $3.88$ & $4.6$ \\
		PFLBM, $D=40$ & & $2.14$ & $3.38$ & $3.92$ & $4.45$ \\
		PFLBM, $D=80$ & & $2.19$ & $3.45$ & $3.95$ & $4.49$ \\
		\bottomrule
	\end{tabular}
	\caption{
		Simulated non-dimensionalized splash crown diameter, $d_{\text{cr}}^{*}(t^{*}) = d_{\text{cr}}(t^{*}) / D$, of the oblique drop impact. 
		The splash crown diameter, $d_{\text{cr}}(t^{*})$, is the crown's inner diameter at the position of the initial liquid surface at time, $t^{*}=0$, measured in a center cross-section with normal in the $x$-direction.
		The results are presented for different dimensionless times, $t^{*}$, and computational resolutions as defined by the initial drop diameter, $D$.
	}
	\label{tab:drop-oblique-cavity}
\end{table}

\FloatBarrier

\subsection{Supplementary material: open source simulation setups} \label{subsec:app-implementation-links}
The following supplementary material is available as part of the online article:
\begin{itemize}
	\item An archive of the \cpp{} source code of the FSLBM and PFLBM as part of the software framework \walberla~\cite{waLBerla}, version
	\url{https://i10git.cs.fau.de/walberla/walberla/-/tree/01a28162ae1aacf7b96152c9f886ce54cc7f53ff}.\\
	The simulation setups are located in the directories \texttt{apps/showcases/FreeSurface} and\\ \texttt{apps/showcases/PhaseFieldAllenCahn/CPU}.\par
	
	\item An archive of the Python source code used for the PFLBM gravity and capillary wave test cases.
	These test cases are provided as Jupyter Notebooks as part of the code generation framework \lbmpy~\cite{lbmpy}, version \url{https://pypi.org/project/lbmpy/1.0.1/}.
	The notebooks are located in the directory \texttt{lbmpy\_tests/full\_scenarios/phasefield\_allen\_cahn}.
\end{itemize}
\FloatBarrier

\section*{Acknowledgments}
The authors C.\ Schwarzmeier and U.\ Rüde are grateful to the Deutsche Forschungsgemeinschaft (DFG, German Research Foundation) for funding project 408062554.\\
This work was supported by the SCALABLE project.
This project has received funding from the European High-Performance Computing Joint Undertaking (JU) under grant agreement No 956000.
The JU receives support from the European Union’s Horizon 2020 research and innovation programme and France, Germany, the Czech Republic.\\
Author T.\ Mitchell acknowledges that this work was supported by resources provided by the Pawsey Supercomputing Centre with funding from the Australian Government and the Government of Western Australia.\\
Author M.\ Lehmann acknowledges funding by the DFG -- project number 391977956 -- SFB 1357 and support from the ENB Biological Physics.\\
The authors gratefully acknowledge the Gauss Centre for Supercomputing e.V. (www.gauss-centre.eu) for funding this project by providing computing time on the GCS Supercomputer SuperMUC at Leibniz Supercomputing Centre (www.lrz.de).\\
The authors gratefully acknowledge the scientific support and HPC resources provided by the Erlangen National High Performance Computing Center (NHR@FAU) of the Friedrich-Alexander-Universität Erlangen-Nürnberg (FAU).

\bibliographystyle{elsarticle-num}
\bibliography{literature.bib}

\end{document}